\documentclass[
reprint,
superscriptaddress,
nofootinbib,
 amsmath,amssymb,
 aps,
pra,
a4paper,twocolumn, floatfix
]{revtex4-2}
\usepackage[english]{babel}
\usepackage{dsfont}
\usepackage{graphicx}
\usepackage{dcolumn}
\usepackage{bm}
\usepackage[hidelinks,colorlinks=true,linkcolor=blue]{hyperref}
\usepackage[normalem]{ulem}
\usepackage{stmaryrd}
\usepackage{amsmath}
\usepackage{mathrsfs}
\usepackage{makecell}
\usepackage{xcolor}
\usepackage{booktabs}
\usepackage{multirow}
\usepackage{amsmath}
\usepackage{amssymb}
\usepackage{physics}
\usepackage{nicefrac}
\usepackage{capt-of}

\hypersetup{
    colorlinks,
    linkcolor={red!50!black},
    citecolor={blue!50!black},
    urlcolor={blue!80!black}
}

\renewcommand{\paragraph}[1]{\textbf{#1}}

\usepackage[normalem]{ulem} 
\usepackage{comment}

\usepackage[colorlinks=true,linkcolor=blue]{hyperref}
\usepackage[percent]{overpic}
\usepackage{ragged2e} 
\usepackage{subcaption}

\DeclareCaptionJustification{realjustified}{\justifying}

\newsavebox{\QELMPanelImage}
\newcommand{\QELMPanel}[5]{%
  \phantomsection\phantomsubcaption\label{#3}%
  \sbox{\QELMPanelImage}{\includegraphics[width=#1]{#2}}%
  \begingroup
    \setlength{\unitlength}{0.01\dimexpr#1\relax}%
    \rlap{\usebox{\QELMPanelImage}}%
    \rlap{%
      \hspace*{#4\unitlength}%
      \raisebox{#5\unitlength}[0pt][0pt]{%
        \makebox[0pt][r]{\small\bfseries(\thesubfigure)}%
      }%
    }%
    \phantom{\usebox{\QELMPanelImage}}%
  \endgroup
}
\usepackage{tikz}
\usetikzlibrary{matrix,backgrounds,quantikz}
\pgfdeclarelayer{myback}
\pgfsetlayers{myback,background,main}

\usepackage{booktabs}
\usepackage{multirow}
\renewcommand*{\arraystretch}{1.3}
\usepackage{mathtools}

\usepackage{amsthm}
\theoremstyle{plain}

\newtheorem*{theorem*}{Theorem}
\newtheorem{theorem}{Theorem}
\newtheorem*{observation*}{Observation}
\newtheorem*{lemma*}{Lemma}
\newtheorem{lemma}{Lemma}
\newtheorem*{definition*}{Definition}
\newtheorem{definition}{Definition}
\newtheorem{proposition}{Proposition}
\newtheorem*{proposition*}{Proposition}
\newtheorem{corollary}{Corollary}
\newtheorem*{corollary*}{Corollary}
\newtheorem{assumption}{Assumption}
\newtheorem*{assumption*}{Assumption}

\let\oldqed\qed
\renewcommand{\qed}{\oldqed\bigskip} 

\newcommand{\nocontentsline}[3]{}
\let\origcontentsline\addcontentsline
\newcommand\stoptoc{\let\addcontentsline\nocontentsline} 
\newcommand\resumetoc{\let\addcontentsline\origcontentsline} 

\begin{document}

\renewcommand{\tocname}{Supplementary Material} 
\stoptoc

\preprint{APS/123-QED}
\title{Shaping double descent in quantum extreme learning through model design
}

\author{Sabri Meyer}
\thanks{These authors contributed equally to this work.}
\affiliation{Department of Mathematics and Computer Science,
University of Basel, Switzerland}
\author{Ege Yilmaz}
\thanks{These authors contributed equally to this work.}
\affiliation{Department of Mathematics and Computer Science,
University of Basel, Switzerland}
\affiliation{IBM Research, 8803 R\"uschlikon, Switzerland}
\author{Aurelien Lucchi}
\affiliation{Department of Mathematics and Computer Science,
University of Basel, Switzerland}
\author{Francesco Tacchino}
\affiliation{IBM Research, 8803 R\"uschlikon, Switzerland}
\author{Francesco Scala}
\affiliation{Department of Mathematics and Computer Science,
University of Basel, Switzerland}
\date{\today}

\begin{abstract}

Increasing capacity can induce a machine learning model to generalize worse until it perfectly interpolates the training data, and better once it moves beyond interpolation. 
This inversion, known as \emph{double descent}, is well understood for independent features. For correlated features, its shape is governed by the spectrum of the feature covariance, but existing analyses typically yield only implicit equations.
We first close this gap for classical linear regression, analytically characterizing the entire risk curve through two quantities, namely the condition number of the feature population covariance and a spectral measure of correlation strength. 
Then, we apply this theory on quantum models, focusing on quantum extreme learning machines (QELMs). 
We show that three tunable design knobs, \textit{i.e.,} encoding strength, input reuploads, and measurement shots, each regularize the QELM output covariance, and derive conditions under which the interpolation threshold remains stable. 
Finite measurement shot noise can be leveraged to guarantee this stability.
Numerical experiments confirm each prediction, yielding concrete design principles for QELMs that generalize well beyond interpolation.
\end{abstract}

\maketitle

\section{Introduction}

\begin{figure*}
    \centering
    \includegraphics[width=\linewidth]{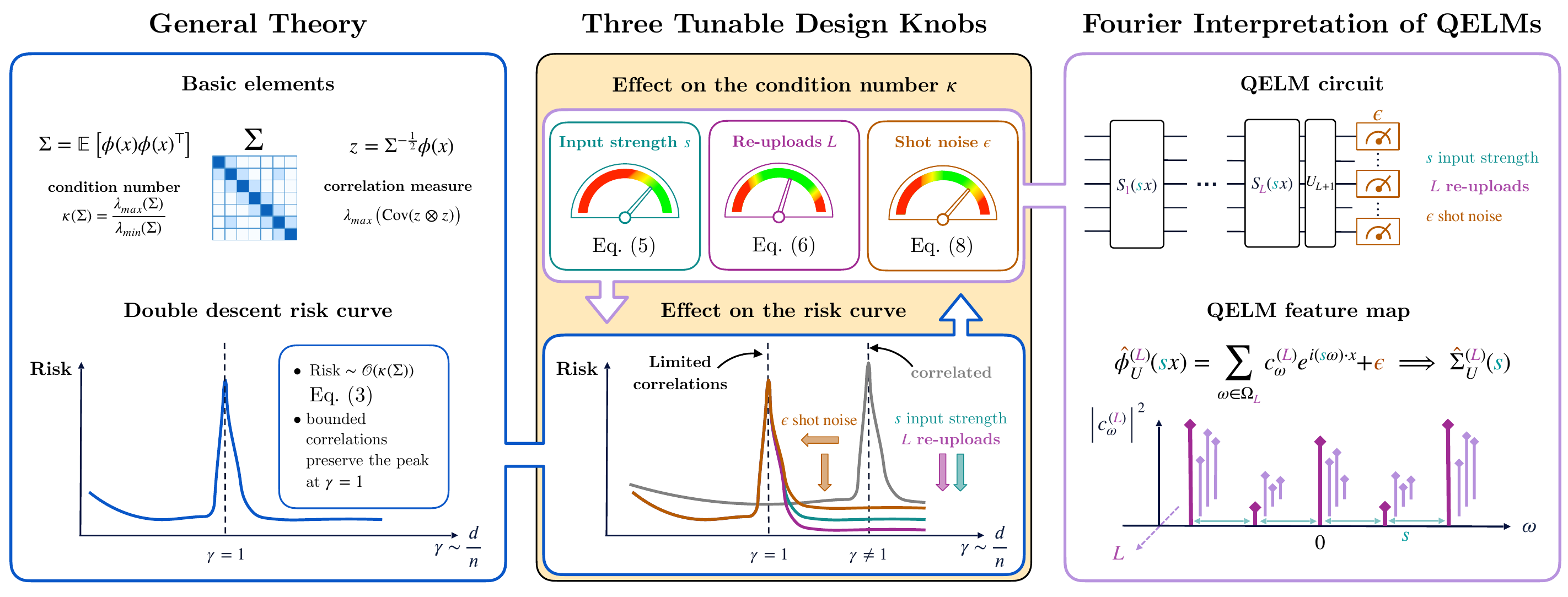}
    \caption{A graphical overview of our approach. First, we establish general results about double descent in linear regression models with correlated features \textbf{(left panel)}. Then, we leverage these results in the context of QELMs. By representing the population covariance in the Fourier picture \textbf{(right panel)}, we identify three tunable design-specific knobs that determine the shape of the double descent curve \textbf{(central panel)}. Input strength, data re-uploadings and shot noise can reduce the risk. In addition, shot noise can also reduce feature correlations, making the interpolation threshold predictable.}
    \label{fig:main-overview}
\end{figure*}

Machine learning models routinely operate in the \emph{over-parametrized} regime, where the number of trainable parameters exceeds the number of training samples. 
Classical theory predicts that such models should overfit catastrophically, yet in practice they often generalize well. 
The phenomenon that reconciles these two facts is \emph{double descent}~\cite{dobriban2018high, belkin2019reconciling, mei2022generalization, hastie2022surprises, bartlett2020benign, bartlett2021deep, tsigler2023benign, schaeffer2023double}: 
as model capacity grows, test error first follows the textbook U-shaped curve, spikes sharply at the \emph{interpolation threshold}, the precise point at which the model has just enough capacity to fit the training data exactly, and then, falls again as capacity increases further. 
Classical generalization bounds that grow with the complexity of the hypothesis class can substantially overestimate the generalization gap in this regime~\cite{mohri2012foundations}, failing to capture the decrease in test error beyond the interpolation threshold.

Understanding this second descent can inform how practitioners size and design over-parametrized models.

Despite its importance, the theoretical foundations of double descent remain confined to specific statistical regimes.
Many study use random matrix theory and Stieltjes transform techniques~\cite{bai2010spectral} to characterize the generalization risk in linear models~\cite{dobriban2018high, mei2022generalization, hastie2022surprises}.
However, these results often assume of isotropic features or independent coordinates, limiting their applicability to realistic correlated data. 
While more recent work has extended these bounds to handle arbitrary sub-Gaussian correlations~\cite{bartlett2020benign, bartlett2021deep, kuzborskij2021role, tsigler2023benign}, it remains poorly understood how these correlations influence the shape of the risk curve.    

Linear models provide an analytically tractable setting for identifying principles that also govern more complex architectures, including nonlinear neural networks~\cite{adlam2020understanding, mel2021anisotropic, mei2022generalization, heckel2020early, kuzborskij2021role, hessian2023sabri, curth2023u, li2024least, mallinar2024minimum, moakher2025generalization}. Their generalization properties and double descent behavior are now well characterized theoretically~\cite{dobriban2018high, bartlett2020benign, bartlett2021deep, adlam2020understanding, mei2022generalization, hastie2022surprises, schaeffer2023double, tsigler2023benign, moakher2025generalization}, while analogous behavior has been observed empirically in deep neural networks~\cite{belkin2019reconciling, heckel2020early, nakkiran2021deep, liu2023understanding}. These results point to the spectral properties of the learning problem as a key factor. In particular, the condition number of a linear system can itself exhibit a double descent profile~\cite{henighan2023double}, and the conditioning of the feature covariance has been linked to the height and geometry of the risk peak~\cite{kuzborskij2021role, rangamani2023interpolating, liu2023understanding}. In addition, complementary studies show that accounting for the effective number of parameters is also essential for predicting shifts in the location of this peak~\cite{curth2023u, li2024least}. 
However, these observations do not yet provide a simple quantitative link between feature correlations, spectral conditioning, model architecture, and the full generalization curve.
It therefore remains unclear whether covariance conditioning quantifies generalization, whether feature correlations explain shifts in the risk peak, and how architectural hyperparameters control the full double descent curve. This will be the focus of this work.

These questions are becoming increasingly salient in quantum machine learning (QML)~\cite{schuld2015introduction, biamonte2017quantum, cerezo2022challenges, chang2025primer} as well. 
An early attempt to address double descent was provided by Hu and Gao~\cite{hu2024generalization}, who employed statistical mechanics to analyse the impact of measurement shot noise on models with fixed parameters. Their analysis shows that measurement noise need not be detrimental, but focuses on the under-parametrized regime and does not address the role of model architecture beyond interpolation.

The non-monotonic double descent behaviour of test error around the interpolation threshold has been recently established in quantum kernel models~\cite{schuld2019quantum, kempkes2026double, kamisoyama2026double} and parametrized quantum circuits~\cite{kempkes2026cautious}.
Kempkes et al.~\cite{kempkes2026double, kempkes2026cautious} demonstrated that quantum models exhibit a universal double descent curve governed by the Marchenko-Pastur law. 
Furthermore, Kamisoyama et al.~\cite{kamisoyama2026double} provided an analytic characterization of test risk for quantum kernel ridge regression. 
However, such methods rely on universal spectral laws or highly non-trivial and implicit equations~\cite{bai2010spectral}, which do not directly link the curve to the circuit architecture.
How architectural choices affect generalization in the over-parametrized regime therefore remains open.

We focus on quantum extreme learning machines (QELMs)~\cite{ghosh2019quantum, innocenti2023potential, suprano2024experimental, xiong2025fundamental, xiong2025role, vetrano2025state, zia2025quantum, kawanabe2026efficient, dao2026breaking}, which train a classical linear readout on features extracted from a quantum circuit.
Their separation between feature generation and fitting makes circuit statistics directly accessible to analysis.

Our main contributions are threefold, which we summarize in Figure~\ref{fig:main-overview}. 
First, we revisit double descent for classical linear regression with generic, correlated random features and prove generalization bounds that depend only on the \emph{population covariance} of the features. This matrix captures, on average, how the features co-vary across the input distribution. We identify two governing quantities: the \emph{condition number}, the ratio of its largest to its smallest eigenvalue, and a spectral measure of correlations. 
We show that a well-conditioned covariance improves generalization and, when correlations satisfy a stated growth condition, preserves the canonical interpolation threshold. These results quantify the over-parametrized risk and the conditions under which correlations shift the risk peak [Equations~\eqref{eq:main-correlation-measure} and~\eqref{eq:main-generalization-bound}].

Second, we specialize this general theory to QELM feature maps by representing the circuit readout as a discrete Fourier series of its inputs, with frequencies dictated by the eigenvalue spectra of the data-encoding generators~\cite{schuld2021effect, xiong2025fundamental, mhiri2025constrained}. 
This representation links the population covariance to three controllable design hyperparameters: the input encoding strength, the number of input reuploads (\textit{i.e.}, repeated encoding across layers), and the number of measurement shots used to estimate each feature.

Third, we prove that each hyperparameter can regularize the QELM population covariance [Equation~\eqref{eq:main-input-strength},~\eqref{eq:main-reuploads} and~\eqref{eq:main-shot-noise-regularization}] and test these predictions numerically [Figure~\ref{fig:experiments}].
Stronger input encoding and more reuploads each expand the accessible Fourier spectrum of the model, which improves the conditioning of the covariance and therefore the risk bound in the over-parametrized regime. 
Finite measurement shot noise instead plays a constructive double role: it adds a diagonal regularizing term that improves the conditioning of the covariance and suppresses feature correlations that shift the interpolation threshold
Taken together, these results provide the first set of theoretically grounded, concrete design principles connecting QELM architectural choices to the shape of the generalization curve.

\section{Results}

\paragraph{Learning setup and the min-norm interpolator.}
We consider a supervised learning problem with training set $S=\{(x_\ell,y_\ell)\}_{\ell=1}^n$ drawn i.i.d. from a joint distribution $\mathcal{D}$ on $\mathcal{X}\times\mathcal{Y}\subseteq\mathbb{R}^p\times\mathbb{R}$. We assume that labels are generated as $y=f_*(x)+\delta$ for an unknown $f_*$ and Gaussian label noise $\delta\sim\mathcal{N}(0,\sigma^2)$. 
Given a feature map $\phi:\mathcal{X}\to\mathbb{R}^d$ and feature matrix $\Phi=[\phi(x_1),\ldots,\phi(x_n)]$, we fit a linear model $f_w(x)=w^\top\phi(x)$ by minimizing the empirical mean squared loss $\mathcal{L}_n(w)=(2n)^{-1}\|\Phi^\top w - Y\|_2^2$ with weight vector $w\in\mathbb{R}^d$ and label vector $Y=[y_1,\dots,y_n]$.
We generally assume that there exists a true weight vector $w_*\in\mathbb{R}^d$ such that $f_{w_*}(x)=f_*(x)$.
In the under-parametrized regime ($d<n$) the loss $\mathcal{L}_n(w)$ has a unique minimizer. 
In the over-parametrized regime ($d>n$), infinitely many weight vectors interpolate the training data exactly. 
We select the minimum $\ell_2$-norm interpolator $\hat w=\arg\min_w \lVert w\rVert^2$ subject to zero training error, which coincides with the solution reached by gradient descent from zero initialization. 
Generalization error is measured by the expected risk $\mathcal{R}(\hat w)=\mathbb{E}_{(x,y)\sim\mathcal{D}}[(\hat w^\top\phi(x)-y)^2]$, and its asymptotic value $\mathcal{R}(\gamma)$ is studied as $n,d\to\infty$ with $d/n\to\gamma$. The \emph{population covariance} is the deterministic matrix defined as $\Sigma := \mathbb{E}_x[\phi(x)\phi(x)^\top]$.

\smallskip
\par
\paragraph{The canonical double descent curve.} We begin working in the classical linear regression setting, since QELM readouts are ultimately combined with a classical linear layer for prediction~\cite{xiong2025fundamental}. 
For the textbook case of independent, identically distributed (i.i.d.) Gaussian features, \textit{i.e.}\ when $\Sigma = \sigma_X^2I_d$ for some $\sigma_X>0$, the asymptotic expected risk $\mathcal{R}(\gamma)$ diverges as $\gamma \to 1$ from the left and then decays for $\gamma > 1$, yielding the double descent curve~\cite{dobriban2018high, belkin2019reconciling, mei2022generalization, hastie2022surprises, bai2010spectral}. 
More explicitly we have a risk curve obeying
\begin{equation}
    \mathcal{R}(\gamma) = \begin{cases}
        \frac{\sigma^2}{1-\gamma},&\text{if $\gamma<1$,}
        \\ \sigma_X^2\left(1-\frac{1}{\gamma}\right)\,\lim_{d\to\infty}\|w_*\|^2 + \frac{\gamma\sigma^2}{\gamma-1}, &\text{if $\gamma>1$,} \label{eq:canonical-DD-curve}
    \end{cases}
\end{equation}
recalling that $\sigma^2$ is the variance of label noise.
The Marchenko-Pastur law yields this expression in the limit $n,d\to\infty$~\cite{bai2010spectral}. We use this curve as the standard reference and call it the \emph{canonical double descent curve}. Note that its only feature-map dependence is through $\sigma_X^2$.

\smallskip
\par
\paragraph{Double descent with correlated features.} 
Realistic feature maps are rarely i.i.d. Gaussian and often exhibit correlations between feature coordinates. We now determine when the canonical curve survives these correlations. We refer to the Methods section for more details on the mathematical derivation of these results.

In the under-parametrized regime ($\gamma<1$), the canonical curve $\mathcal{R}(\gamma)=\sigma^2/(1-\gamma)$ and its interpolation threshold at $\gamma=1$ are preserved if a high-dimensional spectral measure of correlation (which we call \emph{quadratic feature covariance measure}),
\begin{equation}
    \lambda_{\max}(\mathrm{Cov}(z\otimes z))=o(d) \label{eq:main-correlation-measure}
\end{equation}
grows sub-linearly as $d\to\infty$, with $z=\Sigma^{-1/2}\phi(x)$ denoting the ``whitened'' features, which may be non-centred in general. 
If this bound is violated, the location of the divergence can shift unpredictably.
We additionally prove that this condition can equivalently be phrased in terms of the sub-Gaussian norm of $z$~\cite{vershynin2020high}.

In the over-parametrized regime ($\gamma>1$), the assumption that almost surely $\lVert z\rVert=\sqrt{d}$, yields the generalization bound,
\begin{equation}
    \mathcal{R}(\gamma) = \sigma^2 + \mathcal{O}\!\left(\lVert w_*\rVert^2\,\lambda_{\max}(\Sigma) \;+\; \frac{\sigma^2\,\kappa(\Sigma)}{(\sqrt{\gamma}-C_K)^2}\right) \label{eq:main-generalization-bound}
\end{equation}
as $\gamma\to\infty$, in terms of the population covariance condition number $\kappa(\Sigma)=\lambda_{\max}(\Sigma)/\lambda_{\min}(\Sigma)$. Here, $C_K>0$ denotes a constant depending on the sub-Gaussian norm~\cite{vershynin2020high} of the whitened features $z$.
Equation~\eqref{eq:main-generalization-bound} shows that the risk in the over-parametrized regime depends on the extreme eigenvalues of the population covariance: 
a well-conditioned $\Sigma$ tightens the bound while an exploding condition number inflates it, independently of the specific correlation structure. 
We therefore use $\kappa(\Sigma)$ to connect model design to over-parameterized generalization.
Equation~\eqref{eq:main-generalization-bound} exhibits a singularity controlled by the constant $C_K$, near which the upper bound deteriorates sharply. Since the sub-Gaussian norm $K$ of $z$ is directly related to the spectral correlation measure introduced in Equation~\eqref{eq:main-correlation-measure}, provided that $\kappa(\Sigma)$ and $\lambda_{\max}(\Sigma)$ remain uniformly bounded as $d\to\infty$, stronger feature correlations, corresponding to faster growth of $C_K$ with $d$, lead to a looser upper bound in Equation~\eqref{eq:main-generalization-bound}.

\smallskip
\par
\paragraph{Quantum extreme learning machines.} A QELM encodes input data $x$ into a quantum state through a fixed circuit built from alternating data-independent, untrained unitaries and data-dependent encoding layers.
We apply a data-uploading unitary $U(x)=U_{L+1}\prod_{\ell=1}^{L}\{G_\ell(x)U_\ell\}$ (with $L$ \emph{reuploads}) to an initial state $\rho_0$, where $U_1,\ldots,U_{L+1}$ are fixed reservoir unitaries and each $G_\ell(x)=\bigotimes_{\eta=1}^p e^{-ix_\eta H_\eta^{(\ell)}}$ is a data-encoding layer generated by Hermitian operators $H_\eta^{(\ell)}$, as shown in the central panel of Figure~\ref{fig:main-overview}. 
The circuit is then measured to produce the \emph{readout}: a vector of quantum expectation value features $\phi_U(x) = (\langle O_1\rangle_x, \ldots, \langle O_d\rangle_x)$ with a chosen set of observables $O_k$, where $\phi_U(x)_k=\mathrm{Tr}(O_k\rho_U(x))$.
Only a classical linear layer on top of these features, i.e., a classical weight vector $w$, is trained with respect to the mean-squared loss. 
Training only the readout avoids optimizing quantum circuit parameters and the associated barren-plateau problem~\cite{mcclean2018barren, larocca2025barren, arrasmith2022equivalence, meyer2026gradient}, provided that measure concentration remains controlled~\cite{xiong2025fundamental,dao2026breaking}.

\smallskip
\par
\paragraph{Double descent in QELMs.}
The QELM model fits directly into the double descent framework outlined above, with the classical feature map $\phi$ replaced by the quantum readout $\phi_U$ and the classical population covariance replaced by its quantum counterpart, $\Sigma_U := \mathbb{E}_x[\phi_U(x)\phi_U(x)^\top]$.

The simplest theoretical benchmark is a circuit sampled from a unitary 2-design~\cite{mele2024introduction}, for which $\phi_U(x)$ becomes a zero-mean Gaussian process~\cite{garcia2025quantum} and $\Sigma_U$ is exactly isotropic. In this ideal case double descent follows the canonical curve in Equation~\eqref{eq:canonical-DD-curve} with $\sigma_X^2 = \Theta(2^{-N})$. 
As expected from a 2-design, the readout variance $\sigma_X^2$ decays exponentially in the number of qubits~\cite{xiong2025fundamental}. 
This idealized limit does not scale in practice.
Countering concentration is known to require exponentially many measurement shots, but we show that representing a non-concentrating target requires also exponentially many observables.
Practical QELMs therefore depart from the 2-design limit, at the cost of introducing feature correlations and a generally non-isotropic, non-trivially conditioned $\Sigma_U$~\cite{dao2026breaking, fischer2026dynamical}. 
Understanding and controlling $\kappa(\Sigma_U)$ for realistic, non-2-design circuits is thus the key point in which  our theory goes beyond the established literature. 

Our key technical tool is the Fourier interpretation of the quantum outputs~\cite{schuld2021effect, xiong2025fundamental, mhiri2025constrained, kang2026fourier}. Because each data encoding layer applies input-dependent phases determined by the eigenvalues of Hermitian generators, the output of \emph{any} QELM circuit can be written exactly as a finite trigonometric polynomial in the input~\cite{schuld2021effect},
\begin{equation}
    \phi_U(x) \;=\; \sum_{\omega \in \Omega} c_\omega\, e^{i\,\omega \cdot x}, \label{eq:main-Fourier-decomposition}
\end{equation}
where the discrete frequency set $\Omega$ is built from spectral differences accumulated along all pairs of ``quantum spectral tree paths''~\cite{mhiri2025constrained} through the generator eigenspaces across layers, and $c_\omega$ is obtained from trace expression involving the reservoir unitaries and observables.
This representation reduces the conditioning of $\Sigma_U$ to the geometry of the frequency set $\Omega$ and the coefficients $c_\omega$, that we can analyse explicitly as a function of architectural choices. We are particularly interested in three control knobs: input strength, number of reuploads, and effect of finite sampling (shot noise). 

\smallskip
\par
\paragraph{The effect of input strength.} We first consider \emph{input strength}, also known as ``kernel bandwidth'' which was introduced as a tunable parameter in quantum kernel methods by Shaydulin et al.~\cite{shaydulin2022importance}.
Rescaling the input according to $x\mapsto sx$, with $s>0$, before it enters the encoding layers rescales every frequency by the same factor. Consequently, the frequency set transforms as $\Omega\mapsto s\Omega$, while its cardinality remains unchanged. The Fourier coefficients are also unchanged in value. Only their frequency labels are relabelled, so that the coefficient associated with $\omega$ becomes associated with $s\omega$.
For Pauli encoding strategies~\cite{xiong2025fundamental, mhiri2025constrained} we have $\Omega=[-sL,sL]^p\cap s\mathbb{Z}^p$.

The two limits $s \to 0$ and $s \to \infty$ have opposite effects on conditioning. 
As $s\to 0$, the condition number diverges at least as fast as $s^{-2}$. Small input strengths therefore worsen the risk bound in Equation~\eqref{eq:main-generalization-bound} at least quadratically.

As $s \to \infty$, the rapidly oscillating cross-frequency terms of the covariance average out (a Riemann-Lebesgue-type effect~\cite{bochner1949fourier}), so that $\Sigma_U(s)$ converges to a fixed limiting matrix, $\Sigma_{\infty}=\sum_{\omega} c_\omega c_\omega^{\dagger}$.
If the input density has Sobolev order $K$, the cross-frequency contributions vanish as $s^{-K}$, which gives
\begin{equation}
    \kappa\big(\Sigma_U(s)\big) \;\le\; \kappa(\Sigma_\infty) + \mathcal{O}(s^{-K}), \label{eq:main-input-strength}
\end{equation}
where $K$ dictates the convergence rate with a clean power law. As a result, for large $s$ the quality of the bound in Equation~\eqref{eq:main-generalization-bound} mainly depends on the conditioning of the design-dependent matrix $\Sigma_\infty$.

Thus, weak input encoding produces poor conditioning, whereas strong input encoding drives the condition number down to a fixed, architecture-dependent floor at a rate controlled by the smoothness of the data distribution.
More details on the analytical derivations of these results can be found in Methods.

\smallskip
\par
\paragraph{The effect of increased reuploads.} The second knob is the number of \emph{reuploads} $L$, i.e., the number of times the input is re-encoded across the circuit's layers. 
Additional reuploads improve conditioning when they inject high-frequency content into the weakest eigenspace of the existing covariance.
Increasing $L \mapsto L + \Delta L$ enlarges the accessible Fourier frequency set through a \emph{Minkowski sum} over layers~\cite{holzer2026spectral}, so that $\Omega_L \subseteq \Omega_{L+\Delta L}$. In contrast to the input-strength case, we show that this also fundamentally changes the Fourier coefficients associated with the pre-existing frequencies via an affine transformation.

Writing $\Sigma_L$ for the base frequency covariance contribution and $\Sigma_{\Delta L}$ for the contribution of the newly accessible frequencies, the smallest eigenvalue increases under the mentioned injection condition, $\lambda_{\min}(\Sigma_L + \Sigma_{\Delta L}) \;\ge\; \lambda_{\min}(\Sigma_L) + \eta_{\Delta L}\,\delta(\Delta L)$, for a high-frequency injection parameter $\delta(\Delta L)>0$ and an efficiency factor $\eta_{\Delta L} \in (0,1]$ that depends on the spectral gap of $\Sigma_L$.
Intuitively, if the original architecture leaves some direction of feature space poorly represented, additional reuploads can fill in exactly that direction.
Since the largest eigenvalue of the readout stays of constant order $\mathcal{O}_{\Delta L}(1)$, this yields a decreasing upper bound on the condition number as reuploads accumulate:
\begin{equation}
    \kappa\bigl(\Sigma_{L+\Delta L}(s)\bigr) \leq \frac{\mathcal{O}_{\Delta L}(1)}{\lambda_{\min}\bigl(\Sigma_L\bigr)+ \eta_{\Delta L}\delta(\Delta L) - \mathcal{O}_s(s^{-K})}, \label{eq:main-reuploads}
\end{equation}
where $\mathcal{O}_s(s^{-K})$ is the previously discussed Sobolev power law due to input strength. Under the injection condition, additional reuploads therefore tighten the QELM risk bound in Equation~\eqref{eq:main-generalization-bound}.
It is worth mentioning that encoding strategies such as \emph{exponential encoding}~\cite{shin2023exponential, xiong2025fundamental, mhiri2025constrained} can reach high Fourier frequencies with significantly fewer reuploads.
We refer to the Methods section for details on these derivations.

\begin{figure*}[t]
  \centering
  \setcounter{subfigure}{0}

  \begingroup
  \setlength{\unitlength}{1bp}%
  \resizebox{\linewidth}{!}{%
    \begin{picture}(532.800000,216.000000)
      \put(11.520000,124.800000){\QELMPanel{102.480000bp}{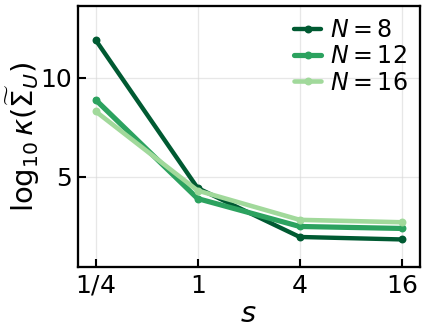}{fig:qelm-a}{11.709602}{75.644028}}
      \put(134.160000,124.800000){\QELMPanel{84.960000bp}{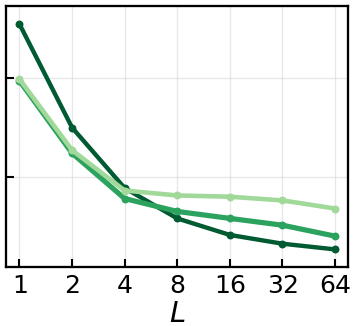}{fig:qelm-b}{-0.847458}{91.242938}}
      \put(238.320000,123.360000){\QELMPanel{87.840000bp}{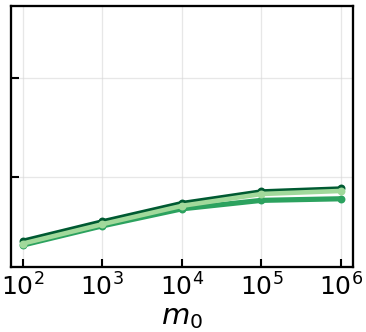}{fig:qelm-c}{0.546448}{89.890710}}

      \put(11.760000,12.000000){\QELMPanel{102.240000bp}{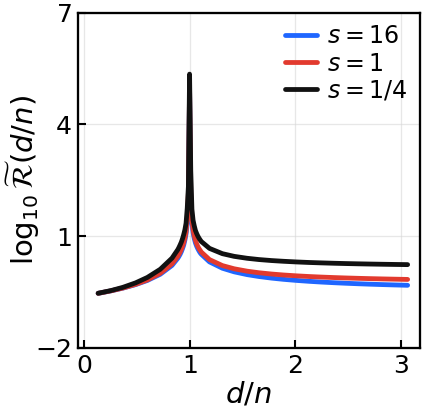}{fig:qelm-d}{11.971831}{93.896714}}
      \put(134.160000,12.000000){\QELMPanel{84.960000bp}{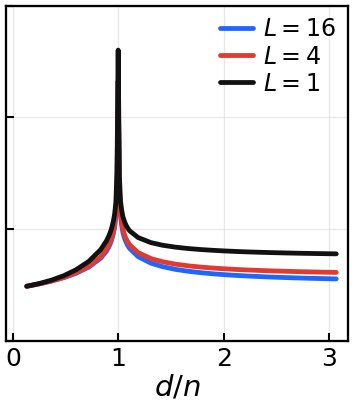}{fig:qelm-e}{-0.847458}{112.994350}}
      \put(239.520000,12.000000){\QELMPanel{84.960000bp}{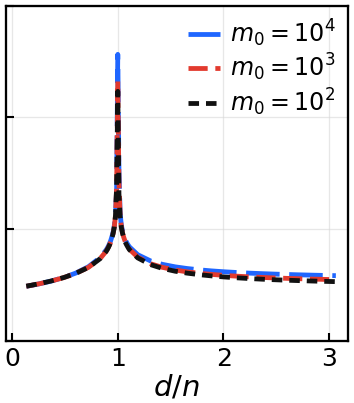}{fig:qelm-f}{-0.847458}{112.994350}}

      \put(344.880000,12.000000){\QELMPanel{174.960000bp}{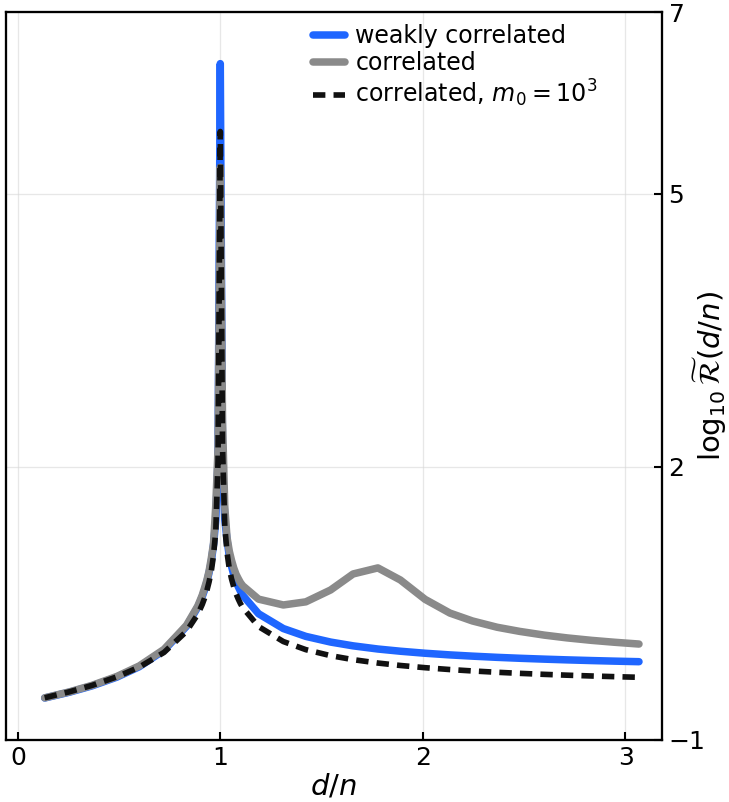}{fig:qelm-g}{-0.411523}{108.779150}}
    \end{picture}%
  }%
  \endgroup

\caption{Conditioning and empirical risk of the QELM features. Panels
 \subref{fig:qelm-a}--\subref{fig:qelm-c} show
 $\log_{10}\kappa(\widetilde{\Sigma}_U)$ for different $N$, with each
 distinct empirical estimate of the population covariance computed from
 $8{'}192$ randomized Sobol input points. Panel
 \subref{fig:qelm-a} varies $s$ at $L=4$, while panel
 \subref{fig:qelm-b} varies $L$ at $s=1$.
 Panel \subref{fig:qelm-c} varies
 $m_0$ at $s=1$ and $L=4$ by reusing
 the corresponding $8{'}192$-sample matrix and applying the analytical
 shot-noise transformation. Panels \subref{fig:qelm-d}--\subref{fig:qelm-f}
 show $\log_{10}\widetilde{\mathcal{R}}(d/n)$ for $N=16$, varying input
 strength, number of input reuploads, and number of measurement shots,
 respectively. All empirical-risk panels use the Moore--Penrose
 minimum-norm least-squares estimator. Panel \subref{fig:qelm-g} compares
 weakly and strongly correlated architectures. For the strongly correlated one,
 exact quantum features produce a secondary peak away from $d/n=1$;
 finite-shot features suppress this secondary peak while preserving the
 interpolation peak at $d/n=1$.}
\label{fig:experiments}

\end{figure*}

\smallskip
\par
\paragraph{The effect of finite shot noise.} The third hyperparameter we consider is \emph{finite measurement shot noise}. 
In experimental setups, each feature $\langle O_k \rangle_x$ is estimated from $m_0$ repeated quantum measurements, $\hat\phi_U(x)_k=m_0^{-1}\sum_{m=1}^{m_0}\mu_k^{(m)}(x)$, rather than computed exactly.
This yields a noisy estimator $\hat\phi_U(x) = \phi_U(x) + \epsilon(x)$ with conditionally centered, independent-across-observables noise $\epsilon(x)$. 
The resulting noisy population covariance decomposes exactly as
\begin{equation}
    \hat\Sigma_U \;:=\; \mathbb{E}_{x,\text{shot}}[\hat\phi_U(x)\hat\phi_U(x)^\top] \;=\; \Sigma_U + D_{\mathrm{shot}} \label{eq:main-noisy-covariance}
\end{equation}
with diagonal matrix $D_{\mathrm{shot}} = \tfrac{1}{m_0}\,\mathrm{diag}(\sigma_1^2,\ldots,\sigma_d^2)$ where $\sigma_k^2 := \mathbb{E}_x[\langle O_k^2\rangle_x - \langle O_k\rangle_x^2]$ is the input-averaged quantum measurement variance of observable $k$.
Shot noise therefore behaves as a diagonal regularizer on the population covariance, analogous to ridge regularization in classical statistics~\cite{dobriban2018high}.
Combining this decomposition with Weyl's eigenvalue inequalities~\cite{vershynin2020high} yields the bound
\begin{equation}
    \kappa(\hat\Sigma_U) \;\leq\; \frac{\lambda_{\max}(\Sigma_U) + \sigma^2_{\max}/m_0}{\lambda_{\min}(\Sigma_U) + \sigma^2_{\min}/m_0}, \label{eq:main-shot-noise-regularization}
\end{equation}
which shows that finite shot noise strictly \emph{improves} the noiseless condition number whenever $\kappa(\Sigma_U) > \sigma^2_{\max}/\sigma^2_{\min}$, when the noiseless covariance is ill-conditioned relative to the uniformity of measurement noise across observables. 
Fewer shots (\textit{i.e.}, stronger statistical noise) help, only while the regularization is needed. 
Beyond that, shot noise is detrimental.

Finite shot noise also limits high-dimensional feature correlations, thereby helping to preserve the canonical interpolation threshold at $\gamma=1$ in the under-parametrized regime. 
Under mild and physically standard assumptions on the observable set, this correlation-limiting condition is guaranteed to hold whenever the number of shots satisfies $m_0 = o(\sqrt{d})$ as the feature dimension $d$ grows. 
Thus, it is sufficient for the shot budget to grow more slowly than the square root of the number of measured observables. 
A moderate level of measurement noise can therefore restore the canonical threshold in architectures whose noiseless correlations shift it.

\smallskip
\par
\paragraph{Numerical experiments.}
We study the three design parameters primarily using a QELM architecture consistent with our correlation assumptions, and use a strongly correlated adversarial architecture, for which our sufficient correlation condition is not satisfied, to highlight the resulting change in behavior. For the weakly correlated model, panels~\subref{fig:qelm-a}--\subref{fig:qelm-c} of Figure~\ref{fig:experiments} show the condition numbers of empirical estimates of the population covariance, while panels~\subref{fig:qelm-d}--\subref{fig:qelm-f} show the corresponding empirical risk curves.

For all three values of $N$, increasing the input strength reduces the
condition number by several orders of magnitude before it approaches an
$N$-dependent plateau [Figure~\ref{fig:experiments}\subref{fig:qelm-a}].
Increasing the number of reuploads produces the same qualitative
improvement [Figure~\ref{fig:experiments}\subref{fig:qelm-b}].
For the $N=16$ model, stronger inputs and more reuploads also lower the
empirical risk in the over-parametrized regime
[Figure~\ref{fig:experiments}\subref{fig:qelm-d} and~\subref{fig:qelm-e}].
Thus, the empirical risk in the over-parametrized regime follows the dependence on $\kappa(\Sigma_U)$ predicted by Equation~\eqref{eq:main-generalization-bound}, while input strength and reuploads influence the risk by changing $\kappa(\Sigma_U)$ according to Equations~\eqref{eq:main-input-strength} and~\eqref{eq:main-reuploads}.

Figure~\ref{fig:experiments}\subref{fig:qelm-c} shows that the condition number increases
toward its noiseless value as the number of shots $m_0$ grows.
Equivalently, finite shot noise regularizes the poorly conditioned
population covariance.
The same ordering appears in the empirical risk curves: decreasing $m_0$
lowers the peak at $d/n=1$ and the risk throughout the over-parametrized
regime [Figure~\ref{fig:experiments}\subref{fig:qelm-f}].
Together, panels \subref{fig:qelm-a}--\subref{fig:qelm-f} connect each of the three design knobs to both
the conditioning of the population covariance and the empirical risk in
the over-parametrized regime.

Figure~\ref{fig:experiments}\subref{fig:qelm-g} shows that, for the strongly correlated architecture, noiseless features produce a second peak away from $d/n=1$, while a sufficiently small shot budget suppresses this peak, consistent with our theoretical result that finite-shot noise limits feature correlations.
Supplementary Section~\ref{appendix:numerical-experiments} directly quantifies these correlations numerically, distinguishing the weakly and strongly correlated regimes and confirming that finite-shot noise suppresses correlations as predicted by Theorem~\ref{thm:finite-shot-noise-correlation}. Furthermore, we test these conclusions on three real regression datasets, California Housing, Diabetes, and AAPL limit-order-book data. Across circuit depths, all three datasets exhibit interpolation peaks near $d/n=1$, while their exact feature maps show different trends in the computed correlation measure as $d$ increases, and finite-shot readout systematically reduces this measure.

These experiments support the predicted effects of all three design parameters in realistic circuits beyond the 2-design limit.

\section{Discussion}

Our results connect the double descent curve of linear models to the random feature population covariance, extending previous generalization analyses~\cite{dobriban2018high, bartlett2020benign, bartlett2021deep, mei2022generalization, hastie2022surprises, tsigler2023benign}.
In the under-parametrized regime, we derived a spectral measure of correlations that may be used to predict shifts in the interpolation threshold.
In the over-parametrized regime, we established a generalization bound whose quality is fully determined by the covariance condition number and the largest eigenvalue.
Consequently, our results advance the theory of double descent by establishing a direct analytical link between the statistical properties of random features and the shape of the risk curve.

For quantum machine learning models, double descent has until now been documented focusing mainly on the interpolation threshold or relying on empirical observation~\cite{hu2024generalization, kempkes2026double, kamisoyama2026double}.
In QELMs, we used the Fourier interpretation~\cite{schuld2021effect, xiong2025fundamental, mhiri2025constrained} to analytically characterize the quantum readout population covariance in terms of three knobs that are already standard in QELM design:
how strongly the data are encoded, how many times they are re-encoded across layers, and how many measurement shots are spent estimating each feature.
This gives practitioners concrete, testable guidance rather than a purely qualitative recommendation to over-parametrize. Our results suggest to increase input encoding strength and the number of reuploads as far as circuit depth and coherence budgets allow, and tuning the measurement-shot budget to balance estimation cost and implicit regularization. We emphasize that our findings do not guarantee that the second descent reaches a lower risk than the underparameterized regime, as this depends on the interplay between the dataset and the model architecture. Nevertheless, they provide concrete design principles for steering QELMs toward more favorable generalization.

Finite sampling noise can be constructive because it acts analogously to ridge regularization. 
A similar beneficial role for shot noise was observed numerically, though without an analytic derivation, in the double descent curves reported for quantum kernel methods by Kempkes et al.~\cite{kempkes2026double}.
This also revisits the earlier statistical-mechanics analysis of Hu and Gao~\cite{hu2024generalization}, extending it into an explicit, architecture-level mechanism that also governs the over-parametrized regime.
Additionally, we show that introducing controlled shot noise stabilizes the generalization curve by suppressing pathological feature correlations, preventing the interpolation threshold from shifting into the over-parametrized regime.
We stress that shot budgets should therefore be tuned to trade off estimation accuracy and implicit regularization in regimes where the covariance is poorly conditioned.

Anisotropic feature covariances with significantly distinct eigenvalues have been associated with ``multiple descent,'' in which the risk has several peaks~\cite{mel2021theory, mel2021anisotropic}. Under certain assumptions, recent work links this phenomenon to covariance correlations~\cite{hanhuang2026zero}. Our results show that sufficiently strong finite shot noise reduces these correlations and restores a single peak.

Our analysis opens several important questions that merit further investigation. The over-parametrized regime bound relies on a geometric regularity condition on the whitened features that we justify explicitly for 2-design circuits and, more generally, in a high-dimensional limit under a Gaussian input assumption.
Extending this justification to broader, non-Gaussian input distributions and to finite, moderate dimensions remains open. 
Our theory also does not explain the emergence of second peaks in the over-parametrized regime, leaving the ``multiple descent'' mechanism open~\cite{mel2021theory, mel2021anisotropic, hanhuang2026zero}.

In the quantum context, our analysis of data reuploading identifies when additional layers improve conditioning, but does not establish a general, non-perturbative guarantee that the required injection condition is satisfied for arbitrary reservoir unitaries.
Characterizing the reservoir architectures for which it holds automatically, rather than case by case, would sharpen the practical design recipe considerably. 
Finally, our theory addresses ridgeless linear regression on top of a fixed quantum feature map: whether an analogous covariance conditioning perspective extends to trained variational quantum models, to non-linear quantum kernel methods more broadly, or to classification losses, is an interesting direction with potential impact well beyond the QELM setting considered here.

In general, the population covariance perspective developed in this work, connecting an entirely classical, statistical quantity to concrete, physically meaningful circuit parameters via a Fourier representation, offers a blueprint that other quantum machine learning generalization questions can build on.

\section{Methods}

\paragraph{Sub-Gaussian whitened features and correlation.} Our analysis holds true as long as the whitened features $z=\Sigma^{-1/2}\phi(x)$ are sub-Gaussian~\cite{vershynin2012introduction, vershynin2020high} (cf. Supplementary Section~\ref{subsubsection:high-dim-probability}), that is,
\begin{equation}
     \sup_{v\in\mathbb{S}^{d-1}}\mathbb{P}\left(|\langle z, v\rangle| \geq t \right) \leq 2\exp\left(-\frac{t^2}{K^2}\right)
\end{equation}
uniformly in $t\geq0$ for some constant $K>0$ generally depending on $d$. The smallest constant $K$ satisfying this property is known as the \emph{sub-Gaussian norm} of $z$. This property is key in our argumentation because it essentially allows us to control and quantitatively describe the eigenspectrum of random matrices despite correlated entries~\cite{bryson2021marchenko, vershynin2012introduction, vershynin2020high}.

\smallskip
\par
\paragraph{Bias-variance decomposition with isotropic features.} Given a training set $\mathcal{S}:=\{(x_\ell,y_\ell)\}_{\ell=1}^n$, and assuming $Y = \Phi^\top w_* + \mathcal{E}$ for some Gaussian noise $\mathcal{E}\sim\mathcal{N}(0,\sigma^2 I_n)$, we aim to derive a closed-form expression for the expected risk,
\begin{align}
    \mathcal{R}(\hat{w}) &= \mathbb{E}_{\mathcal{S}}\left[\mathbb{E}_{( x, y)\sim \mathcal{D}}\left[ (\hat{w}^\top\phi(x) -  y)^2 \mid\mathcal{S}\right]\right] \nonumber
    \\ &= \mathbb{E}_{\mathcal{S}}\left[\|\hat{w}-w_*\|_{ \Sigma}^2\right] + \sigma^2,
\end{align}
where $\hat{w}$ denotes the minimal $\ell_2$-norm interpolator. 

In the textbook case of independent Gaussian features $\phi(x)\sim\mathcal{N}(0,\sigma_X^2 I_d)$ one has $\Sigma = \sigma_X^2 I_d$, reducing the problem to solving expressions of the form $\mathbb{E}_x[S_n(0)]$, where $S_n(0)$ denotes the Stieltjes transform evaluated at the origin of a random matrix for the under- or over-parametrized regime (cf. Supplementary Section~\ref{appendix:Stieltjes}).
By the Marchenko-Pastur law~\cite{bai2010spectral}, the limiting Stieltjes transform $S(z) = \lim_{n\to\infty}S_N(z)$ is governed by the quadratic equation,
\begin{equation}
    \beta z\sigma_X^2 S(z)^2 + (z+\sigma_X^2(\beta-1))S(z) + 1 = 0, \label{eq:main-MP}
\end{equation}
where $\beta\in\{\gamma,\gamma^{-1}\}$, depending on whether we are in the under- or over-parametrized regimes, respectively. 
Supplementary Section~\ref{appendix:gaussian-features} derives the canonical risk curve for $\mathcal{R}(\gamma)$.

\smallskip
\par
\paragraph{Bias-variance decomposition with correlated features.}
If $\Sigma\not\propto I_d$, i.e. the covariance is not isotropic, the same calculation is more difficult.
While Stieltjes techniques have previously been applied to derive the risk of linear models with anisotropic features~\cite{dobriban2018high, mei2022generalization, bai2010spectral}, the solution is generally determined by an implicit, highly non-trivial fixed point equation with closed-form solutions only in special cases, such as, if $\Sigma\propto I_d$.
For that reason, we follow an alternative approach relying on sub-Gaussian features~\cite{bartlett2020benign, bartlett2021deep, tsigler2023benign, vershynin2012introduction,vershynin2020high}.
By assuming that the whitened features $z=\Sigma^{-1/2}\phi(x)$ are sub-Gaussian, we can reduce the under-parametrized problem to Equation~\eqref{eq:main-MP} by leveraging Theorem~1.7 of Bryson et al.~\cite{bryson2021marchenko} at the cost of additionally assuming the growth condition
\begin{equation}
     \max_{j\in[n]} \sup_{\substack{A\in\mathbb{R}^{d\times d} \\ \|A\|=\mathcal{O}_d(1)}} \operatorname{Var}(z_j^\top A z_j) = o(d^2) \label{eq:main-correlation-bound}
\end{equation}
as $d\to\infty$. 
Hence, despite correlations, in the under-parametrized regime we get the closed-form canonical curve $\sigma^2(1-\gamma)^{-1}$, governed by the Marchenko-Pastur law, provided that Equation~\eqref{eq:main-correlation-bound} is satisfied, see Supplementary Section~\ref{sec:non-isotropic-analysis}.
In Supplementary Section~\ref{subsubsection:high-dim-probability} we show that it is sufficient to assume $\lambda_{\max}(\operatorname{Cov}(z\otimes z))=o(d)$ or $K=o(d^{1/4})$, where $K$ is the sub-Gaussian norm of $z$ in order for Equation~\eqref{eq:main-correlation-bound} to hold.

In the over-parametrized regime using the explicit form of the min-norm interpolator, $\hat{w}=\Phi(\Phi^\top\Phi)^{-1}Y$, gives the \emph{bias-variance decomposition}~\cite{belkin2019reconciling, adlam2020understanding, mel2021anisotropic}:
\begin{align}
    &\mathbb{E}_\mathcal{E}\left[\|\hat{w}-w_*\|^2_{ \Sigma}\right] = \underbrace{\|(I-\Phi(\Phi^\top\Phi)^{-1}\Phi^\top)w_*\|^2_{ \Sigma}}_{=:\operatorname{Bias}_\Phi(w_*, \Sigma)^2} \nonumber
    \\ & \, \qquad+\,\,\underbrace{\sigma^2\operatorname{Tr}\left( \Sigma\,\Phi(\Phi^\top\Phi)^{-2}\Phi^\top\right)}_{=:\operatorname{Var}_\Phi( \Sigma)}\,. \label{eq:main-bias-variance-decomposition}
\end{align}
When the whitened features $z$ are sub-Gaussian and $\|z\|=\sqrt{d}$ almost surely, Theorem~5.58 of Vershynin~\cite{vershynin2012introduction} bounds the lowest singular value of the random matrix $Z=\Sigma^{-1/2}\Phi$ used to bound the variance term $\operatorname{Var}_\Phi(\Sigma)$ in Equation~\eqref{eq:main-bias-variance-decomposition}. 
Similar techniques have been employed in previous works~\cite{kuzborskij2021role}.
Further bounds on the variance lead to the condition number $\kappa(\Sigma)$ in Equation~\eqref{eq:main-generalization-bound}. The bias term $\operatorname{Bias}_\Phi(w_*,\Sigma)^2=\mathcal{O}(\|w_*\|^2\|\Sigma\|)$ follows from basic linear algebra. Supplementary Section~\ref{sec:non-isotropic-analysis} gives the full derivation.

Importantly, to ensure that the theories of the under- and over-parametrized regimes are consistent, we have to check that the assumption $\|z\|=\sqrt{d}$ and the growth condition in Equation~\eqref{eq:main-correlation-bound} do not contradict each other. We do this in Supplementary Section~\ref{appendix:RMT-assumptions}. 
There, we explicitly show, using Lipschitz concentration on the sphere~\cite{vershynin2020high}, that if $z\sim\operatorname{Unif}(\sqrt{d}\mathbb{S}^{d-1})$ (in particular $\|z\|=\sqrt{d}$), then the growth condition of the sub-Gaussian norm $K=o(d^{1/4})$ of $z$ is satisfied.
Consequently, Equation~\eqref{eq:main-correlation-bound} holds true.

\smallskip
\par
\paragraph{Applicability to QELMs and Fourier interpretation.} For unitary 2-design circuits, the QELM readout is a zero-mean Gaussian process with isotropic population covariance, so Equation~\eqref{eq:canonical-DD-curve} applies with $\sigma_X^2=\Theta(2^{-N})$. Supplementary Section~\ref{appendix:2-designs} proves this result and gives the corresponding shot and observable scaling. Supplementary Section~\ref{appendix:quantum-RMT-assumptions} verifies the random matrix assumptions for 2-designs and for models with anisotropic $\Sigma_U$ in the high-dimensional limit.
The QELM readout admits the finite Fourier decomposition in Equation~\eqref{eq:main-Fourier-decomposition}, where the frequency set $\Omega$ contains accumulated spectral differences and the coefficients $c_\omega$ encode the reservoir unitaries, input state, and observables. Supplementary Section~\ref{appendix:fourier-interpretation} derives this representation using generalized quantum spectral tree paths. 

\smallskip
\par
\paragraph{The effect of input strength, characteristic function and Sobolev spaces.} To analyse the effect of increasing input strength, we combine the Fourier decomposition in Equation~\eqref{eq:main-Fourier-decomposition} with the multi-variate \emph{characteristic function}~\cite{billingsley2017probability} $\varphi_x:\mathbb{R}^p\to\mathbb{C}$, defined as $\varphi_x(v) = \mathbb{E}_x[e^{i\langle v,x\rangle}]$, which yields
\begin{align}
    \Sigma_U(s) &= \mathbb{E}_x[\phi_U(sx)\phi_U(sx)^\top] \nonumber
    \\ &= \sum_{\omega,\omega'\in\Omega} \varphi_{ x}\left(s(\omega+\omega')\right)   c_\omega  c_{\omega'}^\top,
\end{align}
i.e., a weighted Fourier decomposition with cross-frequency rank-1 outer products $c_\omega c_{\omega'}^\top$. Exploiting the regularity properties of the characteristic function, we may Taylor-expand each weight $\varphi_x(s(\omega+\omega'))$ at $s=0$ in order to derive
\begin{equation}
    \Sigma_U(s) = \sum_{m\geq0}\frac{(is)^m}{m!}\sum_{\omega,\omega'\in\Omega} M_{ x}^{(m)}(\omega+\omega') \, c_\omega c_{\omega'}^\top \label{eq:main-moment-expansion-covariance}
\end{equation}
with projected Fourier frequency moments $M_{ x}^{(m)}(\omega+\omega') := \mathbb{E}_{ x}[((\omega+\omega')\cdot x)^m]$. 
Taking $s\to0$ and using $\varphi_x(0)=1$ reveals the singular limit $\Sigma_U(0) =\phi_U(0)\phi_U(0)^\top$. 
A more careful analysis of Equation~\eqref{eq:main-moment-expansion-covariance} then shows the claimed bound $\kappa(\Sigma_U(s))=\Omega(s^{-2})$ as $s\to0$.

To bound $\kappa(\Sigma_U(s))$ as $s\to\infty$, we use \emph{Sobolev spaces}~\cite{adams2003sobolev}: we assume that the input density $f:\mathcal{X}\to[0,1]$ is a $K$-order Sobolev function, requiring that
\begin{equation}
    \int_{\mathcal{X}}|D_{ x}^\alpha f(x)|^q\,dx<\infty
\end{equation}
for all multi-indices $\alpha=(\alpha_1,\dots,\alpha_p)\in\mathbb{N}_0^p$ with $\|\alpha\|_1\leq K$, where $D_x^\alpha f$ denotes the $\alpha$-order \emph{weak derivative}~\cite{adams2003sobolev, wheeden1977measure} of $f$. 
We use the \emph{integration by parts} formula, which states that if $f$ belongs to the $K$-order Sobolev space, then
\begin{equation}
    \mathbb{E}_x[D_{ x}^\alpha \varphi( x)] = (-1)^K\int_{\mathcal{X}}\varphi( x)D_{ x}^\alpha f( x)dx
\end{equation}
for all smooth functions $\varphi:\mathcal{X}\to\mathbb{R}$ compactly supported in $\mathcal{X}$ and for all multi-indices $\alpha\in\mathbb{N}_0^p$ with $\|\alpha\|_1=K$. This trick then allows us to expand the rescaled population covariance in the limit $s\to\infty$, showing that
\begin{equation}
     \Sigma(s) = \sum_{\omega\in\Omega}c_\omega c_\omega^\dagger + \Delta(s) \label{eq:main-large-input-strength-covariance}
\end{equation}
with a cross-frequency, asymmetric remainder term explicitly determined by
\begin{equation}
    \Delta(s) = \sum_{\substack{\omega,\omega'\in\Omega \\ \omega+\omega'\neq0}}\left(\frac{i}{s\|\omega+\omega'\|^2}\right)^K I_K(\omega+\omega')\,c_\omega c_{\omega'}^\top \label{eq:main-asymmetric-remainder}
\end{equation}
for a certain, uniformly bounded integral $I_K(\omega+\omega')=\mathcal{O}_s(1)$, provided that $f$ belongs to the $K$-order Sobolev space.
The proof of this technical statement, from which it follows that $\Delta(s) = \mathcal{O}(s^{-K})$,  is fully elaborated in Supplementary Section~\ref{appendix:Sobolev-spaces} and involves the construction of bump functions and the Lebesgue dominated convergence theorem to justify exchanging limits and integration~\cite{wheeden1977measure}.
There, we also demonstrate how using Equations~\eqref{eq:main-large-input-strength-covariance} and~\eqref{eq:main-asymmetric-remainder} leads to the claimed bound for the conditioning $\kappa(\Sigma_U(s))$ in the large input strength limit of Equation~\eqref{eq:main-input-strength}.
Supplementary Section~\ref{subsubsec:input-strength} collects the small- and large-$s$ proofs.

\smallskip
\par
\paragraph{Fourier coefficients under reuploads.} The bound in Equation~\eqref{eq:main-input-strength} shows that the condition number $\kappa(\Sigma(s))$ converges to a stable, $s$-independent limit $\kappa(\Sigma_\infty)$, introducing $\Sigma_{\infty}=\sum_{\omega} c_\omega c_\omega^{\dagger}$ as $s\to\infty$.
To connect architecture design to this limit, we track the Fourier coefficients $c_\omega$ in Equation~\eqref{eq:main-Fourier-decomposition} as one modifies the architecture. Specifically, we increase the number of reuploads from $L$ to $L+\Delta L$.
In Supplementary Section~\ref{appendix:fourier-interpretation}, we derive the Fourier decomposition of Equation~\eqref{eq:main-Fourier-decomposition} and express the Fourier coefficients as the discrete double path integral
\begin{equation}
    c_\omega^k = \sum_{\substack{\gamma,\gamma'\in([2^{n_1}]\times\cdots\times[2^{n_p}])^L \\ \Lambda_{\gamma'}-\Lambda_{\gamma}=\omega}} \operatorname{Tr}(O_kU_{L+1}W_\gamma\rho_0 W_{\gamma'}^\dagger U_{L+1}^\dagger). \label{eq:main-Fourier-coefficients}
\end{equation}
The Fourier frequency space $\Omega$ consists of all possible elements of the form $\omega = \Lambda_{\gamma} - \Lambda_{\gamma'}$, using
\begin{equation}
    \Lambda_\gamma := \sum_{\ell=1}^L\left(\begin{matrix}  \lambda_{\gamma_\ell,1}^{(\ell)} \\ \vdots \\ \lambda_{\gamma_\ell,p}^{(\ell)}\end{matrix}\right)  \in \sum_{\ell=1}^L\prod_{\eta=1}^p\operatorname{Spec}(H_\eta^{(\ell)})\subset\mathbb{R}^p
\end{equation}
to calculate the spectral accumulation of the generators $H_\eta^{(\ell)}$ along the generalized quantum spectral tree path $\gamma=(\gamma_{\ell,\eta})$~\cite{mhiri2025constrained} (cf. Supplementary Section~\ref{appendix:Fourier-analysis}).
We further introduced the linear operators
\begin{equation}
    W_\gamma=P_{\gamma_L}U_LP_{\gamma_{L-1}} U_{L-1}\cdots P_{\gamma_1}U_1
\end{equation}
for each quantum spectral tree path $\gamma=(\gamma_{\ell,\eta})$ with spectral projectors $P_{\gamma_\ell}:=\bigotimes_{\eta=1}^p |\gamma_{\ell,\eta}\rangle\langle\gamma_{\ell,\eta}|$.
The linear operators $W_\gamma$ are important to account for the distribution of all input generators $H_\eta^{(\ell)}$ used across the circuit $U(x)$. This is also why we generalized the spectral quantum tree path formalism in Supplementary Section~\ref{appendix:fourier-interpretation}.
Each generator $H_\eta^{(\ell)}$ acts on a subsystem of $n_\eta$ qubits such that $n_1+\cdots +n_p = N$ matches the total number of qubits of the system.
This framework fully accounts for the quantum architecture because the Fourier coefficients in Equation~\eqref{eq:main-Fourier-coefficients} depend explicitly on the choice of the unitaries $U_\ell$, the observables $O_k$, the input state $\rho_0$, and the distribution of the input generators captured by $W_\gamma$.

Returning to the problem of identifying $\Sigma_\infty$ in Equation~\eqref{eq:main-input-strength} in terms of the change $L\mapsto L+\Delta L$, the key idea is therefore to study
\begin{equation}
    \Sigma_U^{(L+\Delta L)}(s) = \underbrace{\sum_{\omega\in\Omega_{L+\Delta L}}c_{\omega}^{(L+\Delta L)} \left(c_{\omega}^{(L+\Delta L)}\right)^\dagger}_{\equiv \Sigma_\infty} + \Delta^{(L+\Delta L)}(s), \label{eq:main-large-reuploads-covariance}
\end{equation}
as per Equation~\eqref{eq:main-large-input-strength-covariance} where, similar to Equation~\eqref{eq:main-asymmetric-remainder}, $\Delta^{(L+\Delta L)}=\mathcal{O}(s^{-K})$ again defines an asymmetric remainder term that vanishes for strong inputs. 
First, we leverage the Minkowski additivity $\Omega_{L+\Delta L} = \Omega_L + \Omega_{\Delta L}$ of the Fourier frequency space~\cite{holzer2026spectral}, where the Minkowski addend $\Omega_{\Delta L}= \prod_{\eta=1}^p\left(\sum_{\ell=L+1}^{L+\Delta L}\operatorname{Spec}(H_\eta^{(\ell)})\right)$ contains the incremental high frequencies.
Next, we use this insight to decompose $\Sigma_\infty = \Sigma_L + \Sigma_{\Delta L}$ in Equation~\eqref{eq:main-large-reuploads-covariance} into two terms where we separately sum over $\Omega_L$ and $\Omega_{\Delta L}$, respectively.
Note that $\Sigma_U^{(L+\Delta L)}(s)$ is a generalization of $\Sigma(s)$ in Equation~\eqref{eq:main-large-input-strength-covariance} which highlights the dependency on the architecture $U$ and the total number $L+\Delta L$ of reuploads. 
In contrast, $\Sigma_L$ and $\Sigma_{\Delta L}$ are $s$-independent matrices specifically used to highlight the Fourier structure in $\Sigma_\infty$ due to the distinct frequency space components $\Omega_L$ and $\Omega_{\Delta L}$, respectively.
Under the conditions of Equations~\eqref{eq:main-large-input-strength-covariance} and~\eqref{eq:main-asymmetric-remainder}, this establishes $\lambda_{\min}(\Sigma_U^{(L+\Delta L)}(s)) \geq \lambda_{\min}\left(\Sigma_L + \Sigma_{\Delta L}\right) - \mathcal{O}_s\left(s^{-K}\right)$ and since $\lambda_{\max}(\Sigma_U^{(L+\Delta L)}(s))= \mathcal{O}_{\Delta L}(1)$, it suffices to prove that $\lambda_{\min}(\Sigma_L + \Sigma_{\Delta L}) \geq  \lambda_{\min}(\Sigma_L) + \eta_{\Delta L}\delta(\Delta L)$ in order to conclude with Equation~\eqref{eq:main-reuploads}.
We refer to the details in Supplementary Section~\ref{subsubsec:input-reuploads}. The most important take-away is that the high-frequency injection $\delta(\Delta L)$ is explicitly determined by the infimum
\begin{equation}
    \delta(\Delta L) = \inf_{\substack{v\in E_{\min}(\Sigma_L)\\ \|v\|=1}}
    v^\dagger \Sigma_{\Delta L}v \label{eq:main-energy-injection}
\end{equation}
over the linear eigenspace $E_{\min}(\Sigma_L) := \{v\in\mathbb{C}^d: \Sigma_L v = \lambda_{\min}(\Sigma_L) v\}$ corresponding to the lowest energy of $\Sigma_L$, and efficiency factor $\eta_{\Delta L} := \frac{\Delta \lambda}{\Delta \lambda + \delta(\Delta L) + \|\Sigma_{\Delta L}\|}$ assuming a positive spectral gap $\Delta\lambda=\lambda_2(\Sigma_L)-\lambda_{\min}(\Sigma_L)>0$.
Notably, we may express $\delta(\Delta L)$ in Equation~\eqref{eq:main-energy-injection} more explicitly using Equation~\eqref{eq:main-Fourier-coefficients}.
All those details are found in Supplementary Section~\ref{subsubsec:input-reuploads}. In order to account for the correct base $\Sigma_U^{(L)}\leftrightarrow \Sigma_L$ (or equivalently, $c_\omega^{(L)}\leftrightarrow c_\omega^{(L+\Delta L)}$) in Equation~\eqref{eq:main-reuploads}, we also show there that $\Sigma_L = T_0\,\Sigma_U^{(L)}(s\to\infty)\,T_0^\dagger + \tilde{\mathcal{E}}^\perp$ for certain, explicitly determined matrices $T_0,\tilde{\mathcal{E}}^\perp\in\mathbb{C}^{d\times d}$ that depend on the deeper architecture due to $\Delta L$.

\smallskip
\par
\paragraph{Finite shot noise fixes double descent peaks.} Previous work revealed that finite shot noise can help suppress the double descent peak~\cite{kempkes2026double}. 
In our contribution, we analytically show that shot noise can improve conditioning and undo shifted peaks, essentially re-establishing the canonical curve $\mathcal{R}(\gamma) = \sigma^2(1-\gamma)^{-1}$ for $\gamma<1$. 
To derive the improved conditioning as per Equation~\eqref{eq:main-shot-noise-regularization}, we replace $\phi_U(x)$ with the noisy features $\hat{\phi}_U(x)$, and $\Sigma_U$ with $\hat{\Sigma}_U$ in Equation~\eqref{eq:main-noisy-covariance}, respectively, and bound as discussed in the Results.
Then, to prove that shot noise limits correlations, one has to compute the variance in Equation~\eqref{eq:main-correlation-bound} with respect to the joint probability space~\cite{billingsley2017probability} taking into account the new randomness due to the noisy vector $\epsilon(x)=\hat{\phi}_U(x) - \phi_U(x)$.
The core idea is the following: if the peak of the noiseless curve is shifted, then the correlation growth condition in Equation~\eqref{eq:main-correlation-bound} must be violated.
Intuitively, to restore the validity, it is sufficient to assume $m_0=o(\sqrt{d})$ so that the noise is large enough to weaken the noiseless correlations.
Since the proof of this fact is very convoluted, we refer to the thorough elaboration in Supplementary Sections~\ref{subsec:finite-shot-noise} and~\ref{appendix:techniques}.

\smallskip
\par
\paragraph{Numerical QELM setup.}

To conduct our numerical experiments, we use two QELM architectures. As a weakly correlated reference, we adopt the kicked-Ising QELM architecture of Dao et al.~\cite{dao2026breaking}, for which the reservoir parameters are sampled independently as $b\sim\mathcal{N}(0.707,0.031)$, $h\sim\mathcal{N}(0.683,0.034)$, and $J\sim\mathcal{N}(0.237,0.038)$. 
Here, $J$, $h$, and $b$ denote the Ising coupling, longitudinal-field, and transverse-kick strengths, respectively. Each circuit block receives one independent draw from the corresponding distribution, which is then held fixed.
To construct a feature map with stronger correlations, we linearly interpolate each independently sampled baseline triple $\boldsymbol{\theta}_{0}=(J,h,b)$ toward a dual-unitary kicked-Ising configuration based on Fischer et al.~\cite{fischer2026dynamical}, chosen here as $\boldsymbol{\theta}_{\mathrm{DU}}=(\pi/4,7\pi/32,\pi/4)$, according to
\begin{equation}
\boldsymbol{\theta}_t=(1-t)\boldsymbol{\theta}_{0}+t\boldsymbol{\theta}_{\mathrm{DU}},
\end{equation}
and choose $t=0.998$. This places the model very close to the dual-unitary regime while retaining a small deviation from it. We evaluate this configuration using both exact expectation values and finite-shot readout. The $t=0$ configuration produces weaker feature correlations, whereas the $t=0.998$ configuration generates substantially stronger ones and serves as a counterexample. We use $t=0.998$ rather than $t=1$, which corresponds exactly to the dual-unitary point where circuit symmetries force many feature-map entries to vanish, resulting in an artificially sparse representation. Supplementary Section~\ref{appendix:numerical-experiments} provides a direct characterization of the feature correlations. In particular, we evaluate the normalized quadratic feature covariance measure $\lambda_{\max}(\operatorname{Cov}(z\otimes z))/d$, which clearly separates the weakly correlated reference architecture from the more strongly correlated near-dual-unitary regime.

For a system of $N$ qubits, we draw input vectors
$x\sim\operatorname{Unif}([0,1)^p)$ with $p=N/2$ and assign one input
component to each qubit pair.
The same scaled input $sx$ is encoded in each of the $L$ data-encoding
layers.
For each curve in panels \subref{fig:qelm-a}--\subref{fig:qelm-f} of Figure~\ref{fig:experiments}, we use fixed reservoir unitaries and a
fixed observable pool while varying the displayed design knob.
The observable pool contains all weight-one and weight-two Pauli
observables supported on eight fixed qubits, giving $276$ observables for
every value of $N$.

\smallskip
\par
\paragraph{Population covariance estimation.}
For panels \subref{fig:qelm-a}--\subref{fig:qelm-c} of Figure~\ref{fig:experiments}, we estimate the population covariance using exact
QELM features obtained by statevector simulation.
For every distinct architecture setting, we evaluate four independently
scrambled Sobol sequences of $2{'}048$ inputs each~\cite{owen1998scrambling}
and pool the resulting $8{'}192$ inputs to form
\begin{equation}
    \widetilde{\Sigma}_U
    :=
    \frac{1}{8{'}192}
    \sum_{j=1}^{8{'}192}
    \phi_U(x_j)\phi_U(x_j)^\top. \label{eq:empirical-covariance}
\end{equation}
These inputs are separate from the training and test sets used for the
empirical risk curves.
Panel \subref{fig:qelm-a} uses $N\in\{8,12,16\}$, $L=4$, and
$s\in\{1/4,1,4,16\}$.
Panel \subref{fig:qelm-b} uses the same values of $N$, $s=1$, and
$L\in\{1,2,4,8,16,32,64\}$.
For panel \subref{fig:qelm-c}, all shot counts share the corresponding $s=1$, $L=4$
estimate.
For panel~\subref{fig:qelm-c}, we model each Pauli expectation value as
being estimated from an independent batch of $m_0$ ideal measurement
shots. This gives
\begin{equation}
 \widetilde{\Sigma}_{U,m_0}
 =
 \widetilde{\Sigma}_{U}
 +
 \frac{1}{m_0}
 \operatorname{Diag}\!\left(
 \mathbf{1}_d-\operatorname{diag}\widetilde{\Sigma}_{U}
 \right).
\end{equation}
To see this, write $\phi_U(x)_k$ for the $k$th component of the feature
vector. A single ideal measurement of the corresponding Pauli observable
returns $\pm1$ with mean $\phi_U(x)_k$. The average of $m_0$ independent
shots therefore has conditional mean $\phi_U(x)_k$ and conditional
variance $(1-\phi_U(x)_k^2)/m_0$. Because our model uses independent shot
batches for distinct feature coordinates, shot noise contributes only the
diagonal correction above. Averaging this correction over the sampled
inputs gives $\widetilde{\Sigma}_{U,m_0}$. We evaluate the expectation over
the binomial shot noise analytically rather than drawing finite-shot
outcomes, thereby avoiding additional shot-sampling fluctuations.
Panel \subref{fig:qelm-c} evaluates this expression for
$m_0\in\{10^2,10^3,10^4,10^5,10^6\}$ without drawing additional input
samples.
Thus, the shared notation $\widetilde{\Sigma}_U$ in
Figure~\ref{fig:experiments} denotes the empirical estimate of the population
covariance before the transformation in panels \subref{fig:qelm-a} and \subref{fig:qelm-b}, and after the
transformation in panel \subref{fig:qelm-c}.

\smallskip
\par
\paragraph{Empirical risk estimation.}
For panels \subref{fig:qelm-d}--\subref{fig:qelm-g} of Figure~\ref{fig:experiments}, each architecture is evaluated on $n=90$ training
inputs and $n_{\mathrm{test}}=1000$ independently drawn test inputs.
For each architecture, the reservoir unitaries and the training and test
inputs are held fixed.
For learned weights $\hat w$, let $f_{\hat w}$ denote the fitted predictor
formed from the exact or finite-shot QELM features used for the
corresponding curve.
The plotted empirical risk is
\begin{equation}
    \widetilde{\mathcal{R}}(d/n)
    :=
    \frac{1}{n_{\mathrm{test}}}
    \sum_{j=1}^{n_{\mathrm{test}}}
    \left(f_{\hat w}(x_j)-y_j\right)^2.
\end{equation}
At every plotted feature count $d\leq276$, we perform $3{'}200$
subset--teacher realizations. In each realization, we select $d$
observables uniformly without replacement from the pool and independently
draw a unit-norm linear teacher $w_*$
and generate labels from the exact QELM features according to
\begin{equation}
    y=w_*^\top\phi_U(x)+\delta,
    \qquad
    \delta\sim\mathcal{N}(0,0.5^2).
\end{equation}
Independent noise drawn from the same Gaussian distribution is added to
the training and test labels.
The linear readout is computed as the Moore--Penrose minimum-norm
least-squares solution. For each training design matrix $\Phi^\top$, we
compute a compact singular-value decomposition in double precision.
Singular values not exceeding
$\max(n,d)\epsilon_{\mathrm{mach}}\sigma_{\max}$ are treated as zero,
where $\epsilon_{\mathrm{mach}}$ is the double-precision machine epsilon
and $\sigma_{\max}$ is the largest singular value of $\Phi^\top$.
For each plotted point, we average the test mean-squared error over these $3{'}200$ subset--teacher realizations. 
For each realization, we average over $800$ antithetic label-noise realizations, comprising $400$
independent Gaussian draws and their negatives. Thus, each plotted point
averages $2{'}560{'}000$ test-error evaluations.

Panel \subref{fig:qelm-d} uses $N=16$, $L=4$, noiseless features, and
$s\in\{1/4,1,16\}$.
Panel \subref{fig:qelm-e} uses $N=16$, $s=1$, noiseless features, and
$L\in\{1,4,16\}$.
Panel \subref{fig:qelm-f} uses $N=16$, $s=1$, $L=4$, and
$m_0\in\{10^2,10^3,10^4\}$.
For finite $m_0$, the features supplied to the linear readout are sampled
as
\begin{equation}
    \hat{\phi}_U(x)_k
    =
    \frac{1}{m_0}
    \sum_{m=1}^{m_0}\mu_k^{(m)}(x),
\end{equation}
where the independent outcomes $\mu_k^{(m)}(x)\in\{-1,1\}$ satisfy
\begin{equation}
    \mathbb{P}\!\left(\mu_k^{(m)}(x)=1\mid x\right)
    =
    \frac{1+\phi_U(x)_k}{2}.
\end{equation}
The outcomes are sampled independently across observables and input
samples.
Thus, $m_0$ denotes the number of shots per observable and input.
Labels remain generated from the exact QELM features $\phi_U(x)$.
One realization of the finite-shot training and test features is held
fixed while averaging over feature subsets and label-noise
realizations.
Panel \subref{fig:qelm-g} uses $N=8$, $s=1$, and $L=4$ for both reservoir settings; its
finite-shot curve uses $m_0=10^3$.

\section*{Acknowledgements}

We thank Tin Sum Cheng for his valuable feedback including insightful discussions on generalization and high-dimensional probability theory.
S.M., E.Y., F.S. and A.L. acknowledge the support from SNSF grant No. 214919. F.T.\
acknowledges support from the Swiss National
Science Foundation through grant 225229 (RESQUE).

\bibliography{bibliography}

\resumetoc
\newpage
\onecolumngrid
\appendix

{
  \tableofcontents
}

\section{Notations} \label{appendix:notations}

\textbf{Mathematical Fundamentals.} The set $\mathbb{N}=\{1,2,3,\dots\}$ denotes the natural numbers and we explicitly exclude the zero element unless when we specify $\mathbb{N}_0:=\mathbb{N}\cup\{0\}$. For each $p\in\mathbb{N}$ we use the notations $[p]:=\{1,\dots,p\}$ for indexing purposes. The real numbers are denoted by $\mathbb{R}$ and the complex numbers by $\mathbb{C}$, respectively. 
For each $z\in\mathbb{C}$ we decompose $z = \operatorname{Re}(z) + i\operatorname{Im}(z)$ into real and imaginary parts, where $i$ denotes the imaginary unit. Also, $z^*:=\operatorname{Re}(z)-i\operatorname{Im}(z)$ denotes the complex conjugate of $z\in\mathbb{C}$. By linear algebra, it is well known that the $p$-fold Cartesian product $\mathbb{R}^p=\mathbb{R}\times \cdots\times\mathbb{R}$ (analogously for complex generalizations $\mathbb{C}^p$) defines a $p$-dimensional vector space with Euclidean norm $\|v\|^2:=\sum_{k=1}^p|v_k|^2$ for each $v=(v_1,\dots,v_p)\in\mathbb{R}^p$. This motivates the definition of the unit sphere $\mathbb{S}^{p-1}:=\{v\in\mathbb{R}^p:\|v\|=1\}$. An inner product may also be defined on Euclidean space via $\langle v,w\rangle := \sum_{k=1}^p v_kw_k$ for all $v,w\in\mathbb{R}^p$. 
The Cauchy Schwarz inequality states that we generally have $|\langle v,w\rangle|\leq \|v\|\|w\|$ for all $v,w\in\mathbb{R}^p$. Sometimes we write $v^\top w$ instead of $\langle v,w\rangle$, where $v^\top\in\mathbb{R}^{1\times p}$ is the transpose of the column vector $v\in\mathbb{R}^{p\times 1}$ (i.e. $v^\top$ is a row vector). Linear transformations $\mathbb{R}^{n}\to\mathbb{R}^d$ between Euclidean spaces are determined using matrices $A\in\mathbb{R}^{d\times n}$ with entries $A_{i,j}\in\mathbb{R}$ via matrix-vector product mappings $x\mapsto Ax$. Once again, the space of matrices defines a Euclidean space and in the case of square matrices one may define norms on it. For each square matrix $A\in\mathbb{R}^{d\times d}$ we define the operator norm (or spectral norm) $\|A\|:=\sup_{v\in\mathbb{S}^{d-1}}\|Av\|$. For two square matrices $A,B\in\mathbb{R}^{d\times d}$ we generally have the sub-multiplicativity property $\|AB\|\leq\|A\|\|B\|$. Another norm of our interest is the Hilbert-Schmidt norm $\|A\|_{\operatorname{HS}}^2:=\operatorname{Tr}(AA^\top)$ where $\operatorname{Tr}(B):=\sum_{k=1}^dB_{k,k}$ denotes the trace of a square matrix $B\in\mathbb{R}^{d\times d}$. 
The $d\times d$ identity matrix is denoted by $I_d:=(\delta_{i,j})$, where $\delta_{i,j}$ is the Kronecker delta (i.e., $\delta_{i,j}=1$ if $i=j$ and $0$ otherwise). Considering rectangular matrices, we use the notations $s_k(A)$ to specify the singular values of a rectangular matrix $A\in\mathbb{R}^{d\times n}$ with $s_{\min}(A)$ and $s_{\max}(A)$ highlighting the smallest and the largest singular value of $A$, respectively (note that ordering eigenvalues by size is only meaningful for real eigenspectra). Analogously, we use $\lambda_k(A)$ to denote the eigenvalues of a square matrix $A\in\mathbb{R}^{d\times d}$ with $\lambda_{\min}(A)$ and $\lambda_{\max}(A)$ specifying the extreme eigenvalues of $A$. Note that we generally have
\begin{equation}
    \|A\| := \sup_{v \in \mathbb{S}^{d-1}} \|Av\| = \sqrt{\lambda_{\max}(A^\top A)} = s_{\max}(A)
\end{equation}
for all $A\in\mathbb{R}^{d\times d}$. Thus, $\|A\|=\max_{k\in[d]}|\lambda_k(A)|$ if $A$ is symmetric, i.e., $A=A^\top$ (or if $A$ is Hermitian in the complex case). In particular, $\|A\|=\lambda_{\max}(A)=\sup_{v\in\mathbb{S}^{d-1}}v^\top A v$ if $A$ is both symmetric (or Hermitian in the complex case) and positive semi-definite (the latter meaning $\lambda_k(A)\geq0$ for all $k\in[d]$). A subset $U\subseteq\mathbb{R}^p$ is open if for every $x\in U$ there exists a ball $B_r(x):=\{y\in\mathbb{R}^p:\|x-y\|< r\}$ of radius $r>0$ centered at $x$ such that $B_r(x)\subseteq U$. A subset $A\subseteq\mathbb{R}^p$ is closed if the complement $\mathbb{R}^p\setminus A$ is open. A subset $B\subset\mathbb{R}^p$ is bounded if it is contained in a ball $B\subseteq B_r(0)$ for some $r>0$. A subset $K\subseteq\mathbb{R}^p$ is compact if $K$ is closed and bounded. For a $\alpha_k$-fold partially differentiable function $f:U\to\mathbb{R}$ in the $k$-th coordinate on an open domain $U\subseteq \mathbb{R}^p$ we use $\partial_{x_k}^{\alpha_k}f$ to denote the $\alpha_k$-th order partial derivative of $f$ along the $k$-th coordinate. The support of a function $f:U\subseteq\mathbb{R}^p\to\mathbb{R}$ is denoted $\operatorname{supp}(f):=\overline{\{x\in U:f(x)\neq 0\}}$ where $\overline{U}$ denotes the topological closure of $U\subseteq\mathbb{R}^p$ with respect to the standard Euclidean topology.

\smallskip
\par
\textbf{Quantum.} Let $N\in\mathbb{N}$ denote the number of qubits describing a complex $2^N$-dimensional Hilbert space $\mathcal{H}(2^N)$ to represent quantum computation. For a thorough introduction to the topic we recommend Ref.~\cite{nielsen2010quantum}. A normalized element of $\mathcal{H}(2^N)$ is called a (pure) quantum state and represented with the Dirac notations $|\psi\rangle$. The computational basis of $\mathcal{H}(2^N)$ is given by $\{|b\rangle:b\in\{0,1\}^N\}$ where we conventionally abbreviate $|0\rangle\equiv|0\rangle^{\otimes N}\in\mathcal{H}(2^N)$. For any complex operator $A\in\mathbb{C}^{2^N\times 2^N}$ acting on the Hilbert space $\mathcal{H}(2^N)$ we let $A^\dagger:=(A^*)^T$ denote the complex conjugate transpose (or adjoint) of $A$. A general quantum state at the beginning of the computation is represented by an element of the space $\mathcal{D}(2^N)$ of positive semi-definite, self-adjoint, unit-trace operators also known as density matrices. Density matrices generalize pure quantum states since they can also account for mixtures of the form $\rho=\sum_jp_j|\psi_j\rangle\langle\psi_j|$ for an orthonormal basis $\{|\psi_j\rangle\}_j$ of $\mathcal{H}(2^N)$ and probabilities $p_j\in[0,1]$ such that $\sum_jp_j=1$. In particular, $\operatorname{Tr}(\rho)=1$ for all $\rho\in\mathcal{D}(2^N)$. To describe the computation itself we process an input $\rho\in\mathcal{D}(2^N)$ with a unitary evolution with the help of the unitary group $\operatorname{U}(2^N):=\{U\in\mathbb{C}^{2^N\times 2^N}:U^{-1}=U^\dagger\}$. An operator $U\in\operatorname{U}(2^N)$ fully determines the quantum computation since the result of the mapping $\rho\mapsto \rho_U:=U\rho\,U^\dagger$ defines the quantum state after the computation $U$ (hence unitary evolution). However, one has to perform measurement in order to extract information from the output system. For that purpose we consider the real Hilbert-Schmidt space $\operatorname{Herm}(2^N):=\{H\in\mathbb{C}^{2^N\times 2^N}: H=H^\dagger\}$ of Hermitian operators whose elements are also known as observables (or Hamiltonians). Expectation values of the form
\begin{equation}
    \langle O\rangle := \operatorname{Tr}(O\rho_U)
\end{equation}
are used to define the average outcome of measuring the observable $O\in\operatorname{Herm}(2^N)$ when the quantum system $\mathcal{H}(2^N)$ is prepared in the state $\rho\in\mathcal{D}(2^N)$ and is usually defined as the processable output of a quantum computation ready for classical post-processing. Furthermore, it is a well-known fact that every unitary operator $U\in\operatorname{U}(2^N)$ may be represented as an exponential mapping of a Hamiltonian, i.e., $U=e^{iH}$ for some $H\in\operatorname{Herm}(2^N)$ which naturally gives rise to a physical model of computation. This highlights that Hermitian operators can be used in two different contexts, namely as ``observables'' in measurement processes and ``Hamiltonians'' for describing the unitary evolution of quantum states. For a sequence $U_1,\dots,U_L$ of individual layers of unitary operations we define the product unitary $\prod_{\ell=1}^L U_\ell \equiv U_L\cdots U_1$ which reversed ordering by convention. It is worth noting that finding the Hamiltonian $H\in\operatorname{Herm}(2^N)$ such that $\prod_{\ell=1}^Le^{iH_\ell}=e^{iH}$ for some given $H_\ell\in\operatorname{Herm}(2^N)$ is, in general, a very challenging problem with complexity highly depending on the nested commutation structures $[H_\ell,H_k]:=H_\ell H_k - H_k H_\ell$. We refer to the Baker-Campbell-Hausdorff formula and the underlying Lie theory~\cite{gilmore2006lie} for readers who are interested in the details.

\smallskip
\par
\textbf{Asymptotics and Bounds.} We use the following Landau notations to characterize asymptotic behaviour of a function $f:\mathbb{R}\to\mathbb{R}$ in terms of some reference function $g:\mathbb{R}\to\mathbb{R}$, as $x\to x_0\in\mathbb{R}\cup\{\pm\infty\}$:
\begin{align}
    {{f(x)}} &= {{\mathcal{O}(g(x)) \quad:\iff\quad \limsup_{x\to x_0}\left|\frac{f(x)}{g(x)}\right|\in[0,\infty)}} \nonumber
    \\ {{f(x)}} &= {{\Omega(g(x)) \quad:\iff\quad \liminf_{x\to x_0} \left|\frac{f(x)}{g(x)}\right| \in (0,\infty]}} \nonumber
    \\ {{f(x)}} &= \Theta(g(x)) \quad:\iff f(x) = \mathcal{O}(g(x))\quad\text{and}\quad f(x)=\Omega(g(x))\quad\text{as $x\to x_0$} \nonumber
    \\ {f(x)} &= {o(g(x))\quad:\iff\quad\lim_{x\to\ x_0}\frac{f(x)}{g(x)}=0,} \nonumber
    \\ {f(x)} &= {\omega(g(x))\quad:\iff\quad\lim_{x\to\ x_0}\left|\frac{f(x)}{g(x)}\right|=\infty}
    \label{eq:landau-notations}
\end{align}
For a multivariate function $f(x,y)$, the notations $\mathcal{O}_x, \Omega_x, \Theta_x,$ and $o_x$ denote the asymptotic behaviour purely with respect to $x$ as $x \to x_0$, where any expressions involving $y$ are treated as constants (meaning the implicit bounding constants may depend on $y$). Furthermore, we say that $f(x)$ and $g(x)$ are \emph{asymptotically equivalent} as $x\to x_0$ if the much stronger condition,
\begin{equation}
    {{\lim_{x\to x_0}\frac{f(x)}{g(x)} = 1,}} \label{eq:asymptotic-equivalence}
\end{equation}
holds true. In that case, we write $f(x)\sim g(x)$ as $x\to x_0$.

\smallskip
\par
\paragraph{Probability Theory Foundations.} To maintain focus on the geometric and computational aspects, we assume an underlying probability space $(\Omega, \mathcal{F}, \mathbb{P})$ exists, but we operate directly with the random objects and their induced distributions on Euclidean spaces. A real-valued random variable $X$ is a measurable mapping $X: \Omega \to \mathbb{R}$, whose probability distribution is uniquely characterized by its Cumulative Distribution Function (CDF) $F(t) = \mathbb{P}(X \leq t)$ for $t \in \mathbb{R}$. The CDF is right-continuous, non-decreasing, and satisfies $\lim_{t \to -\infty} F(t) = 0$ and $\lim_{t \to \infty} F(t) = 1$. For a thorough introduction to measure-theoretic probability, we recommend Refs.~\cite{wheeden1977measure, billingsley2017probability}. Two random variables $X$ and $Y$ defined on the same probability space are said to be equal \emph{almost surely} if $\mathbb{P}(X=Y) = 1$. A weaker notion of equivalence is equality in distribution, denoted $X \stackrel{d}{=} Y$, meaning their induced probability measures coincide on the Borel $\sigma$-algebra $\mathcal{B}(\mathbb{R})$ (i.e., they share the same CDF). For absolutely continuous random variables, this is equivalent to stating that their probability density functions are equal \emph{almost everywhere} with respect to the Lebesgue measure.

In what follows, let $\mathcal{X}\subseteq\mathbb{R}^p$ be an open subset of $p$-dimensional Euclidean space for some $p\in\mathbb{N}$. Now, we consider random vectors $x$ with probability distributions $\mathbb{P}:\mathcal{B}(\mathcal{X})\to[0,1]$ defined on the Borel $\sigma$-algebra on $\mathcal{X}$. In this work, we frequently consider scalar mappings of random vectors of the form $X\equiv g(x)$ where $g:\mathcal{X}\to\mathbb{R}$ is a measurable function. If $g$ is additionally integrable with respect to the measure $\mathbb{P}$, we may define its expectation value via
\begin{equation}
    \mathbb{E}[h(x)] := \int_{\mathcal{X}}h(x)\,d\mathbb{P}(x). \label{eq:Lebesgue-integral}
\end{equation}
The integral in Eq.~\eqref{eq:Lebesgue-integral} may be understood as a standard Lebesgue integral if $x$ has a continuous distribution by noting that $d\mathbb{P}(x) = f(x)\,dx$ is the differential element of the measure $\mathbb{P}$, and $f$ denotes the Radon-Nikodym derivative of the measure $\mathbb{P}$ with respect to the Lebesgue measure (in other words, $f(x)$ defines the probability density of $x$). In particular, we generally have $\int_{\mathcal{X}}d\mathbb{P}(x) = \mathbb{P}(\mathcal{X}) = 1$. For a non-negative random variable $X\in[0,\infty)$, we recall \emph{Markov's inequality} stating that
\begin{equation}
    \mathbb{P}(X\geq t) \leq \frac{\mathbb{E}[X]}{t} \label{eq:Markov-inequality}
\end{equation}
for all $t>0$. Random matrices $Z=(z_{ij})$ may be bijectively represented as a random vectorization where each component $z_{ij}$ defines a random variable on the same probability space. We sometimes use the notations $X\in\mathbb{R}$, $x\in\mathbb{R}^p$, and $\Phi\in\mathbb{R}^{d\times n}$ to specify the dimension of random variables $X$, random vectors $x$ and random matrices $\Phi$, respectively. That is, using these notations, each component $z_{ij}\in\mathbb{R}$ is to be understood as a measurable function $z_{ij}:\Omega\to\mathbb{R}$ defined on the same underlying Kolmogorovian probability space $(\Omega,\mathcal{F},\mathbb{P})$. Returning to the scalar case $p=1$, the variance of a random variable $X$ (for which the second moment $\mathbb{E}[X^2]<\infty$ exists) may be defined as $\operatorname{Var}(X) := \mathbb{E}[(X-\mathbb{E}[X])^2] = \mathbb{E}[X^2]-\mathbb{E}[X]^2$. It is worth observing that the variance of the sum of two random variables generally satisfies the following bound:

\begin{lemma} \label{lemma:bound-variance-sum-of-RV}
    Let $X,Y\in\mathbb{R}$ be two random variables with finite second moments, i.e., $\mathbb{E}[X^2],\mathbb{E}[Y^2]<\infty$. Then,
    \begin{equation*}
        \operatorname{Var}(X+Y) \leq 2\operatorname{Var}(X)+ 2\operatorname{Var}(Y).
    \end{equation*}
\end{lemma}

The covariance of two random variables $X,Y\in\mathbb{R}$ with finite second moments $\mathbb{E}[X^2],\mathbb{E}[Y^2]<\infty$ (not necessarily defined on the same probability space) is used to quantify correlations between their underlying randomness and is defined as $\operatorname{Cov}(X,Y) := \mathbb{E}[XY]-\mathbb{E}[X]\mathbb{E}[Y]$. We say that two such random variables $X,Y$ are correlated if and only if $\operatorname{Cov}(X,Y) \neq 0$.
In particular, independent random variables are never correlated.
Moreover, we generally have $\operatorname{Cov}(X,X)=\operatorname{Var}(X)$.

\section{Overview}

\begin{figure}
    \captionsetup{justification=raggedright, singlelinecheck=false}
    \centering
    \includegraphics[width=0.99\linewidth]{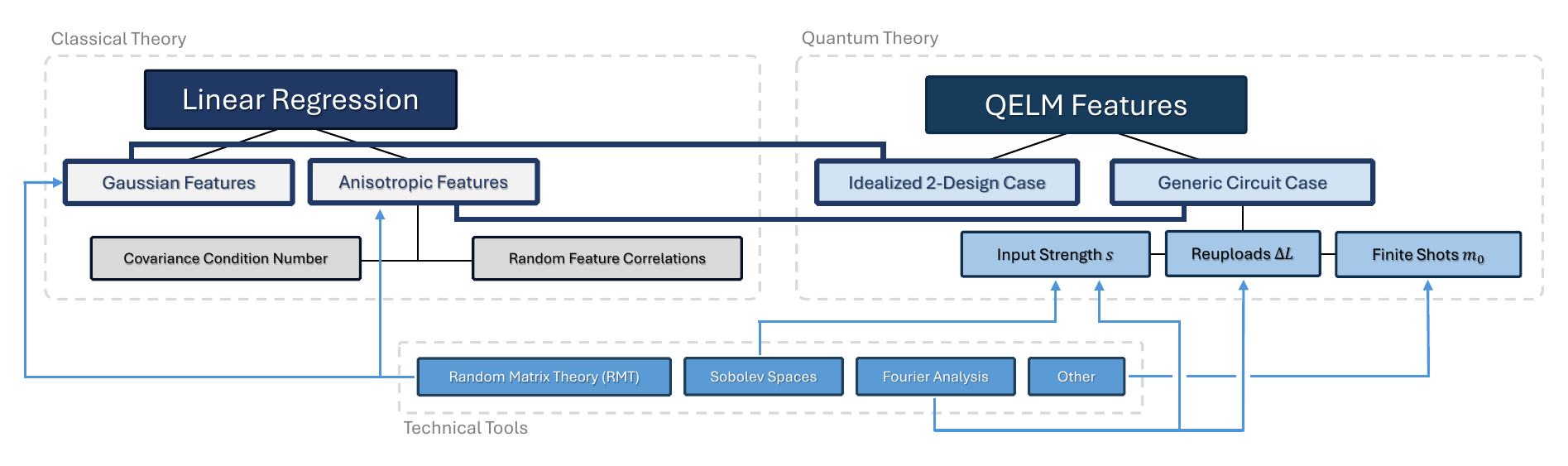}

    \caption{A flowchart representing the structure in the appendix. First, we classically analyse the expected risk of arbitrary linear regression models. Then, we specifically leverage these results to QELM feature maps. The technical details used to prove our results are compiled at the end of the appendix.}
    \label{fig:abstract-flowchart}
\end{figure}

The appendix is roughly organized into three principal parts: Classical theory (cf. Supplementary Section~\ref{appendix:double_descent_linear2}), quantum theory (cf. Supplementary Section~\ref{appendix:QELM-math},~\ref{appendix:2-designs} and~\ref{appendix:non-concentrating-quantum}), and technical tools (cf. Supplementary Section~\ref{appendix:techniques}). A flowchart representation of this organization is shown in Figure~\ref{fig:abstract-flowchart}.

\smallskip
\par
\paragraph{Classical Theory.} Although the study of double descent in linear regression models with Gaussian features are already well-known~\cite{dobriban2018high, mei2022generalization, hastie2022surprises}, we provide complementary details for the sake of completeness.
In Section~\ref{appendix:linear-regression}, we begin by analytically bounding the expected risk of linear regression models with generic feature maps through the lens of classical random matrix theory.
In the special case of Gaussian random features one may derive an exact, analytical description of the risk, effectively proving the existence of double descent, as elaborated in detail in Section~\ref{appendix:gaussian-features}.
Subsequently, we continue with our contributions in this work: in Section~\ref{sec:non-isotropic-analysis}, we extend the risk analysis to arbitrary, non-isotropic features. In the under-parametrized regime, we recover the exact representation of the previous Gaussian case under certain limitations on the random feature correlation in order to ensure the diverging part of the double descent curve (cf. Proposition~\ref{prop:over-fitting}). In the over-parametrized regime, we prove asymptotic upper bounds showing that the curve is decreasing for larger number of parameters and that the generalization performance is determined by the covariance condition number (cf. Theorem~\ref{thm:risk-bound-nonisotropic}).

\smallskip
\par
\paragraph{Quantum Theory.} Supplementary Section~\ref{appendix:QELM-math} begins by introducing the QELM framework. There, we define the quantum version of the features maps and population covariance.
In Supplementary Section~\ref{appendix:2-designs} we prove in Corollary~\ref{corr:double-descent-2design} the existence of double descent in the idealized 2-design case which is equivalent to using isotropic, Gaussian features~\cite{garcia2025quantum} from the previous, classical theory. 
For more generic quantum circuits (e.g., non-concentrating or chaotic correlators~\cite{dao2026breaking, fischer2026dynamical}), we leverage the classical results the we derived in Section~\ref{sec:non-isotropic-analysis} covering anisotropic features and characterize double descent for non-concentrating and correlating quantum features in Supplementary Section~\ref{appendix:non-concentrating-quantum}.
We further analyse how architectural and physical quantities, namely input strength (cf. Theorem~\ref{thm:condition-number-input-strength} in Section~\ref{subsubsec:input-strength}), number of reuploads (cf. Theorem~\ref{thm:condition-number-reuploads} in Section~\ref{subsubsec:input-reuploads}) and finite shot noise (cf. Theorem~\ref{thm:condition-number-improves-with-shots} in Section~\ref{subsec:finite-shot-noise}), each affect the covariance condition number and thus the behaviour of the risk curve in detail.

\smallskip
\par
\paragraph{Technical Tools.} Finally, in Supplementary Section~\ref{appendix:techniques}, we compile all technical details that are not essential for understanding of the key argumentation used in the previous two parts.
The purpose of this part is mainly to ensure mathematically rigorous integrity of our theoretical results. 
We also include a very brief introduction to all necessary concepts used in our proofs with further literature for the interested reader. 
We categorize the technical details into four mathematical ``flavors''. 
First, we employ tools from random matrix theory (cf. Section~\ref{appendix:RMT}) and high-dimensional probability theory~\cite{vershynin2012introduction, vershynin2020high}, including Stieltjes methods for the spectral analysis of large-dimensional random matrices~\cite{bai2010spectral}, to prove our results in the classical theory. 
Specifically, the risk curve of linear regression models is explicitly determined with the help of Stieltjes techniques (cf. Subsection~\ref{appendix:Stieltjes}) of Gaussian features. Other tools from high-dimensional probability theory (cf. Subsection~\ref{subsubsection:high-dim-probability}) are used in the case of anisotropic features.
In the quantum realm, the key argument used in our analysis is the Fourier interpretation of quantum models, thoroughly discussed in Section~\ref{appendix:Fourier-analysis}. It is utilized to analytically describe how the Fourier coefficients transform as the number of reuploads are added, and bound the condition number of the population covariance in terms of input strength and the number of reuploads. However, the analysis with respect to the input strength requires a particularly refined argumentation using Sobolev spaces, as elaborated in Section~\ref{appendix:Sobolev-spaces}. Last but not least, Section~\ref{appendix:others} contains additional computations and technicalities, including arguments needed to derive our results regarding finite shot noise and other useful properties.

\section{Classical Double Descent in Linear Regression} \label{appendix:double_descent_linear2}

This section studies classical double descent of ridgeless linear regression models through the lens of random matrix theory and high-dimensional probability theory. We separate the special case of isotropic, Gaussian random features and more generic, anisotropic features.

\subsection{Linear Regression with Classical Random Features} \label{appendix:linear-regression}

\paragraph{Supervised Learning.} We are interested in solving a supervised learning problem. That is, we are given a training set $\mathcal{S}:=\{(x_\ell,y_\ell)\}_{\ell=1}^n\subset\mathcal{X}\times\mathcal{Y}$ of labelled data sampled from an unknown distribution on $\mathcal{X}\times\mathcal{Y}$, where $\mathcal{X}\subset\mathbb{R}^p$ and $\mathcal{Y}\subset\mathbb{R}$. Let $\varepsilon>0$ be an arbitrary error tolerance. Specifically, the goal is to find an output function $f:\mathcal{X}\to\mathcal{Y}$ such that
\begin{equation}
    \mathbb{P}(|f(x)-y|<\varepsilon)\geq \frac{2}{3}. \label{eq:supervised-learning}
\end{equation}
In other words, we would like to find a function $f$ that captures the input-output relation with sufficiently high accuracy and high probability with respect to all randomness.

\smallskip
\par
\paragraph{Linear Regression Model.} Consider an input data matrix $X:=[x_1,\dots,x_n]\in\mathbb{R}^{p\times n}$ and let $
Y:=(y_1,\dots,y_n)\in\mathbb{R}^{n\times 1}$ be the target vector. Let further $w\in\mathbb{R}^{d\times 1}$ be a weight vector. Assume that $d/n\to\gamma$ as $n\to\infty$ for some constant $\gamma>0$ called the \emph{overparameterization ratio}. The goal of this section is to prove the existence of classical double descent with interpolation threshold at $\gamma=1$ for the linear regression model,
\begin{equation}
    f_w(X) := w^\top\phi(X) \in \mathbb{R}^{1\times n}, \label{eq:linear-regression}
\end{equation}
where $\phi:\mathbb{R}^{p\times n}\to\mathbb{R}^{d\times n}$ is an arbitrary, column-wise feature and we shall use the abbreviation
\begin{equation}
    \Phi := \phi(X)\in\mathbb{R}^{d\times n}, \label{eq:random-features}
\end{equation}
which we call the \emph{random features} of the input. With ``column-wise'' we formally mean $\Phi_j = \phi(x_j)$ for all $j\in[n]$. An \emph{empirical loss} $\mathcal{L}_n(w)$ then simply measures the distance between our deterministic model $f_w(X)$ and the training labels $Y$. By choosing a square loss, this means that
\begin{equation}
    \mathcal{L}_n(w) := \frac{1}{2n}\|f_w(X)^\top-Y\|^2=\frac{1}{2n}\sum_{j=1}^n\left(w^\top\phi(x_j)-y_j\right)^2,
\end{equation}
where $\phi:\mathbb{R}^{p}\to\mathbb{R}^d$ may also be understood as a feature map acting on a single column.

\smallskip
\par
\paragraph{Generalization.} In general, one may assume that the true labels are generated by an unknown  function $f_*:\mathcal{X}\to\mathcal{Y}$ via
\begin{equation}
     y := f_*( x) + \delta, \label{eq:ground-truth}
\end{equation}
where $\delta\sim\mathcal{N}(0,\sigma^2)$ is some Gaussian noise. Let $\mathcal{D}$ be the unknown joint distribution on $\mathcal{X}\times\mathcal{Y}$ giving rise to the function we are trying to learn in Eq.~\eqref{eq:ground-truth}. We define the \emph{expected risk} (or generalization error) of a given weight vector $w$ as
\begin{equation}
    \mathcal{R}(w) := \mathbb{E}_{( x, y)\sim \mathcal{D}}\left[ (w^\top\phi( x) -  y)^2 \right]. \label{eq:expected-risk}
\end{equation}
The generalization gap is the difference between this expected risk $\mathcal{R}(w)$ and the empirical training loss $\mathcal{L}_n(w)$. The model is said to generalize better the faster its generalization gap decreases with the number of training samples $n$.

\smallskip
\par
\paragraph{Classical Double Descent.} To define double descent, we must specify exactly which weight vector $\hat{w}$ the model converges to. In the \emph{under-parametrized regime} ($d < n$), the empirical loss $\mathcal{L}_n(w)$ typically has a unique global minimum. However, in the \emph{over-parametrized regime} ($d > n$), there are infinitely many weight vectors that may perfectly interpolate the training data (i.e., $\mathcal{L}_n(w) = 0$ for $d$ large enough). We define the \emph{min-norm interpolator} $\hat{w}$ as the solution of
\begin{equation}
    \hat{w} := \arg\min_{w \in \arg\min\mathcal{L}_n(w)} \|w\|_2 \quad. \label{eq:min-norm-interpolator}
\end{equation}

With the expected risk $\mathcal{R}(\hat{w})$ and the overparameterization ratio $\gamma = d/n$ defined, we can informally state a simple case of the double descent phenomenon. Let $\mathcal{R}(\gamma) := \lim_{n,d\to\infty} \mathcal{R}(\hat{w})$ denote the asymptotic expected risk. The double descent phenomenon is characterized by the behaviour of $\mathcal{R}(\gamma)$ across three distinct regimes:

\begin{itemize}
    \item[(i)] \emph{Under-parametrized Regime} ($\gamma < 1$): As the ratio $\gamma$ increases, the model capacity grows. The asymptotic expected risk $\mathcal{R}(\gamma)$ follows the classical U-shape, initially decreasing but eventually increasing as the model begins to overfit the noise $z$ in the training labels.

    \item[(ii)] \emph{Interpolation Threshold} ($\gamma\to 1$): The asymptotic expected risk $\mathcal{R}(\gamma)$ exhibits a singularity where it diverges to infinity. At this exact point, there is only one solution that interpolates the data, and it is highly sensitive to the label noise, causing variance to explode.

    \item[(iii)] \emph{Over-parametrized Regime} ($\gamma>1$): As $\gamma$ continues to grow beyond $1$, the model has excess capacity. Because we select the min-norm solution $\hat{w}$, this excess capacity is used to smoothly interpolate between the noisy data points, reducing the variance of the estimator. Consequently, $\mathcal{R}(\gamma)$ monotonically decreases, forming the ``second descent''.
\end{itemize}

Let $\hat{w}\in\mathbb{R}^d$ be the min-norm interpolator defined in Eq.~\eqref{eq:min-norm-interpolator} and assume that there exists a true, unknown weight vector $w_* \in \mathbb{R}^{d\times 1}$ such that the labels $ y$ are generated by $ y = w^\top_* \phi( x) + \delta$ with $\delta\sim\mathcal{N}(0,\sigma^2)$ for some variance $\sigma^2>0$ (as per Eq.~\eqref{eq:ground-truth}). Consequently, the training labels $Y \in \mathbb{R}^{n\times 1}$ are generated via
\begin{equation}
    Y = \Phi^\top w_* + \mathcal{E}, \label{eq:training-labels}
\end{equation}
for some noisy column vector $\mathcal{E}\sim \mathcal{N}(0,\sigma^2 I_n)$, where $\Phi \in \mathbb{R}^{d\times n}$ denotes the random features introduced in Eq.~\eqref{eq:random-features}. Then, at least for linear models, the parameter error $\|\hat{w}-w_*\|$ is directly linked to the prediction error $|\hat{w}^\top\phi(x)- y|$. Indeed, in that case the expected risk $\mathcal{R}(\hat{w}\mid\mathcal{S})$ at the learned weights $\hat{w}$, conditioned on the training set $\mathcal{S}$ reads
\begin{align}
    \mathcal{R}(\hat{w}\mid\mathcal{S}) &= \mathbb{E}_{( x, y)\sim \mathcal{D}}\left[ (\hat{w}^\top\phi(x) -  y)^2 \mid\mathcal{S}\right] = \mathbb{E}_{ x,\delta}\left[ \left(\hat{w}^\top\phi(x) - (w_*^\top\phi(x) + \delta)\right)^2\mid \mathcal{S} \right] \nonumber
    \\ &= \mathbb{E}_{ x,\delta}\left[\left((\hat{w} - w_*)^\top\phi(x)-\delta\right)^2\mid\mathcal{S}\right] \nonumber
    \\ &= \mathbb{E}_{ x}\left[(\hat{w}-w_*)^\top \phi(x)\phi(x)^\top (\hat{w}-w_*)\mid\mathcal{S}\right] + \underbrace{\mathbb{E}_{\delta}[\delta^2]}_{=\sigma^2} - 2\underbrace{\mathbb{E}_{\delta}[\delta]}_{=0}\,\mathbb{E}_{ x}\left[(\hat{w}-w_*)^\top \phi(x)\right] \nonumber
    \\ &= (\hat{w}-w_*)^\top \mathbb{E}_{ x}\left[\phi(x)\phi(x)^\top\right](\hat{w}-w_*) + \sigma^2 = \|\hat{w}-w_*\|_{ \Sigma}^2 +\sigma^2,
\end{align}
recalling that the randomness in $ y$ arises from the noise $\delta$. We also have the $ \Sigma$-norm $\|\hat{w}-w_*\|^2_{ \Sigma}=(\hat{w}-w_*)^\top  \Sigma (\hat{w}-w_*)$ with the deterministic covariance matrix
\begin{equation}
     \Sigma:=\mathbb{E}_{ x}\left[\phi( x)\phi( x)^\top\right]\in\mathbb{R}^{d\times d}. \label{eq:feature-covariance}
\end{equation}
In order to properly describe the risk conditionally on the training set $\mathcal{S}$, it is worth noting the fact that the test input $ x$, drawn from the marginal distribution of $\mathcal{D}$, is independent of the training set $\mathcal{S}$: by definition, a test point $( x,  y)$ is an $(n+1)$-th sample drawn from that exact same joint distribution $\mathcal{D}$. Because it is a newly drawn sample under the i.i.d. framework, the random vector $( x,  y)$ is independent of the random variables that make up the training set $\mathcal{S}$.

By the law of total expectation, we thus find by averaging over the training set,
\begin{equation}
    \mathcal{R}(\hat{w}) = \mathbb{E}_{\mathcal{S}}\left[\mathcal{R}(\hat{w}\mid\mathcal{S})\right] = \mathbb{E}_{\mathcal{S}}\left[\|\hat{w}-w_*\|_{ \Sigma}^2\right] + \sigma^2. \label{eq:risk-formula}
\end{equation}
In particular, if we additionally assume that $ \Sigma=\sigma_X^2I_d$ for some data driven variance $\sigma_X^2>0$, then the expected risk reduces to the Euclidean parameter error:
\begin{equation}
    \mathcal{R}(\hat{w}) = \sigma_X^2\mathbb{E}_{\mathcal{S}}\left[\|\hat{w}-w_*\|^2\right] + \sigma^2. \label{eq:expected-risk-isotropic-case}
\end{equation}

\subsection{Proving Double Descent for Gaussian Random Features} \label{appendix:gaussian-features}

We prove double descent for linear regression models with Gaussian features $\Phi_{i,j}\stackrel{\text{i.i.d.}}{\sim}\mathcal{N}(0,\sigma_X^2)$ for some variance $\sigma_X^2>0$. For that purpose, we are going to make the following assumptions on our model:

\begin{assumption} \label{assumption:gaussian-features} Suppose that:
    \begin{itemize}
        \item[(i)] The matrix elements $\Phi_{ij}$ of the random features $\Phi$ in Eq.~\eqref{eq:random-features} are i.i.d. with $\mathbb{E}_{X}[\Phi_{ij}]=0$ and $\mathbb{E}_X[\Phi_{ij}^2]=\sigma_X^2<\infty$ for all $(i,j)\in[d]\times[n]$. We also assume isotropy $ \Sigma=\sigma_X^2 I_d$ of the random features covariance in Eq.~\eqref{eq:feature-covariance}.
        
        \item[(ii)] The random matrices $Q\Phi\stackrel{\text{d}}{=}\Phi$ are equal in distribution (cf. Supplementary Section~\ref{appendix:notations}) for all orthogonal matrices $Q\in\operatorname{O}(d)$. Furthermore, we suppose that $\operatorname{rank}(\Phi\Phi^\top)=d$ if $\gamma<1$, and $\operatorname{rank}(\Phi^\top\Phi)=n$ if $\gamma>1$, respectively.
        
        \item[(iii)] There exists a Stieltjes transform $m(z)$ such that $m_{\frac{1}{n}\Phi\Phi^\top}(z)\to m(z)$ almost surely as $n\to\infty$ for all $z\in\mathbb{C}_+$, where $m_{\frac{1}{n}\Phi\Phi^\top}(z)$ denotes the Stieltjes transform of the empirical spectral distribution of the random matrix $\frac{1}{n}\Phi\Phi^\top$ (cf. Definitions~\ref{def:stieltjes-transform} and~\ref{def:empirical-spectral-measure})
    \end{itemize}
\end{assumption}

By invoking Maxwell's theorem~\cite{herschel2025essays}, items (i) and (ii) collectively necessitate that $\Phi_{ij}\stackrel{\text{i.i.d.}}{\sim}\mathcal{N}(0,\sigma_X^2)$, which effectively treats our random features as white noise. While such features often lack practical utility in modern machine learning applications, we adopt this assumption for didactic clarity before transitioning to a more technical generalization that relaxes item (i). This specific constraint allows for a direct application of the standard Marchenko-Pastur law, resulting in a streamlined self-consistent equation for the Stieltjes transform. Conversely, item (ii) remains an indispensable requirement for which we currently lack a relaxation method, whereas item (iii) constitutes a standard mathematical regularity condition that typically holds in practice and is provable in the context of the Marchenko-Pastur law (cf. Theorem~\ref{thm:marchenko-pastur}). We now proceed to state the following, well-known theorem, which characterizes the double descent phenomenon for Gaussian features specifically.

\begin{theorem}[Double Descent with Gaussian Features] \label{thm:double-descent-gaussian-features}
Assume that there exists a true, unknown weight vector $w_* \in \mathbb{R}^{d\times 1}$ such that the labels $ y$ are generated by $ y = w^\top_* \phi(x) + \delta$ with $\delta\sim\mathcal{N}(0,\sigma^2)$ for some variance $\sigma^2>0$. Suppose further that $\frac{d}{n}\to\gamma\in(0,\infty)$ as $n,d\to\infty$. Under the conditions on the data in Assumption~\ref{assumption:gaussian-features}, we have that the asymptotic expected risk $\mathcal{R}(\gamma)$ of the linear regression model in Eq.~\eqref{eq:linear-regression} satisfies
\begin{equation*}
    \mathcal{R}(\gamma) = \begin{cases}
        \frac{\sigma^2}{1-\gamma},&\text{if $\gamma<1$,}
        \\ \sigma_X^2\left(1-\frac{1}{\gamma}\right)\,\lim_{d\to\infty}\|w_*\|^2 + \frac{\gamma\sigma^2}{\gamma-1}, &\text{if $\gamma>1$.}
    \end{cases}
\end{equation*}
    
\end{theorem}

\noindent
\emph{Proof.} By Assumption~\ref{assumption:gaussian-features}(i) we are given that $ \Sigma=\sigma_X^2I_d$ so that Eq.~\eqref{eq:expected-risk-isotropic-case} yields
\begin{equation}
    \mathcal{R}(\gamma)=\lim_{d\to\infty}\mathcal{R}(\hat{w}) = \sigma_X^2\lim_{d\to\infty}\mathbb{E}_{\mathcal{S}}\left[\|\hat{w}-w_*\|^2\right] + \sigma^2.
\end{equation}
The proof of double descent is split into two cases: the under-parametrized regime ($\gamma<1$) and the over-parametrized regime ($\gamma>1$), respectively. We shall show both separately and begin with the simpler, under-parametrized case.

\subsubsection{Under-parametrized Regime}

By Assumption~\ref{assumption:gaussian-features}(ii), we are guaranteed that $\operatorname{rank}(\Phi\Phi^\top)=d$ so that the random matrix $\Phi\Phi^\top$ is invertible. By Lemma~\ref{lemma:least-square-solution}, we know that $\hat{w}=(\Phi\Phi^\top)^{-1}\Phi Y$ defines the least square solution that minimizes the norm $\|Y-\Phi^\top w\|$ across all weights $w\in\mathbb{R}^d$. Therefore, using that $Y=\Phi^\top w_*+\mathcal{E}$ as per Eq.~\eqref{eq:training-labels},

\begin{align}
    \|\hat{w}-w_*\|^2 &= \|(\Phi\Phi^\top)^{-1}\Phi Y - w_*\|^2 = \|(\Phi\Phi^\top)^{-1}\Phi (\Phi^\top w_*+\mathcal{E}) - w_*\|^2 \nonumber
    \\ &= \|w_*+(\Phi\Phi^\top)^{-1}\Phi \mathcal{E} - w_*\|^2 = \|(\Phi\Phi^\top)^{-1}\Phi \mathcal{E}\|^2 = \mathcal{E}^\top\Phi^\top(\Phi\Phi^\top)^{-2}\Phi \mathcal{E}, \label{eq:parameter-distance-eq1}
\end{align}
noting that $(\Phi\Phi^\top)^{-1}$ is symmetric. Taking the expectation over the training set $\mathcal{S}$ (whose randomness is determined by the independent quantities $X$ and $\mathcal{E}$) results in
\begin{align}
    \mathbb{E}_{\mathcal{S}}\left[\|\hat{w}-w_*\|^2\right] &= \mathbb{E}_{X,\mathcal{E}}\left[\mathcal{E}^\top\Phi^\top(\Phi\Phi^\top)^{-2}\Phi \mathcal{E}\right] = \sigma^2\,\mathbb{E}_X\left[\operatorname{Tr}\left(\Phi^\top(\Phi\Phi^\top)^{-2}\Phi\right)\right] \nonumber
    \\ &= \sigma^2\,\mathbb{E}_X\left[\operatorname{Tr}\left((\Phi\Phi^\top)^{-1}\right)\right] = \frac{\sigma^2}{n}\mathbb{E}_X\left[\operatorname{Tr}\left(\left(\frac{1}{n}\Phi\Phi^\top\right)^{-1}\right)\right] \nonumber
    \\ &= \frac{\sigma^2}{n}\mathbb{E}_X\left[\operatorname{Tr}Q_{\frac{1}{n}\Phi\Phi^\top}(0)\right] = \sigma^2\,\frac{d}{n}\mathbb{E}_X\left[\frac{1}{d}\operatorname{Tr}Q_{\frac{1}{n}\Phi\Phi^\top}(0)\right] = \sigma^2\,\frac{d}{n} \,\mathbb{E}_X\left[S_n(0)\right], \label{eq:risk1}
\end{align}
where in the second equality we applied Lemma~\ref{lemma:expected-gaussian-quadratic-form} with $\mathcal{E}\sim\mathcal{N}(0,\sigma^2 I_n)$ being independent of $X$ and thus $\Phi$, and the cyclic property of the trace in the third step. We also employed the definition of the resolvent $Q_{\frac{1}{n}\Phi\Phi^\top}$ of $\frac{1}{n}\Phi\Phi^\top$ and the Stieltjes transform $S_n(z) := m_{\frac{1}{n}\Phi\Phi^\top}(z)$ of the empirical spectral measure of $\frac{1}{n}\Phi\Phi^\top$ (cf. Definition~\ref{def:resolvent} and Lemma~\ref{lemma:Stieltjes-transform}). Note that $S_n(0)$ is well-defined since $\Phi\Phi^\top$ is invertible. Because of Assumption~\ref{assumption:gaussian-features}(iii) the empirical Stieltjes transform $S_n(z)\to S(z)$ converges almost surely as $n\to\infty$ to a deterministic Stieltjes transform $S$. Since Assumption~\ref{assumption:gaussian-features}(i) and~(ii) together imply that $\Phi_{ij}\stackrel{\text{i.i.d.}}{\sim}\mathcal{N}(0,\sigma_X^2)$. The Marchenko-Pastur law (cf. Theorem~\ref{thm:marchenko-pastur}) thus applies which states that the limiting Stieltjes transform $S$ obeys the quadratic equation
\begin{equation}
    \gamma z\sigma_X^2 S(z)^2 + (z+\sigma_X^2(\gamma-1))S(z) + 1 = 0.
\end{equation}
for all $z\in\mathbb{C}_+$. In particular,
\begin{equation}
    S(0) = \lim_{\substack{z\to0 \\ \operatorname{Im}(z)>0}} S(z) = \frac{1}{\sigma_X^2(1-\gamma)}
\end{equation}
is deterministic. Taking the limit $\frac{d}{n}\to\gamma$, we thus conclude with
\begin{align}
    \mathcal{R}(\gamma) &= \lim_{d\to\infty}\mathcal{R}(\hat{w}) = \sigma_X^2\lim_{d\to\infty}\mathbb{E}_{\mathcal{S}}\left[\|\hat{w}-w_*\|^2\right] + \sigma^2 \stackrel{\eqref{eq:risk1}}{=} \sigma^2\sigma_X^2\,\lim_{d\to\infty}\frac{d}{n}\,\mathbb{E}_{X}\left[S_n(0)\right] + \sigma^2 \nonumber
    \\&= \sigma^2\sigma_X^2\gamma\,S(0)+\sigma^2 = \sigma^2\gamma\left(\frac{\sigma_X^2}{\sigma_X^2(1-\gamma)}\right)+\sigma^2 = \sigma^2\left(1+\frac{\gamma}{1-\gamma}\right) = \frac{\sigma^2}{1-\gamma}.
\end{align}
This covers the under-parametrized case, $\gamma<1$.

\subsubsection{Over-parametrized Regime}

Again, Assumption~\ref{assumption:gaussian-features}(ii) ensures that $\operatorname{rank}(\Phi^\top\Phi)=n$ so that the random matrix $\Phi^\top\Phi$ is invertible.  By Lemma~\ref{lemma:existence-interpolator} we further know that $\hat{w}:=\Phi(\Phi^\top\Phi)^{-1}Y$ defines the min-norm interpolator as per Eq.~\eqref{eq:min-norm-interpolator}. By Proposition~\ref{prop:bias-variance-decomposition}, we have the bias-variance decomposition,
\begin{equation}
    \|\hat{w}-w_*\|^2 = \underbrace{\|(I-\Phi(\Phi^\top\Phi)^{-1}\Phi^\top)w_*\|^2}_{=:\operatorname{Bias}_\Phi(w_*)^2} + \underbrace{\|\Phi(\Phi^\top\Phi)^{-1}\mathcal{E}\|^2}_{=:\operatorname{Var}_\Phi(\mathcal{E})}. \label{eq:bias-variance1}
\end{equation}

\smallskip
\par
\paragraph{Analysis of the Variance.} Averaging over the Gaussian noise,
\begin{align}
    \mathbb{E}_\mathcal{E}\left[\operatorname{Var}_\Phi(\mathcal{E})\right] &= \mathbb{E}_\mathcal{E}\left[\mathcal{E}^\top(\Phi^\top\Phi)^{-1}\Phi^\top\Phi(\Phi^\top\Phi)^{-1}\mathcal{E}\right] = \mathbb{E}_\mathcal{E}\left[\mathcal{E}^\top(\Phi^\top\Phi)^{-1}\mathcal{E}\right] = \sigma^2\operatorname{Tr}\left((\Phi^\top\Phi)^{-1}\right) \nonumber
    \\ &= \frac{\sigma^2}{d}\operatorname{Tr}\left(\left(\frac{1}{d}\Phi^\top\Phi\right)^{-1}\right) = \frac{\sigma^2}{d}\operatorname{Tr}Q_{\frac{1}{d}\Phi^\top\Phi}(0) = \sigma^2\frac{n}{d}\,\frac{1}{n}\operatorname{Tr}Q_{\frac{1}{d}\Phi^\top\Phi}(0) = \sigma^2\frac{n}{d}\,S_d(0),
\end{align}
where we employed Lemma~\ref{lemma:expected-gaussian-quadratic-form} in the third equality with $\mathcal{E}\sim\mathcal{N}(0,\sigma^2I_n)$ being independent of $\Phi$. Similarly, we utilized the resolvent $Q_{\frac{1}{d}\Phi^\top\Phi}$ of $\frac{1}{d}\Phi^\top\Phi$ and the Stieltjes transform $S_d(z) := m_{\frac{1}{d}\Phi^\top\Phi}(z)$ of the empirical spectral measure of $\frac{1}{d}\Phi^\top\Phi$ (cf. Definition~\ref{def:resolvent} and Lemma~\ref{lemma:Stieltjes-transform}). Again, using the fact that $\Phi_{ij}\stackrel{\text{i.i.d.}}{\sim}\mathcal{N}(0,\sigma_X^2)$, the Marchenko-Pastur law  (cf. Theorem~\ref{thm:marchenko-pastur}) applies (but this time with the roles of $n$ and $d$ being reversed), almost surely giving the limiting Stieltjes transform $S_d(z)\to S(z)$ such that
\begin{equation}
    \frac{1}{\gamma} z\sigma_X^2 S(z)^2 + \left(z+\sigma_X^2\left(\frac{1}{\gamma}-1\right)\right)S(z) + 1 = 0.
\end{equation}
for all $z\in\mathbb{C}_+$. Hence,
\begin{equation}
    S(0) = \lim_{\substack{z\to0 \\ \operatorname{Im}(z)>0}} S(z) = \frac{1}{\sigma_X^2\left(1-\frac{1}{\gamma}\right)}.
\end{equation}
Taking the limit,
\begin{equation}
    \lim_{d\to\infty}\mathbb{E}_z\left[\operatorname{Var}_\Phi(z)\right] = \sigma^2\lim_{d\to\infty}\frac{n}{d}S_d(0) = \sigma^2\frac{1}{\gamma}S(0) = \sigma^2\frac{1}{\gamma\sigma_X^2\left(1-\frac{1}{\gamma}\right)} = \frac{\sigma^2}{\sigma_X^2(\gamma-1)}. \label{eq:variance1}
\end{equation}

\smallskip
\par
\paragraph{Analysis of the Bias.} Let us begin by abbreviating the column vector $b_\Phi(w_*) := (I-P_\Phi)w_*$, where $P_\Phi:= \Phi(\Phi^\top\Phi)^{-1}\Phi^\top$. In particular, $\operatorname{Bias}_\Phi(w_*)^2=\|b_\Phi(w_*)\|^2$. First, we make the observation that $P_\Phi$ defines a symmetric projector, that is $P_\Phi^2=P_\Phi$. Indeed, $P_\Phi^\top=P_\Phi$ is straightforward, and
\begin{align}
    P_\Phi^2 = \Phi(\Phi^\top\Phi)^{-1}\underbrace{\Phi^\top \Phi(\Phi^\top\Phi)^{-1}}_{=I_n}\Phi^\top = \Phi(\Phi^\top\Phi)^{-1}\Phi^\top = P_\Phi. \label{eq:projector}
\end{align}
As a result,
\begin{align}
    \operatorname{Bias}_\Phi(w_*)^2 &= b_\Phi(w_*)^\top b_\Phi(w_*) = w_*^\top (I-P_\Phi)^\top(I-P_\Phi)w_* \nonumber
    \\ &= \|w_*\|^2 + w_*^\top P_\Phi^2 w_* -2w_*^\top P_\Phi w_* = \|w_*\|^2 - w_*^\top P_\Phi w_*. \label{eq:bias-projection1}
\end{align}

Next, we show that
\begin{equation}
    u^\top P_\Phi u \stackrel{\text{d}}{=} v^\top P_\Phi v
\end{equation}
are equal in distribution for all unit vectors $u,v\in\mathbb{S}^{d-1}$. This statement relies on the rotational invariance in Assumption~\ref{assumption:gaussian-features}(ii), i.e., $Q\Phi\stackrel{\text{d}}{=}\Phi$ are equal in distribution for all orthogonal matrices $Q\in\operatorname{O}(d)$. Indeed, for any pair $u,v\in\mathbb{S}^{d-1}$ of unit vectors there exists an orthogonal matrix $Q\in\operatorname{O}(d)$ such that $Q^\top v=u$. As a result, using $Q^\top Q=I_d$ due to orthogonality,
\begin{align}
    u^\top P_\Phi u &= v^\top Q P_\Phi Q^\top v = v^\top Q\,\Phi(\Phi^\top\Phi)^{-1}\Phi^\top Q^\top v = v^\top Q\Phi(\Phi^\top \underbrace{Q^\top Q}_{=I_d}\Phi)^{-1}\Phi Q^\top v \nonumber
    \\ &= v^\top(Q\Phi)\left((Q\Phi)^\top(Q\Phi)\right)^{-1}(Q\Phi)^\top = v^\top P_{Q\Phi} v \stackrel{\text{d}}{=} v^\top P_\Phi v,
\end{align}
by the assumption that $Q\Phi\stackrel{\text{d}}{=}\Phi$ (which particularly implies $P_{Q\Phi}\stackrel{\text{d}}{=} P_\Phi$). Specifically, we observe that the distribution of the bias
\begin{equation}
    \operatorname{Bias}_\Phi(w_*)^2 \stackrel{\eqref{eq:bias-projection1}}{=} \|w_*\|^2 - w_*^\top P_\Phi w_* = \|w_*\|^2\left(1-\frac{w_*^\top}{\|w_*\|}P_\Phi\frac{w_*}{\|w_*\|}\right) \stackrel{\text{d}}{=} \|w_*\|^2(1 - v^\top P_\Phi v) \label{eq:bias1}
\end{equation}
for any random unit vector $v\in\mathbb{S}^{d-1}$ (in particular independent of $X$).

Additionally, $P_{Q\Phi}\stackrel{\text{d}}{=} P_\Phi$ also results in
\begin{equation}
    \mathbb{E}_X[P_\Phi] = \mathbb{E}_X[P_{Q\Phi}] = \mathbb{E}_X[Q\Phi(\Phi^\top Q^\top Q\Phi)^{-1}\Phi^\top Q^\top] = \mathbb{E}_X[QP_\Phi Q^\top] = Q\,\mathbb{E}_X[P_\Phi]Q^\top,
\end{equation}
where the first identity arises from $P_\Phi$ and $P_{Q\Phi}$ sharing the same distribution with respect to the randomness in $X$. This establishes $[\mathbb{E}_X[P_\Phi],Q]=0$ for all orthogonal matrices $Q\in\operatorname{O}(d)$. This allows us to apply Lemma~\ref{lemma:orthogonal-group-commutant}, telling us that $\mathbb{E}_X[P_\Phi]=\lambda I_d$ is a multiple of the identity for some scalar $\lambda\in\mathbb{R}$. Taking the expectation, since $v$ is any random unit vector independent of $X$ by the previous observation, this establishes
\begin{align}
    \mathbb{E}_X[\operatorname{Tr}(v^\top P_\Phi v)] = \mathbb{E}_X[\operatorname{Tr}(P_\Phi vv^\top)] = \operatorname{Tr}(\underbrace{\mathbb{E}_X[P_\phi]}_{=\lambda I_d}vv^\top) = \lambda\operatorname{Tr}(vv^\top) = \lambda \underbrace{v^\top v}_{=1} = \lambda
\end{align}
by the cyclic property of the trace and recalling $v^\top v = \|v\|^2=1$. Note that $\mathbb{E}_X[P_\Phi]=\lambda I_d$ additionally yields $\operatorname{Tr}(\mathbb{E}_X[P_\Phi])=\lambda\operatorname{Tr}(I_d) = \lambda\,d$ so that, in summary,
\begin{align}
    \lambda &= \frac{1}{d}\operatorname{Tr}(\mathbb{E}_X[P_\Phi]) = \frac{1}{d}\,\mathbb{E}_X[\operatorname{Tr}(P_\Phi)] = \frac{1}{d}\mathbb{E}_X\,[\operatorname{Tr}(\Phi(\Phi^\top\Phi)^{-1}\Phi^\top)] \nonumber
    \\ &= \frac{1}{d}\mathbb{E}_X[\operatorname{Tr}((\Phi^\top\Phi)^{-1}\Phi^\top\Phi)] = \frac{1}{d}\,\mathbb{E}_X[\underbrace{\operatorname{Tr}(I_n)}_{=n}] = \frac{n}{d},
\end{align}
again by the cyclic property of the trace, hence proving $\mathbb{E}_X[\operatorname{Tr}(v^\top P_\Phi v)]=\frac{n}{d}$. Finally, taking the expectation and the limit yields
\begin{equation}
    \lim_{d\to\infty}\mathbb{E}_X[\operatorname{Bias}_\Phi(w_*)^2] \stackrel{\eqref{eq:bias1}}{=} \lim_{d\to\infty}\|w_*\|^2(1-v^\top\mathbb{E}_X[P_\Phi]v) = \left(1-\frac{1}{\gamma}\right)\,\lim_{d\to\infty}\|w_*\|^2, \label{eq:bias1'}
\end{equation}
recalling $\frac{d}{n}\to\gamma$ as $d\to\infty$.

Putting everything together, and recalling that the randomness in the training set $\mathcal{S}$ arises from the independent random quantities $X$ and $z$,
\begin{align}
    \mathcal{R}(\gamma) &= \lim_{d\to\infty}\mathcal{R}(\hat{w}) = \sigma_X^2\lim_{d\to\infty}\mathbb{E}_{\mathcal{S}}\left[\|\hat{w}-w_*\|^2\right] + \sigma^2 \stackrel{\eqref{eq:bias-variance1}}{=} \sigma_X^2\lim_{d\to\infty}\mathbb{E}_X[\operatorname{Bias}_\Phi(w_*)^2] + \sigma_X^2\lim_{d\to\infty}\mathbb{E}_z\left[\operatorname{Var}_\Phi(z)\right] + \sigma^2 \nonumber
    \\ &= \sigma_X^2\left(1-\frac{1}{\gamma}\right)\,\lim_{d\to\infty}\|w_*\|^2 + \frac{\sigma_X^2\sigma^2}{\sigma_X^2(\gamma-1)} + \sigma^2 = \sigma_X^2\left(1-\frac{1}{\gamma}\right)\,\lim_{d\to\infty}\|w_*\|^2 + \frac{\gamma\sigma^2}{\gamma-1} \label{eq:risk-bound-isotropic}
\end{align}
by Eq.~\eqref{eq:bias1'} and~\eqref{eq:variance1}, respectively. This concludes the proof of the theorem.
\qed

\subsection{Analysis of the Non-Isotropic Case} \label{sec:non-isotropic-analysis}

In the previous section we have shown Theorem~\ref{thm:double-descent-gaussian-features}, which provides an analytic description of the linear regression expected risk if the random features are sampled from an isotropic Gaussian. This section extends the description of the risk to more general, anisotropic random features. While standard double descent theory typically assumes independent structures, isotropic features, or rely on highly non-trivial Stieltjes transform solutions assuming large signal-to-noise ratios~\cite{bartlett2020benign, bartlett2021deep, tsigler2023benign, dobriban2018high, mei2022generalization, bai2010spectral},
a key novelty of our work lies in extending these guarantees by bounding random feature correlations, and by conditioning a non-singular population covariance  $\Sigma:=\mathbb{E}_x[\phi(x)\phi(x)^\top]$.
In the under-parametrized regime ($\gamma < 1$), we show that the classic risk divergence $\mathcal{R}(\gamma) = \sigma^2(1-\gamma)^{-1}$ is preserved at the interpolation threshold ($\gamma=1$) if the feature correlations remain limited.
If these correlations are too excessive, the exact behaviour at the interpolation threshold becomes unpredictable. Regarding the risk decay in the over-parametrized regime ($\gamma > 1$), we establish a generalization bound characterized by the condition number $\kappa(\Sigma)$ of the population covariance. The strategy is to separate the cases of the under- and over-parametrized regimes, respectively. However, instead of using Stieltjes techniques (cf. Subsection~\ref{appendix:Stieltjes}), we rely on tools of high-dimensional probability theory~\cite{vershynin2012introduction, vershynin2020high} (cf. Subsection~\ref{subsubsection:high-dim-probability}).

\subsubsection{Under-parametrized Regime}

The first step is to relax the condition of isotropy $ \Sigma=\sigma_X^2I_d$ of the deterministic feature covariance defined in Eq.~\eqref{eq:feature-covariance}. As a result, one requires a different analysis since the parameter distance $\|\hat{w}-w_*\|_{ \Sigma}$ in Eq.~\eqref{eq:risk-formula} does not coincide with the Euclidean distance any more. However, the assignment $v\mapsto \|v\|_{ \Sigma}$ is a well-defined functions which implies that
\begin{equation}
    \hat{w}-w_* = (\Phi\Phi^\top)^{-1}\Phi \mathcal{E} \implies \|\hat{w}-w_*\|_{ \Sigma} = \|(\Phi\Phi^\top)^{-1}\Phi \mathcal{E}\|_{ \Sigma}
\end{equation}
as derived from Eq.~\eqref{eq:parameter-distance-eq1}. As a result, we similarly obtain
\begin{equation}
    \|\hat{w}-w_*\|^2_{ \Sigma} = \mathcal{E}^\top \Phi^\top (\Phi\Phi^\top)^{-1} \Sigma\,(\Phi\Phi^\top)^{-1}\Phi \mathcal{E}
\end{equation}
and we may apply Lemma~\ref{lemma:expected-gaussian-quadratic-form} since $\mathcal{E}\sim\mathcal{N}(0,\sigma^2 I_n)$ is Gaussian:
\begin{align}
    \mathbb{E}_{\mathcal{S}}\left[\|\hat{w}-w_*\|^2_{ \Sigma}\right] &= \mathbb{E}_{X,\mathcal{E}}\left[\mathcal{E}^\top \Phi^\top (\Phi\Phi^\top)^{-1} \Sigma\,(\Phi\Phi^\top)^{-1}\Phi \mathcal{E}\right] = \sigma^2\,\mathbb{E}_X\left[\operatorname{Tr}\left(\Phi^\top (\Phi\Phi^\top)^{-1} \Sigma\,(\Phi\Phi^\top)^{-1}\Phi\right)\right] \nonumber 
    \\ &= \sigma^2\,\mathbb{E}_X\left[\operatorname{Tr}\left( (\Phi\Phi^\top)^{-1} \Sigma\,(\Phi\Phi^\top)^{-1}\Phi\Phi^\top\right)\right] = \sigma^2\,\mathbb{E}_X\left[\operatorname{Tr}\left( (\Phi\Phi^\top)^{-1} \Sigma\right)\right], \label{eq:risk-underparametrized1}
\end{align}
where we applied the cyclic property of the trace. 

\smallskip
\par
\paragraph{Standard Model of Random Matrix Theory (RMT).} Assume that
\begin{equation}
    \Phi =  \Sigma^{\frac{1}{2}}Z \label{eq:standard-RMT-model}
\end{equation}
for some random matrix $Z=[z_1|\cdots|z_n]\in\mathbb{R}^{d\times n}$ with independent columns $z_j\in\mathbb{R}^{d\times 1}$. We suppose that the lowest eigenvalue $\lambda_{\min}(\Sigma)>0$ of the population covariance $\Sigma$ is non-zero so that we may represent the columns of $Z$ via $z_j = \Sigma^{-\frac{1}{2}}\Phi_j$ with $\Phi_j$ denoting the column of $\Phi$, respectively. Since $\Phi_j$ is equal in distribution to $\phi(x)$ for each $j\in[n]$, we find
\begin{equation}
    \mathbb{E}[z_jz_j^\top] = \Sigma^{-\frac{1}{2}}\mathbb{E}[\Phi_j\Phi_j^\top] \Sigma^{-\frac{1}{2}} = \Sigma^{-\frac{1}{2}}\underbrace{\mathbb{E}_x[\phi(x)\phi(x)^\top]}_{=\Sigma} \Sigma^{-\frac{1}{2}} = I_d \label{eq:isotropic-columns}
\end{equation}
for all $j\in[n]$, showing that the columns $z_j$ of $Z$ are isotropic.

\par
Utilizing the cyclic property of the trace, the population covariance matrix perfectly cancels out of the expression:
\begin{align}
    \operatorname{Tr}\left((\Phi\Phi^\top)^{-1} \Sigma\right) &\stackrel{\eqref{eq:standard-RMT-model}}{=} \operatorname{Tr}\left( \left(\Sigma^{\frac{1}{2}} Z Z^\top \Sigma^{\frac{1}{2}}\right)^{-1} \Sigma \right) = \operatorname{Tr}\left( \Sigma^{-1/2} (Z Z^\top)^{-1} \Sigma^{-1/2} \Sigma \right) = \operatorname{Tr}\left( (Z Z^\top)^{-1} \Sigma^{-1/2} \Sigma \Sigma^{-1/2} \right) \nonumber
    \\ &= \operatorname{Tr}\left( (Z Z^\top)^{-1} \right) = \frac{1}{n} \operatorname{Tr}(W_n^{-1}), \label{eq:trace-to-stieltjes1}
\end{align}
where $W_n := \frac{1}{n}ZZ^\top \in \mathbb{R}^{d \times d}$ denotes the empirical sample covariance matrix of the isotropic columns of $Z$. To evaluate $\frac{1}{n} \operatorname{Tr}(W_n^{-1})$ under the relaxed assumption of independent columns, we appeal to the following generalization of the Marchenko-Pastur law (cf. Theorem~\ref{thm:marchenko-pastur}) which is originally established in Ref.~\cite{bai2008large}:

\begin{lemma}[Bryson et al.~\cite{bryson2021marchenko}, Theorem~1.7] \label{lemma:generalized-marchenko-pastur}
    Let $Z=[z_1|\cdots|z_n]\in\mathbb{R}^{d\times n}$ be a random matrix with independent, isotropic columns $z_j\in\mathbb{R}^{d\times 1}$ such that
    \begin{equation*}
        \max_{j\in[n]}\operatorname{Var}(z_j^\top A z_j) = o(d^2)
    \end{equation*}
    for all deterministic $A\in\mathbb{R}^{d\times d}$ with uniformly bounded operator norm $\|A\|=\mathcal{O}(1)$ as $n,d\to\infty$. Then, the Stieltjes transform $m_{W_n}(z)$ of the empirical spectral measure of $W_n$ converges almost surely to the Stieltjes transform $S$ satisfying the Marchenko-Pastur equation
    \begin{equation*}
        \gamma z S(z)^2 + (z + (\gamma-1))S(z) + 1 = 0. \label{eq:quadratic-MP-equation}
    \end{equation*}
\end{lemma}

Recalling the definition of the Stieltjes transform $m_{W_n}(z) = \frac{1}{d} \operatorname{Tr}\left((W_n - z I_d)^{-1}\right)$, Lemma~\ref{lemma:generalized-marchenko-pastur} implies that $m_{W_n}(z) \to S(z)$ almost surely so that,
\begin{equation}
    \frac{1}{n}\operatorname{Tr}(W_n^{-1}) = \frac{d}{n}\lim_{\substack{z\to 0 \\ \operatorname{Im}(z)>0}} m_{W_n}(z) \to \underbrace{\lim_{n,d\to\infty}\frac{d}{n}}_{=\gamma} \, \lim_{\substack{z\to 0 \\ \operatorname{Im}(z)>0}} \, \underbrace{\lim_{n,d\to\infty}m_{W_n}(z)}_{=S(z)} = \gamma S(0) = \frac{\gamma}{1-\gamma} \label{eq:trace-to-stieltjes2}
\end{equation}
as $n,d\to\infty$, where the last equality follows from Eq.~\eqref{eq:quadratic-MP-equation} as $z\to0$ (assuming $\operatorname{Im}(z)>0$). In conclusion,
\begin{align}
     \mathcal{R}(\gamma)=\lim_{n,d\to\infty}\mathcal{R}(\hat{w}) &\stackrel{\eqref{eq:risk-formula}}{=} \sigma^2 + \lim_{n,d\to\infty} \mathbb{E}_{\mathcal{S}}\left[\|\hat{w}-w_*\|_{ \Sigma}^2\right] \stackrel{\eqref{eq:risk-underparametrized1}}{=} \sigma^2 + \lim_{n,d\to\infty}\sigma^2\,\mathbb{E}_X\left[\operatorname{Tr}\left( (\Phi\Phi^\top)^{-1} \Sigma\right)\right] \nonumber
     \\ &\stackrel{\eqref{eq:trace-to-stieltjes1}}{=} \sigma^2 + \sigma^2\lim_{n,d\to\infty}\frac{1}{n} \operatorname{Tr}(W_n^{-1}) \stackrel{\eqref{eq:trace-to-stieltjes2}}{=} \sigma^2 + \sigma^2 \left(\frac{\gamma}{1-\gamma}\right) = \sigma^2 \left(1 + \frac{\gamma}{1-\gamma}\right) = \frac{\sigma^2}{1-\gamma}. \label{eq:unit-interpolation-threshold}
\end{align}
As $\gamma\uparrow 1$, this perfectly demonstrates the double-descent spike from the left side. Observe the fact that this bound remains independent of $\Sigma$ and coincides with the isotropic case in Theorem~\ref{thm:double-descent-gaussian-features}. This establishes the following result in the under-parametrized regime:

\begin{proposition}[Re-established Double Descent Curve] \label{prop:over-fitting}
    Let $\gamma<1$ and assume that
    \begin{equation}
        \max_{j\in[n]}\sup_{\substack{A\in\mathbb{R}^{d\times d}\\\|A\|=\mathcal{O}_d(1)}}\operatorname{Var}_x(\phi_U(x_j)^\top A^{(\Sigma)}\,\phi_U(x_j))=o(d^2) \label{eq:variance-bound1i}
    \end{equation}
    as $d\to\infty$, where $A^{(\Sigma)}:=\Sigma^{-\frac{1}{2}}A\Sigma^{-\frac{1}{2}}$. Then, $\mathcal{R}(\gamma) = \frac{\sigma^2}{1-\gamma}$.
\end{proposition}

In fact, we explicitly prove in Lemma~\eqref{eq:second-order-correlation-measure} of Section~\ref{subsubsection:high-dim-probability} that the variance bound in Eq.~\eqref{eq:variance-bound1i} is equivalent to 
\begin{equation}
    \operatorname{Var}(z^\top A z) = \operatorname{vec}(A)^\top \operatorname{Cov}(z\otimes z)\operatorname{vec}(A) = \mathcal{O}_d\left( d \,\lambda_{\max}(\operatorname{Cov}(z \otimes z))\right)
\end{equation}
as $d\to\infty$, where $z:=\Sigma^{-\frac{1}{2}}\phi_U(x)$. The latter bound follows from Proposition~\ref{prop:spectral-variance-bound} of Section~\ref{subsubsection:high-dim-probability}.
Therefore, the condition in Eq.~\eqref{eq:variance-bound1i} is automatically satisfied if
\begin{equation}
    \lambda_{\max}(\operatorname{Cov}(z \otimes z)) = o(d)
\end{equation}
as $d\to\infty$, which may thus be interpreted as a spectral measure of second moment random feature correlations. If it is bounded by $o(d)$, then the standard double descent curve with an interpolation threshold at $\gamma=1$ is guaranteed.

\subsubsection{Over-parametrized Regime}

We begin by establishing a noise averaged bias-variance decomposition for the non-isotropic case. By Proposition~\ref{prop:general-bias-variance-decomposition}, we have the identity
\begin{equation}
    \mathbb{E}_\mathcal{E}\left[\|\hat{w}-w_*\|^2_{ \Sigma}\right] = \underbrace{\|(I-\Phi(\Phi^\top\Phi)^{-1}\Phi^\top)w_*\|^2_{ \Sigma}}_{=:\operatorname{Bias}_\Phi(w_*, \Sigma)^2} + \, \underbrace{\sigma^2\operatorname{Tr}\left( \Sigma\,\Phi(\Phi^\top\Phi)^{-2}\Phi^\top\right)}_{=:\operatorname{Var}_\Phi( \Sigma)}.
\end{equation}

The next step is to upper bound both bias and variance separately by an expression that converges to a small risk as $\gamma\to\infty$.

\smallskip
\par
\paragraph{Bounding the Bias.} Abbreviate $P_\Phi:=\Phi(\Phi^\top\Phi)^{-1}\Phi^\top$, which defines a projector as observed in Eq.~\eqref{eq:projector}. By sub-multiplicativity of the norm we directly estimate
\begin{align}
    \operatorname{Bias}_\Phi(w_*, \Sigma)^2 = \|(I-P_\Phi)w_*\|^2_{ \Sigma} \leq \lambda_{\max}( \Sigma)\|(I-P_\Phi)w_*\|^2 \leq \lambda_{\max}( \Sigma)\underbrace{\|I-P_\Phi\|^2}_{=1}\|w_*\|^2 = \lambda_{\max}( \Sigma)\|w_*\|^2,
\end{align}
using that the population covariance $ \Sigma$ is positive semi-definite, and the fact that $\|I-P_\Phi\|=1$ due to
\begin{equation}
    I-P_\Phi = I-2P_\Phi + \underbrace{P_\Phi}_{P_\Phi^2} = (I-P_\Phi)^2
\end{equation}
defining a non-trivial projector which has eigenvalues $0$ and $1$. Therefore,
\begin{equation}
    \operatorname{Bias}_\Phi(w_*, \Sigma)^2 = \mathcal{O}\left(\|w_*\|^2 \lambda_{\max}( \Sigma)\right) = \mathcal{O}\left(\|w_*\|^2 \|\Sigma\|\right)
\end{equation}
as $d\to\infty$, which also shows why it is desirable to minimize the norm $\|w_*\|$. Furthermore, note that the bias term is independent of $y$.

\smallskip
\par
\paragraph{Bounding the Variance.} Let us now upper bound the variance term. We again assume that the lowest eigenvalue $\lambda_{\min}(\Sigma)>0$ of the population covariance $\Sigma$ is non-zero. We first observe with the help of Proposition~\ref{prop:general-bias-variance-decomposition} that 
\begin{align}
    \frac{1}{\sigma^2}\operatorname{Var}_\Phi( \Sigma) = \operatorname{Tr}\left( \Sigma\,\Phi(\Phi^\top\Phi)^{-2}\Phi^\top\right) \leq \| \Sigma\|\operatorname{Tr}\left(\Phi(\Phi^\top\Phi)^{-2}\Phi^\top\right) = \| \Sigma\|\operatorname{Tr}\left((\Phi^\top\Phi)^{-2}\Phi^\top\Phi\right) = \| \Sigma\|\operatorname{Tr}\left((\Phi^\top\Phi)^{-1}\right) \label{eq:var-bound1}
\end{align}
by the cyclic property of the trace. It thus remains to upper bound the factor $\operatorname{Tr}\left((\Phi^\top\Phi)^{-1}\right)$. Since the columns of $\Phi$ now generally have dependent coordinates, we cannot apply the previous Marchenko-Pastur technique. Instead, we are going to employ high dimensional probability theory~\cite{vershynin2012introduction, vershynin2020high} to derive an expression in terms of the population covariance $\Sigma$, despite correlations among the feature coordinates.
We begin by noting $\Sigma\succeq \lambda_{\min}(\Sigma)I_d$, where ``$\succeq$'' denotes the Loewner order which defines the partial order $A\succeq B$ for two matrices $A,B\in\mathbb{R}^d$ if and only if $A-B$ is positive semi-definite (using ``$\succ$'' for strict positive definiteness, respectively). Similar to the standard model of RMT in Eq.~\eqref{eq:standard-RMT-model}, we assume $\Phi =  \Sigma^{\frac{1}{2}}Z$ for some random matrix $Z=[z_1|\cdots|z_n]\in\mathbb{R}^{d\times n}$ with independent columns $z_j\in\mathbb{R}^{d\times 1}$. Then, we particularly have $\Phi^\top \Phi = (\Sigma^{\frac{1}{2}}Z)^\top (\Sigma^{\frac{1}{2}}Z) = Z^\top \Sigma Z$ and, by using the representation $z_j=\Sigma^{-\frac{1}{2}} \Phi_j$ for each column $\Phi_j$ of $\Phi$, we find that $z_j$ is isotropic for all $j\in[n]$ as previously shown in Eq.~\eqref{eq:isotropic-columns}.

Moreover, as a result of $A\succeq B\succ 0\implies B^{-1} \preceq A^{-1}$, we find
\begin{equation}
    \Phi^\top\Phi = Z^\top\Sigma Z \succeq \lambda_{\min}(\Sigma)Z^\top Z \implies (\Phi^\top\Phi)^{-1} = (Z^\top \Sigma Z)^{-1} \preceq \frac{1}{\lambda_{\min}(\Sigma)}(Z^\top Z)^{-1}, \label{eq:tracebound1}
\end{equation}
where the implication is due to the assumption $\operatorname{rank}(\Phi^\top \Phi)=n$, in particular $\lambda_{\min}(\Sigma)>0$, and under the extra assumption that the random matrix $Z^\top Z\in\mathbb{R}^{n\times n}$ is almost surely invertible. Since we have $A \preceq B \implies \operatorname{Tr}A \leq \operatorname{Tr}B$ we particularly have
\begin{equation}
    \operatorname{Tr}\left((\Phi^\top\Phi)^{-1}\right) \stackrel{\eqref{eq:tracebound1}}{\leq} \frac{1}{\lambda_{\min}(\Sigma)}\operatorname{Tr}\left((Z^\top Z)^{-1}\right) \label{eq:tracebound-1.5}
\end{equation}
meaning that we have reduced the problem to upper bounding
\begin{equation}
    \operatorname{Tr}\left((Z^\top Z)^{-1}\right) \leq n\,\lambda_{\max}\left((Z^\top Z)^{-1}\right) = \frac{n}{\lambda_{\min}(Z^\top Z)}, \label{eq:tracebound-2}
\end{equation}
recalling that the trace of a matrix $A$ equals to the sum of its eigenvalues $\lambda_j(A)$, and that $\lambda_{\max}(A^{-1})=\lambda_{\min}(A)^{-1}$ for any positive semi-definite matrix $A$, respectively. Equivalently, we need to lower bound the eigenvalue $\lambda_{\min}(Z^\top Z)$. In order to do so, inspired by the approach of Ref.~\cite{kuzborskij2021role}, we shall make use of the following random matrix theory statement:

\begin{lemma}[Vershynin~\cite{vershynin2012introduction}, Theorem~5.58] \label{lemma:singular-value-lower-bound}
    Assume $d\geq n$. Let $Z=[z_1|\cdots|z_n]\in\mathbb{R}^{d\times n}$ be a random matrix with independent, sub-Gaussian and isotropic columns $z_j\in\mathbb{R}^{d\times 1}$ such that $\|z_j\|=\sqrt{d}$ almost surely for all $j\in[n]$. Then, for all $t\geq0$, the smallest singular value $s_{\min}(Z)$ of $Z$ satisfies the lower bound
    \begin{equation*}
        s_{\min}(Z) \geq \sqrt{d} - C_K\sqrt{n}-t
    \end{equation*}
    with probability at least $1-2e^{-ct^2}$ with absolute constant $C_K\geq 1$ that only depends on the sub-Gaussian norm $K:=\max_{j\in[n]}\|z_j\|_{\psi_2}$ of the columns of $Z$.
\end{lemma}

Note that $\lambda_{\min}(Z^\top Z) = s_{\min}(Z)^2$ since the singular values of $Z$ and $Z^\top$ coincide. As a result, under the additional assumption that $Z$ has sub-Gaussian columns, applying the previous lemma yields the bound
\begin{align}
    \operatorname{Tr}\left((Z^\top Z)^{-1}\right)  \stackrel{\eqref{eq:tracebound-2}}{\leq} \frac{n}{\lambda_{\min}(Z^\top Z)} \leq \frac{n}{\left(\sqrt{d}-C_K\sqrt{n}-t\right)^2} = \frac{1}{\left(\frac{\sqrt{d}}{\sqrt{n}}-C_K-\frac{t}{\sqrt{n}}\right)^2} \stackrel{\gamma\sim\frac{d}{n}}{\sim} \frac{1}{(\sqrt{\gamma} - C_K)^2} \label{eq:tracebound-3}
\end{align}
as $n,d\to\infty$ with probability at least $1-2e^{-ct^2}$ for every $t\geq0$. Note that by taking $t=\Theta(n^{\frac{1}{2}-\varepsilon})$ for some $\varepsilon>0$ we will have $\frac{t}{\sqrt{n}}\to0$ as $n\to\infty$. This means that the non-asymptotic bound holds true almost surely in the limit $n,d\to\infty$.

Finally, it follows that, almost surely as $n,d\to\infty$,
\begin{align}
    \frac{1}{\sigma^2}\operatorname{Var}_\Phi(\Sigma) \stackrel{\eqref{eq:var-bound1}}{\leq} \| \Sigma\|\operatorname{Tr}\left((\Phi^\top\Phi)^{-1}\right) \stackrel{\eqref{eq:tracebound-1.5}}{\leq} \underbrace{\frac{\|\Sigma\|}{\lambda_{\min}(\Sigma)}}_{=\kappa(\Sigma)}\operatorname{Tr}\left((Z^\top Z)^{-1}\right) \stackrel{\eqref{eq:tracebound-3}}{=} \mathcal{O}\left(\frac{\kappa(\Sigma)}{(\sqrt{\gamma}-C_K)^2}\right), \label{eq:variance-bound-nonisotropic}
\end{align}
where
\begin{equation}
    \kappa(\Sigma) := \frac{\lambda_{\max}(\Sigma)}{\lambda_{\min}(\Sigma)} = \frac{\|\Sigma\|}{\lambda_{\min}(\Sigma)}
\end{equation}
denotes the condition number of the population covariance $\Sigma$.

In summary, we have established with the help of Proposition~\ref{prop:general-bias-variance-decomposition} that
\begin{equation}
    \mathcal{R}(\gamma)=\lim_{n,d\to\infty}\mathcal{R}(\hat{w}) \stackrel{\eqref{eq:risk-formula}}{=} \sigma^2 + \lim_{n,d\to\infty} \mathbb{E}_{\mathcal{S}}\left[\|\hat{w}-w_*\|_{ \Sigma}^2\right] = \sigma^2 + \lim_{n,d\to\infty}\mathbb{E}_{\mathcal{S}}\left[\operatorname{Bias}_\Phi(w_*)^2 + \operatorname{Var}_\Phi( \Sigma)\right].
\end{equation}
In combination with the previously derived upper bounds for the bias and variance terms, we may thus conclude with the following asymptotic bound of the risk in the over-parametrized regime:

\begin{theorem} \label{thm:risk-bound-nonisotropic}
     Let $\Phi\in\mathbb{R}^{d\times n}$ denote the random features in Eq.~\eqref{eq:random-features} and assume that $\operatorname{rank}(\Phi^\top\Phi) = n$. Let $\Sigma=\mathbb{E}_{ x}\left[\phi(x)\phi(x)^\top\right]\in\mathbb{R}^{d\times d}$ denotes the population covariance in Eq.~\eqref{eq:feature-covariance} and suppose that $Z:=\Sigma^{-\frac{1}{2}}\Phi$ has independent, sub-Gaussian columns $z_j\in\mathbb{R}^{d\times 1}$ such that $\|z_j\|=\sqrt{d}$ almost surely for all $j\in[n]$. Let further $Y\in\mathbb{R}^{n\times 1}$ be the training labels and suppose that there exists $w_*\in\mathbb{R}^d$ satisfying $Y=\Phi^\top w_*+\mathcal{E}$ for some Gaussian noise $\mathcal{E}\sim\mathcal{N}(0,\sigma^2I_n)$. Then, as $\gamma\to\infty$, the risk in Eq.~\eqref{eq:risk-formula} satisfies the upper bound
    \begin{equation*}
        \mathcal{R}(\gamma) = \sigma^2 + \mathcal{O}\left(\|w_*\|^2\| \Sigma\| + \frac{\sigma^2\kappa(\Sigma)}{(\sqrt{\gamma}-C_K)^2}\right),
    \end{equation*}
    where $C_K\geq1$ is an absolute constant that only depends on the sub-Gaussian norm $K:=\max_{j\in[n]}\|z_j\|_{\psi_2}$ of the columns of $Z$.
\end{theorem}

It is worth noting that this bound is consistent with Theorem~\ref{thm:double-descent-gaussian-features}. Indeed, if $\Sigma = \sigma_X^2I_d$, then $\|\Sigma\|=\sigma_X^2$ and $\kappa(\Sigma)=1$. Also, we have the bound $1-\frac{1}{\gamma}\leq 1$ if $\gamma>1$ which cover the bias term. For the variance we find
\begin{align}
    \operatorname{Var}_\Phi(\Sigma) \stackrel{\eqref{eq:variance-bound-nonisotropic}}{=} \mathcal{O}\left(\frac{\sigma^2}{(\sqrt{\gamma}-C_K)^2}\right) = \mathcal{O}\left(\frac{\sigma^2}{\gamma -2C_K\sqrt{\gamma}+C_K^2}\right) = \Theta\left(\frac{1}{\gamma}\right)
\end{align}
as $\gamma\to\infty$. On the other hand, in the context of Theorem~\ref{thm:double-descent-gaussian-features}, a comparison with Eq.~\eqref{eq:risk-bound-isotropic} yields that the variance term in the isotropic case obeys
\begin{equation}
    \operatorname{Var}_\Phi(\sigma_X^2 I_d) = \sigma_X^2\lim_{d\to\infty}\mathbb{E}_\mathcal{E}\left[\operatorname{Var}_\Phi(\mathcal{E})\right] \stackrel{\eqref{eq:risk-bound-isotropic}}{=} \Theta\left(\frac{\sigma^2}{\gamma-1}\right) = \Theta\left(\frac{1}{\gamma}\right)
\end{equation}
as $\gamma\to\infty$. This demonstrates that the bounds we have established in the non-isotropic case are formally consistent with in the matrix limit $\Sigma\to \sigma_X^2 I_d$.

\smallskip
\par
\paragraph{A Remark on the Random Matrix Theory Assumptions.} The key of our anisotropic covariance analysis was the application of two lemmata from high-dimensional probability theory: Lemma~\ref{lemma:generalized-marchenko-pastur}, used to cover the under-parametrized regime, imposes the bound $\operatorname{Var}(z_j^\top A z_j)=o(d^2)$ as $d\to\infty$ for all $j\in[n]$ and $A\in\mathbb{R}^{d\times d}$ with $\|A\|=\mathcal{O}_d(1)$, where $z_j=\Sigma^{-\frac{1}{2}}\phi(x_j)$. Lemma~\ref{lemma:singular-value-lower-bound}, used for the over-parametrized regime on the other hand, requires the assumption that $\|z_j\|=\sqrt{d}$ almost surely for all $j\in[n]$. In Subsection~\ref{appendix:RMT-assumptions} of Supplementary Section~\ref{appendix:techniques} we show that these two assumptions do not contradict each other, making the theory between the under- and over-parametrized regimes compatible, and thus allowing us to prove double descent. Furthermore, below in Section~\ref{appendix:quantum-RMT-assumptions} of Supplementary Section~\ref{appendix:QELM-math}, we also justify these random matrix theory assumptions explicitly in the context of quantum feature maps.

\section{Mathematical Analysis of Quantum Extreme Learning Machines (QELMs)} \label{appendix:QELM-math}

Previously, in Supplementary Section~\ref{appendix:double_descent_linear2}, we have proven in Proposition~\ref{prop:over-fitting} that the expected risk of a generic, ridgeless linear regression model obeys the double descent curve if certain random feature correlations are limited. Furthermore, Theorem~\ref{thm:risk-bound-nonisotropic} establishes a generalization bound in terms of the population covariance condition number in the over-parametrized regime, regardless of the strength of correlations. This naturally motivates the next theoretical question: How can we precisely quantify the the condition number $\kappa( \Sigma)$ and the maximum eigenvalue $\lambda_{\max}(\Sigma)$ of the population covariance $ \Sigma$ in the context of random features arising from Quantum Extreme Learning Machines (QELMs)~\cite{xiong2025fundamental}? Before we analytically demonstrate double descent in QELMs, we shall first define their overarching mathematical structure.

\subsection{The Framework of QELMs}

In a nutshell, the QELM framework generally proceeds as follows~\cite{xiong2025fundamental}: first, there is a \emph{data uploading} stage, where the data in $\mathcal{X}$ is encoded into a parametrized unitary which acts as a feature map on an ''accessible`` subsystem. In the next stage, one applies \emph{reservoir evolution} which is achieved by evolving the entire-system input state (including the subsystem feature map) through a unitary operator known as the ''quantum reservoir``. Afterwards, one performs a series of measurements in the \emph{readout} stage. In the final \emph{classical post-processing} stage, these measurement results are used to train the model with linear regression using the labels in $\mathcal{Y}$.

For the purpose of studying QELMs more rigorously, we shall first introduce all necessary objects to define the QELM output function. Let $w=(w_1,\dots w_d)\in\mathbb{R}^d$ be a trainable weight vector in $d=\mathcal{O}(\operatorname{Poly}(N))$ dimensions, and $O_1,\dots,O_d\in\operatorname{Herm}(2^N)$ arbitrary observables. Let $U_1,\dots,U_{L+1}\in\operatorname{U}(2^N)$ be arbitrary unitary operators acting on all qubits, which we call the \emph{quantum reservoirs} that could be implemented by Parametrized Quantum Circuits (PQCs). 
Decompose $N=n_a+n_h$ for some non-negative integers $n_a,n_h\leq N$. The \emph{accessible qubits} are defined by the indices in $\{1,\dots,n_a\}$ while the \emph{hidden qubits} by the remaining $n_h$ indices in $\{n_a+1,\dots,N\}$, respectively. 
Consider an input density matrix $\rho_0^{(a)}\in\mathcal{D}(2^{n_a})$ acting on the accessible qubits, and let $\rho_0:=\rho_0^{(a)}\otimes |0\rangle\langle0|\in\mathcal{D}(2^N)$ denote the input density matrix of the entire $N$-qubit system. 
Finally, let $\mathcal{G}_1,\dots,\mathcal{G}_L:\mathcal{X}\to\operatorname{U}(2^{n_a})$ be $L\in\mathbb{N}$ arbitrary PQCs which we call the (angle encoding) \emph{feature maps} due to their role in encoding the data $x$ into the unitary rotation angles. 
The number $L$ is also referred to as the \emph{number of reuploads} (or number of layers). The pre-measurement state of the system shall be denoted
\begin{equation}
    \rho_U(x) := U(x)\rho_0 {U(x)}^\dagger,
\end{equation}
where we defined the entire unitary
\begin{equation}
    U(x):= U_{L+1}\prod_{\ell=1}^L \left\{\left( \mathcal{G}_\ell(x) \otimes I_{2^{n_h}} \right)U_\ell \right\} \label{eq:data-uploading}
\end{equation}
of the QELM model. We shall sometimes use the abbreviation $S_\ell(x):=\mathcal{G}_\ell(x)\otimes I_{2^{n_h}}$ for each layer $\ell\in[L]$ which may be seen as a data-encoding feature map acting on the full $N$-qubit space. The output function of the QELM is subsequently defined as the Euclidean inner product

\begin{equation}
    f_w(x) = w^\top \phi_U(x) = \sum_{j=1}^d w_j\operatorname{Tr}(O_j\rho_U(x)), \label{eq:quantum-regression-model}
\end{equation}
between the real-valued readout vector $\phi_U(x) := \left(\operatorname{Tr}(O_j\rho_U(x))\right)_{j=1}^d\in\mathbb{R}^d$ and the weight vector $w\in\mathbb{R}^{d}$. Further below we prove that the readout mapping $\phi_U:\mathcal{X}\to\mathbb{R}^d$ is Lipschitz continuous with a Lipschitz constant that is generally independent from the dimension $d$ which makes it possible to apply useful concentration inequalities such as Lipschitz concentration~\cite{vershynin2020high}.

In order to effectively solve the supervised learning task in Eq.~\eqref{eq:supervised-learning}, we rely on optimizing the empirical mean squared error (MSE) loss
\begin{equation}
    \mathcal{L}_w(x) := \frac{1}{2n}\sum_{\ell=1}^n(f_w(x_\ell)-y_\ell)^2
\end{equation}
with linear regression with respect to the weight vector $w$.

\subsection{Quantum Features in QELMs}

The next step is to identify the structure of the random features $\Phi\in\mathbb{R}^{d\times n}$ in the QELM setting analogous to the classical linear regression model as per Eq.~\eqref{eq:random-features}. In other words, we would like to find a correspondence $\phi(x) \equiv \phi_U(x)$ with $\phi_U(x)$ as defined in Eq~\eqref{eq:quantum-regression-model}. First, recall that
\begin{equation}
    \phi_U(x) = \left(\operatorname{Tr}(O_k\rho_U(x))\right)_{k=1}^d = \left(\begin{matrix}
        \operatorname{Tr}(O_1\,\rho_U(x)) \\ \vdots \\ \operatorname{Tr}(O_d\,\rho_U(x))
    \end{matrix}\right) = \left(\begin{matrix}
        \langle O_1\rangle_{x} \\ \vdots \\ \langle O_d\rangle_{x}
    \end{matrix}\right) \in \mathbb{R}^d. \label{eq:quantum-readout}
\end{equation}
by introducing the standard quantum notations $\langle O_k\rangle_{x}:= \operatorname{Tr}(O_k\rho_U(x))$ for each $k\in[d]$. Instead of a classical feature map $\Phi$, we now introduce \emph{quantum} features:

\begin{equation}
    \Phi_U := \left(\phi_U(x_1)\mid\dots\mid \phi_U(x_n)\right) = \left(\begin{matrix}\langle O_1\rangle_{x_1} & \cdots & \langle O_1\rangle_{x_n} \\ \vdots & \ddots & \vdots \\ \langle O_d\rangle_{x_1} & \cdots & \langle O_d\rangle_{x_n} \end{matrix}\right) \in \mathbb{R}^{d\times n}, \label{eq:quantum-feature-sample-covariance}
\end{equation}
where we again abbreviated the expectation values $\langle O_k\rangle_{x_j}:=\operatorname{Tr}(O_k\rho_U(x_j))$ for each $k\in[d]$ and $j\in[n]$. In other words, the column-wise feature map in Eq.~\eqref{eq:random-features} is now explicitly given by the quantum readout mapping $\phi\equiv \phi_U(\cdot)$. This quantum reinterpretation forces us to adapt the classical deterministic feature covariance defined in Eq.~\eqref{eq:feature-covariance} accordingly, where we are required to replace $\phi(x)$ by $\phi_U(x)$. This modification results in the following \emph{quantum} deterministic feature covariance:
\begin{equation}
     \Sigma \equiv \Sigma_U := \mathbb{E}_{ x}\left[\phi_U( x)\,\phi_U( x)^\top\right] \in \mathbb{R}^{d\times d}. \label{eq:quantum-feature-covariance}
\end{equation}

Other than these modification to obtain the quantum feature objects $\Phi_U$ and $ \Sigma_U$, the proof of Theorem~\ref{thm:double-descent-gaussian-features} is completely identical and relies solely on the Gaussian features condition imposed by Assumption~\ref{assumption:gaussian-features}. This raises the question of whether the quantum features satisfy the Gaussian assumptions. The answer is yes: If the model unitary $U$ defines a so-called \emph{2-design} as shown in Ref.~\cite{garcia2025quantum}. We elaborate the details further below in Supplementary Section~\ref{appendix:2-designs}. In a subsequent analysis, we treat the non-2-design case separately which relies on the Fourier interpretation of the readout~\cite{schuld2021effect, xiong2025fundamental, mhiri2025constrained}.

\subsection{Compatibility with the Classical RMT Assumptions} \label{appendix:quantum-RMT-assumptions}

In Supplementary Section~\ref{appendix:double_descent_linear2} we have established bounds for the expected risk $\mathcal{R}(\gamma)$ defined in Eq.~\eqref{eq:expected-risk} in the under- and over-parametrized regimes, respectively. For the over-parametrized regime, we specifically had to assume the strong condition that $\|z\|=\sqrt{d}$ almost surely, where $z:=\Sigma^{-\frac{1}{2}}\phi(x)$ with population covariance $\Sigma=\mathbb{E}_x[\phi(x)\phi(x)^\top]$ and $\phi(x)$ denoting the random feature vector. In order to apply the results in the context of QELMs, we need to justify that it is physically possible to achieve $\|\Sigma^{-\frac{1}{2}}\phi_U(x)\|=\sqrt{d}$ almost surely, where $\Sigma$ and $\phi$ are replaced by their quantum counterparts, $\Sigma_U$ and $\phi_U(x)$, defined in Eqs.~\eqref{eq:quantum-feature-covariance} and~\eqref{eq:quantum-readout}, respectively. We consider two concrete examples: first, the 2-design limit where we fix $d$ and show that $\|z\|=\sqrt{d}$ holds true deterministically and is furthermore compatible with the results in Supplementary Section~\ref{appendix:2-designs}. In the second example, where we assume Gaussian distributed inputs $x\sim\mathcal{N}(0,I_p)$, we recover that $\|z\|=\sqrt{d}$ holds true almost surely in the thermodynamic limit $\frac{d}{n}\to\gamma$ as $n,d\to\infty$.

\smallskip
\par
\paragraph{The 2-Design Limit.} Let us first demonstrate the possibility of $\|z\|=\sqrt{d}$ holding true deterministically. Suppose that the observables $\{O_k:k\in[d]\}=\left\{P:P\in\mathcal{P}_N\setminus\{I\}\right\}$ span all non-identity $N$-qubit Pauli strings (where $\mathcal{P}_N$ denotes the set of all $N$-qubit Pauli strings) so that, together with the identity element $I\in\mathcal{P}_N$, it defines an orthogonal basis of the Hilbert-Schmidt space $\operatorname{Herm}(2^N)$ due to $\operatorname{Tr}(O_kO_{k'})=2^N\delta_{k,k'}$ for all $k,k'\in[d]$. In particular, $d=2^{2N}-1$ distinct observables are required to implement this QELM model. Regardless of the architecture $U\in\operatorname{U}(2^N)$ and the input distribution $x\sim\mathbb{P}$, we always have $\rho_U(x)\in\operatorname{Herm}(2^N)$ so that the Pauli basis choice $\{O_k\}_{k\in[d]}$ yields
\begin{equation}
    \rho_U(x) = \frac{1}{2^N}I+\sum_{k=1}^d c_kO_k \label{eq:Pauli-decomposition1}
\end{equation}
for some unique, real coefficients $c_k\in\mathbb{R}$. Here, we used $\operatorname{Tr}\rho_U(x)=1$ to get the $2^{-N}$ factor for the identity term (observe further that $\operatorname{Tr}O_k=0$ for all $k\in[d]$ since any non-identity Pauli string is traceless). Furthermore, we find for each $k'\in[d]$ that
\begin{align}
    \langle O_{k'}\rangle_x = \operatorname{Tr}\left(O_{k'}\rho_U(x)\right) \stackrel{\eqref{eq:Pauli-decomposition1}}{=} \operatorname{Tr}\left(\frac{1}{2^N}O_{k'}+\sum_{k=1}^d c_k\,O_kO_{k'}\right) = \frac{1}{2^N}\underbrace{\operatorname{Tr}O_{k'}}_{=0} + \sum_{k=1}^d c_k\,\underbrace{\operatorname{Tr}\left(O_kO_{k'}\right)}_{=2^N\delta_{k,k'}} = 2^Nc_{k'}. \label{eq:Pauli-coefficient1}
\end{align}
As a result, assuming that $\rho_U(x)$ is pure,
\begin{align}
    1 &\stackrel{\eqref{eq:Pauli-decomposition1}}{=} \operatorname{Tr}\left(\rho_U(x)^2\right) = \operatorname{Tr}\left(\left(\frac{1}{2^N}I+\sum_{k=1}^d c_kO_k\right)^2\right) = \operatorname{Tr}\left(\frac{1}{2^{2N}}I + \frac{2}{2^{N}}\sum_{k=1}^d c_kO_k + \sum_{k,k'=1}^d c_kc_{k'}\,O_kO_{k'}\right) \nonumber
    \\ &= \frac{1}{2^{2N}}\underbrace{\operatorname{Tr}(I)}_{=2^N} + \frac{1}{2^{N-1}}\sum_{k=1}^d c_k\,\underbrace{\operatorname{Tr}O_k}_{=0} + \sum_{k,k'=1}^d c_kc_{k'}\,\underbrace{\operatorname{Tr}\left(O_kO_{k'}\right)}_{=2^N\delta_{k,k'}} = \frac{1}{2^N}+2^N\sum_{k=1}^d c_k^2 \stackrel{\eqref{eq:Pauli-coefficient1}}{=} \frac{1}{2^N} + \frac{1}{2^N}\underbrace{\sum_{k=1}^d \langle O_k\rangle_x^2}_{\|\phi_U(x)\|^2},
\end{align}
recalling the definition of $\phi_U(x)$ in Eq.~\eqref{eq:quantum-readout}, which immediately implies that 
\begin{equation}
    \|\phi_U(x)\|^2 = 2^N-1 \label{eq:constant-readout1}
\end{equation}
is constant. In other words, the readout vector $\phi_U(x)$ lies on the sphere $r\,\mathbb{S}^{d-1}$ of radius $r=\sqrt{2^N-1}$.

If we now assume that $\Sigma = \sigma^2I_d$ is isotropic, then
\begin{equation}
    d \stackrel{!}{=} \|z\|^2 = \left\|\Sigma^{-\frac{1}{2}}\phi_U(x)\right\|^2 = \phi_U(x)^\top \Sigma^{-1}\, \phi_U(x) = \frac{1}{\sigma^2} \phi_U(x)^\top\phi_U(x) = \frac{1}{\sigma^2}\|\phi_U(x)\|^2 \stackrel{\eqref{eq:constant-readout1}}{=} \frac{2^N-1}{\sigma^2}
\end{equation}
if and only if
\begin{equation}
    \sigma^2 = {\frac{2^N-1}{d}} = \frac{2^N}{2^{2N}-1}\left(1-\frac{1}{2^N}\right), \label{eq:isotropic-concentration1}
\end{equation}
recalling that $d=2^{2N}-1$. Remarkably, we have derived the exact same expression (cf. Lemma~1 in Ref.~\cite{garcia2025quantum}) for the unitary 2-designs variance $\sigma_N^2$ as elaborated in Lemma~\ref{lemma:quantum-covariance} below of Supplementary Section~\ref{appendix:2-designs} (without using representation theory) which requires at least $d=\Omega(2^N\operatorname{Poly}(N)^{-1})$ distinct readout components in order to being able to express a non-concentrating target function (i.e., $\mathbb{E}_x[f_*(x)^2]=\Omega(\operatorname{Poly}(N)^{-1})$). Our construction above specifically relies on using $d=2^{2N}-1$ distinct readout components.

\smallskip
\par
\paragraph{The High-Dimensional Limit.} Suppose that $d=\mathcal{O}(\operatorname{Poly}(N))$. The idea here is to prove
\begin{equation}
    \lim_{d\to\infty}\left|\|z\|/{\sqrt{d}}-1\right| = 0 \label{eq:almost-sure-limit-readout}
\end{equation}
almost surely, to justify the condition asymptotically in the high-dimensional limit $d\to\infty$. 
For pure input states, $\rho_0=|0\rangle\langle 0|$, Ref.~\cite{berberich2024training} established that each component $\phi_U(x)_k$ of the readout map is Lipschitz continuous as a scalar function from $\mathcal{X}$ to $\mathbb{R}$. Here, to prove Eq.~\eqref{eq:almost-sure-limit-readout}, we pursue the stronger result that the full vector-valued map $\phi_U:\mathcal{X}\to\mathbb{R}^d$ is Lipschitz continuous with a dimension-independent constant, $L_{\phi_U}=\Theta_d(1)$.
We refer to Lemma~\ref{lemma:lipschitz-readout} in Section~\ref{appendix:others} for a formal proof of this statement. Notably, this holds true for arbitrary density matrices $\rho_0\in\mathcal{D}(2^N)$ and holds true under the additional assumption that the system $\{O_k:k\in[d]\}$ of observables defines a \emph{1-tight frame system}, that is,
\begin{equation}
    \sum_{k=1}^d\left|\operatorname{Tr}\left(O_kA\right)\right|^2 \leq \|A\|_{\operatorname{HS}}^2
\end{equation}
for all Hermitian operators $A\in\operatorname{Herm}(2^N)$, where $\|A\|_{\operatorname{HS}}:=\sqrt{\langle A,A\rangle_{\operatorname{HS}}}$ denotes the Schatten 2-norm induced by the Hilbert-Schmidt inner product $\langle A,B\rangle_{\operatorname{HS}}:=\operatorname{Tr}\left(A^\dagger B\right)$. In Lemma~\ref{lemma:1-tight-system} of Section~\ref{appendix:others} we explicitly show that any orthonormal system $\{O_k:k\in[d]\}$ with $\langle O_k,O_{k'}\rangle_{\operatorname{HS}}=\delta_{k,k'}$ (e.g., normalized Pauli strings) defines a 1-tight frame system as per Eq.~\eqref{eq:1-tight-frame}.

We proceed by making two extra assumptions, namely that $x\sim\mathcal{N}(0,I_p)$ is normal, allowing us to apply concentration results from the classical literature~\cite{vershynin2020high}, and that the growth of the smallest eigenvalue $\lambda_{\min}(\Sigma)=\omega\left(\frac{\log d}{d}\right)$ of the population covariance $\Sigma_U$ is restricted as $d\to\infty$. In particular, $\lambda_{\min}(\Sigma)>0$. The idea is to use the previous Lipschitz continuity of the readout to show that the norm $f(x):=\|\Sigma^{-\frac{1}{2}}\phi_U(x)\|$ of the whitened readout is Lipschitz as a scalar function $f:\mathcal{X}\to\mathbb{R}$. Then, we use the classical concentration result in order to conclude with the almost sure limit in Eq.~\eqref{eq:almost-sure-limit-readout} after an application of the first Borel-Cantelli lemma (cf. Theorem~4.3 in Ref.~\cite{billingsley2017probability}).

Indeed, using $L_{\phi_U}=\mathcal{O}_d(1)$ from above, the reverse triangle inequality and sub-multiplicativity of the Euclidean norm yield
\begin{align}
    |f(x)-f(x')| &= \left|\|\Sigma^{-\frac{1}{2}}\phi_U(x)\|-\|\Sigma^{-\frac{1}{2}}\phi_U(x')\|\right| \leq \|\Sigma^{-\frac{1}{2}}\phi_U(x)-\Sigma^{-\frac{1}{2}}\phi_U(x')\|\leq \|\Sigma^{-\frac{1}{2}}\|\,\|\phi_U(x)-\phi_U(x')\| \nonumber
    \\ & \stackrel{\eqref{eq:readout-Lipschitz-bound}}{\leq} \frac{L_{\phi_U}}{\sqrt{\lambda_{\min}(\Sigma)}}\|x-x'\| = \frac{\mathcal{O}_d(1)}{\sqrt{\lambda_{\min}(\Sigma)}}\|x-x'\|
\end{align}
for all inputs $x,x'\in\mathcal{X}$, where we also used $\|\Sigma^{-\frac{1}{2}}\| = \lambda_{\max}(\Sigma^{-\frac{1}{2}})=\lambda_{\min}(\Sigma)^{-\frac{1}{2}}$ since the population covariance $\Sigma$ is positive definite. This confirms that $f:\mathcal{X}\to\mathbb{R}$ if $L_f$-Lipschitz with
\begin{equation}
    L_f = \frac{\mathcal{O}_d(1)}{\sqrt{\lambda_{\min}(\Sigma)}}.
\end{equation}
which depends on the smallest eigenvalue $\lambda_{\min}(\Sigma)$ of the population covariance $\Sigma$. By Theorem~5.2.3 of Ref.~\cite{vershynin2020high}, since we assumed $x\sim\mathcal{N}(0,I_p)$, we have Gaussian concentration in the sense that 
\begin{equation}
    \|f(x)-\mathbb{E}_x[f(x)]\|_{\psi_2} \leq C_0 L_f
\end{equation}
for some absolute constant $C_0>0$, recalling the definition of the sub-Gaussian norm $\|\cdot\|_{\psi_2}$ in Definition~\eqref{eq:orlicz-space} of Section~\ref{appendix:others}. By Lemma~\ref{lemma:subgaussian-characterization} in Section~\ref{appendix:others} we have for all $t\geq0$ that
\begin{equation}
    \mathbb{P}\left(|f(x)-\mathbb{E}_x[f(x)]|\geq t\right) \leq 2\exp\left(-\frac{t^2}{C_1L_f^2}\right) \label{eq:Gaussian-concentration1}
\end{equation}
for some other absolute constant $C_1>0$. We can leverage this to show that $\|z\|$ concentrates exponentially around $\sqrt{d}$ in high dimensions. Indeed, with the help of Lemma~\ref{lemma:lipschitz-concentration-quantum} in Section~\ref{appendix:others}, Eq.~\eqref{eq:Gaussian-concentration1} allows us to prove the concentration inequality
\begin{align}
    \mathbb{P}\left(\left|\frac{\|z\|}{\sqrt{d}}-1\right|\geq\varepsilon\right) = \mathcal{O}_d\left(e^{-{d\lambda_{\min}(\Sigma)}}\right)
\end{align}
as $d\to\infty$, which holds true for arbitrary $\varepsilon>0$. By the first Borel-Cantelli lemma (cf. Theorem~4.3 in Ref.~\cite{billingsley2017probability}) we must thus have that
\begin{equation}
    \lim_{d\to\infty}\left|\frac{\|z\|}{\sqrt{d}}-1\right| = 0
\end{equation}
almost surely.

\section{Double Descent in QELMs with Haar Random Unitaries} \label{appendix:2-designs}

This section is dedicated to proving double descent in the setting of QELMs under the assumption of unitaries sampled from a 2-design~\cite{mele2024introduction}.

\subsection{Interlude: Unitary Designs}

We begin by briefly reminding the reader the notion of a 2-design. For that purpose, we need to define the \emph{Haar measure} on the unitary group $\operatorname{U}(d)$. We refer to Ref.~\cite{mele2024introduction} for additional details.

\begin{definition}[Haar Measure]
    Let $d\in\mathbb{N}$. The Haar measure $\mu$ on the unitary group $\operatorname{U}(d)$ is defined as the unique probability measure $\mu$ satisfying
    \begin{equation*}
        \int_{\operatorname{U}(d)}f(U)\,d\mu(U) = \int_{\operatorname{U}(d)}f(UV)\,d\mu(U) = \int_{\operatorname{U}(d)}f(VU)\,d\mu(U),
    \end{equation*}
    element-wise for all integrable functions $f:\operatorname{U}(d)\to\operatorname{Mat}_{d_1\times d_2}(\mathbb{C})$ and all unitaries $V\in\operatorname{U}(d)$. We say that the Haar measure is left- and right invariant under group multiplication, respectively.
\end{definition}

A unitary $t$-design is subsequently defined as a probability measure on a subset of the unitary group $\operatorname{U}(d)$ such that the $t$-th moment operator matches the corresponding Haar integral:

\begin{definition}[Unitary Design] \label{def:2-design}
    Let $t\in\mathbb{N}$. A unitary $t$-design is a probability measure $\nu$ on a subset $S\subseteq\operatorname{U}(d)$ of the unitary group such that
    \begin{equation*}
        \int_S V^{\otimes t}O{V^\dagger}^{\otimes t}\, d\nu(V) = \int_{\operatorname{U}(d)}U^{\otimes t}O{U^\dagger}^{\otimes t}\,d\mu_d(U)
    \end{equation*}
    for all linear operators $O:(\mathbb{C}^d)^{\otimes t}\to\mathbb{C}^d)^{\otimes t}$.
\end{definition}

With the help of the vectorization formalism~\cite{mele2024introduction} one may equivalently formulate the notion of unitary $t$-designs by the condition
\begin{equation}
    \int_S V^{\otimes t}\otimes {V^*}^{\otimes t}\, d\nu(V) = \int_{\operatorname{U}(d)}U^{\otimes t}\otimes {U^*}^{\otimes t}\, d\mu_d(U),
\end{equation}
where $V\mapsto V^*$ denotes element-wise complex conjugation.

\par
Now, having briefly introduced the concept of 2-designs, we proceed by proving the existence of double descent in QELMs with unitaries sampled from the Haar measure~\cite{mele2024introduction}.

\subsection{Gaussian Processes in QELMs}

In fact, the QELM in the idealized 2-design case is mathematically equivalent to the linear regression model with Gaussian features. We elaborate this in detail in the following by leveraging results in previous the literature showing that quantum neural networks are Gaussian processes~\cite{garcia2025quantum}.

Consider the vectorization
\begin{equation}
    \operatorname{vec}{\Phi} = (\langle O_1\rangle_{x_1},\dots,\langle O_d\rangle_{x_1},\dots,\langle O_{1}\rangle_{x_{n}},\dots, \langle O_d\rangle_{x_n}) \in \mathbb{R}^{dn}
\end{equation}
of the quantum random features $\Phi\equiv\Phi_U$ defined in Eq.~\eqref{eq:quantum-feature-sample-covariance} recalling that $\langle O_k\rangle_{x_j}=\operatorname{Tr}(O_k\rho_U(x_j))$. We say that $U(x)$ in Eq.~\eqref{eq:data-uploading} is sampled from a unitary 2-design if the ensemble $\{U(x):x\sim \mathbb{P}_x\}$ defines a unitary 2-design as per Definition~\ref{def:2-design} where $\mathbb{P}_x$ denotes the probability distribution of the inputs $x$. Note that if $U(x)$ is sampled from a 2-design, then the integrals of the form $\mathbb{E}_x[f(U(x))]$ for certain integrable functions $f:\operatorname{U}(2^N)\to\mathbb{C}$ match the integration over the Haar measure $\mathbb{E}_{U\sim\mu_{\text{Haar}}}[f(U))]$. In more formal words,
\begin{equation}
    \mathbb{E}_{x}[f(U(x))] = \int_{\operatorname{U}(2^N)} f(U)\,d\mu_{\text{Haar}}(U) \label{eq:2-design-input-sampling}
\end{equation}
for all degree-2 homogeneous polynomials $f\in\mathbb{C}[\{(U_{i,j},U^*_{i,j}):i,j\in[2^N]\}]_2$ in the entries of $U$ and $U^*$, respectively. Therefore, we are able to apply the following insight from previous literature: if $U(x)$ is sampled from a unitary 2-design, then it has been shown by means of representation theory~\cite{mele2024introduction} that the readout vector defines a Gaussian process~\cite{garcia2025quantum}. More specifically, we shall apply the following results in the context of our theoretical framework:

\begin{lemma}[Ref.~\cite{garcia2025quantum}, Lemma~1] \label{lemma:quantum-covariance}
    Let $O_1,\dots,O_d\in\operatorname{Herm}(2^N)$ be traceless observables satisfying $O_k^2 = I_{2^N}$ and $\operatorname{Tr}(O_k O_{k'})=\delta_{k,k'}\, 2^N$ for all $k,k'\in[d]$ (e.g. Pauli strings). Assume that $U(x)\in\operatorname{U}(2^N)$ in Eq.~\eqref{eq:data-uploading} is sampled from a unitary 2-design. Then, we have $\mathbb{E}_x[\langle O_k\rangle_{x_j}] = 0$ for all $k\in[d]$ and $j\in[n]$. Furthermore, we have 
    \begin{equation*}
        \operatorname{Cov}(\langle O_k\rangle_{x_j},\langle O_{k'}\rangle_{x_{j'}}) = \begin{cases}
            0, & k\neq k', 
            \\  \frac{2^N}{2^{2N}-1}\left(\operatorname{Tr}\left(\rho(x_j)\rho(x_{j'})\right)-\frac{1}{2^N}\right) = \Theta(2^{-N}),& k=k',
        \end{cases}
    \end{equation*}
    for all $k,k'\in[d]$ and all $j,j'\in[n]$.
\end{lemma}

The previous lemma thus precisely quantifies the covariance between the different readout elements $\{\langle O_k\rangle_{x_j}\}_{k\in[d]}$ of $\phi_U(x_j)$ in Eq.~\eqref{eq:quantum-readout}. The following lemma shows that these elements obey a Gaussian law with a variance that decays exponentially in the number $N$ of qubits:

\begin{lemma}[Ref.~\cite{garcia2025quantum}, Corollary~1] \label{lemma:quantum-gaussian-process}
    Let $O_1,\dots,O_d\in\operatorname{Herm}(2^N)$ be traceless observables satisfying $O_k^2 = I_{2^N}$ and $\operatorname{Tr}(O_k O_{k'})=\delta_{k,k'}\, 2^N$ for all $k,k'\in[d]$ (e.g. Pauli strings). If $U(x)$ is sampled from a unitary 2-design, then $\langle O_k\rangle_{x_j}\sim\mathcal{N}(0,\sigma_N^2)$ for all $k\in[d]$ and $j\in[n]$ where $\sigma_N^2 = \Theta(2^{-N})$ as $N\to\infty$.
\end{lemma}

\subsection{Gaussian QELMs exhibit Standard Double Descent}

To prove double descent in the quantum unitary 2-design case, we need to introduce the risk function in the setting of QELMs. Define,
\begin{equation}
    \mathcal{R}_U(\gamma) := \mathbb{E}_{( x, y)\sim \mathcal{D}}\left[ (w^\top \phi_U( x) -  y)^2 \right], \label{eq:quantum-risk}
\end{equation}
where $\phi_U( x)$ denotes the quantum readout vector defined in Eq.~\eqref{eq:quantum-readout}. Note that we have to make the following standard assumptions in order to apply our previous classical result with Gaussian features (cf. Theorem~\ref{thm:double-descent-gaussian-features}):
\begin{itemize}
    \item[(i)] There exists a true, unknown weight vector $w_*\in\mathbb{R}^{d\times 1}$ such that the labels $ y$ are generated by $ y=w_*^\top \phi_U( x) + \delta$ with $\delta\sim\mathcal{N}(0,\sigma^2)$ for some variance $\sigma^2$.
    \item[(ii)] The quantum features $\Phi=\Phi_U$ satisfy $\operatorname{rank}(\Phi\Phi^\top)=d$ if $\gamma<1$ and $\operatorname{rank}(\Phi^\top\Phi)=n$ if $\gamma>1$, respectively.
\end{itemize}
These assumptions allow us to establish our first result regarding double descent with quantum features:

\begin{corollary}\label{corr:double-descent-2design}
     Let $O_1,\dots,O_d\in\operatorname{Herm}(2^N)$ be traceless observables satisfying $O_k^2 = I_{2^N}$ and $\operatorname{Tr}(O_k O_{k'})=\delta_{k,k'}\, 2^N$ for all $k,k'\in[d]$ (e.g. Pauli strings). Assume that $U(x)$ in Eq.~\eqref{eq:data-uploading} is sampled from a 2-design and suppose that $\operatorname{Tr}\left(\rho_1^2\right) \neq 2^{-N}$ (e.g., if $\rho_1$ is pure) and $\operatorname{Tr}\left(\rho(x_j)\rho(x_{j'})\right)=2^{-N}$ for all $j,j'\in[n]$ with $j\neq j'$. Then, we have that the asymptotic expected quantum risk $\mathcal{R}_U(\gamma)$ in Eq.~\eqref{eq:quantum-risk} satisfies
    \begin{equation*}
        \mathcal{R}_U(\gamma) = \begin{cases}
            \frac{\sigma^2}{1-\gamma},&\text{if $\gamma<1$,}
            \\ \Theta(2^{-N})\left(1-\frac{1}{\gamma}\right)\,\lim_{d\to\infty}\|w_*\|^2 + \frac{\gamma\sigma^2}{\gamma-1}, &\text{if $\gamma>1$,}
        \end{cases}
    \end{equation*}
    as $N\to\infty$.
\end{corollary}

\noindent
\emph{Proof.} First, recall the definitions of $\Phi=\Phi_U$ in Eq.~\eqref{eq:quantum-feature-sample-covariance} and the population covariance $\Sigma=\Sigma_U$ in Eq.~\eqref{eq:quantum-feature-covariance}. Observe that, since the inputs $x_j\sim\mathbb{P}_x$ are sampled i.i.d. from its underlying probability distribution we can just fix the first sample $x_1$ (i.e., $x\stackrel{d}=x_j$ are equal in distribution for all $j\in[n]$ so we may fix $j=1$ without loss of generality). As a result,
\begin{align}
    \Sigma = \mathbb{E}_x[\phi_U(x)\phi_U(x)^\top] = \mathbb{E}_x[\phi_U(x_1)\phi_U(x_1)^\top] = \mathbb{E}_x[\Phi_1\Phi_1^\top],
\end{align}
where $\Phi_1\in\mathbb{R}^{d\times 1}$ denotes the first column of $\Phi$. In particular, the population covariance $\Sigma\in\mathbb{R}^{d\times d}$ has entries
\begin{equation}
    \Sigma_{k,k'} = \mathbb{E}_x\left[(\Phi_1\Phi_1^\top)_{k,k'}\right] = \mathbb{E}_x[\Phi_{k,1}\Phi_{k',1}] = \mathbb{E}_x\left[\langle O_k\rangle_{x_1}\langle O_{k'}\rangle_{x_1}\right].
\end{equation}
By applying Lemma~\ref{lemma:quantum-covariance} in the special case $j=j'=1$, we know that $\mathbb{E}_x[\langle O_k\rangle_{x_1}]=0$ for all $k\in[d]$ and that the covariance matrix $\operatorname{Cov}(\langle O_k\rangle_{x_1},\langle O_{k'}\rangle_{x_{1}}) = \delta_{k,k'}\cdot \sigma_N^2$ is diagonal, where
\begin{equation}
    \sigma_N^2 := \frac{2^N}{2^{2N}-1}\left(\operatorname{Tr}\left(\rho(x_1)^2\right)-\frac{1}{2^N}\right) = \Theta(2^{-N}) \label{eq:unitary-2-design-concentration}
\end{equation}
as $N\to\infty$ since $\operatorname{Tr}(\rho(x_1)^2)\neq 2^{-N}$ by assumption. As a result, we have
\begin{equation}
    \Sigma_{k,k'} = \mathbb{E}_x\left[\langle O_k\rangle_{x_1}\langle O_{k'}\rangle_{x_1}\right] = \mathbb{E}_x\left[\langle O_k\rangle_{x_1}\langle O_{k'}\rangle_{x_1}\right] - \underbrace{\mathbb{E}_x\left[\langle O_k\rangle_{x_1}\right]\mathbb{E}_x\left[\langle O_{k'}\rangle_{x_1}\right]}_{=0} = \operatorname{Cov}(\langle O_k\rangle_{x_1},\langle O_{k'}\rangle_{x_{1}}) = \delta_{k,k'}\cdot \sigma_N^2, \label{eq:covariance-entries-2design}
\end{equation}
which proves that $\Sigma = \sigma_N^2 I_d$. Because of Lemma~\ref{lemma:quantum-gaussian-process} we further know that $\langle O_k\rangle_{x_j}\sim\mathcal{N}(0,\sigma_N^2)$ for all $k\in[d]$ and $j\in[n]$. Note that Lemma~\ref{lemma:quantum-covariance} also implies that $\Phi$ has uncorrelated entries $\Phi_{k,j}=\langle O_k\rangle_{x_j}$ since we additionally assumed that $\operatorname{Tr}\left(\rho(x_j)\rho(x_{j'})\right)=2^{-N}$ for all $j,j'\in[n]$ with $j\neq j'$. This shows that the entries $\Phi_{k,j}$ of $\Phi$ are i.i.d. In particular, items (i) and (ii) of Assumption~\ref{assumption:gaussian-features} are satisfied. The third item (iii) of Assumption~\ref{assumption:gaussian-features} follows from the Marchenko-Pastur law (cf. Theorem~\ref{thm:marchenko-pastur}). Therefore, Theorem~\ref{thm:double-descent-gaussian-features} applies directly and immediately yields the corollary by substituting the variance $\sigma_X^2=\sigma_N^2$ accordingly.
\qed

Even though double descent exists in QELM with 2-design unitaries, the bottleneck is that this model is not scalable. 
If the ground truth target function is concentrating exponentially with the model size, then the problem one tries to solve is non-scalable in the first place.
A more nuanced situation is the case where the target is non-concentrating but the 2-design readout is:
not only do 2-designs result in exponential concentration $\phi_U(x)_k=\mathcal{O}(\operatorname{Exp}(N)^{-1})$ of the readouts~\cite{xiong2025fundamental, xiong2025role}, and thus non-scalable practical applications, but one might require an exponential number $d=\Omega(\operatorname{Exp}(N))$ of observables in order to learn a non-concentrating ground truth $f_*:\mathcal{X}\to\mathbb{R}$.
In other words, even if one has oracle access to the readout components, effectively bypassing the exponential complexity of estimating the expectation value, one still requires an exponential number of expectation values.

To demonstrate a generic situation where this happens, suppose that the non-concentrating ground truth is directly given by $f_*(x) := w_*^\top \phi_U(x)$ with optimal weight configuration $w_*\in\mathbb{R}^d$. 
Non-concentration of the target $f_*$ means that $\mathbb{E}_x [f_*(x)^2] = \Omega(\operatorname{Poly}(N)^{-1})$ as $N\to\infty$ which is closely related to the barren plateau phenomenon~\cite{mcclean2018barren, larocca2025barren, arrasmith2022equivalence, meyer2026gradient}. 
However, as demonstrated below, this is only possible if the optimal weight vector obeys $\|w_*\|^2 = \Omega(2^N\operatorname{Poly}(N)^{-1})$ as $N\to\infty$, which also implies a non-concentrating bias term $\Theta(2^{-N})\lim_{d\to\infty}\|w_*\|^2=\Omega(\operatorname{Poly}(N)^{-1})$ as $N\to\infty$. 

To see why we indeed require $\|w_*\|^2 = \Omega(2^N\operatorname{Poly}(N)^{-1})$ as $N\to\infty$ for $\mathbb{E}_x [f_*(x)^2] = \Omega(\operatorname{Poly}(N)^{-1})$ to hold true, we directly compute
\begin{align}
    \mathbb{E}_x [f_*(x)^2] &= \mathbb{E}_x \left[ \left( \sum_{k=1}^d w_{*,k} \phi_U(x)_k \right)^2 \right] = \sum_{k,k'=1}^d w_{*,k} w_{*,k'} \underbrace{\mathbb{E}_x [\phi_U(x)_k \phi_U(x)_{k'}]}_{\stackrel{\eqref{eq:covariance-entries-2design}}{=}\delta_{k,k'}\cdot \sigma_N^2} \nonumber
    \\ &= \sigma_N^2\sum_{k=1}^d w_{*,k}^2 = \sigma_N^2\|w_*\|^2 = \Theta\left(\frac{\|w_*\|^2}{2^N}\right)
\end{align}
recalling that $\sigma_N^2 = \Theta(2^{-N})$ as $N\to\infty$. Therefore, $\mathbb{E}_x [f_*(x)^2] = \Omega(\operatorname{Poly}(N)^{-1})$ if and only if $\|w_*\|^2 = \Omega(2^N\operatorname{Poly}(N)^{-1})$ as $N\to\infty$.

In most practical QELM or quantum random feature settings, the number of measurement operators (features) $d$ is restricted to a polynomial size relative to the number of qubits, i.e., $d = \mathcal{O}(\operatorname{Poly}(N))$, to ensure efficient readout. Assuming by contradiction that $d = \mathcal{O}(\operatorname{Poly}(N))$ and that the optimal weights $w_{*,k}=\mathcal{O}(1)$ are physically bounded as $N\to\infty$ for all $k\in[d]$, then the maximum possible squared norm is strictly bounded by the feature dimension:
\begin{equation}
    \|w_*\|^2 = \sum_{k=1}^d \underbrace{w_{*,k}^2}_{=\mathcal{O}(1)} = \mathcal{O}(d) = \mathcal{O}(\operatorname{Poly}(N))
\end{equation}
as $N\to\infty$. This creates a mathematical impossibility: $\|w_*\|^2$ can never scale as $\Omega(2^N \operatorname{Poly}(N)^{-1})$ if $d$ is polynomial in the system size $N$. As a result, any 2-design QELM utilizing only a polynomial number of readout components is fundamentally incapable of expressing a non-concentrated target function (i.e., $\mathbb{E}_x [f_*(x)^2] = \Omega(\operatorname{Poly}(N)^{-1})$ as $N\to\infty$). This proves that double descent is not scalable in the number $N$ of qubits if $U(x)$ is sampled from a unitary 2-design. 

\section{Double Descent for Non-Concentrating Quantum Models} \label{appendix:non-concentrating-quantum}

Now that we have treated the concentrating Gaussian 2-design case, we shift our attention to more general models. This requires a more delicate analysis and we rely on the Fourier interpretation of quantum models~\cite{schuld2021effect, mhiri2025constrained} as the basis of our main theoretical framework for the quantum part. To analyse generalization performance in the over-parametrized regime, by Theorem~\ref{thm:risk-bound-nonisotropic} in Section~\ref{sec:non-isotropic-analysis}, it suffices to separately study how input strength (cf. Section~\ref{subsubsec:input-strength}), the number of reuploads (cf. Section~\ref{subsubsec:input-reuploads}) and finite shot noise (cf. Section~\ref{subsec:finite-shot-noise}) each affect the condition number of the quantum population covariance. Regarding the under-parametrized regime, we explicitly show in Section~\ref{subsec:finite-shot-noise} when using finite sampling shot noise guarantees the limited correlations required in Proposition~\ref{prop:over-fitting} in Section~\ref{sec:non-isotropic-analysis}, which ultimately re-establishes the double descent curve in the context of QELM feature maps.

\subsection{Interlude: Fourier Interpretation of Quantum Models} \label{appendix:fourier-interpretation}

\paragraph{Spectral Accumulation along Quantum Circuit Paths.} Recall our unitary model assumption from Eq.~\eqref{eq:data-uploading}, namely that
\begin{equation}
    U(x):= U_{L+1}\prod_{\ell=1}^L \left\{\left( \mathcal{G}_\ell(x) \otimes I_{2^{n_h}} \right)U_\ell \right\}
\end{equation}
We generally suppose ``time-evolution'' encoding of the input data: For each input $x=(x_1,\dots,x_p)\in\mathcal{X}\subseteq\mathbb{R}^p$, we shall make the assumption that, for each reupload $\ell\in[L]$,
\begin{equation}
    \mathcal{G}_\ell(x) = e^{-ix_1H_1^{(\ell)}}\otimes\cdots\otimes e^{-ix_pH_p^{(\ell)}} \label{eq:separable-encoding}
\end{equation}
for some generating Hamiltonians $H_1^{(\ell)},\dots,H_p^{(\ell)}$ each with $H_\eta^{(\ell)}\in\operatorname{Herm}(2^{n_\eta})$ with block sizes $n_\eta\in\mathbb{N}$ acting on the accessible qubits such that
\begin{equation}
    n_a=\sum_{\eta=1}^pn_\eta. \label{eq:n_a-restriction}
\end{equation}
In other words, each generator $H_\eta^{(\ell)}$ is acting on an $n_\eta$-qubit subsystem of the accessible qubits, respectively.

\begin{figure}
    \captionsetup{justification=raggedright, singlelinecheck=false}
    \centering
    \includegraphics[width=0.80\linewidth]{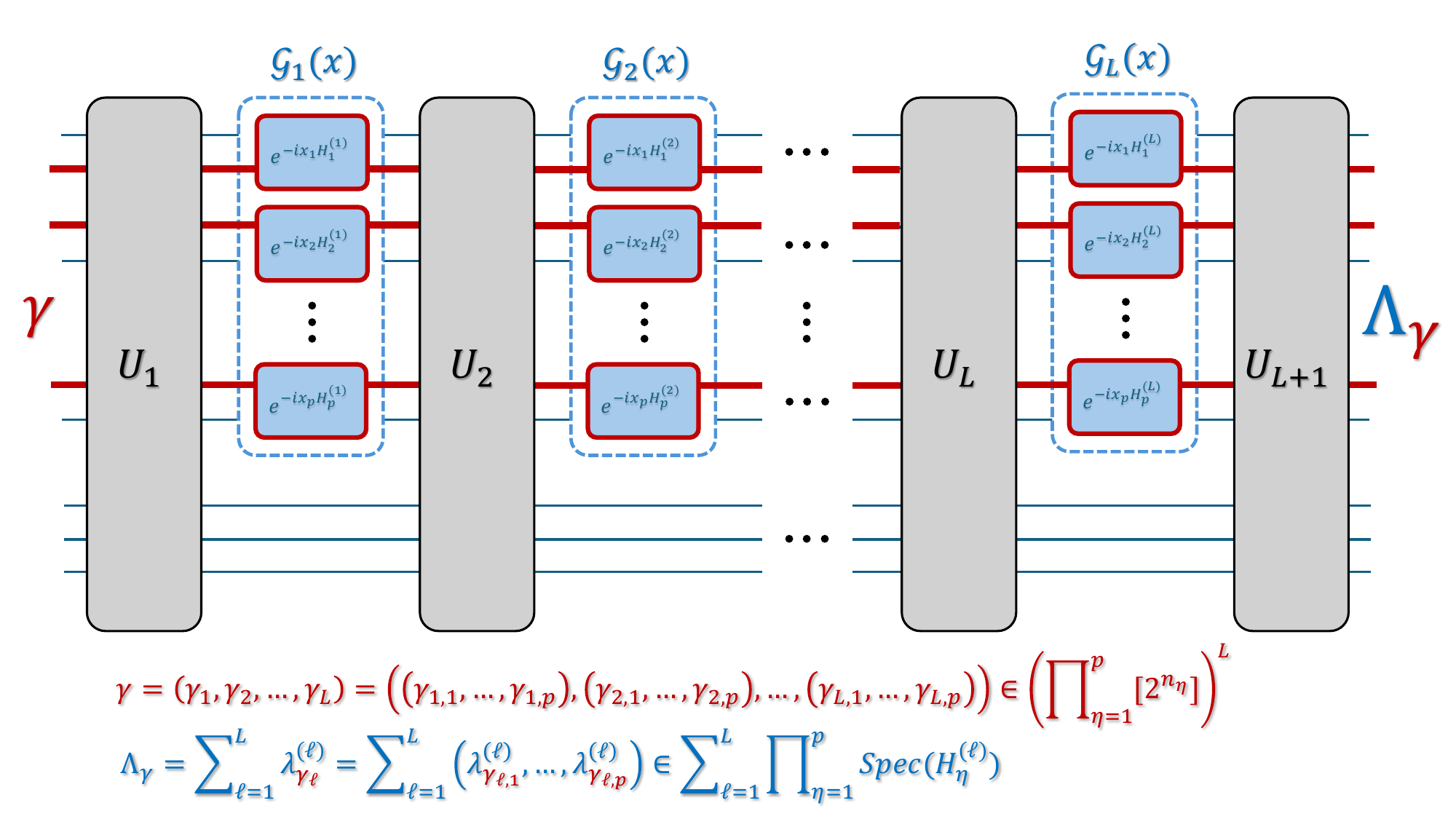}

    \caption{A path $\gamma$ propagates through the accessible qubits of the unitary $U(x)$ in a layer-wise. Along this path, we calculate the vector-valued, spectral accumulation $\Lambda_\gamma$ of the matrix $\{H_\eta^{(\ell)}:(\eta,\ell)\in[p]\times [L]\}$ of generators by summing all eigenvalues $\lambda_{\gamma_{\ell,1}}^{(\ell)}\in\operatorname{Spec}(H_\eta^{(\ell)})$ over the layers $\ell\in[L]$, component-wise in the coordinates $\eta\in[p]$. Notably, the path is used to index this accumulation in a very certain manner.}
    \label{fig:QELM-unitary-path}
\end{figure}

Specifically, in Ref.~\cite{mhiri2025constrained} the authors introduced the quantum spectrum tree which contains of paths representing different choices of eigenvalues in the generator eigenspectrum. We adapt this idea to our more general context and introduce a formalism with paths propagating layer-wise through the input encoding Hamiltonian generators in the quantum circuit.
For each layer $\ell\in[L]$, we first introduce a corresponding path component $\gamma_{\ell} =(\gamma_{\ell,1},\dots,\gamma_{\ell,p})\in[2^{n_1}]\times\cdots\times[2^{n_p}]$ to quantify what will happen in each of these generator blocks. It is worth noting the dimensional consistency
\begin{equation}
    \#([2^{n_1}]\times\cdots\times[2^{n_p}]) = \prod_{\eta=1}^p\#[2^{n_\eta}] = \prod_{\eta=1}^p 2^{n_\eta} = 2^{\sum_{\eta=1}^p n_\eta} = 2^{n_a}, \label{eq:accessible-dimension}
\end{equation}
which is satisfied due to the restriction in Eq.~\eqref{eq:n_a-restriction}. Equivalently, the path component $\gamma_\ell$ exactly covers all possible combinations in the $2^{n_a}$-dimensional Hilbert space of accessible qubits used in the input-reuploading process in the $\ell$-th layer due to the generators $H_1^{(\ell)},\dots, H_p^{(\ell)}$. The rigorous introduction of the layer-wise path components $\gamma_\ell$ allows us define the full path
\begin{equation}
    \gamma := (\gamma_1,\dots,\gamma_L)=((\gamma_{1,1},\dots,\gamma_{1,p}),\dots,(\gamma_{L,1},\dots,\gamma_{L,p})) \in ([2^{n_1}]\times\cdots\times[2^{n_p}])^L \label{eq:general-quantum-spectrum-tree}
\end{equation}
which takes into account all possible paths in the QELM model among all input reuploading layers. This constitutes to indexing a $2^{Ln_a}$-dimensional space by Eq.~\eqref{eq:accessible-dimension}. Along such a path $\gamma$, the key idea is to layer-wise accumulate the discrete eigenspectrum $\operatorname{Spec}(H_\eta^{(\ell)}):=\{\lambda_{j,\eta}^{(\ell)}\}_{j\in[2^{n_\eta}]}$ of the generators $H_1^{(\ell)},\dots, H_p^{(\ell)}$. Without loss of generality, we may assume that all generators $H_\eta^{(\ell)}=\operatorname{diag}\left(\lambda_{1,\eta}^{(\ell)},\dots,\lambda_{2^{n_\eta},\eta}^{(\ell)}\right)$ are diagonal to simplify our formalism. Otherwise, we may absorb the diagonalizing transformation unitaries into the quantum reservoirs. The aforementioned spectral accumulation along the full path $\gamma$ is formally determined by the vector summation
\begin{equation}
    \Lambda_\gamma := \sum_{\ell=1}^L \lambda_{\gamma_\ell}^{(\ell)} \in \sum_{\ell=1}^L\prod_{\eta=1}^p\operatorname{Spec}(H_\eta^{(\ell)})\subset\mathbb{R}^p, \label{eq:generator-spectrum}
\end{equation}
where $\lambda_{\gamma_\ell}^{(\ell)} := (\lambda_{\gamma_{\ell,1}}^{(\ell)},\dots,\lambda_{\gamma_{\ell,p}}^{(\ell)})\in\mathbb{R}^p$ for each reuploading layer $\ell\in[L]$. Intuitively, the vector $\Lambda_\gamma$ captures the total phase accumulation along the path $\gamma$ through all input encoding generators across the $p$ subsystems of size $2^{n_\eta}$. Figure~\ref{fig:QELM-unitary-path} graphically depicts the concept of paths $\gamma$ propagating through the circuit, and how the spectral accumulation of the generators along this path is calculated. Observe that for each layer $\ell\in[L]$, the elements $\gamma_{\ell,\eta}$ path component $\gamma_\ell=(\gamma_{\ell,1},\dots,\gamma_{\ell,p})$ is used specifically to choose one of the $2^{n_\eta}$ possible eigenvalues of the generator $H_\eta^{(\ell)}$. On a high level, all the path $\gamma$ may look like the same as depicted in red in Figure~\ref{fig:QELM-unitary-path}, however, each path-component $\gamma_\ell$ is the key which determines which eigenvalue to choose for each generator in that layer. For a fixed generator $H_\eta^{(\ell)}$, the process of choosing such an eigenvalue is represented as a branch in the previously mentioned quantum spectrum tree~\cite{mhiri2025constrained}. The main technical difference compared to Ref.~\cite{mhiri2025constrained} is the fact that we allow for distinct generators $H_1^{(\ell)},\dots, H_p^{(\ell)}$ in each layer $\ell\in[L]$, which is the reason why we introduced our generalized path framework above.

\smallskip
\par
\paragraph{Fourier Decomposition of the Readout.} According to Refs.~\cite{schuld2021effect, xiong2025fundamental, mhiri2025constrained, wiedmann2025fourier}, the readout vector $\phi_U(x)$ in Eq.~\eqref{eq:quantum-readout} may then be decomposed as a truncated Fourier series of the general form
\begin{equation}
    \phi_U(x) = (\langle O_k\rangle_{x})_{k\in[d]} = \left(\sum_{\omega\in\Omega} c_\omega^k e^{i\omega\cdot x}\right)_{k\in[d]} = \sum_{\omega\in\Omega} c_\omega \, e^{i\omega\cdot x}\in\mathbb{R}^d \label{eq:Fourier-readout}
\end{equation}
for very certain Fourier coefficient vectors $ c_\omega:=( c_\omega^1,\dots, c_\omega^d)\in\mathbb{C}^d$. Here, $\Omega$ denotes the Fourier spectrum with vector-valued frequencies $\omega=(\omega_1,\dots,\omega_p)\in\mathbb{R}^p$ whose geometry is fully determined by the input generator spectrum. More precisely, each Fourier frequency $\omega\in\Omega$ is given by the relative spectral accumulation $\omega=\Lambda_\gamma-\Lambda_{\gamma'}$ for two paths $\gamma,\gamma'\in\Lambda$ in the sense discussed above. In particular $0\in\Omega$ by choosing two identical paths $\gamma=\gamma'$. The Fourier coefficients $ c_\omega^k$, on the other hand, are non-trivially determined by the reservoir architecture $U_1,\dots,U_{L+1}$, the input generators $\{H_\eta^{(1)},\dots,H_\eta^{(L)}\}_{\eta\in[p]}$, and the choice of measurement observables $\{O_k\}_{k\in[d]}$ as shown, for example, in Ref.~\cite{kang2026fourier}. We explicitly demonstrate below that the Fourier coefficients $c^k_\omega$ may be represented as a restricted double path integral with respect to the spectral accumulation paths $\gamma$ introduced above. The double path integral representation is very convenient to identify symmetries among the Fourier coefficients that, in turn, are useful for studying double descent as elaborated subsequently in Sections~\ref{subsubsec:input-strength} and~\ref{subsubsec:input-reuploads}.

If the inputs $\mathcal{X}\subseteq\mathbb{R}$ are scalar (i.e., $p=1$) then Refs.~\cite{schuld2021effect, mhiri2025constrained} explicitly prove Eq.~\eqref{eq:Fourier-readout} for arbitrary $L\geq 1$. Otherwise, if $L=1$, then Ref.~\cite{xiong2025fundamental} provide an explicit proof for arbitrary vector-valued inputs $x$ in the context of QELMs. In order to fully justify our use of the Fourier decomposition, we shall rigorously re-establish the existence of the discrete Fourier decomposition as per Eq.~\eqref{eq:Fourier-readout} in the context of our QELM framework:

\begin{theorem}[Fourier Interpretation of Quantum Readouts] \label{thm:general-Fourier-decomposition}
    Let $\mathcal{X}\subseteq\mathbb{R}^p$ and the number $L\geq1$ of reuploads be arbitrary. For each path $\gamma=(\gamma_1,\dots,\gamma_L)\in([2^{n_1}]\times\cdots\times[2^{n_p}])^L$ consider the operator
    \begin{equation*}
        W_\gamma\equiv W_{\gamma_1,\dots,\gamma_L} := P_{\gamma_L}^{(a)}U_LP_{\gamma_{L-1}}^{(a)} U_{L-1}\cdots P_{\gamma_1}^{(a)}U_1,
    \end{equation*}
    introducing the spectral projector $P_{\gamma_\ell}^{(a)}:=\left(\bigotimes_{\eta=1}^p |\gamma_{\ell,\eta}\rangle\langle\gamma_{\ell,\eta}|\right)\otimes I_{2^{n_h}}$ acting on the $n_a\leq N$ accessible qubits for all layers $\ell\in[L]$, recalling that $N=n_a+n_h$ with $n_h$ denoting the number of hidden qubits. Then, Eq.~\eqref{eq:Fourier-readout} holds true and for all $k\in[d]$ we explicitly have
    \begin{equation}
         c_\omega^k = \sum_{\substack{\gamma,\gamma'\in([2^{n_1}]\times\cdots\times[2^{n_p}])^L \\ \Lambda_{\gamma'}-\Lambda_\gamma=\omega}} \operatorname{Tr}(O_kU_{L+1}W_\gamma\rho_0 W_{\gamma'}^\dagger U_{L+1}^\dagger) \label{eq:Fourier-coefficients-vector-frequencies}
    \end{equation}
    with Fourier frequency space $\Omega = \{\Lambda_{\gamma'}-\Lambda_\gamma:\gamma,\gamma'\in([2^{n_1}]\times\cdots\times[2^{n_p}])^L\}$. In particular, $|\Omega| = \mathcal{O}\left((2^{Ln_a})^2\right)=\mathcal{O}\left(4^{Ln_a}\right)$.
\end{theorem}

\noindent
\emph{Proof.} Let $\ell\in[L]$ be arbitrary. Due to our assumption that $H^{(\ell)}_\eta=\operatorname{diag}(\lambda_{1,\eta}^{(\ell)},\dots,\lambda_{2^{n_\eta},\eta}^{(\ell)})$ is diagonal for all $\eta\in[p]$, we have the identity
\begin{align}
    S_\ell(x) &= \mathcal{G}_\ell(x)\otimes I_{2^{n_h}} \stackrel{\eqref{eq:separable-encoding}}{=} \left(\bigotimes_{\eta=1}^pe^{-ix_\eta H^{(\ell)}_\eta}\right)\otimes I_{2^{n_h}} = \bigotimes_{\eta=1}^p \left(\left\{\sum_{\gamma_{\ell,\eta}=1}^{2^{n_\eta}}e^{-ix_\eta\lambda_{\gamma_{\ell,\eta}}^{(\ell)}}|\gamma_{\ell,\eta}\rangle\langle\gamma_{\ell,\eta}|\right\}\right)\otimes I_{2^{n_h}} \nonumber
    \\ &= \sum_{\gamma_{\ell,1}=1}^{2^{n_1}}\cdots\sum_{\gamma_{\ell,p}=1}^{2^{n_p}}\left(\bigotimes_{\eta=1}^p e^{-ix_\eta\lambda_{\gamma_{\ell,\eta}}^{(\ell)}}|\gamma_{\ell,\eta}\rangle\langle\gamma_{\ell,\eta}|\right)\otimes I_{2^{n_h}} \nonumber
    \\ &= \sum_{\gamma_\ell\in[2^{n_1}]\times\cdots\times[2^{n_p}]} \underbrace{\left(\prod_{\eta=1}^p e^{-ix_\eta\lambda_{\gamma_{\ell,\eta}}^{(\ell)}}\right)}_{=\exp\left(-i\sum_{\eta=1}^px_\eta\lambda_{\gamma_{\ell,\eta}}^{(\ell)}\right)}\underbrace{\left(\bigotimes_{\eta=1}^p|\gamma_{\ell,\eta}\rangle\langle\gamma_{\ell,\eta}|\right)\otimes I_{2^{n_h}}}_{=P_{\gamma_\ell}^{(a)}}  \nonumber
    \\ &= \sum_{\gamma_\ell\in[2^{n_1}]\times\dots\times[2^{n_p}]}e^{-ix\cdot\lambda_{\gamma_\ell}^{(\ell)}} P_{\gamma_\ell}^{(a)}
    \label{eq:fourier-decomposition-step1-generalized}
\end{align}
using the notations $x=(x_1,\dots,x_p)\in\mathcal{X}\subseteq\mathbb{R}^p$, $\gamma_\ell=(\gamma_{\ell,1},\dots,\gamma_{\ell,p})\in[2^{n_1}]\times\cdots\times[2^{n_p}]$ and $\lambda_{\gamma_\ell}^{(\ell)}=(\lambda_{\gamma_{\ell,1}}^{(\ell)},\dots,\lambda_{\gamma_{\ell,p}}^{(\ell})\in\mathbb{R}^p$, respectively.
Accordingly, for each $\eta\in[p]$, the projector $|\gamma_{\ell,\eta}\rangle\langle\gamma_{\ell,\eta}|$ acts on the $2^{n_\eta}$-dimensional subsystem of the accessible qubits so that the total spectral projector $P_{\gamma_\ell}^{(a)}=\left(\bigotimes_{\eta=1}^p|\gamma_{\ell,\eta}\rangle\langle\gamma_{\ell,\eta}|\right)\otimes I_{2^{n_h}}$ acts on the the $n_a$ accessible qubits by Eq.~\eqref{eq:accessible-dimension}. As a result, we may use this identity for each layer $\ell\in[L]$ in order to find
\begin{align}
    U(x) &\stackrel{\eqref{eq:data-uploading}}{=} U_{L+1}\prod_{\ell=1}^L\{S_\ell(x)U_\ell\} \stackrel{\eqref{eq:fourier-decomposition-step1-generalized}}{=} U_{L+1}\prod_{\ell=1}^L \left(\sum_{\gamma_\ell\in[2^{n_1}]\times\dots\times[2^{n_p}]}e^{-ix\cdot\lambda_{\gamma_\ell}^{(\ell)}} P_{\gamma_\ell}^{(a)}U_\ell\right) \nonumber
    \\ &= \sum_{\gamma_1,\dots,\gamma_L\in[2^{n_1}]\times\cdots\times[2^{n_p}]}\underbrace{\left(\prod_{\ell=1}^L e^{-ix\cdot\lambda_{\gamma_\ell}^{(\ell)}}\right)}_{=\exp\left(-ix\cdot\sum_{\ell=1}^L \lambda_{\gamma_\ell}^{(\ell)}\right)}U_{L+1} \underbrace{(P_{\gamma_L}^{(a)}U_L)(P_{\gamma_{L-1}}^{(a)}U_{L-1})\cdots(P_{\gamma_1}^{(a)}U_1)}_{=W_{\gamma_1,\dots,\gamma_L}} \nonumber
    \\ &\stackrel{\eqref{eq:generator-spectrum}}{=} \sum_{\gamma\in([2^{n_1}]\times\cdots\times[2^{n_p}])^L}e^{-ix\Lambda_\gamma} U_{L+1}W_\gamma, \label{eq:fourier-decomposition-step2-generalized}
\end{align}
where we applied the notations $\gamma=(\gamma_1,\dots,\gamma_L)\in([2^{n_1}]\times\cdots\times[2^{n_p}])^L$. Thus,
\begin{align}
     \phi_U(x)_k &= \langle O_k\rangle_x = \operatorname{Tr}(O_k\rho_U(x)) = \operatorname{Tr}(O_kU(x)\rho_0 U(x)^\dagger) \nonumber
     \\ &\stackrel{\eqref{eq:fourier-decomposition-step2-generalized}}{=} \operatorname{Tr}\left(O_k\left\{\sum_{\gamma\in([2^{n_1}]\times\cdots\times[2^{n_p}])^L}e^{-ix\Lambda_\gamma}U_{L+1}W_\gamma\right\}\rho_0\left\{\sum_{\gamma'\in([2^{n_1}]\times\cdots\times[2^{n_p}])^L}e^{ix\Lambda_{\gamma'}}W_{\gamma'}^\dagger U_{L+1}^\dagger\right\}\right) \nonumber
     \\ &= \sum_{\gamma,\gamma'\in([2^{n_1}]\times\cdots\times[2^{n_p}])^L} e^{ix(\Lambda_{\gamma'}-\Lambda_{\gamma})}\operatorname{Tr}(O_kU_{L+1}W_\gamma\rho_0W_{\gamma'}^\dagger U_{L+1}^\dagger) \nonumber
     \\ &= \sum_{\omega\in\Omega}\sum_{\substack{\gamma,\gamma'\in([2^{n_1}]\times\cdots\times[2^{n_p}])^L \\ \Lambda_{\gamma'}-\Lambda_\gamma = \omega}} e^{ix\omega} \operatorname{Tr}(O_k U_{L+1}W_\gamma\rho_0W_{\gamma'}^\dagger U_{L+1}^\dagger) \nonumber
     \\ &= \sum_{\omega\in\Omega}e^{ix\omega} \underbrace{\sum_{\substack{\gamma,\gamma'\in([2^{n_1}]\times\cdots\times[2^{n_p}])^L \\ \Lambda_{\gamma'}-\Lambda_\gamma = \omega}}\operatorname{Tr}(O_k U_{L+1}W_\gamma\rho_0W_{\gamma'}^\dagger U_{L+1}^\dagger)}_{ =c_\omega^k}  =  \sum_{\omega\in\Omega} c_\omega^k e^{ix\omega},
\end{align}
which indeed completes the proof since this holds true regardless of $k\in[d]$.
\qed

It is worthwhile to note that the Fourier frequency space $\Omega = \{\Lambda_{\gamma'} - \Lambda_\gamma : \gamma, \gamma' \in ([2^{n_1}]\times \cdots \times [2^{n_p}])^L\}$ is symmetric under reflection, meaning
\begin{equation}
    -\Omega = \Omega. \label{eq:symmetric-Fourier-frequency-space}
\end{equation}
In other words, we have $-\omega\in\Omega$ for all $\omega\in\Omega$. In particular, we always have $0\in\Omega$. Indeed, let $\omega \in \Omega$. By definition, there exist paths $\gamma, \gamma'$ such that $\omega = \Lambda_{\gamma'} - \Lambda_\gamma$. Inverting the sign gives $-\omega = -(\Lambda_{\gamma'} - \Lambda_\gamma) = \Lambda_{\gamma} - \Lambda_{\gamma'}$. Since $\gamma, \gamma'$ belong to the same path space, swapping their roles shows that $-\omega \in \Omega$. Thus, $-\Omega = \Omega$ as claimed.

Next, we shall prove the following subtle symmetry of the Fourier coefficients which will be useful in our subsequent analysis:

\begin{corollary} \label{corollary:symmetric-Fourier-coefficients}
    For every vector frequency $\omega \in \Omega$, the Fourier coefficient vector $c_\omega = (c_\omega^1, \dots, c_\omega^d)^\top \in \mathbb{C}^d$ defined in Eq.~\eqref{eq:Fourier-coefficients-vector-frequencies} satisfies the symmetry
    \begin{equation*}
        c_{-\omega} = c_\omega^*.
    \end{equation*}
\end{corollary}

\noindent
\emph{Proof.} Let $\omega\in\Omega$ and $k\in[d]$ be arbitrary. Recalling the observables $O_k\in\operatorname{Herm}(2^N)$ and operators $W_\gamma,U_\ell$ appearing inside in the expression of the Fourier coefficients in Eq.~\eqref{eq:Fourier-coefficients-vector-frequencies}
, we use the identity $\operatorname{Tr}(A)^* = \operatorname{Tr}(A^\dagger)$ for any linear operator $A$ in order to compute the adjoint of the operator inside the trace via
\begin{equation}
    \left(O_k U_{L+1} W_\gamma \rho_0 W_{\gamma'}^\dagger U_{L+1}^\dagger\right)^\dagger = (U_{L+1}^\dagger)^\dagger (W_{\gamma'}^\dagger)^\dagger \rho_0^\dagger W_\gamma^\dagger U_{L+1}^\dagger O_k^\dagger 
    = U_{L+1} W_{\gamma'} \rho_0 W_\gamma^\dagger U_{L+1}^\dagger O_k. \label{eq:operator-identity1}
\end{equation}
We use this identity to directly evaluate the complex conjugate of the $k$-th component $c_\omega^k$ as per Eq.~\eqref{eq:Fourier-coefficients-vector-frequencies}:
\begin{align}
    (c_\omega^k)^* &= \sum_{\substack{\gamma,\gamma'\in([2^{n_1}]\times\cdots\times[2^{n_p}])^L \\ \Lambda_{\gamma'}-\Lambda_\gamma=\omega}} \operatorname{Tr}\left(O_k U_{L+1} W_\gamma \rho_0 W_{\gamma'}^\dagger U_{L+1}^\dagger\right)^* \stackrel{\eqref{eq:operator-identity1}}{=} \sum_{\substack{\gamma,\gamma'\in([2^{n_1}]\times\cdots\times[2^{n_p}])^L \\ \Lambda_{\gamma'}-\Lambda_\gamma=\omega}} \operatorname{Tr}\left(U_{L+1} W_{\gamma'} \rho_0 W_\gamma^\dagger U_{L+1}^\dagger O_k\right) \nonumber
    \\ &= \sum_{\substack{\gamma,\gamma'\in([2^{n_1}]\times\cdots\times[2^{n_p}])^L \\ \Lambda_{\gamma'}-\Lambda_\gamma=\omega}} \operatorname{Tr}\left(O_k U_{L+1} W_{\gamma'} \rho_0 W_\gamma^\dagger U_{L+1}^\dagger\right) = \sum_{\substack{\alpha,\alpha'\in([2^{n_1}]\times\cdots\times[2^{n_p}])^L \\ \Lambda_{\alpha'}-\Lambda_\alpha=-\omega}} \operatorname{Tr}\left(O_k U_{L+1} W_{\alpha} \rho_0 W_{\alpha'}^\dagger U_{L+1}^\dagger\right) = c_{-\omega}^k,
\end{align}
where we also employed the cyclic property of the trace in the third equality.
\qed

\subsection{The Effect of Input Strength} \label{subsubsec:input-strength}

In Section~\ref{appendix:fourier-interpretation} we have rigorously covered the fundamental theoretical framework for the quantum part, namely, the Fourier interpretation of the readout $\phi_U(x)$ in Eq.~\eqref{eq:quantum-readout}. This tool allows us to precisely describe the behaviour of the expected risk $\mathcal{R}_U(\gamma)$ in Eq.~\eqref{eq:quantum-risk} in terms of quantities of interest. One such quantity is the \emph{input strength}, also known as ``kernel bandwidth''~\cite{shaydulin2022importance}, which is defined as a fixed scalar $s>0$ that rescales the input via $x\mapsto sx$. This naturally introduces a rescaled version of the expected risk:
\begin{equation}
    \mathcal{R}_U^{(s)}(\gamma) := \mathbb{E}_{( x, y)\sim \mathcal{D}}\left[ (w^\top \phi_U(sx) -  y)^2 \right],
\end{equation}
Focusing on the over-parametrized regime, $\gamma>1$, and leveraging the classical bound in Theorem~\ref{thm:risk-bound-nonisotropic} of Supplementary Section~\ref{appendix:double_descent_linear2}, it suffices to upper bound the condition number of the rescaled population covariance,
\begin{equation}
     \Sigma(s) \equiv  \Sigma_U(s) := \mathbb{E}_{ x}\left[\phi_U(s x)\,\phi_U(s x)^\top\right], \label{eq:rescaled-population-covariance}
\end{equation}
with respect to the input strength. This will particularly help us answer the question of how input strength affects the behaviour of the double descent curve.

\smallskip
\par
\paragraph{The Utility of the Characteristic Function.} Leveraging the Fourier interpretation in Theorem~\ref{thm:general-Fourier-decomposition}, we make the observation that the population covariance is explicitly determined by the Fourier coefficients and the characteristic function (cf. Section~\ref{appendix:Sobolev-spaces}) of the input data:

\begin{lemma} \label{lemma:input-strength-covariance}
    Let $s>0$. Then,
    \begin{equation*}
         \Sigma(s) = \sum_{\omega,\omega'\in\Omega} \varphi_{ x}\left(s(\omega+\omega')\right)   c_\omega  c_{\omega'}^\top
    \end{equation*}
    defines a linear combination of rank-1 matrices $ c_\omega c_{\omega'}^\top$, where $c_\omega=(c_\omega^1,\dots, c_\omega^d)\in\mathbb{C}^d$ are explicitly determined in Eq.~\eqref{eq:Fourier-coefficients-vector-frequencies} and independent of the input strength $s>0$, and $\varphi_{ x}(v) = \mathbb{E}_{ x}[e^{iv\cdot x}]$ denotes the characteristic function of the random vector $ x$.
\end{lemma}

\noindent
\emph{Proof.} Let $k,k'\in[d]$ be arbitrary coordinates of the readout $\phi_U(x)$. Then, by the Fourier interpretation of quantum readouts (cf. Theorem~\ref{thm:general-Fourier-decomposition}), we directly may directly represent the entries
\begin{align}
     \Sigma_{k,k'} &= \mathbb{E}_{ x}\left[(\phi_U(s x)\, \phi_U(s x)^{\top})_{k,k'}\right] = \mathbb{E}_{ x}\left[\phi_U(s x)_k\, \phi_U(s x)_{k'}\right] = \mathbb{E}_{ x}\left[\left(\sum_{\omega\in\Omega} c_\omega^k e^{is\omega\cdot  x}\right)\left(\sum_{\omega'\in\Omega}c_{\omega'}^{k'} e^{is\omega'\cdot  x}\right)\right] \nonumber
    \\ &= \sum_{\omega,\omega'\in\Omega}  c_\omega^k c_{\omega'}^{k'}\,\underbrace{\mathbb{E}_{ x}\left[e^{is(\omega+\omega')\cdot x}\right]}_{\varphi_{ x}(s(\omega+\omega'))} = \sum_{\omega,\omega'\in\Omega} \varphi_{ x}(s(\omega+\omega'))\,  c_\omega^kc_{\omega'}^{k'}.
\end{align}
\qed

The role of the characteristic function is also to precisely quantify the asymptotic behaviour of the scaled population covariance $ \Sigma(s)$ as a function of the input strength $s>0$. As an example, we compare the case of high-dimensional uniform and Gaussian inputs (cf. Table~\ref{tab:characteristic-function} in Section~\ref{appendix:Sobolev-spaces}). For the $p$-dimensional uniform case $ x\sim\operatorname{Unif}([-1,1]^p)$, the characteristic function reads
\begin{align}
    \varphi_{ x}(s(\omega+\omega')) &= \prod_{\eta=1}^p\frac{e^{is(\omega_\eta+\omega'_\eta)}-e^{-is(\omega_\eta+\omega'_\eta)}}{2is(\omega_\eta+\omega'_\eta)} =  \prod_{\eta=1}^p \frac{\sin(s(\omega_\eta+\omega'_\eta))}{s(\omega_\eta+\omega'_\eta)} = \prod_{\eta=1}^p \left( 1 - \frac{s^2(\omega_j+\omega'_j)^2}{6} + \mathcal{O}(s^4) \right) \nonumber
    \\ &= 1 - \sum_{\eta=1}^p \frac{s^2(\omega_\eta+\omega'_\eta)^2}{6} + \mathcal{O}(s^4) = 1 - \frac{s^2}{6} \|\omega+\omega'\|^2 + \mathcal{O}(s^4), \label{eq:characteristic-function-uniform}
\end{align}
applying the approximations $\sin(x) = \frac{e^{ix}-e^{-ix}}{2i}=x-\frac{x^3}{6}+\mathcal{O}(x^5)$ and $\prod_j (1 - \varepsilon_\eta) \approx 1 - \sum_\eta \varepsilon_\eta$ for small $\varepsilon_\eta$. On the other hand, the $p$-dimensional normal case $ x\sim\mathcal{N}(0,I_p)$ yields
\begin{align}
    \varphi_{ x}(s(\omega+\omega')) = e^{-\frac{s^2}{2}\|\omega+\omega'\|^2} = 1 - \frac{s^2}{2}\|\omega+\omega'\|^2 + \mathcal{O}(s^4)
\end{align}
using $e^{-x}=1-x+\mathcal{O}(x^2)$. Comparing the two bounds shows that the characteristic function with Gaussian data converges faster for sufficiently small input strengths because $\frac{1}{2} > \frac{1}{6}$.

Let us proceed by applying this idea to analytically study the scaled population covariance $ \Sigma(s)$ in the limits $s\to0$ and $s\to\infty$, respectively.

\smallskip
\par
\paragraph{Analysis of the Limit $s\to0$.} Recall from Eq.~\eqref{eq:Fourier-readout} that the Fourier readout vector with zero inputs,
\begin{equation}
    \phi_U(0)=\sum_{\omega\in\Omega}  c_\omega,
\end{equation}
is obtained by summing $ c_\omega$ over all Fourier frequencies. From basic probability theory it is well-known that the characteristic function $\varphi_{ x}:\mathbb{R}^p\to[-1,1]$ is uniformly continuous and satisfies $\varphi_{ x}(0)=1$. In particular, by Lemma~\ref{lemma:input-strength-covariance} we have that
\begin{align}
    \lim_{s\to0} \Sigma(s) = \sum_{\omega,\omega'\in\Omega}   c_\omega  c_{\omega'}^\top \underbrace{\lim_{s\to0}\varphi_{ x}\left(s(\omega+\omega')\right)}_{=\varphi_{ x}(0)=1} = \sum_{\omega,\omega'\in\Omega}  c_\omega c_{\omega'}^\top = \left(\sum_{\omega\in\Omega} c_\omega\right)\left(\sum_{\omega'\in\Omega} c_{\omega'}\right)^\top = \phi_U(0)\,\phi_U(0)^\top
\end{align}
collapses to a rank-1 matrix. As a result, $\lambda_{\min}( \Sigma(s))\to0$ and $\lambda_{\max}( \Sigma(s))\to \operatorname{Tr}(\phi_U(0)\phi_U(0)^\top) = \|\phi_U(0)\|^2$ which causes the condition number to explode, i.e., $\kappa( \Sigma(s))\to\infty$ as $s\to0$.

\smallskip
\par
\paragraph{Analysis of the Limit $s\to\infty$.} Next, we consider the other limit. By the Riemann-Lebesgue lemma~\cite{bochner1949fourier} we know that the characteristic function obeys $\varphi_{ x}(v)\to 0$ as $\|v\|\to\infty$. In particular,
\begin{equation}
    \varphi_{ x}(s(\omega+\omega')) \to 0\quad\text{as $s\to\infty$}
\end{equation}
if and only if $\omega+\omega'\neq0$ due to the previous remark that $\varphi_{ x}(0)=1$. This means that only Fourier frequency pairs $(\omega,\omega')$ with $\omega'=-\omega$ contribute to the sum, i.e.,
\begin{align}
     \lim_{s\to \infty} \Sigma(s) = \sum_{\omega,\omega'\in\Omega}   c_\omega  c_{\omega'}^\top \underbrace{\lim_{s\to\infty}\varphi_{ x}\left(s(\omega+\omega')\right)}_{=\delta_{\{\omega+\omega'=0\}}} = \sum_{\omega\in\Omega} c_\omega c_{-\omega}^\top = \sum_{\omega\in\Omega} c_\omega( c_{\omega})^\dagger = CC^\dagger,
\end{align}
where we used the fact that $ c_{-\omega}= c_\omega^*$ by Corollary~\ref{corollary:symmetric-Fourier-coefficients} and introduced the matrix $C\in\mathbb{C}^{d\times|\Omega|}$ collecting the Fourier coefficients obtained by the arranging the vectors $ c_\omega$ as columns. In particular, since $\operatorname{rank}(CC^\dagger)=\operatorname{rank}(C)$, the population covariance is guaranteed to exhibit full rank in the limit $s\to\infty$ if and only if the columns of $C$ span the image space, i.e.,
\begin{equation}
    \operatorname{span}_{\mathbb{C}}\{ c_\omega:\omega\in\Omega\} = \mathbb{C}^d.
\end{equation}
This requires that at least $d$ many elements in $\{ c_\omega\}_{\omega\in\Omega}$ are linearly independent, in particular, we require that $|\Omega|>d$.

In summary, we have established the following theoretical insight:

\begin{proposition} \label{prop:asymptotic-population-covariance}
    Let $s>0$ and $ \Sigma(s)=\mathbb{E}_{ x}\left[\phi_U(s x)\,\phi_U(s x)^\top\right]$ denote the scaled population covariance of the QELM model in Eq.~\eqref{eq:quantum-readout}. Then,
    \begin{equation*}
         \Sigma(s) \to \begin{cases}
            \phi_U(0)\phi_U(0)^\top,&\text{as $s\to0$}, \\ \sum_{\omega\in\Omega} c_\omega c_\omega^\dagger,& \text{as $s\to\infty$},
        \end{cases}
    \end{equation*}
    where $\phi_U(0)=\sum_{\omega\in\Omega}  c_\omega$ with explicitly determined $c_\omega=(c_\omega^1,\dots, c_\omega^d)\in\mathbb{C}^d$ in Eq.~\eqref{eq:Fourier-coefficients-vector-frequencies} and independent of the input strength $s>0$. In particular, the condition number $\kappa( \Sigma(s))=\frac{\lambda_{\max}( \Sigma(s))}{\lambda_{\min}( \Sigma(s))}$ explodes as $s\to 0$ and $\lambda_{\min}( \Sigma(s))>0$ is bounded away from zero as $s\to\infty$ if $\operatorname{span}_{\mathbb{C}}\{ c_\omega:\omega\in\Omega\}=\mathbb{C}^d$.
\end{proposition}

In retrospect, Proposition~\ref{prop:asymptotic-population-covariance} tells us about the explicit structure of the rescaled population covariance $\Sigma(s)$ in the input strength limits $s\to0$ and $s\to\infty$, respectively.  In fact, since we are only interested in the rescaled condition number
\begin{equation}
    \kappa(\Sigma(s)) = \frac{\lambda_{\max}(\Sigma(s))}{\lambda_{\min}(\Sigma(s))}
\end{equation}
of $\Sigma(s)$, it is already sufficient to establish corresponding bounds for the smallest and largest eigenvalues $\lambda_{\min}(\Sigma(s))$ and $\lambda_{\max}(\Sigma(s))=\|\Sigma(s)\|$, respectively.

\subsubsection{Condition Number for Small Input Strengths}

By Proposition~\ref{prop:asymptotic-population-covariance} we already know that $\kappa(\Sigma(s))$ explodes as $s\to0$, however, it does not tell us how fast. By combining the Fourier interpretation of quantum models (cf. Section~\ref{appendix:fourier-interpretation}) with fundamental properties of the characteristic function (cf. Section~\ref{appendix:Sobolev-spaces}) we can quantify this explosion more precisely in what follows.

\smallskip
\par
\paragraph{Lower Bound for $\|\Sigma(s)\|$ as $s\to0$.} Let us begin by lower bounding the largest eigenvalue of $\Sigma(s)$ as the input strength approaches zero. As a corollary of the aforementioned techniques, we may prove the following lower bound:

\begin{corollary} \label{corollary:population-operatornorm-lowerbound}
    Let $s>0$ and suppose that $\phi_U(0)=\sum_{\omega\in\Omega} c_\omega\neq0$. Then, if $\mathbb{E}_x[x]=0$, we have that
    \begin{equation*}
        \| \Sigma(s)\| \geq \left|\left|\sum_{\omega\in\Omega} c_\omega\right|\right|^2 - \mathcal{O}_s(s^2)
    \end{equation*}
    as $s\to0$.
\end{corollary}

\noindent
\emph{Proof.} For any normalized test vector $v \in \mathbb{R}^d$ with $\|v\|=1$, we have
\begin{equation}
    \| \Sigma(s)\| = \sup_{u\in\mathbb{S}^{d-1}}u^\top \Sigma(s)u \geq v^\top  \Sigma(s)v. \label{eq:max-eigenvalue-lowerbound1}
\end{equation}
We choose the test vector aligned with the zero-input readout
\begin{equation}
    v:=\frac{\phi_U(0)}{\|\phi_U(0)\|} = \frac{1}{\|\phi_U(0)\|}\sum_{\omega\in\Omega}  c_\omega\in\mathbb{S}^{d-1}
\end{equation}
as per Eq.~\eqref{eq:Fourier-readout}, where we also made use of the assumption that $\sum_{\omega\in\Omega} c_\omega\neq0$. Starting from the Fourier representation established in Lemma~\ref{lemma:input-strength-covariance},
\begin{equation}
    \Sigma(s) = \sum_{\omega,\omega'\in\Omega} \varphi_{ x}\left(s(\omega+\omega')\right)   c_\omega  c_{\omega'}^\top
\end{equation}
we may use the moments property of the characteristic function (cf. Eq.~\eqref{eq:moments-property} in Section~\ref{appendix:Sobolev-spaces}). Indeed, by directly substituting the exact series expansion from Proposition \ref{prop:covariance-moment-series}, independently shown in Section~\ref{appendix:Fourier-analysis}, into the quadratic form
\begin{align}
    v^\top  \Sigma(s) v &= v^\top \left( \sum_{m\geq0}\frac{(is)^m}{m!}\sum_{\omega,\omega'\in\Omega} M_{ x}^{(m)}(\omega+\omega')\,  c_\omega c_{\omega'}^\top \right) v = \sum_{m\geq0}\frac{(is)^m}{m!}\sum_{\omega,\omega'\in\Omega} M_{ x}^{(m)}(\omega+\omega') (v^\top  c_\omega)( c_{\omega'}^\top v) \nonumber
    \\ &= \frac{1}{\|\phi_U(0)\|^2}\sum_{m\geq0}\frac{(is)^m}{m!}\sum_{\omega,\omega'\in\Omega}M_{ x}^{(m)}(\omega+\omega')\,\pi_\omega\pi_{\omega'} \label{eq:moment-expansion1}
\end{align}
observing that
\begin{equation}
    v^\top c_\omega = \frac{\phi_U(0)^\top  c_\omega}{\|\phi_U(0)\|} = \frac{\pi_\omega}{\|\phi_U(0)\|}
\end{equation}
for all $\omega\in\Omega$.

The matrices $\sum_{\omega,\omega'\in\Omega}M_{ x}^{(m)}(\omega+\omega')\, c_\omega c_{\omega'}^\top$ represent the contribution of the $m$-th order frequency projection moments of the the input data to the population covariance, where, more precisely, $M_{ x}^{(m)}(\omega+\omega')$ defines the $m$-th moment of the projection $(\omega+\omega')\cdot  x$ of the input $ x$ onto the Fourier frequency $\omega+\omega'$, which demonstrates in a very precise manner how the input data interacts with the Fourier frequencies of the QELM model as $s\to0$. Note that it is possible that $\omega+\omega'\notin\Omega$ despite $\omega,\omega'\in\Omega$.

As a result of Eq.~\eqref{eq:max-eigenvalue-lowerbound1}, we thus find
\begin{equation}
    \| \Sigma(s)\| \geq v^\top  \Sigma(s)v \stackrel{\eqref{eq:moment-expansion1}}{=} \frac{1}{\|\phi_U(0)\|^2}\sum_{m\geq0}\frac{(is)^m}{m!} \sum_{\omega,\omega'\in\Omega} M_{ x}^{(m)}(\omega+\omega')\,\pi_\omega\pi_{\omega'}. \label{eq:moment-expansion1i}
\end{equation}
We proceed by noting that, by definition, the zeroth moment is $M_{ x}^{(0)}(\omega+\omega')=\mathbb{E}_{ x}[1]=1$. Thus, the $m=0$ term in Eq.~\eqref{eq:moment-expansion1i} becomes
\begin{align}
    \frac{1}{\|\phi_U(0)\|^2} \sum_{\omega,\omega'\in\Omega} \pi_\omega \pi_{\omega'} &= \frac{1}{\|\phi_U(0)\|^2} \left( \sum_{\omega\in\Omega} \pi_\omega \right) \left( \sum_{\omega'\in\Omega} \pi_{\omega'} \right) = \frac{1}{\|\phi_U(0)\|^2} \left( \phi_U(0)^\top\sum_{\omega\in\Omega}  c_\omega \right) \left(\phi_U(0)^\top\sum_{\omega'\in\Omega}  c_{\omega'} \right)
    \nonumber
    \\ &= \frac{1}{\|\phi_U(0)\|^2} \left(\phi_U(0)^\top \sum_{\omega\in\Omega}  c_\omega\right)^2 = \frac{1}{\|\phi_U(0)\|^2} \left(\phi_U(0)^\top \phi_U(0)\right)^2 \nonumber
    \\ &= \frac{1}{\|\phi_U(0)\|^2}\,\|\phi_U(0)\|^4 = \|\phi_U(0)\|^2,
\end{align}
using $\phi_U(0)=\sum_{\omega\in\Omega} c_\omega$. The first moment is $M_{ x}^{(1)}(\omega+\omega') = \mathbb{E}_{ x}[(\omega+\omega')\cdot x] = (\omega+\omega') \cdot \mathbb{E}_{ x}[ x]$. Since the data is centered ($\mathbb{E}_{ x}[ x] = 0$), this term strictly vanishes in the expansion. The second moment expands via the input covariance matrix $\Sigma_{ x}$ as
\begin{align}
    M_{ x}^{(2)}(\omega+\omega') &= \mathbb{E}_{ x}\left[((\omega+\omega')\cdot x)^2\right] = \mathbb{E}_{ x}[(\omega+\omega')^\top x x^\top(\omega+\omega')] = (\omega+\omega')^\top \mathbb{E}_{ x}[ x x^\top] (\omega+\omega') \nonumber
    \\ &= (\omega+\omega')^\top \Sigma_{ x} (\omega+\omega') = \|\omega+\omega'\|^2_{\Sigma_{ x}}.
\end{align}
Substituting this into the $m=2$ component yields the term
\begin{equation}
    \frac{(is)^2}{2!} \frac{1}{\|\phi_U(0)\|^2} \sum_{\omega,\omega'\in\Omega} \|\omega+\omega'\|^2_{\Sigma_{ x}}\,\pi_\omega\pi_{\omega'} = 
    -\frac{s^2}{2\|\phi_U(0)\|^2} \sum_{\omega,\omega'\in\Omega} \|\omega+\omega'\|^2_{\Sigma_{ x}}\,\pi_\omega\pi_{\omega'}.
\end{equation}
This completes the proof of the corollary.
\qed

It is worth remarking that the only assumption $\phi_U(0)=\sum_{\omega\in\Omega}c_\omega=0$ made to prove this bound is only violated in very special cases for quantum architectures with perfect symmetry. By slightly modifying the architecture one may easily break this symmetry to ensure the validity of the assumption. Furthermore, Proposition~\ref{prop:covariance-moment-series} in Section~\ref{appendix:Sobolev-spaces} indeed generalizes the asymptotic limit $s\to0$ of Proposition~\ref{prop:asymptotic-population-covariance}. To see this, observe that
\begin{align}
    \lim_{s\to0} \Sigma(s) &= \lim_{s\to0}\sum_{m\geq0}\frac{(is)^m}{m!}\sum_{\omega,\omega'\in\Omega} M_{ x}^{(m)}(\omega+\omega') \, c_\omega c_{\omega'}^\top = \sum_{\omega,\omega'\in\Omega}\underbrace{M_{ x}^{(0)}(\omega+\omega')}_{=1}\, c_\omega c_{\omega'}^\top \nonumber 
    \\ &= \sum_{\omega,\omega'\in\Omega}\, c_\omega c_{\omega'}^\top = \left(\sum_{\omega\in\Omega} c_\omega\right)\left(\sum_{\omega'\in\Omega} c_{\omega'}\right)^\top = \phi_U(0)\phi_U(0)^\top,
\end{align}
recalling that $\phi_U(0)=\sum_{\omega\in\Omega} c_\omega$.

\smallskip
\par
\paragraph{Upper Bound for $\lambda_{\min}(\Sigma(s))$ as $s\to0$.} Proposition~\ref{prop:covariance-moment-series} in Section~\ref{appendix:Sobolev-spaces} is furthermore useful to obtain upper bounds of the smallest eigenvalue $\lambda_{\min}( \Sigma(s))$ in the asymptotic neighborhood $s\to0$, as demonstrated by the next corollary:

\begin{corollary} \label{corr:upper-min-eigenvalue-bound-population}
    Let $s>0$ and suppose that $\phi_U(0)=\sum_{\omega\in\Omega} c_\omega\neq0$. Then, if $\mathbb{E}_x[x]=0$, we have that $\lambda_{\min}( \Sigma(s)) = \mathcal{O}_s(s^2)$ is decays to zero as $s\to0$.
\end{corollary}

\noindent
\emph{Proof.} Let $s>0$ be arbitrary. Since $ \Sigma(s)$ defines a symmetric positive semi-definite matrix, we have the bound
\begin{equation}
    \lambda_{\min}( \Sigma(s)) = \inf_{u\in\mathbb{S}^{d-1}} u^\top  \Sigma(s)u \leq v^\top  \Sigma(s)v
\end{equation}
for all unit vectors $v\in\mathbb{S}^{d-1}$. Since any specific choice of a unit vector provides a valid upper bound for $\lambda_{\min}( \Sigma(s))$, we restrict our choice to the subspace of unit vectors orthogonal to the zero-input readout vector $\phi_U(0)=\sum_{\omega\in\Omega}c_\omega$. That is, let $v\in\mathbb{S}^{d-1}$ such that
\begin{equation}
    0=\langle v,\phi_U(0)\rangle = v^\top \phi_U(0) = v^\top \sum_{\omega\in\Omega}  c_\omega = \sum_{\omega\in\Omega} \underbrace{v^\top c_\omega}_{=:v_\omega} = \sum_{\omega\in\Omega} v_\omega \label{eq:orthogonality-condition1}
\end{equation}
Substituting Proposition \ref{prop:covariance-moment-series} in Section~\ref{appendix:Sobolev-spaces} into the variational equation using this test vector $v$ gives the upper bound
\begin{align}
    \lambda_{\min}( \Sigma(s)) \leq v^\top  \Sigma(s) v &= v^\top \left( \sum_{m\geq0}\frac{(is)^m}{m!}\sum_{\omega,\omega'\in\Omega} M_{ x}^{(m)}(\omega+\omega')\,c_\omega c_{\omega'}^\top \right)v \nonumber
    \\ &= \sum_{m\geq0}\frac{(is)^m}{m!}\sum_{\omega,\omega'\in\Omega} M_{ x}^{(m)}(\omega+\omega') \underbrace{(v^\top  c_\omega)( c_{\omega'}^\top v)}_{=v_\omega v_{\omega'}} = \sum_{m\geq0}\frac{(is)^m}{m!}\sum_{\omega,\omega'\in\Omega} M_{ x}^{(m)}(\omega+\omega')\,v_\omega v_{\omega'}. \label{eq:upper-bound-expansion1}
\end{align}
Taking the infimum over all $v\in\mathbb{S}^{d-1}$ with $\langle v,\phi_U(0)\rangle=0$ yields
\begin{equation}
    \lambda_{\min}( \Sigma(s)) \leq \inf_{\substack{v\in\mathbb{S}^{d-1}\\\langle v,\phi_U(0)\rangle=0}}\sum_{m\geq0}\frac{(is)^m}{m!}\sum_{\omega,\omega'\in\Omega}M_{ x}^{(m)}(\omega+\omega')\,v_\omega v_{\omega'}.
\end{equation}
Now we evaluate the individual terms of the expansion as $s\to0$. First, we know that the zeroth frequency projection moment equals $M_{ x}^{(0)}(\omega+\omega')= \mathbb{E}_{ x}[1]=1$. Therefore, the constant term $m=0$ in the expansion in Eq.~\eqref{eq:upper-bound-expansion1} simplifies via
\begin{align}
    \sum_{\omega,\omega'\in\Omega}\underbrace{M_{ x}^{(0)}(\omega+\omega')}_{=1}\,v_\omega v_{\omega'} = \sum_{\omega,\omega'\in\Omega}v_\omega v_{\omega'} = \left(\sum_{\omega\in\Omega} v_\omega\right)^2 \stackrel{\eqref{eq:orthogonality-condition1}}{=}0.
\end{align}
Here we made use of the orthogonality restriction in Eq.~\eqref{eq:orthogonality-condition1}. Next, because of the assumption that $\mathbb{E}_{ x}[ x]=0$ we find $M_{ x}^{(1)}(\omega+\omega') = (\omega+\omega') \cdot \mathbb{E}_{ x}[ x]=0$ so that the linear term $m=1$ also vanishes in the expansion in Eq.~\eqref{eq:upper-bound-expansion1}. Finally, we evaluate the quadratic term $m=2$ in the expansion:
\begin{align}
    \sum_{\omega,\omega'\in\Omega}\underbrace{M_{ x}^{(2)}(\omega+\omega')}_{\|\omega+\omega'\|^2_{\Sigma_{ x}}}\,v_\omega v_{\omega'} &= \sum_{\omega,\omega'\in\Omega} (\omega+\omega')^\top \Sigma_{ x} (\omega+\omega')\,v_\omega v_{\omega'} =  \sum_{\omega,\omega'\in\Omega} \left( \omega^\top \Sigma_{ x} \omega + 2\omega^\top \Sigma_{ x} \omega' + \omega'^\top \Sigma_{ x} \omega' \right) \, v_\omega v_{\omega'} \nonumber
    \\ &= \sum_{\omega\in\Omega} v_\omega\,\omega^\top \Sigma_{ x}\omega \underbrace{\sum_{\omega'\in\Omega}v_{\omega'}}_{\stackrel{\eqref{eq:orthogonality-condition1}}{=}0} + 2\sum_{\omega,\omega'\in\Omega}v_\omega v_{\omega'}\,\omega^\top \Sigma_{ x} \omega' + \sum_{\omega'\in\Omega} v_{\omega'}\, \omega'^\top \Sigma_{ x} \omega' \underbrace{\sum_{\omega\in\Omega} v_{\omega}}_{\stackrel{\eqref{eq:orthogonality-condition1}}{=}0} \nonumber
    \\ &= 2\sum_{\omega,\omega'\in\Omega}v_\omega v_{\omega'}\,\omega^\top\Sigma_{ x}\omega' = 2\left( \sum_{\omega\in\Omega} v_\omega\, \omega \right)^\top \Sigma_{ x} \left( \sum_{\omega'\in\Omega} v_{\omega'} \,\omega' \right) = 2 \left\|\sum_{\omega\in\Omega}v_\omega\,\omega\right\|^2_{\Sigma_{ x}}. \label{eq:second-moment-term1}
\end{align}
again applying the orthogonality restriction in Eq.~\eqref{eq:orthogonality-condition1}. We continue by examining the vector $\sum_{\omega\in\Omega}v_\omega\,\omega$ in more detail. By Corollary~\ref{corollary:symmetric-Fourier-coefficients} in Section~\ref{appendix:fourier-interpretation} we know that $c_\omega^* = c_{-\omega}$ for all $\omega\in\Omega$. This immediately implies that
\begin{equation}
    v_\omega^* = (v^\top c_\omega)^* = v^\top(c_\omega)^* = v^\top c_{-\omega} = v_{-\omega} \label{eq:symmetric-weights1}
\end{equation}
since $v\in\mathbb{S}^{d-1}$ is real. By Eq.~\eqref{eq:symmetric-Fourier-frequency-space} we know that the negation $\omega\mapsto-\omega$ is well-defined for all Fourier frequencies $\omega\in\Omega$. In particular, under negation $\omega \mapsto -\omega$, Eq.\eqref{eq:symmetric-weights1} yields the parity properties $\operatorname{Re}(v_{-\omega}) = \operatorname{Re}(v_\omega)$ and $\operatorname{Im}(v_{-\omega}) = -\operatorname{Im}(v_\omega)$ thus implying
\begin{equation}
    \sum_{\omega\in\Omega}\operatorname{Re}(v_\omega)\,\omega = 0
\end{equation}
by the symmetry $-\Omega=\Omega$ so that, as a result,
\begin{equation}
    \sum_{\omega\in\Omega}v_\omega\,\omega = \sum_{\omega\in\Omega} \left(\operatorname{Re}(v_\omega) + i\operatorname{Im}(v_\omega)\right)\omega = \sum_{\omega\in\Omega}\operatorname{Re}(v_\omega)\,\omega + i\sum_{\omega\in\Omega}\operatorname{Im}(v_\omega)\,\omega = i\sum_{\omega\in\Omega}\operatorname{Im}(v_\omega)\,\omega\in i\mathbb{R}^p \label{eq:purely-imaginary1}
\end{equation}
is purely imaginary. Substituting back into Eq.~\eqref{eq:second-moment-term1}, and recalling that the $m=0,1$ terms vanish, results in the conclusion that
\begin{align}
    \lambda_{\min}( \Sigma(s)) &\stackrel{\eqref{eq:upper-bound-expansion1}}{\leq} \sum_{m\geq0}\frac{(is)^m}{m!}\sum_{\omega,\omega'\in\Omega} M_{ x}^{(m)}(\omega+\omega')\,v_\omega v_{\omega'} = \frac{(is)^2}{2!}\sum_{\omega,\omega'\in\Omega} M_{ x}^{(2)}(\omega+\omega')\,v_\omega v_{\omega'} + \mathcal{O}(s^3) \nonumber
    \\ &\stackrel{\eqref{eq:second-moment-term1}}{=} -\frac{s^2}{2}\cdot 2 \left\|\sum_{\omega\in\Omega}v_\omega\,\omega\right\|^2_{\Sigma_{ x}} + \mathcal{O}(s^3) \stackrel{\eqref{eq:purely-imaginary1}}{=} s^2 \underbrace{\left\|\sum_{\omega\in\Omega}\operatorname{Im}(v_\omega)\,\omega\right\|_{\Sigma_{ x}}^2}_{=\mathcal{O}_s(1)} = \mathcal{O}_s(s^2).
\end{align}
as $s\to0$, thus concluding the proof.
\qed

It is possible to relax the assumption that $x\sim\mathbb{P}$ is centered at the cost of making proof more complicated. Nevertheless, any random input may always be assumed to be centered by an affine transformation. Thus, it is not restrictive for practical applications. We are now ready to lower bound the condition number of the rescaled population covariance $\Sigma(s)$ in the small input strength neighborhood:

\begin{theorem}
    Let $s>0$ and suppose that $\phi_U(0)=\sum_{\omega\in\Omega} c_\omega\neq0$. Then, if $\mathbb{E}_x[x]=0$ and
    $$\inf_{\substack{v\in\mathbb{S}^{d-1} \\ \langle v,\phi_U(0)\rangle=0}} \left\|\sum_{\omega\in\Omega}\operatorname{Im}(v^\top c_\omega)\,\omega\right\|_{\Sigma_{ x}}^2 > 0,$$
    we have that the condition number $\kappa(\Sigma(s)) = \Omega(s^{-2})$ diverges to infinity as $s\to 0$.
\end{theorem}

\noindent
\emph{Proof.} Let $s>0$ be arbitrary. By the proofs of Corollaries~\ref{corollary:population-operatornorm-lowerbound} and~\ref{corr:upper-min-eigenvalue-bound-population} we have the bounds
\begin{equation}
    \|\Sigma(s)\| \geq \|\phi_U(0)\|^2 - C_1s^2 + \mathcal{O}(s^3)\quad\text{and}\quad \lambda_{\min}(\Sigma(s)) \leq C_0 s^2 + \mathcal{O}(s^3) \label{eq:s-zero-bounds}
\end{equation}
as $s\to0$, where the constants
\begin{equation}
    C_0 := \inf_{\substack{v\in\mathbb{S}^{d-1} \\ \langle v,\phi_U(0)\rangle=0}} \left\|\sum_{\omega\in\Omega}\operatorname{Im}(v^\top c_\omega)\,\omega\right\|_{\Sigma_{ x}}^2 \quad\text{and}\quad C_1 := \frac{1}{2\|\phi_U(0)\|^2}\sum_{\omega,\omega'\in\Omega}\|\omega+\omega'\|^2_{\Sigma_{ x}}\,\pi_\omega\pi_{\omega'} \label{eq:constants-input-strength-proof}
\end{equation}
are positive and independent of the input strength $s>0$. By combining these bounds, we find
\begin{align}
    \kappa(\Sigma(s)) &= \frac{\|\Sigma_U(s)\|}{\lambda_{\min}(\Sigma_U(s))} \stackrel{\eqref{eq:s-zero-bounds}}{\geq} \frac{\|\phi_U(0)\|^2 - C_1s^2 + \mathcal{O}(s^3)}{C_0s^2 + \mathcal{O}(s^3)} = \frac{\|\phi_U(0)\|^2 \left(1 - \frac{C_1s^2}{\|\phi_U(0)\|^2} + \mathcal{O}(s^3)\right)}{C_0s^2 \left(1 + \frac{\mathcal{O}(s^3)}{C_0s^2}\right)} = \frac{\|\phi_U(0)\|^2}{C_0s^2} \, \frac{1 + \mathcal{O}(s^2)}{1 + \mathcal{O}(s)} \nonumber \\
    &= \frac{\|\phi_U(0)\|^2}{C_0s^2} (1 + \mathcal{O}(s^2))(1 + \mathcal{O}(s)) = \frac{\|\phi_U(0)\|^2}{C_0s^2} (1 + \mathcal{O}(s))= \frac{\|\phi_U(0)\|^2}{C_0s^2} + \mathcal{O}(s^{-1}) = \Omega(s^{-2}),
\end{align}
where we applied the Taylor expansion $(1+x)^{-1} = 1 + \mathcal{O}(x)$ as $x\to0$ to the denominator's parenthesis, and used the facts that $\|\phi_U(0)\|^2,C_0,C_1 \in \{\Theta_s(1)\}$ are positive constants independent of $s$. Since the leading term scales as $s^{-2}$ and dominates as $s \to 0$, this yields the desired lower bound $\Omega(s^{-2})$ and completes the proof.
\qed

We again highlight the fact that the assumption of centered inputs $x\sim\mathbb{P}$ may be relaxed at a mild cost: If the inputs are not centered, then it may happen that the constant $C_0=0$ vanishes in Eq.~\eqref{eq:constants-input-strength-proof} above. Hence, we see from Eq.~\eqref{eq:s-zero-bounds} that the smallest eigenvalue $\lambda_{\min}(\Sigma(s))=o(s^2)$ may decays faster than $s^2$ as $s\to0$. This means that one has to do a more refined analysis by identifying the smallest, non-zero constant $C_3>0$ in the moments expansion in the proof of Corollary~\ref{corr:upper-min-eigenvalue-bound-population} which will ultimately result in a faster explosion $\kappa(\Sigma(s))=\Omega(s^{-P})$ for some $P>2$ as $s\to0$.

\subsubsection{Condition Number for Large Input Strengths}

Since $\Sigma(0)$ is singular being a rank-1 matrix, meaning that the condition number $\kappa(\Sigma(s))$ explodes as $s\to\infty$, the more interesting case is the limit $s\to\infty$ because this is the only way the condition number could improve with the input strength. Instead of studying the limit, we are more interested in an approximation of the form, as $s\to\infty$
\begin{equation}
    \Sigma(s) = \sum_{\omega\in\Omega}c_\omega c_\omega^\dagger + o_s(1)
\end{equation}
with some vanishing error term $o_s(1)$ that we would like to quantify more precisely. Once again, it suffices to bound the smallest and the largest eigenvalues of $\Sigma(s)$ since we are only interested in upper bounding the rescaled population covariance in the large input strength limit. However, the analysis is more technical and we need to leverage tools from Sobolev theory (cf. Section~\ref{appendix:Sobolev-spaces}).

\smallskip
\par
\paragraph{The Asymmetric Remainder $\Delta(s)$.} In order to show that the population covariance improves as $s\to\infty$, particularly by lower bounding $\lambda_{\min}( \Sigma(s))$ away from zero, we shall again employ Lemma~\ref{lemma:input-strength-covariance} as a basis for this refined analysis:
\begin{equation}
     \Sigma(s) = \sum_{\omega,\omega'\in\Omega} \varphi_{ x}\left(s(\omega+\omega')\right)  c_\omega  c_{\omega'}^\top.
\end{equation}
From here, we continue by making use of the Fourier symmetry in Corollary~\ref{corollary:symmetric-Fourier-coefficients} which states $c_{-\omega}^\top = (c_{\omega}^*)^\top = c_\omega^\dagger$ for all $\omega\in\Omega$ so that
\begin{align}
     \Sigma(s) &= \sum_{\omega,\omega'\in\Omega} \varphi_{ x}\left(s(\omega+\omega')\right)  c_\omega  c_{\omega'}^\top = \sum_{\substack{\omega,\omega'\in\Omega \\ \omega + \omega' = 0}} \underbrace{\varphi_{ x}\left(s(\omega+\omega')\right)}_{=\varphi_{ x}(0)=1}  c_\omega  c_{\omega'}^\top + \sum_{\substack{\omega,\omega'\in\Omega\\ \omega + \omega'\neq0}} \varphi_{ x}\left(s(\omega+\omega')\right)  c_\omega  c_{\omega'}^\top \nonumber
    \\ &= \sum_{\omega\in\Omega}c_{\omega}\underbrace{c_{-\omega}^\top}_{= c_\omega^\dagger} + \,\Delta(s) = \sum_{\omega\in\Omega} c_\omega c_\omega^\dagger + \Delta(s) \label{eq:population-covariance-decomposition}
\end{align}
for all input strengths $s>0$, recalling $\Omega=-\Omega$ by Eq.~\eqref{eq:symmetric-Fourier-frequency-space}, where we also introduced the asymmetric remainder term
\begin{equation}
    \Delta(s) := \sum_{\substack{\omega,\omega'\in\Omega\\ \omega + \omega'\neq0}} \varphi_{ x}\left(s(\omega+\omega')\right)  c_\omega  c_{\omega'}^\top. \label{eq:asymmetric-remainder-term}
\end{equation}
According to Proposition~\ref{prop:asymptotic-population-covariance} in Section~\ref{appendix:Sobolev-spaces} we can already tell that the asymmetric remainder must vanish as $s\to\infty$, i.e., $\|\Delta(s)\|\to 0$. However, the goal of the following refined analysis is to quantify precisely how fast this convergence happens in terms of increasing input strength $s>0$.

First, we demonstrate that it is possible to explicitly compute $\Delta(s)$ in Eq.~\eqref{eq:asymmetric-remainder-term} by a case-by-case analysis for each input distribution. For example, when considering high-dimensional uniform inputs $ x\sim\operatorname{Unif}([-1,1]^p)$. In that case, with the help of Table~\ref{tab:characteristic-function} in Section~\ref{appendix:Sobolev-spaces}, we have for all $\omega,\omega'\in\Omega$ with $\omega+\omega'\neq 0$ that
\begin{align}
    \varphi_{ x}(s(\omega+\omega')) &= \prod_{\eta=1}^p \frac{\sin(s(\omega_\eta+\omega'_\eta))}{s(\omega_\eta+\omega'_\eta)}  \leq \max_{\substack{\omega,\omega'\in\Omega \\ \omega+\omega'\neq 0}}\prod_{\eta=1}^p\frac{|\sin(s(\omega_\eta+\omega'_\eta))|}{s|\omega_\eta+\omega'_\eta|} = \mathcal{O}_s\left(\max_{\substack{\omega,\omega'\in\Omega \\ \omega+\omega'\neq 0}}{s^{-P(\omega,\omega')}}\right)
\end{align}
with polynomial degree $P(\omega,\omega'):=\#\{\eta\in[p]:\omega_\eta+\omega'_\eta\neq0\}\leq p$ notably being non-zero due to $\omega+\omega'\neq0$. Indeed, if $\omega_\eta+\omega'_\eta=0$ we need to employ the limit $\frac{\sin(x)}{x}\to1$ as $x\to0$ and only the factors with $\omega_\eta+\omega'_\eta\neq0$ remain in the product. For these, we use $|\sin(x)|\leq 1$ and the fact that $\omega,\omega'\in\Omega$ are independent of $s>0$.

Therefore, since $c_\omega$ is independent of $s$ for all $\omega\in\Omega$, we have the upper bound
\begin{equation}
    \|\Delta(s)\| \leq \underbrace{|\Omega|^2\max_{\substack{\omega,\omega'\in\Omega\\ \omega+\omega'\neq0}}\|c_\omega c_{\omega'}^\top\|}_{\text{independent of $s$}}\cdot|\varphi_{ x}(s(\omega+\omega'))| = \mathcal{O}_s\left(\max_{\substack{\omega,\omega'\in\Omega\\ \omega+\omega'\neq0}}{s^{-P(\omega,\omega')}}\right) \to 0
\end{equation}
as $s\to\infty$ since $P(\omega,\omega')\neq0$, thus confirming convergence to zero.

On the other hand, if $x\sim\mathcal{N}(0,I_p)$ is normal, then again using Table~\ref{tab:characteristic-function} yields
\begin{equation}
    \|\Delta(s)\| = C_{\Omega,U}\max_{\omega,\omega'\in\Omega}|\varphi_{ x}(s(\omega+\omega'))| = \mathcal{O}_s\left(\max_{\substack{\omega,\omega'\in\Omega\\ \omega+\omega'\neq0}}e^{-\frac{s^2}{2}\|\omega+\omega'\|^2}\right) \to 0 \label{eq:Gaussian-decay1}
\end{equation}
as $s\to\infty$ for some constant $C_{\Omega,U}>0$ that is independent of the input strength $s$. This example demonstrates faster convergence for normal inputs as $s\to\infty$ compared to the previous uniform case.

\smallskip
\par
\paragraph{Upper Bound for $\|\Sigma(s)\|$ as $s\to\infty$.} It turns out that under certain regularity assumptions on the probability density of the input $x$, one may prove a very explicit formula for the asymmetric remainder term $\Delta(s)$ in Eq.~\eqref{eq:asymmetric-remainder-term} which, in turn, precisely quantifies the speed of the convergence $\|\Delta(s)\|\to0$. In fact, the more regular the density of $ x$, the faster the convergence which is formally characterized by the notion of Sobolev spaces (cf. Definition~\ref{def:Sobolev-space} in Section~\ref{appendix:Sobolev-spaces}). The important consequence is the following bound:

\begin{corollary} \label{corr:input-strength-upperbound-infty}
    Let $s>0$. Assume that the input $ x$ has a density function $f( x)$ belonging to the $K$-th order Sobolev space $W^{K,1}(\mathcal{X})$ for some $K\in\mathbb{N}$. Then, we have that
    \begin{equation*}
        \|\Sigma(s)\| \leq \left\|\sum_{\omega\in\Omega}c_\omega c_{\omega}^\dagger\right\| + \mathcal{O}_s\left(s^{-K}\right)
    \end{equation*}
    as $s\to\infty$.
\end{corollary}

\noindent
\emph{Proof.} Let $s>0$ and $\omega,\omega'\in\Omega$ be arbitrary with $\omega+\omega'\neq0$. From Eq.~\eqref{eq:population-covariance-decomposition} we recall that
\begin{equation}
     \Sigma(s) = \sum_{\omega\in\Omega}c_\omega c_\omega^\dagger + \Delta(s)\quad\text{with}\quad \Delta(s) = \sum_{\substack{\omega,\omega'\in\Omega\\ \omega + \omega'\neq0}} \varphi_{ x}\left(s(\omega+\omega')\right)  c_\omega  c_{\omega'}^\top. \label{eq:population-decomposition1}
\end{equation}
By using the assumption that $f\in W^{K,1}(\mathcal{X})$, Proposition~\ref{prop:regularity-bound}, shown independently with the help of Sobolev techniques in Supplementary Section~\ref{appendix:Sobolev-spaces}, implies the refined representation
\begin{equation}
    \Delta(s) = \sum_{\substack{\omega,\omega'\in\Omega \\ \omega+\omega'\neq0}}\left(\frac{i}{s\|\omega+\omega'\|^2}\right)^K \int_{\mathcal{X}}e^{is(\omega+\omega')\cdot x}D_{\omega+\omega'}^K f( x)\,d x\,c_\omega c_{\omega'}^\top
\end{equation}
with directional derivatives defined as
\begin{equation}
    D_{\omega+\omega'}^K f( x) := \sum_{\eta_1,\dots,\eta_K=1}^p \prod_{\kappa =1}^K (\omega_{\eta_\kappa}+\omega'_{\eta_\kappa})\frac{\partial^K f( x)}{\partial_{ x_{\eta_1}}\cdots\,\,\partial_{ x_{\eta_K}}}
\end{equation}
for each $\omega,\omega'\in\Omega$ with $\omega+\omega'\neq0$, respectively. Therefore, we may directly bound using the triangle inequality,
\begin{align}
    \|\Delta(s)\| &\leq \sum_{\substack{\omega,\omega'\in\Omega \\ \omega+\omega'\neq0}} \left| \frac{i}{s\|\omega+\omega'\|^2} \right|^K  \int_{\mathcal{X}}\underbrace{|e^{is(\omega+\omega')\cdot x}|}_{=1}\left|D_{\omega+\omega'}^K f( x)\,d x\right|\, \|c_\omega c_{\omega'}^\top\| \nonumber
    \\ &\leq \frac{1}{s^K} \underbrace{\sum_{\substack{\omega,\omega'\in\Omega \\ \omega+\omega'\neq 0}} \frac{1}{\|\omega+\omega'\|^{2K}} \int_{\mathcal{X}}  \left| D_{\omega+\omega'}^K f( x) \right|\,d x \, \|c_\omega c_{\omega'}^\top\|}_{\text{independent of $s$}} = \mathcal{O}_s\left(s^{-K}\right) \label{eq:final-input-strength-bound}
\end{align}
as $s\to\infty$, and the corollary follows because of Eq.~\eqref{eq:population-decomposition1}, i.e.,
\begin{equation}
    \|\Sigma(s)\| \leq \left\|\sum_{\omega\in\Omega}c_\omega c_\omega^\dagger\right\| + \|\Delta(s)\| \stackrel{\eqref{eq:final-input-strength-bound}}{=} \left\|\sum_{\omega\in\Omega}c_\omega c_\omega^\dagger\right\| + \mathcal{O}_s\left(s^{-K}\right).
\end{equation}
\qed

Note that if we suppose $ x$ to have a Schwartz regular density $f\in\mathcal{S}(\mathcal{X})$, where
\begin{equation}
    \mathcal{S}(\mathcal{X}):=\left\{g\in\mathcal{C}^\infty(\mathcal{X}):\sup_{x\in\mathcal{X}}|x^\alpha(D_x^\beta g)(x)|<\infty\text{ for all $\alpha,\beta\in\mathbb{N}_0^p$}\right\},
\end{equation}
such as a Gaussian, then
\begin{equation}
    f\in\bigcap_{K\in\mathbb{N}}W^{K,1}(\mathcal{X})
\end{equation}
automatically defines a Sobolev function for all orders $K$. As a result, Proposition~\ref{prop:regularity-bound} in Section~\ref{appendix:Sobolev-spaces} implies that $\|\Delta(s)\| = \mathcal{O}_s(s^{-K})$ for all $K\in\mathbb{N}$ and thus, in particular, $\|\Delta(s)\| = \mathcal{O}_s(\operatorname{SuperPoly}(s)^{-1})$ decays super-polynomially in the input strength. This observation is consistent with our previous example in Eq.~\eqref{eq:Gaussian-decay1}.

\smallskip
\par
\paragraph{Lower Bound for $\lambda_{\min}(\Sigma(s))$ as $s\to\infty$.} By leveraging the exact decomposition $ \Sigma(s) = \sum_{\omega\in\Omega} c_\omega c_\omega^\dagger + \Delta(s)$ in Eq.~\eqref{eq:population-covariance-decomposition} and the bound of the asymmetric remainder term $\Delta(s)$ as per Corollary~\ref{corr:input-strength-upperbound-infty} above, it is straightforward to derive a lower bound for $\lambda_{\min}( \Sigma(s))$ in the asymptotic regime $s\to\infty$ as it converges to the limiting eigenvalue $\lambda_{\min}\left(\sum_{\omega\in\Omega} c_\omega c_\omega^\dagger\right)$.

\begin{corollary} \label{corr:lower-eigenvalue-bound-population}
Let $s>0$ and assume that the input $ x$ has a density function $f( x)$ belonging to the $K$-th order Sobolev space $W^{K,1}(\mathcal{X})$ for some $K\in\mathbb{N}$. Then, we have the lower bound
\begin{equation*}
    \lambda_{\min}( \Sigma(s)) \geq \lambda_{\min}\left(\sum_{\omega\in\Omega}c_\omega c_\omega^\dagger\right) - \mathcal{O}_s\left(s^{-K}\right),
\end{equation*}
as $s\to\infty$, where $c_\omega=(c_\omega^1,\dots,c_\omega^d)\in\mathbb{C}^d$ are explicitly determined in Eq.~\eqref{eq:Fourier-coefficients-vector-frequencies} and independent of the input strength $s>0$.
\end{corollary}

\noindent
\emph{Proof.} By Proposition~\ref{prop:regularity-bound} we have
\begin{equation}
         \Sigma(s) = \sum_{\omega\in\Omega}c_\omega c_\omega^\dagger + \Delta(s) \quad\text{with}\quad \|\Delta(s)\|=\mathcal{O}_s\left(s^{-K}\right)
\end{equation}
as $s\to\infty$. Note that the matrix $\sum_{\omega\in\Omega}c_\omega c_\omega^\dagger$ is Hermitian, and so is the real symmetric matrix $ \Sigma(s)$. Hence, $\Delta(s) =  \Sigma(s)-\sum_{\omega\in\Omega}c_\omega c_\omega^\dagger$ is also Hermitian, which allows us to apply Weyl's inequality~\cite{vershynin2020high}:
\begin{align}
    \lambda_{\min}( \Sigma(s)) = \lambda_{\min}\left( \sum_{\omega\in\Omega}c_\omega c_\omega^\dagger + \Delta(s) \right) \geq \lambda_{\min} \left(\sum_{\omega\in\Omega}c_\omega c_\omega^\dagger\right) + \lambda_{\min}(\Delta(s))
\end{align}
for all $s>0$. Next, we recall the general spectral bound $|\lambda_k(\Delta(s))| \leq \|\Delta(s)\|$ for all $k\in[d]$. In particular, $\lambda_{\min}(\Delta(s)) \geq - \|\Delta(s)\|$ from which we obtain
\begin{equation}
    \lambda_{\min}( \Sigma(s)) \geq \lambda_{\min} \left(\sum_{\omega\in\Omega}c_\omega c_\omega^\dagger\right) + \lambda_{\min}(\Delta(s)) \geq \lambda_{\min} \left(\sum_{\omega\in\Omega}c_\omega c_\omega^\dagger\right) - \|\Delta(s)\| = \lambda_{\min} \left(\sum_{\omega\in\Omega}c_\omega c_\omega^\dagger\right) - \mathcal{O}_s\left(s^{-K}\right)
\end{equation}
by the previous bound in combination with the asymptotics $\|\Delta(s)\|=\mathcal{O}_s\left(s^{-K}\right)$ as $s\to\infty$. The corollary now follows.
\qed

Finally, we are fully prepared to upper bound the condition number of the rescaled population covariance $\Sigma(s)$ in the large input strength neighborhood:

\begin{theorem} \label{thm:condition-number-input-strength}
    Let $s>0$ and suppose that $\phi_U(0)=\sum_{\omega\in\Omega} c_\omega\neq0$. Let $s>0$ and assume that the input $ x$ has a density function $f( x)$ belonging to the $K$-th order Sobolev space $W^{K,1}(\mathcal{X})$ for some $K\in\mathbb{N}$. Then,
    \begin{equation*}
        \kappa\left(\Sigma_U(s)\right) \leq
            \kappa\left(\sum_{\omega\in\Omega}c_\omega c_\omega^\dagger\right) + \mathcal{O}(s^{-K})
    \end{equation*}
    as $s\to\infty$.
\end{theorem}

\noindent
\emph{Proof.} Let $s>0$ be arbitrary and define the matrix
\begin{equation}
    \Sigma_\infty:=\sum_{\omega\in\Omega}c_\omega c_\omega^\dagger,
\end{equation}
which is independent of the input strength $s>0$. With the help of Corollaries~\ref{corr:input-strength-upperbound-infty} and~\ref{corr:lower-eigenvalue-bound-population} we establish the bounds
\begin{equation}
    \|\Sigma_U(s)\| \leq \|\Sigma_\infty\| + \mathcal{O}(s^{-K})\quad\text{and}\quad \lambda_{\min}(\Sigma_U(s)) \geq \lambda_{\min}(\Sigma_\infty) - \mathcal{O}(s^{-K}) \label{eq:s-infty-bounds}
\end{equation}
as $s\to\infty$. Therefore, we obtain
\begin{align}
    \kappa(\Sigma_U(s)) &= \frac{\|\Sigma_U(s)\|}{\lambda_{\min}(\Sigma_U(s))} \stackrel{\eqref{eq:s-infty-bounds}}{\leq} \frac{\|\Sigma_\infty\| + \mathcal{O}(s^{-K})}{\lambda_{\min}(\Sigma_\infty) - \mathcal{O}(s^{-K})} = \frac{\|\Sigma_\infty\|\left(1 + \frac{\mathcal{O}(s^{-K})}{\|\Sigma_\infty\|}\right)}{\lambda_{\min}(\Sigma_\infty)\left(1 - \frac{\mathcal{O}(s^{-K})}{\lambda_{\min}(\Sigma_\infty)}\right)} = \kappa(\Sigma_\infty) \, \frac{1 + \mathcal{O}(s^{-K})}{1 - \mathcal{O}(s^{-K})} \nonumber \\
    &= \kappa(\Sigma_\infty) \left(1 + \mathcal{O}(s^{-K})\right)\left(1 + \mathcal{O}(s^{-K})\right) = \kappa(\Sigma_\infty) \left(1 + \mathcal{O}(s^{-K})\right) = \kappa(\Sigma_{\infty}) + \mathcal{O}(s^{-K}),
\end{align}
where we applied the Taylor approximation $(1-x)^{-1} = 1 + \mathcal{O}(x)$ as $x\to0$, and that $\lambda_{\min}(\Sigma_\infty),\|\Sigma_\infty\|,\kappa(\Sigma_\infty)\in\{\Theta_s(1)\}$ are each independent of $s$. This proves the claimed upper bound.
\qed

\subsection{The Effect of Increased Reuploads} \label{subsubsec:input-reuploads}

In this section we are going to study the effect of increasing the number $L$ of reuploads with the help of the Fourier interpretation of quantum models in Supplementary Section~\ref{appendix:non-concentrating-quantum} and show how the number of reuploads affects the condition number of the population covariance and hence the double descent curve. As our first insight, we establish an alternative representation of the Fourier frequency space $\Omega$ defined in Theorem~\ref{thm:general-Fourier-decomposition} of Section~\ref{appendix:fourier-interpretation} which allows us to explicitly describe $\Omega\equiv\Omega_L$ in terms of the number $L$ of reuploads. As shown below, this allows us to understand precisely how increasing reuploads transforms the Fourier coefficients and thus the population covariance $\Sigma_U\equiv \Sigma_U^{(L)}$.

Instead of the generalized quantum spectral tree interpretation of the path space $\Gamma$ as per Eq.~\eqref{eq:general-quantum-spectrum-tree} (cf. Ref.~\cite{mhiri2025constrained}), we bijectively reinterpret the path space $\Gamma$ in order to obtain an alternative Minkowski sum representation. Indeed, Theorem~\ref{thm:equivalent-fourier-space} of Section~\ref{appendix:Fourier-analysis} implies
\begin{equation}
    \Omega_L = \prod_{\eta=1}^p \sum_{\ell=1}^L \Omega_\eta^{(\ell)} \label{eq:Fourier-space-Minkowski-representation}
\end{equation}
is an input coordinate-wise ($\eta\in[p]$) Cartesian product of a layer-wise ($\ell\in[L]$) Minkowski sum, where
\begin{equation}
    \Omega_\eta^{(\ell)}:=\{\lambda-\lambda':\lambda,\lambda'\in\operatorname{Spec}(H_\eta^{(\ell)})\},\quad \eta\in[p],\quad\ell\in[L]
\end{equation}
denotes the set of all possible spectral differences of the generator $H_\eta^{(\ell)}$ with $(\eta,\ell)\in[p]\times[L]$.

\bigskip
\par
\paragraph{Fourier Frequency Space with Pauli Encoding Generators.} As a first, brief demonstration of how Eq.~\eqref{eq:Fourier-space-Minkowski-representation} may be used to analyse the geometry of the Fourier frequency space, suppose that every generator $H_\eta^{(\ell)}$ has an equidistant spectrum such that $\operatorname{Spec}(H_\eta^{(\ell)}) = \{-\frac{s}{2}, \frac{s}{2}\}$ for all $\eta \in [p]$ and $\ell \in [L]$ for some fixed input strength $s>0$ (e.g., if $H_\eta^{(\ell)}=\frac{s}{2}P_{\eta,\ell}$ is a scaled Pauli string $P_{\eta,\ell}\in\mathcal{P}_{n_\eta}$). Then, the layer-wise spectral difference set is strictly given by
\begin{equation}
    \Omega_{\eta}^{(\ell)} = \left\{\lambda-\lambda':\lambda,\lambda'\in\left\{\pm \frac{s}{2}\right\}\right\} = \{-s,0,s\}
\end{equation}
for all $\eta\in[p]$ and $\ell\in[L]$. Consequently, Eq.~\eqref{eq:Fourier-space-Minkowski-representation} ensures that the total frequency space for an $L$-layer model decomposes as
\begin{equation}
    \Omega_L = \prod_{\eta=1}^p\sum_{\ell=1}^L \{-s,0,s\} = \{-sL, -s(L-1), \dots, 0, \dots, s(L-1), sL\}^p = [-sL,sL]^p\cap s\mathbb{Z}^p, \label{eq:Pauli-encoding}
\end{equation}
which defines a $p$-dimensional equidistant lattice of points arranged inside a bounded hypercube of side length $2sL$ which is controlled by the input strength $s>0$. In particular, $|\Omega_L|=(2L+1)^p$ grows polynomially with $L$ with the number $p$ of input features defining the degree of that polynomial growth. Furthermore, we have that $\lim_{L\to\infty}\Omega_L=s\mathbb{Z}^p$ defines a $p$-dimensional free Abelian group that is isomorphic to the standard integer lattice $\mathbb{Z}^p$.

\smallskip
\par
\paragraph{Fourier Frequency Space with Exponential Encoding Generators.} To explore an alternative paradigm, suppose that the exponential scaling factor depends strictly on the layer index $\ell\in[L]$, also known as \emph{exponential encoding}~\cite{shin2023exponential, xiong2025fundamental, mhiri2025constrained}. Specifically, let $H_\eta^{(\ell)} = \frac{s}{2}\beta_{\ell} P_{\eta,\ell}$ under a fixed input strength $s>0$, where $P_{\eta,\ell}\in\mathcal{P}_{n_\eta}$ is a Pauli string and the layer-dependent scaling is chosen as $\beta_{\ell} = 3^{\ell-1}$ for all $\ell \in [L]$ and $\eta \in [p]$. Under this choice, we have $\operatorname{Spec}(H_\eta^{(\ell)})=\{-\frac{s}{2}\beta_\ell,\frac{s}{2}\beta_\ell\} = \{-\frac{s}{2}3^{\ell-1},\frac{s}{2}3^{\ell-1}\}$, thus yielding the layer-wise spectral difference set given by
\begin{equation}
    \Omega_{\eta}^{(\ell)} = \{-s 3^{\ell-1}, 0, s 3^{\ell-1}\} \label{eq:exponential-Fourier-spectral-difference-space}
\end{equation}
for all $\eta\in[p]$ and $\ell\in[L]$.

Let us now evaluate the Minkowski sum over the layers $L$ by inductively proving the identity
\begin{equation}
    \sum_{\ell=1}^L \Omega_\eta^{(\ell)} = \left[-s\frac{3^L-1}{2}, s\frac{3^L-1}{2}\right] \cap s\mathbb{Z} =: \mathcal{I}_{L}.
\end{equation}
For $L=1$, we have $\Omega_\eta^{(1)} = \{-s, 0, s\}$, which trivially satisfies the claim with $\frac{3^1-1}{2} = 1$. Now, assume the relation holds for $L-1$ layers, generating a contiguous integer lattice $\mathcal{I}_{L-1}$ bounded by $\frac{3^{L-1}-1}{2}$.
By the induction hypothesis, and since the set union operation distributes over the Minkowski sum, we find
\begin{align}
    \mathcal{I}_L &= \sum_{\ell=1}^L \Omega_\eta^{(\ell)} = \Omega_{\eta}^{(L)} + \underbrace{\sum_{\ell=1}^{L-1}\Omega_{\eta}^{(\ell)}}_{=\mathcal{I}_{L-1}} \stackrel{\eqref{eq:exponential-Fourier-spectral-difference-space}}{=} \{-s 3^{\ell-1}, 0, s 3^{\ell-1}\} + \mathcal{I}_{L-1} = \left(\{-s3^{L-1}\}\cup \{0\} \cup \{s3^{L-1}\}\right) + \mathcal{I}_{L-1} \nonumber
    \\ &= \left(\{-s3^{L-1}\}+\mathcal{I}_{L-1}\right) \cup \left(\{0\}+\mathcal{I}_{L-1}\right) \cup \left(\{s3^{L-1}\}+\mathcal{I}_{L-1}\right) \nonumber
    \\ &= \left( \left[-s\frac{3^L-1}{2}, -s\frac{3^{L-1}+1}{2}\right] \cup \left[-s\frac{3^{L-1}-1}{2}, s\frac{3^{L-1}-1}{2}\right] \cup \left[s\frac{3^{L-1}+1}{2}, s\frac{3^L-1}{2}\right] \right) \cap s\mathbb{Z} \nonumber
    \\ &= \left[-s\frac{3^L-1}{2}, s\frac{3^L-1}{2}\right] \cap s\mathbb{Z} = \mathcal{I}_{L},
\end{align}
which completes the inductive step.

As a result, Eq.~\eqref{eq:Fourier-space-Minkowski-representation} ensures that the total frequency space decomposes as a symmetric, uniform hypercube lattice:
\begin{equation}
    \Omega_L = \prod_{\eta=1}^p \left( \left[-s\frac{3^L-1}{2}, s\frac{3^L-1}{2}\right] \cap s\mathbb{Z} \right) = \left[-s\frac{3^L-1}{2}, s\frac{3^L-1}{2}\right]^p \cap s\mathbb{Z}^p. \label{eq:exponential-encoding}
\end{equation}
Once again, in the asymptotic limit, we naturally retain $\lim_{L\to\infty}\Omega_L=s\mathbb{Z}^p$, matching the standard free Abelian group structure. However, the convergence is exponentially faster than in the previous, Pauli encoding case.

\smallskip
\par
\paragraph{Comparing Increased Reuploads with Increasing Input Strength.} Increasing the number of reuploads increases the maximal Fourier frequency. Below, we show that increasing the input strength (cf. Section~\ref{subsubsec:input-strength}) has the same effect. However, there is an important difference: Below, we show that increasing the input strength $s>0$ scales the Fourier frequency space while the effect of increasing reuploads $\Delta L >0$ is adding new frequencies by Minkowski additivity. Indeed, the effect of increasing reuploads $\Delta L >0$ with the effect of increasing input strength $s>0$. Furthermore, we show that increasing $\Delta L$ fundamentally transforms the Fourier coefficients $c_\omega\equiv c_\omega^{(L)}$ (cf. Theorem~\ref{thm:general-Fourier-decomposition} in Section~\ref{appendix:fourier-interpretation}), while the Fourier coefficients are left invariant when modifying $s$.

When the number of reuploads is increased via $L\mapsto L+\Delta L$ (where $\Delta L\in\mathbb{N}$), the frequency support expands to $\Omega_L\mapsto \Omega_{L+\Delta L}$. 
With the help of Theorem~\ref{thm:equivalent-fourier-space} in Section~\ref{appendix:Fourier-analysis}, in combination with the distributivity of the Minkowski sum, we can partition this enlarged space into the pre-existing frequency lattice and the newly accessible high-frequency components:
\begin{equation}
    \Omega_{L+\Delta L} = \prod_{\eta=1}^p \sum_{\ell=1}^{L+\Delta L}\Omega_\eta^{(\ell)} = \prod_{\eta=1}^p\left(\sum_{\ell=1}^L\Omega_\eta^{(\ell)} + \sum_{\ell=L+1}^{L+\Delta L}\Omega_\eta^{(\ell)}\right) = \underbrace{\left(\prod_{\eta=1}^p\sum_{\ell=1}^L\Omega_\eta^{(\ell)}\right)}_{=:\Omega_L} + \prod_{\eta=1}^p\left(\sum_{\ell=L+1}^{L+\Delta L}\Omega_{\eta}^{(\ell)}\right), \label{eq:new-high-energy-fourier-frequencies}
\end{equation}
showing that $\Omega_L\subseteq \Omega_{L+\Delta L}$ since $0\in\Omega_\eta^{(\ell)}$ for all $\eta\in[p]$ and $\ell\in[L]$. 
Hence, we may define $\Omega_{\Delta L}:=\Omega_{L+\Delta L}\setminus \Omega_L$ (note that $\Omega_{\Delta L}=\emptyset$ is empty if $\Delta L =0$). In particular $|\Omega_{L+\Delta L}|>|\Omega_L|$. 
On the other hand, when increasing $s>0$, another application of Theorem~\ref{thm:equivalent-fourier-space} yields
\begin{equation}
    \Omega_{L}(s) = \prod_{\eta=1}^p\sum_{\ell=1}^{L}\{\lambda-\lambda':\lambda,\lambda'\in\operatorname{Spec}(s H_{\eta}^{(\ell)})\} = \prod_{\eta=1}^p\sum_{\ell=1}^{L}\{s(\lambda-\lambda'):\lambda,\lambda'\in\operatorname{Spec}(H_{\eta}^{(\ell)})\} = s\Omega_{L}(1)
\end{equation}
so that, after absorbing the input strength into the generators via $e^{i(sx)H_{\eta}^{(\ell)}}=e^{ix(sH_\eta^{(\ell)})}$, we see that increasing the input strength $s>0$ has the effect of uniformly expanding the entire Fourier frequency space. In particular $|\Omega_L|=|s\Omega_L|$ is left invariant.

Next, we study the behaviour of the largest Fourier frequency. Let $\|\cdot\|$ be an arbitrary norm on the Euclidean space $\mathbb{R}^p$. Since $\Omega_L\subset\mathbb{R}^p$, we can study the maximal norm $\max_{\omega\in\Omega_L}\|\omega\|$. In the case of increasing input strength, we have homogeneity by the previous result: 
\begin{equation}
    \max_{\omega\in\Omega_L(s)}\|\omega\| = \max_{\omega\in s\Omega_L(1)}\|\omega\| = \max_{\omega\in \Omega_L(1)}\|s\omega\| = s\max_{\omega\in\Omega_L(1)}\|\omega\|.
\end{equation}
In contrast to scalar multiplicativity, $\Delta L$  scales maximal frequency norm by Minkowski additivity: 
Let $\omega\in \Omega_{L+\Delta L}$ be arbitrary. Then, by the Minkowski additivity in Eq.~\eqref{eq:new-high-energy-fourier-frequencies} there exists a base term $\omega^{(L)}\in\Omega_L$ and an additive term $\omega^{(\Delta L)}\in\prod_{\eta=1}^p\sum_{\ell=L+1}^{L+\Delta L}\Omega_\eta^{(\ell)}$ being element of a high-frequency Minkowski addend such that $\omega=\omega^{(L)}+\omega^{(\Delta L)}$. 
Because of $0\in\Omega_L$ we can choose the path component $\omega^{(L)}=0$ and since $\Omega_{\Delta L}\subseteq\Omega_{L+\Delta L}$, we find
\begin{align}
    \max_{\omega\in\Omega_{L+\Delta L}}\|\omega\| = \max_{\omega\in\Omega_{L+\Delta L}}\|\omega^{(L)}+\omega^{(\Delta L)}\| \geq \max_{\omega^{(\Delta L)}\in\prod_{\eta=1}^p\sum_{\ell=L+1}^{L+\Delta L}\Omega_\eta^{(\ell)}}\|\omega^{(\Delta L)}\|.
\end{align}
Therefore, $\max_{\omega\in\Omega_{L+\Delta L}}\|\omega\|$ is guaranteed to grow with the high frequency contribution $\|\omega^{(\Delta L)}\|$.

The most notable difference between increasing $s$ and $\Delta L$ is the frequency resolution gap:
\begin{equation}
    \min_{\substack{\omega,\omega'\in\Omega_L \\ \omega\neq \omega'}}\|\omega-\omega'\|.
\end{equation}
By the previous arguments, we know that the distance $\|s(\omega-\omega')\|$ between two distinct frequencies increases with $s>0$ which causes the frequency resolution gap to increase. However, if a pair $\omega,\omega'\in\Omega_L$ of distinct frequencies minimizes $\|\omega-\omega'\|$, then increasing $\Delta L$ can only make $\|\omega-\omega'\|$ smaller or leave it invariant since $\Omega_L\subseteq\Omega_{L+\Delta L}$ results in a larger frequency space for all $\Delta L$.

Finally, we analyse the effect on the Fourier coefficients. As dictated by Theorem~\ref{thm:general-Fourier-decomposition} in Section~\ref{appendix:fourier-interpretation}, we have that the Fourier coefficients are explicitly determined by
\begin{equation}
    \left(c_\omega^{(L+\Delta L)}\right)_k = \sum_{\substack{\gamma,\gamma'\in([2^{n_1}]\times\cdots\times[2^{n_p}])^{L+\Delta L} \\ \Lambda_{\gamma'}-\Lambda_\gamma=\omega}} \operatorname{Tr}(O_kU_{L+\Delta L+1}W_\gamma\rho_0 W_{\gamma'}^\dagger U_{L+\Delta L+1}^\dagger), \quad k\in[d],
\end{equation}
which is independent of the input strength $s>0$ since scaling the Fourier frequencies has the equivalent effect of scaling the spectral differences $\Lambda_{\gamma'}-\Lambda_\gamma$. On the other hand, increasing $\Delta L$ fundamentally changes the operators within the trace, and thus, the Fourier coefficients themselves.

\smallskip
\par
\paragraph{Increased Reuploads regularize the Condition Number.} Finally, we shift our attention to our main result, namely showing how increasing the number of reuploads affects the condition number of the population covariance and thus the double descent curve. Let $\Sigma_U^{(L)}(s)$ and $\Sigma_U^{(L+\Delta L)}(s)$ denote the scaled population covariance matrices of the $L$-layer and $(L+\Delta L)$-layer architectures, respectively. Crucially, because appending layers introduces additional reservoir unitaries, the Fourier coefficients $c_\omega^{(L+\Delta L)}$ of the deeper network will generally differ from $c_\omega^{(L)}$ even for the old frequencies $\omega \in \Omega_L$.
Invoking Lemma~\ref{lemma:input-strength-covariance} from Section~\ref{subsubsec:input-strength} and expanding the double summation over the partitioned frequency space $\Omega_{L+\Delta L} = \Omega_L \sqcup \Omega_{\Delta L}$ yields an exact structural decomposition of the updated population covariance matrix into three distinct terms: A term with the base frequencies in $\Omega_L$, a term only taking into account high-frequencies $\Omega_{\Delta L}$, and a symmetric cross term containing frequencies from both $\Omega_L$ and $\Omega_{\Delta L}$. More explicitly,
\begin{align}
    \Sigma_U^{(L+\Delta L)}(s) &= \sum_{\omega,\omega'\in \Omega_{L+\Delta L}} \varphi_x(s(\omega+\omega'))\,c_{\omega}^{(L+\Delta L)} \left(c_{\omega'}^{(L+\Delta L)}\right)^\top \nonumber \\
    &= \sum_{\omega,\omega'\in\Omega_L}\varphi_x(s(\omega+\omega'))\,c_{\omega}^{(L+\Delta L)}\left(c_{\omega'}^{(L+\Delta L)}\right)^\top + \sum_{\omega,\omega'\in\Omega_{\Delta L}}\varphi_x(s(\omega+\omega'))\,c_{\omega}^{(L+\Delta L)}\left(c_{\omega'}^{(L+\Delta L)}\right)^\top \nonumber \\
    &\quad + \underbrace{2\operatorname{Sym}\left(\sum_{\omega\in\Omega_L}\sum_{\omega'\in\Omega_{\Delta L}}\varphi_x(s(\omega+\omega'))\,c_{\omega}^{(L+\Delta L)}\left(c_{\omega'}^{(L+\Delta L)}\right)^\top\right)}_{=: \Sigma_U^{(\text{cross})}(s)},
\end{align}
where $\operatorname{Sym}(A) := \frac{1}{2}(A+A^\top)$. It is worth remarking that $\omega+\omega'\neq 0$ for all $\omega\in\Omega_L$ and $\omega'\in\Omega_{\Delta L}$ because this means that the cross term $\Sigma_U^{(\text{cross})}(s)$ always strictly depends on $s$. This allows us to absorb the cross term into the asymmetric remainder $\Delta(s)$ which will not affect the argumentation of Eq.~\eqref{eq:population-covariance-decomposition} that concludes with $\Sigma(s) =\sum_{\omega\in\Omega} c_\omega c_\omega^\dagger + \Delta(s)$. Hence, we have
\begin{align}
    \Sigma_U^{(L+\Delta L)}(s) &= \underbrace{\sum_{\omega\in\Omega_L}c_{\omega}^{(L+\Delta L)} \left(c_{\omega}^{(L+\Delta L)}\right)^\dagger}_{=:\Sigma_L} + \underbrace{\sum_{\omega\in\Omega_{\Delta L}}c_{\omega}^{(L+\Delta L)} \left(c_{\omega}^{(L+\Delta L)}\right)^\dagger}_{=:\Sigma_{\Delta L}} + \Delta^{(L+\Delta L)}(s), \label{eq:Sigma_L-term}
\end{align}
introducing the $s$-independent terms $\Sigma_L$ and $\Sigma_{\Delta L}$, respectively, where we absorbed the remaining $s$-dependent cross-frequency terms $\Sigma_U^{(\text{cross})}(s)$ into the updated asymmetric remainder
\begin{equation}
    \Delta^{(L+\Delta L)}(s) :=  \sum_{\substack{\omega,\omega'\in\Omega_{L+\Delta L} \\ \omega+\omega'\neq0}}\varphi_x(\omega+\omega')c_{\omega}^{(L+\Delta L)} \left(c_{\omega'}^{(L+\Delta L)}\right)^\top.
\end{equation}
as described above. Assuming that the input density $f\in W^{1,K}(\mathcal{X})$ is sufficiently regular, we may invoke Corollary~\ref{corr:lower-eigenvalue-bound-population} from Section~\ref{subsubsec:input-strength} which guarantees that the lowest eigenvalue of the updated population covariance $\Sigma_U^{(L+\Delta L)}(s)$ obeys the lower bound
\begin{align}
    \lambda_{\min}\left(\Sigma_U^{(L+\Delta L)}(s)\right) \geq \lambda_{\min}\left(\Sigma_L + \Sigma_{\Delta L}\right) - \mathcal{O}_s\left(s^{-K}\right)
\end{align}
as $s\to\infty$. On the other hand, we generally have the upper bound
\begin{align}
    \left\|\Sigma_U^{(L+\Delta L)}(s)\right\| \leq \operatorname{Tr}\left(\Sigma_U^{(L+\Delta L)}(s)\right) = \sum_{k=1}^d\left(\Sigma_U^{(L+\Delta L)}(s)\right)_{k,k} = \sum_{k=1}^d \underbrace{\mathbb{E}_x\left[\left(\phi_U^{(L+\Delta L)}(sx)_{k}\right)^2\right]}_{\leq M^2} \leq d M^2 = \mathcal{O}_{\Delta L}(1)
\end{align}
as $\Delta L\to\infty$ where $M:=\max_{k\in[d]}\|O_k\|=\mathcal{O}_{\Delta L}(1)$. Therefore, by combining the previous two bounds, the condition number of $\Sigma_U^{(L+\Delta L)}(s)$ satisfies
\begin{equation}
    \kappa\left(\Sigma_U^{(L+\Delta L)}(s)\right) = \frac{\lambda_{\max}\left(\Sigma_U^{(L+\Delta L)}(s)\right)}{\lambda_{\min}\left(\Sigma_U^{(L+\Delta L)}(s)\right)} \leq \frac{\mathcal{O}_{\Delta L}(1)}{\lambda_{\min}\left(\Sigma_L + \Sigma_{\Delta L}\right) - \mathcal{O}_s(s^{-K})} \label{eq:condition-number-increased-reupload-bound}
\end{equation}
which decreases with $\Delta L$ if and only if the lowest eigenvalue $\lambda_{\min}(\Sigma_L + \Sigma_{\Delta L})$ increases. Intuitively, this happens if the newly appearing high-frequency term $\Sigma_{\Delta L}$ ``injects'' energy into the specific subspace where $\Sigma_L$ is weak. Intuitively, this injection has the effect of regularizing the condition number of the population covariance. That is, if $\Sigma_{U}^{(L)}$ is ill-conditioned, then increasing $\Delta L$ can help improving the condition number.

As the next step, to make this idea more rigorous, we would like to isolate the the smallest eigenvalue $\lambda_{\min}(\Sigma_L)$ in Eq.~\eqref{eq:condition-number-increased-reupload-bound} and use the term $\Sigma_{\Delta L}$ to construct an explicitly determined measure of how much energy is injected into the weak subspace of $\Sigma_L$. First, we need to define what the weak subspace of $\Sigma_L$ is: For that purpose, we introduce the linear eigenspace
\begin{equation}
    E_{\min}(\Sigma_L) := \{v\in\mathbb{C}^d: \Sigma_L v = \lambda_{\min}(\Sigma_L) v\}
\end{equation}
corresponding to the lowest energy of $\Sigma_L$  (note that $\Sigma_L,\Sigma_{\Delta L}$ define Hermitian operators so that their eigenspectra are real). 
If the eigenvalue of $\lambda_{\min}(\Sigma_L)$ is small, then $\kappa\left(\Sigma_U^{(L+\Delta L)}(s)\right)$ is large if $\Delta L \approx 0$, as evident from Eq.~\eqref{eq:condition-number-increased-reupload-bound}. 
In that sense, $E_{\min}(\Sigma_L)$ is indeed a faithful representation for the weak subspace that we would like to inject energy into by increasing $\Delta L$.  The process of injecting the high-frequency energy into $\Sigma_L$ is formally described by the growth in
\begin{equation}
   \lambda_{\min}(\Sigma_L) = \inf_{\substack{v\in\mathbb{C}^d \\ \|v\|=1}}v^\dagger \Sigma_L v < \inf_{\substack{v\in\mathbb{C}^d \\ \|v\|=1}}\left(v^\dagger \Sigma_Lv+v^\dagger \Sigma_{\Delta L}v\right)  = \inf_{\substack{v\in\mathbb{C}^d \\ \|v\|=1}}\left(v^\dagger (\Sigma_L+\Sigma_{\Delta L})v\right) = \lambda_{\min}(\Sigma_L+\Sigma_{\Delta L}),
\end{equation}
where the strict inequality holds true if and only if
\begin{equation}
    v^\dagger \Sigma_{\Delta L} v > 0 \quad\text{for all $v\in E_{\min}(\Sigma_L)\setminus\{0\}$}. \label{eq:energy-injection-condition1i}
\end{equation}
The latter condition formally means that the high-frequencies due to $\Delta L$ indeed inject positive energy into the weak subspace of $\Sigma_L$ and plays the key role in defining the high-frequency injection denoted by $\delta(\Delta L)$. Specifically, it is natural to define
\begin{equation}
    \delta(\Delta L) = \inf_{\substack{v\in E_{\min}(\Sigma_L)\\ \|v\|=1}}
    v^\dagger \Sigma_{\Delta L}v. \label{eq:injection-definition}
\end{equation}
Having introduced the concepts above, we are now ready to formulate the following theorem:

\begin{theorem} \label{thm:condition-number-reuploads}
    Let $s>0$ be an arbitrary input strength. For each $\Delta L\in\mathbb{N}$, let $\Omega_{\Delta L}:=\Omega_{L+\Delta L}\setminus\Omega_L$ denote the newly accessible high-energy Fourier frequencies as per Eq.~\eqref{eq:new-high-energy-fourier-frequencies}. Assume that the input $ x$ has a density function $f( x)$ belonging to the $K$-th order Sobolev space $W^{K,1}(\mathcal{X})$ for some $K\in\mathbb{N}$. Suppose that the spectral gap $\Delta\lambda := \lambda_2(\Sigma_L)-\lambda_{\min}(\Sigma_L)>0$ of $\Sigma_L:=\sum_{\omega\in\Omega_L}c_{\omega}^{(L+\Delta L)} \left(c_{\omega}^{(L+\Delta L)}\right)^\dagger$ is positive, and define the high-frequency injection
    \begin{equation*}
        \delta(\Delta L) := \inf_{\substack{v\in E_{\min}(\Sigma_L)\\ \|v\|=1}}\sum_{\omega\in\Omega_{\Delta L}}\left|\operatorname{Tr}\left(O(v)^\dagger U_{L+\Delta L + 1} P_\omega U_{L+\Delta L+1}^\dagger\right)\right|^2, \quad P_\omega := \sum_{\substack{\gamma,\gamma'\in \Gamma \\ \Lambda_{\gamma'}-\Lambda_\gamma=\omega}}W_\gamma\rho_0W_{\gamma'}^\dagger,
    \end{equation*}
    where $ E_{\min}(\Sigma_L) := \{v\in\mathbb{C}^d: \Sigma_L v = \lambda_{\min}(\Sigma_L) v\}$ is the eigenspace of $\Sigma_L$ corresponding to the smallest eigenvalue, and $O(v):=\sum_{k=1}^dv_kO_k\neq0$ for all $v\in E_{\min}(\Sigma_L)\setminus\{0\}$ (e.g., if the system $\{O_k\}_{k\in[d]}$ is linearly independent). Then,
    \begin{equation*}
        \kappa\left(\Sigma_U^{(L+\Delta L)}(s)\right) \leq \frac{\mathcal{O}_{\Delta L}(1)}{\lambda_{\min}(\Sigma_L) + \eta_{\Delta L}\delta(\Delta L) - \mathcal{O}_s(s^{-K})},
    \end{equation*}
    with a spectral efficiency factor $\eta_{\Delta L} := \frac{\Delta \lambda}{\Delta \lambda + \delta(\Delta L) + \|\Sigma_{\Delta L}\|}$.
\end{theorem}

\noindent
\emph{Proof.} Le $s>0$ and $\Delta L\in\mathbb{N}$ be arbitrary. We begin with deriving the claimed, explicit expression for the high-frequency injection $\delta(\Delta L)$. Recall from Eq.~\eqref{eq:energy-injection-condition1i} that, by definition, the injection is successful if and only if
\begin{equation}
    v^\dagger \Sigma_{\Delta L} v > 0 \quad\text{for all $v\in E_{\min}(\Sigma_L)\setminus\{0\}$}. \label{eq:energy-injection-condition}
\end{equation}
In order to further analyse this condition, we employ the exact expression of the Fourier coefficient components 
\begin{align}
    \left(c_\omega^{(L+\Delta L)}\right)_k = \sum_{\substack{\gamma,\gamma'\in \Gamma \\ \Lambda_{\gamma'}-\Lambda_\gamma=\omega}} \operatorname{Tr}(O_kU_{L+\Delta L+1}W_\gamma\rho_0 W_{\gamma'}^\dagger U_{L+\Delta L +1}^\dagger) = \operatorname{Tr}\left(O_kU_{L+\Delta L+1}\underbrace{\left(\sum_{\substack{\gamma,\gamma'\in \Gamma \\ \Lambda_{\gamma'}-\Lambda_\gamma=\omega}}W_\gamma\rho_0W_{\gamma'}^\dagger\right)}_{=:P_\omega}U_{L+\Delta L+1}^\dagger\right)  \label{eq:Fourier-coefficient-formula-layer}
\end{align}
as dictated by Theorem~\ref{thm:general-Fourier-decomposition} in Section~\ref{appendix:fourier-interpretation} with an enlarged, updated path space $\Gamma := ([2^{n_1}]\times\cdots\times[2^{n_p}])^{L+\Delta L}$, and introducing operator $P_\omega$ to simplify the notations. 
Then, we may explicitly evaluate the Fourier coefficient projection
\begin{align}
    v^\dagger c_\omega^{(L+\Delta L)} = \sum_{k=1}^dv_k^*\left(c_\omega^{(L+\Delta L)}\right)_k \stackrel{\eqref{eq:Fourier-coefficient-formula-layer}}{=} \sum_{k=1}^dv_k^*\operatorname{Tr}\left(O_kU_{L+\Delta L +1}P_\omega U_{L+\Delta L+1}^\dagger\right) = \operatorname{Tr}\left(\underbrace{\left(\sum_{k=1}^d v_kO_k\right)^\dagger}_{=:O(v)^\dagger} U_{L+\Delta L + 1}P_\omega U_{L+\Delta L + 1}^\dagger\right) \label{eq:Fourier-coefficient-eigenspace-projection}
\end{align}
onto each eigenspace element $v\in E_{\min}(\Sigma_L)\setminus\{0\}$, where we also introduced the linear combination $O(v):=\sum_{k=1}^d v_k O_k$ of the system $\{O_k\}_{k\in[d]}$ of observables with respect to the eigenspace of $\Sigma_L$, and thus we find
\begin{align}
    v^\dagger \Sigma_{\Delta L}v &= v^\dagger\left(\sum_{\omega\in\Omega_{\Delta L}}c_{\omega}^{(L+\Delta L)}\left(c_\omega^{(L+\Delta L)}\right)^\dagger\right)v = \sum_{\omega\in\Omega_{\Delta L}} \left|v^\dagger c_\omega^{(L+\Delta L)}\right|^2 \stackrel{\eqref{eq:Fourier-coefficient-eigenspace-projection}}{=} \sum_{\omega\in\Omega_{\Delta L}}\left|\operatorname{Tr}\left(O(v)^\dagger U_{L+\Delta L + 1} P_\omega U_{L+\Delta L+1}^\dagger\right)\right|^2 \label{eq:injection-condition-architecture}
\end{align}
for all  $v\in E_{\min}(\Sigma_L)\setminus\{0\}$. Therefore, the high-frequency energy injection condition in Eq.~\eqref{eq:energy-injection-condition} is satisfied if and only if there exists $\omega\in\Omega_{\Delta L}$ such that
\begin{equation}
    \operatorname{Tr}\left(O(v)^\dagger U_{L+\Delta L + 1} P_\omega U_{L+\Delta L+1}^\dagger\right) \neq 0
\end{equation}
for all $v\in E_{\min}(\Sigma_L)\setminus\{0\}$. In addition, one needs to impose that the system $\{O_k\}_{k\in[d]}$ of observables is $\mathbb{C}$-linearly independent in order to exclude the pathological case $O(v)=0$ for some $v\in E_{\min}(\Sigma_L)\setminus\{0\}$. Under this standard assumption, Eq.~\eqref{eq:injection-condition-architecture} above becomes
\begin{equation}
    \delta(\Delta L) \stackrel{\eqref{eq:injection-definition}}{=} \inf_{\substack{v\in E_{\min}(\Sigma_L)\\ \|v\|=1}}
    v^\dagger \Sigma_{\Delta L}v  =\inf_{\substack{v\in E_{\min}(\Sigma_L)\\ \|v\|=1}}
    \sum_{\omega\in\Omega_{\Delta L}}\left|\operatorname{Tr}\left(O(v)^\dagger U_{L+\Delta L + 1} P_\omega U_{L+\Delta L+1}^\dagger\right)\right|^2.
\end{equation}
This proves the claimed representation of the high-frequency injection $\delta(\Delta L)$.

Now, we show how we can leverage $\delta(\Delta L)$ to isolate it from the base reference $\lambda_{\min}(\Sigma_L)$ and why the spectral efficiency factor $\eta_{\Delta L}$ formally appears. 
Let $\lambda_2(\Sigma_L) > \lambda_{\min}(\Sigma_L)$ be the second smallest eigenvalue of $\Sigma_L$ (we assume that $\Sigma_L$ is not a multiple of the identity) so that $\Delta\lambda := \lambda_2(\Sigma_L)-\lambda_{\min}(\Sigma_L)$ defines the spectral gap of $\Sigma_L$. 
Any arbitrary unit vector $v \in \mathbb{C}^d$ can be uniquely decomposed into components inside and orthogonal to the weak eigenspace $E_{\min}(\Sigma_L)$ so that $v = \alpha v_{\min} + \beta v_\perp$ where $v_{\min} \in E_{\min}(\Sigma_L)$, $v_\perp \in E_{\min}(\Sigma_L)^\perp$, with $\|v_{\min}\| = \|v_\perp\| = 1$, and the coefficients satisfying $|\alpha|^2 + |\beta|^2 = 1$. Because $v_{\min}$ and $v_\perp$ are orthogonal eigenvectors of $\Sigma_L$, the previous decomposition of $v$ yields
\begin{equation}
    v^\dagger \Sigma_L v = \underbrace{|\alpha|^2}_{=1-|\beta|^2} \lambda_{\min}(\Sigma_L) + |\beta|^2 \lambda_2(\Sigma_L) = \lambda_{\min}(\Sigma_L) + |\beta|^2\left(\lambda_2(\Sigma_L)-\lambda_{\min}(\Sigma_L)\right) = \lambda_{\min}(\Sigma_L) + |\beta|^2 \Delta \lambda.
\end{equation}
Similarly, we expand the high-frequency injection matrix $\Sigma_{\Delta L}$ across these two components,
\begin{equation}
    v^\dagger \Sigma_{\Delta L} v = |\alpha|^2 (v_{\min}^\dagger \Sigma_{\Delta L} v_{\min}) + |\beta|^2 (v_\perp^\dagger \Sigma_{\Delta L} v_\perp) +  2 \operatorname{Re}\left(\alpha^* \beta v_{\min}^\dagger \Sigma_{\Delta L} v_\perp\right),
\end{equation}
which allows us to bound
\begin{align}
    v^\dagger (\Sigma_L + \Sigma_{\Delta L}) v &= \underbrace{\lambda_{\min}(\Sigma_L) + |\beta|^2\Delta\lambda}_{=v^\dagger \Sigma_L v} + |\alpha|^2 (v_{\min}^\dagger \Sigma_{\Delta L} v_{\min}) + |\beta|^2 (v_\perp^\dagger \Sigma_{\Delta L} v_\perp) +  \underbrace{2 \operatorname{Re}\left(\alpha^* \beta v_{\min}^\dagger \Sigma_{\Delta L} v_\perp\right)}_{\geq -|\alpha||\beta||v_{\min}^\dagger\Sigma_{\Delta L}v_\perp|} \nonumber
    \\ &\geq \lambda_{\min}(\Sigma_L) + |\beta|^2\Delta\lambda + |\alpha|^2(v_{\min}^\dagger\Sigma_{\Delta L}v_{\min})+|\beta|^2(v_\perp^\dagger\Sigma_{\Delta L}v_\perp) - 2|\alpha||\beta|\sqrt{(v_{\min}^\dagger\Sigma_{\Delta L}v_{\min})(v_\perp^\dagger\Sigma_{\Delta L}v_\perp)} \label{eq:spectral-shift}
\end{align}
by using the elementary identity $\operatorname{Re}(z) \geq -|z|$ for $z\in\mathbb{C}$ and the Cauchy-Schwarz inequality (recalling that $\Sigma_{\Delta L}$ is PSD) in
\begin{equation}
    |v_{\min}^\dagger\Sigma_{\Delta L}v_\perp| = \langle v_{\min},v_\perp\rangle_{\Sigma_{\Delta L}} \leq \sqrt{\|v_{\min}\|_{\Sigma_{\Delta L}}\|v_\perp\|_{\Sigma_{\Delta L}}} = \sqrt{(v_{\min}^\dagger\Sigma_{\Delta L}v_{\min})(v_\perp^\dagger\Sigma_{\Delta L}v_\perp)}.
\end{equation}
The right-hand side of the inequality in Eq.~\eqref{eq:spectral-shift} can be viewed as a quadratic form perturbation, $\lambda_{\min}(\Sigma_L) + (|\alpha|,|\beta|)^\top A\, (|\alpha|,|\beta|)$ away from $\lambda_{\min}(\Sigma_L)$ in the coordinates $(|\alpha|,|\beta|)$ so that the worst-case lower bound is determined by the lowest eigenvalue $\mu_{\min}(A)$ of the matrix system
\begin{equation}
    A = \begin{pmatrix} v_{\min}^\dagger\Sigma_{\Delta L}v_{\min} & -\sqrt{(v_{\min}^\dagger\Sigma_{\Delta L}v_{\min})(v_\perp^\dagger\Sigma_{\Delta L}v_\perp)} \\ -\sqrt{(v_{\min}^\dagger\Sigma_{\Delta L}v_{\min})(v_\perp^\dagger\Sigma_{\Delta L}v_\perp)} & \Delta \lambda + v_\perp^\dagger\Sigma_{\Delta L}v_\perp \end{pmatrix}
\end{equation}
which is explicitly determined as
\begin{equation}
    \mu_{\min}(A) = \frac{1}{2} \left( (v_{\min}^\dagger\Sigma_{\Delta L}v_{\min} + v_\perp^\dagger\Sigma_{\Delta L}v_\perp + \Delta \lambda) - \sqrt{(v_{\min}^\dagger\Sigma_{\Delta L}v_{\min} + v_\perp^\dagger\Sigma_{\Delta L}v_\perp + \Delta \lambda)^2 - 4 v_{\min}^\dagger\Sigma_{\Delta L}v_{\min} \Delta \lambda} \right)
\end{equation}
By introducing the variables $X_1:=v_{\min}^\dagger\Sigma_{\Delta L}v_{\min}$ and $X_2:=v_\perp^\dagger\Sigma_{\Delta L}v_\perp$ one considers
\begin{equation}
    \mu_{\min}(A)\equiv\mu(X_1,X_2) = \frac{1}{2}\left(X_1 + X_2 + \Delta\lambda - \sqrt{(X_1+X_2+\Delta\lambda)^2 - 4X_1\Delta\lambda}\right)
\end{equation}
as a differentiable function of these variables. Taking the partial derivatives with respect to these variables reveals
\begin{equation}
    \partial_{X_1}\mu = \frac{1}{2} \left( 1 - \frac{X_1+X_2 - \Delta \lambda}{\sqrt{(X_1+X_2+\Delta\lambda)^2 - 4 X_1 \Delta \lambda}} \right)\quad\text{and}\quad \partial_{X_2}\mu = \frac{1}{2} \left( 1 - \frac{X_1+X_2+\Delta\lambda}{\sqrt{(X_1+X_2+\Delta\lambda)^2 - 4 X_1 \Delta \lambda}} \right). \label{eq:eigenspectrum-partial-derivatives}
\end{equation}
In particular $\partial_{X_1}\mu \geq 0$ and $\partial_{X_2}\mu<0$ globally for all $X_1>0$ and $X_2\geq0$ since $\Delta\lambda>0$. This means that, in order to find the minimum possible value of the lower bound for $\lambda_{\min}(\Sigma_L+\Sigma_{\Delta L})$ in Eq.~\eqref{eq:spectral-shift}, we must push $X_1=v_1^\dagger\Sigma_{\Delta L}v_1$ to its minimum possible value of $\delta(\Delta L) = \inf_{v\in E_{\min}(\Sigma_{L}):\|v\|=1}v^\dagger \Sigma_L v$, and $X_2=v_2^\dagger\Sigma_{\Delta L}v_2$ to its maximum possible value of $\|\Sigma_{\Delta L}\| \geq \sup_{v\in E_{\min}(\Sigma_L)^\perp:\|v\|=1}v^\dagger\Sigma_{\Delta L}v$, respectively. 
That is, we generally have the lower bound
\begin{equation}
    \lambda_{\min}(\Sigma_L + \Sigma_{\Delta L}) \geq \lambda_{\min}(\Sigma_L) + \frac{1}{2} \left[((\delta(\Delta L) + \|\Sigma_{\Delta L}\| + \Delta \lambda) - \sqrt{(\delta(\Delta L) + \|\Sigma_{\Delta L}\| + \Delta \lambda)^2 - 4 \delta(\Delta L) \Delta \lambda} \right).
\end{equation}
To convert this exact radical expression into a clean, readable bound, we apply the algebraic inequality $- \sqrt{a^2 - b} \geq -a + \frac{b}{2a}$ (where $a = \delta(\Delta L) + \|\Sigma_{\Delta L}\| + \Delta \lambda$ and $b = 4 \delta(\Delta L) \Delta \lambda$):
\begin{equation}
    \lambda_{\min}(\Sigma_L + \Sigma_{\Delta L}) \geq \lambda_{\min}(\Sigma_L) + \underbrace{\left( \frac{\Delta \lambda}{\Delta \lambda + \delta(\Delta L) + \|\Sigma_{\Delta L}\|} \right)}_{=:\eta_{\Delta L}} \delta(\Delta L)
\end{equation}
where we introduced the high-frequency efficiency factor $\eta_{\Delta L}\in(0,1]$ which is well-defined since the spectral gap $\Delta \lambda >0$ is positive. Hence, we find
\begin{equation}
    \kappa\left(\Sigma_U^{(L+\Delta L)}(s)\right) \stackrel{\eqref{eq:condition-number-increased-reupload-bound}}{\leq} \frac{\mathcal{O}_{\Delta L}(1)}{\lambda_{\min}\left(\Sigma_L + \Sigma_{\Delta L}\right) - \mathcal{O}_s(s^{-K})} \leq \frac{\mathcal{O}_{\Delta L}(1)}{\lambda_{\min}(\Sigma_L) + \eta_{\Delta L}\delta(\Delta L) - \mathcal{O}_s(s^{-K})}.
\end{equation}
This completes the proof.
\qed

Theorem~\ref{thm:condition-number-reuploads} thus confirms that, if the initial architecture is poorly conditioned due to a small $\lambda_{\min}(\Sigma_L)$, increasing the number of reuploads by $\Delta L$ strictly regularizes the spectrum. It injects an energy commodity of scale $\delta(\Delta L)$ directly into the weak subspace, stabilizing the denominator of $\kappa\left(\Sigma_U^{(L+\Delta L)}(s)\right)$ up to a saturation ceiling dictated by the remaining structural properties of the reservoir. It is worthwhile to note that, when employing exponential encoding (cf. Eq.~\eqref{eq:exponential-encoding}), one might expected that the regularization is more effective due to the exponential growth of $\Omega_{\Delta L}$, making it more likely to have a larger injection $\delta(\Delta L)$ in contrast to the linearly growth rate due to the simple Pauli encoding strategy (cf. Eq.~\eqref{eq:Pauli-encoding}).

\smallskip
\par
\paragraph{Transformation to the Proper Population Covariance Base Reference.} At this point one might remark that we are in fact interested in the high-frequency injection into $\Sigma_U^{(L)}(s\to\infty)$ (the covariance matrix when the QELM feature map has $L$ layers) rather than $\Sigma_L$ (the covariance matrix of the first $L$ layers when the QELM feature map has $L+\Delta L$ layers). However, in what follows, we prove in the Fourier picture that they are explicitly related by a affine transformation. Specifically, we establish with the help of Proposition~\ref{prop:Fourier-convolution} in Section~\ref{appendix:Fourier-analysis} that
\begin{equation}
    \Sigma_L = T_0\,\Sigma_U^{(L)}(s\to\infty)\,T_0^\dagger + \tilde{\mathcal{E}}^\perp,
\end{equation}
for certain, explicitly determined matrices $T_0,\tilde{\mathcal{E}}^\perp\in\mathbb{C}^{d\times d}$ that depend on the deeper architecture due to $\Delta L$:

\begin{corollary} \label{corr:transformation-base-reference-layers}
    Assume that $\{O_k\}_{k\in[d]}$ defines an orthogonal system of observables and let $\Delta L\in\mathbb{N}$. Then,
    \begin{equation*}
        \Sigma_L = T_0\,\Sigma_U^{(L)}(s\to\infty)\,T_0^\dagger + \tilde{\mathcal{E}}^\perp,
    \end{equation*}
    where for each $k,k'\in[d]$, we defined transition matrices
    \begin{equation*}
        (T_0)_{k,k'}:=\sum_{\substack{\mu,\mu'\in\Gamma_{\Delta L} \\ \Lambda_{\mu'}-\Lambda_\mu=0}}T_{k,k'}(\mu,\mu') \quad \text{with}\quad T_{k,k'}(\mu,\mu'):= \frac{\operatorname{Tr}\left(O_{k'}^\dagger V_{\mu'}^\dagger U_{L+\Delta L+1}^\dagger O_k U_{L+\Delta L+1}V_{\mu}\right)}{\operatorname{Tr}(O_{k'}^2)},
    \end{equation*}
    where $V_\mu:=P_{\mu_{\Delta L}}^{(a)}U_{L+\Delta L}\cdots P_{\mu_1}^{(a)}$ for each partial path $\mu=(\mu_1,\dots,\mu_{\Delta L})\in\Gamma_{\Delta L}:=([2^{n_1}]\times\cdots\times[2^{n_p}])^{\Delta L}$, and an orthogonal leakage error term
    \begin{equation}
        \tilde{\mathcal{E}}^\perp := \sum_{\omega\in\Omega_L}\left\{T_0\,c_\omega^{(L)}\tilde{\epsilon}_\omega^\dagger + \tilde{\epsilon}_\omega\left(T_0 c_\omega^{(L)}\right)^\dagger + \tilde{\epsilon}_\omega\tilde{\epsilon}_\omega^\dagger\right\} \label{eq:orthogonal-leakage-error}
    \end{equation}
    with $\tilde{\epsilon}_\omega\in\mathbb{C}^d$ defined as in Proposition~\ref{prop:Fourier-convolution}.
\end{corollary}

\noindent
\emph{Proof.} By Proposition~\ref{prop:Fourier-convolution} we have that
\begin{equation}
        c_\omega^{(L+\Delta L)}= T_0 c_\omega^{(L)}+\tilde{\epsilon}_\omega, \label{eq:fourier-convolution-formula}
\end{equation}
where for each $k,k'\in[d]$,
\begin{equation}
    (T_0)_{k,k'}:=\sum_{\substack{\mu,\mu'\in\Gamma_{\Delta L} \\ \Lambda_{\mu'}^{(\Delta L)}-\Lambda_\mu^{(\Delta L)}=0}}T_{k,k'}(\mu,\mu') \quad\text{and}\quad (\tilde{\epsilon}_\omega)_k:=(\epsilon_\omega)_k \, + \sum_{\substack{\mu,\mu'\in\Gamma_{\Delta L} \\ \Lambda_{\mu'}^{(\Delta L)}-\Lambda_\mu^{(\Delta L)}\neq0}}\sum_{k'=1}^d T_{k,k'}(\mu,\mu') \left(c^{(L)}_{\omega-(\Lambda_{\mu}^{(\Delta L)}-\Lambda_{\mu'}^{(\Delta L)})}\right)_{k'}.
\end{equation}
Therefore, recalling the definition of $\Sigma_L$ in Eq.~\eqref{eq:Sigma_L-term} we find
\begin{align}
    \Sigma_L &= \sum_{\omega\in\Omega_L}c_\omega^{(L+\Delta L)}\left(c_\omega^{(L+\Delta L)}\right)^\dagger \stackrel{\eqref{eq:fourier-convolution-formula}}{=} \sum_{\omega\in\Omega_L} \left(T_0 c_\omega^{(L)}+\tilde{\varepsilon}_\omega\right)\left(T_0 c_\omega^{(L)}+\tilde{\varepsilon}_\omega\right)^\dagger \nonumber
    \\ &= \sum_{\omega\in\Omega_L}T_0\,c_\omega^{(L)}\left(c_\omega^{(L)}\right)^\dagger T_0^\dagger + \underbrace{\sum_{\omega\in\Omega_L}\left\{T_0\,c_\omega^{(L)}\tilde{\epsilon}_\omega^\dagger + \tilde{\epsilon}_\omega\left(T_0 c_\omega^{(L)}\right)^\dagger + \tilde{\epsilon}_\omega\tilde{\epsilon}_\omega^\dagger\right\}}_{=:\tilde{\mathcal{E}}^\perp} \nonumber
    \\ &= T_0\underbrace{\left(\sum_{\omega\in\Omega_L}c_\omega^{(L)}\left(c_\omega^{(L)}\right)^\dagger\right)}_{=\Sigma_U^{(L)}(s\to\infty)}T_0^\dagger + \tilde{\mathcal{E}}^\perp = T_0\,\Sigma_U^{(L)}(s\to\infty)\,T_0^\dagger + \tilde{\mathcal{E}}^\perp.
\end{align}

Since the matrices $T_0\Sigma_U^{(L)}(s\to\infty)\,T_0^\dagger$ and $\tilde{\mathcal{E}}^\perp$ are both Hermitian, we can apply Weyl's inequality~\cite{vershynin2020high} so that
\begin{equation}
    \lambda_{\min}(\Sigma_L) = \lambda_{\min}\left(T_0\,\Sigma_U^{(L)}(s\to\infty)\,T_0^\dagger + \tilde{\mathcal{E}}^\perp\right) \geq \sigma_{\min}(T_0)^2\,\lambda_{\min}\left(\Sigma_U^{(L)}(s\to\infty)\right) + \lambda_{\min}(\tilde{\mathcal{E}}^\perp),
\end{equation}
which may directly be substituted into Theorem~\ref{thm:condition-number-reuploads} in order to get apply it with the correct base reference population covariance $\Sigma_U^{(L)}$.
\qed

As a final remark, we observe that the orthogonal leakage error $\tilde{\mathcal{E}}^\perp$ is not positive semi-definite in general due to the mixed cross-terms involving the transition matrix $T_0$ in Eq.~\eqref{eq:orthogonal-leakage-error}. In contrast, the $T_0$-independent term $\sum_{\omega\in\Omega_L}\tilde{\epsilon}_\omega \tilde{\epsilon}_\omega^\dagger$ is positive semi-definite by construction.
This raises the question of under what conditions this positive semi-definite term dominates the mixed terms, ensuring that $\lambda_{\min}(\tilde{\mathcal{E}}^\perp)>-\varepsilon$ for some sufficiently small $\varepsilon>0$ and preventing the upper bound in Theorem~\ref{thm:condition-number-reuploads} from becoming vacuous.
Suppose that $\Delta L$ is large enough, featuring layers with many non-Clifford gates and highly entangling operations that nonetheless do not mix into a unitary 2-design. Intuitively, a simple Pauli operator $O_k$ will quickly scramble as it propagates through these additional layers, developing into an exponentially large superposition of high-weight, non-local Paulis.
If the tracked observable system $\{O_k\}_{k\in[d]}$ only contains a polynomial number of elements, i.e., $d=\mathcal{O}(\operatorname{Poly}(N))$, almost the entirety of this scrambled operator's weight will land in the orthogonal complement $O_k^\perp(\mu, \mu')$ defined in Eq.~\eqref{eq:path-operator-orthogonal-decomposition}.
Under these conditions, one would expect the orthogonal leakage contributions to dominate the cross-terms in Eq.~\eqref{eq:orthogonal-leakage-error}, keeping $\lambda_{\min}(\tilde{\mathcal{E}}^\perp)$ safely bounded above $-\varepsilon$. 
However, a rigorous proof remains highly non-trivial due to the complex, path-dependent structure of the leakage vector $\tilde{\epsilon}_\omega$, and it remains a possibility that the application of Weyl's inequality is too loose. 
Nonetheless, Corollary~\ref{corr:transformation-base-reference-layers} establishes an exact, structurally rigorous transformation formula for $\Sigma_L$ relative to the original population covariance base reference.

\subsection{The Effect of Finite Shot Noise} \label{subsec:finite-shot-noise}

Recall from Eq.~\eqref{eq:quantum-readout} that the readout components $\phi_U(x)_k=\langle O_k\rangle_x=\operatorname{Tr}(O_k\rho_U(x))$ are exact expectation values. In practice, one has to use a (random) finite shot estimator $\hat{\phi}_U(x)$ of $\phi_U$ in order to approximate the exact expectation values. In this section, we are going to elaborate how the induced randomness due to finite shot noise acts as a regularizer of the population covariance condition number, and how it limits high-dimensional feature correlations, thus improving the chance of double descent occurring with an interpolation threshold at $\gamma=1$ as dictated by Proposition~\ref{prop:over-fitting} in Section~\ref{sec:non-isotropic-analysis}.

Let $m_0\in\mathbb{N}$ denote the total number of measurement shots used for each observable in the system $\{O_k:k\in[d]\}$. 
For each $k\in[d]$ and $m\in[m_0]$ let $\mu_k^{(m)}(x)\in\mathbb{R}$ be the random variable defined as the random outcome of the $m$-th independent measurement outcome of the observable $O_k$ when the quantum system is prepared in the state $\rho_U(x)$. 
By the postulates of quantum mechanics, the random variable $\mu_k^{(m)}(x)$ takes values in $\operatorname{Spec}(O_k):=\{\lambda_j(O_k):j\in[2^N]\}$ for all $k\in[d]$ and thus defines a discrete random variable. In particular, $|\mu_k^{(m)}(x)|\leq\|O_k\|$ for all $k\in[d]$ and $m\in[m_0]$. Let $\mathbb{P}_{\operatorname{shot},k}(\,\cdot\mid x)$ denote the conditional probability distribution that described the random process of $\mu_k^{(m)}(x)$, conditioned on the input $x\sim\mathbb{P}$. To formalize the connection between individual measurement outcomes and the ideal quantum features, we invoke the spectral theorem for Hermitian operators. 
Every observable $O_k \in \operatorname{Herm}(2^N)$ uniquely admits a spectral decomposition of the form
\begin{equation}
    O_k = \sum_{\lambda\in\operatorname{Spec}(O_k)}\lambda P_{\lambda}^{(k)}, \label{eq:observable-spectral-decompisition1}
\end{equation}
where $P_{\lambda}^{(k)}\in\operatorname{Herm}(2^N)$ denotes the orthogonal projector onto the eigenspace corresponding to the eigenvalue $\lambda$. More precisely, 
$P_\lambda^{(k)} = \sum_{j=1}^{g_\lambda} |\psi_{\lambda,j}^{(k)}\rangle\langle\psi_{\lambda,j}^{(k)}|$ for some $|\psi_{\lambda,j}^{(k)}\rangle\in\mathcal{H}(2^N)$ such that $O_k|\psi_{\lambda,j}^{(k)}\rangle=\lambda |\psi_{\lambda,j}^{(k)}\rangle$ with $g_\lambda\in\mathbb{N}$ denoting the multiplicity of the eigenvalue $\lambda$. According to Born's rule, when the $N$-qubit system is prepared in the pre-measurement state $\rho_U(x)\in\mathcal{D}(2^N)$, the conditional probability of obtaining the specific measurement outcome $\lambda \in \operatorname{Spec}(O_k)$ during the $m$-th shot is explicitly given by
\begin{equation}
    \mathbb{P}_{\operatorname{shot},k}(\mu_k^{(m)}(x)=\lambda\mid x) = \operatorname{Tr}(P_\lambda^{(k)}\rho_U(x)).
\end{equation}
Consequently, the discrete, conditional expectation value of the random variable $\mu_k^{(m)}(x)$ with respect to the measurement distribution $\mathbb{P}_{\operatorname{shot},k}(\,\cdot\mid x)$ perfectly coincides with the exact readout component $\phi_U(x)_k$:
\begin{align}
    \mathbb{E}_{\operatorname{shot},k}[\mu_k^{(m)}(x)\mid x] &= \sum_{\lambda\in\operatorname{Spec}(O_k)}\lambda\underbrace{\mathbb{P}_{\operatorname{shot},k}(\mu_{k}^{(m)}(x)=\lambda\mid x)}_{=\operatorname{
    Tr}(P_\lambda^{(k)}\rho_U(x))} = \operatorname{Tr}\left(\left(\sum_{\lambda\in\operatorname{Spec}(O_k)}\lambda P^{(k)}_\lambda\right)\rho_U(x)\right) \nonumber
    \\ &\stackrel{\eqref{eq:observable-spectral-decompisition1}}{=} \operatorname{Tr}(O_k\rho_U(x)) = \langle O_k\rangle_x = \phi_U(x)_k. \label{eq:born-expectation}
\end{align}
We now define the finite-shot feature estimator $\hat{\phi}_U(x) \in \mathbb{R}^d$, a random vector defined component-wise by taking the empirical sample mean over the $m_0$ independent quantum measurement shots:
\begin{equation}
    \hat{\phi}_U(x)_k := \frac{1}{m_0} \sum_{m=1}^{m_0} \mu_k^{(m)}(x), \quad k\in[d] \label{eq:finite-shot-estimator-def}
\end{equation}
Observe that the finite-shot feature estimator $\hat{\phi}_U(x)$ defines an unbiased estimator of the ideal feature map $\phi_U(x)$, both conditionally and unconditionally, a property used throughout our computations further below. Indeed, taking the conditional expectation of Eq. \eqref{eq:finite-shot-estimator-def} given an input sample $x$, and applying the linearity of expectation alongside Eq. \eqref{eq:born-expectation}, we find
\begin{align}
    \mathbb{E}_{\operatorname{shot},k}[\hat{\phi}_U(x)_k \mid x] &= \mathbb{E}_{\operatorname{shot},k}\left[ \frac{1}{m_0} \sum_{m=1}^{m_0} \mu_k^{(m)}(x) \;\middle|\; x \right] = \frac{1}{m_0} \sum_{m=1}^{m_0} \underbrace{\mathbb{E}_{\operatorname{shot},k}[\mu_k^{(m)}(x) \mid x]}_{=\phi_U(x)_k} = \frac{1}{m_0} \sum_{m=1}^{m_0} \phi_U(x)_k = \phi_U(x)_k. \label{eq:unbiased-estimator}
\end{align}
Thus, $\mathbb{E}_{\operatorname{shot}}[\hat{\phi}_U(x) \mid x] = \phi_U(x)$ holds, establishing conditional unbiasedness. 
For our subsequent analysis we also need to introduce the following probability measure to account for shot noise. In fact, we can also use it to show that the estimator is unconditionally unbiased: We introduce the joint product measure (cf. Section~18 in Ref.~\cite{billingsley2017probability}),
\begin{equation}
    \mathbb{P}_{x,\operatorname{shot}}\left(x, \mu_1^{(m)}(x),\dots,\mu_d^{(m)}(x)\right) := \mathbb{P}(x)  \underbrace{\prod_{k=1}^d \mathbb{P}_{\operatorname{shot},k}\left(\mu_k^{(m)}(x)\;\middle|\; x\right)}_{=: \mathbb{P}_{\operatorname{shot}}\left(\mu_1^{(m)}(x),\dots,\mu_d^{(m)}(x)\;\middle|\; x\right)} \label{eq:joint-product-measure}
\end{equation}
which, by Fubini's theorem (cf. Theorem~6.1 of Ref.~\cite{wheeden1977measure}), yields for all $k\in[d]$ that
\begin{align}
    \mathbb{E}_{x, \text{shot}}[\hat{\phi}_U(x)] &= \int_{\mathcal{X}} \mathbb{E}_{\operatorname{shot}}[\hat{\phi}_U(x)\mid x]\,d\mathbb{P}(x) = \int_{\mathcal{X}} \mathbb{E}_{\operatorname{shot}}\left[\left(\begin{matrix} \hat{\phi}_U(x)_1 \\ \vdots \\ \hat{\phi}_U(x)_d \end{matrix}\right)\;\middle|\; x\right]\,d\mathbb{P}(x) \stackrel{\eqref{eq:joint-product-measure}}{=} \int_{\mathcal{X}} \left(\begin{matrix} \mathbb{E}_{\operatorname{shot},1}[\hat{\phi}_U(x)_1\mid x] \\ \vdots \\ \mathbb{E}_{\operatorname{shot},d}[\hat{\phi}_U(x)_d \mid x] \end{matrix}\right)\,d\mathbb{P}(x) \nonumber
    \\ &\stackrel{\eqref{eq:unbiased-estimator}}{=} \int_{\mathcal{X}} \left(\begin{matrix} \phi_U(x)_1 \\ \vdots \\ \phi_U(x)_d \end{matrix}\right)\,d\mathbb{P}(x) = \mathbb{E}_x[\phi_U(x)],
\end{align}
which precisely matches the exact feature expectation as claimed. Furthermore, the estimator is uniformly bounded via
\begin{equation}
    |\hat{\phi}_U(x)_k| \leq \frac{1}{m_0}\sum_{m=1}^{m_0}\underbrace{|\mu_k^{(m)}(x)|}_{\leq \|O_k\|} \leq \max_{k\in[d]}\|O_k\|. \label{eq:bounded-estimator}
\end{equation}
Because the $m_0$ measurements shots are mutually independent conditionally on $x$, the variance scales inversely with the number of shots:
\begin{equation}
    \operatorname{Var}_{\operatorname{shot},k}(\hat{\phi}_U(x)_k\mid x) = \frac{1}{m_0^2}\sum_{m=1}^{m_0}\operatorname{Var}_{\operatorname{shot},k}(\mu_k^{(m)}\mid x) = \frac{1}{m_0}\left(\langle O_k^2\rangle_x-\langle O_k\rangle_x^2\right),
\end{equation}
where the last identity is derived using Eq.~\eqref{eq:born-expectation} so that
\begin{equation}
    \operatorname{Var}_{\operatorname{shot},k}(\mu^{(m)}_k(x)\mid x)= \underbrace{\mathbb{E}_{\operatorname{shot},k}[\mu_k^{(m)}(x)^2\mid x]}_{=\langle O_k^2\rangle_x} - \underbrace{\mathbb{E}_{\operatorname{shot},k}[\mu_k^{(m)}(x)\mid x]^2}_{=\langle O_k\rangle_x^2} = \langle O_k^2\rangle_x-\langle O_k\rangle_x^2. \label{eq:quantum-variance}
\end{equation}
This also demonstrates why concentrating readouts are problematic for scalable applications (cf. Supplementary Section~\ref{appendix:2-designs}): if the readout concentrates exponentially in the number $N$ of qubits then at least an exponential number $m_0=\Omega(\operatorname{Exp}(N))$ is required in order to achieve a good estimate $\hat{\phi}_U(x)$ of the exact readout $\phi_U(x)$.

\smallskip
\par
\paragraph{Finite Shot Noise Population Covariance.} 
When relying on a finite shot estimator $\hat{\phi}_U(x)$ of the exact readout $\phi_U(x)$ one inevitable induces finite sampling
shot noise. Before we show that this has the effect of regularizing the condition number of the population covariance, we first have to determine the structure of the population covariance when working in the joint probability space in Eq.~\eqref{eq:joint-product-measure}. We shall model the shot noise $\epsilon(x)\in\mathbb{R}^d$ as a random vector conditioned on the input $x\sim\mathbb{P}$ defined via
\begin{equation}
    \hat{\phi}_U(x) = \phi_U(x)+\epsilon(x). \label{eq:shot-noise-vector}
\end{equation}
Observe that the shot noise vector is conditionally centered, i.e.,
\begin{equation}
    \mathbb{E}_{\operatorname{shot}}[\epsilon(x)\mid x] \stackrel{\eqref{eq:shot-noise-vector}}{=} \mathbb{E}_{\operatorname{shot}}[\hat{\phi}_U(x)-\phi_U(x)|x] = \underbrace{\mathbb{E}_{\operatorname{shot}}[\hat{\phi}_U(x)\mid x]}_{\stackrel{\eqref{eq:unbiased-estimator}}{=}\phi_U(x)} - \underbrace{\mathbb{E}_{\operatorname{shot}}[\phi_U(x)\mid x]}_{=\phi_U(x)} = 0 \label{eq:centered-shot-noise}
\end{equation}
since $\hat{\phi}_U(x)$ is an unbiased estimator of the readout $\phi_U(x)$ as per Eq.~\eqref{eq:unbiased-estimator}. 
In particular, applying Fubini's theorem again (cf. Theorem~6.1 of Ref.~\cite{wheeden1977measure}), 
\begin{equation}
    \mathbb{E}_{x,\operatorname{shot}}[\epsilon(x)] \stackrel{\eqref{eq:joint-product-measure}}{=} \mathbb{E}_x\left[\mathbb{E}_{\operatorname{shot}}[\epsilon(x)\mid x]\right] \stackrel{\eqref{eq:centered-shot-noise}}{=} \mathbb{E}_x[0] = 0
\end{equation}
showing that the shot noise $\epsilon(x)$ is also unconditionally centered. Furthermore, we have for each $k\in[d]$ that
\begin{equation}
    \epsilon(x)_k \stackrel{\eqref{eq:shot-noise-vector}}{=} \hat{\phi}_U(x)_k - \underbrace{\phi_U(x)_k}_{=\langle O_k\rangle_x} \stackrel{\eqref{eq:finite-shot-estimator-def}}{=}  \frac{1}{m_0} \left(\sum_{m=1}^{m_0} \mu_k^{(m)}(x)\right) - \langle O_k\rangle_x = \sum_{m=1}^{m_0}\underbrace{\frac{1}{m_0}\left\{\mu_k^{(m)}(x) - \langle O_k \rangle_x\right\}}_{=:X_m^{(k)}} \label{eq:shot-noise-components}
\end{equation}
defines a sum of $m_0$ independent, bounded random variables $X_m^{(k)}$ with $|X_m^{(k)}|\leq\frac{2\|O_k\|}{m_0}$ for all $m\in[m_0]$. 
By Hoeffding's inequality (cf. Theorem~2.2.6 in Ref.~\cite{vershynin2020high}), and since $\mathbb{E}_{\operatorname{shot},k}[X_m^{(k)}\mid x]=0$ due to $\mu_k^{(m)}(x)$ being an unbiased estimator of $\langle O_k\rangle_x$  by Eq.~\eqref{eq:unbiased-estimator}, we have that
\begin{align}
    \mathbb{P}_{\operatorname{shot},k}\left(\epsilon(x)_k\geq t\;\middle|\;x\right) &= \mathbb{P}_{\operatorname{shot},k}\left(\sum_{m=1}^{m_0}\left\{X_m^{(k)}- \mathbb{E}_{\operatorname{shot},k}[X_m^{(k)}\mid x]\right\}\geq t\;\middle|\;x\right) \nonumber
    \\ &\leq \exp\left(-\frac{2t^2}{\sum_{m=1}^{m_0}\left(\frac{4\|O_k\|}{m_0}\right)^2}\right) = \exp\left(-\frac{m_0 t^2}{8\|O_k\|^2}\right)
\end{align}
for all $t\geq0$. This establishes that the components $\epsilon(x)_k$ of the shot noise vector $\epsilon(x)$ concentrate around zero with sub-Gaussian tails depending directly on the total number of measurement shots $m_0$. 
That is, the larger the number of shots, the faster the estimator $\hat{\phi}_U(x)$ concentrates around the exact noiseless quantum readout $\phi_U(x)$. As a matter of fact, by Lemma~\ref{lemma:subgaussian-characterization} in Subsection~\ref{subsubsection:high-dim-probability} we know that each component $\epsilon(x)_k$ is sub-Gaussian conditionally on $x$ with a conditional sub-Gaussian norm of $\|\epsilon(x)_k \mid x\|_{\psi_2} = \mathcal{O} \left( \frac{\|O_k\|}{\sqrt{m_0}} \right)$. 

Now that we have introduced shot noise and studied its fundamental properties, we are interested in assessing how the newly appeared stochasticity affects the population covariance $\Sigma_U:=\mathbb{E}_{ x}\left[\phi_U( x)\,\phi_U( x)^\top\right] \in \mathbb{R}^{d\times d}$ defined in Eq.~\eqref{eq:quantum-feature-covariance}.  Under finite shot noise, the deterministic covariance matrix $\Sigma_U$ must be replaced by the joint expectation over the data and measurement distributions as per Eq.~\eqref{eq:joint-product-measure}:
\begin{equation}
    \hat{\Sigma}_U := \mathbb{E}_{x, \operatorname{shot}} \left[ \hat{\phi}_U(x)\hat{\phi}_U(x)^\top \right] \in \mathbb{R}^{d\times d}. \label{eq:noisy-population-covariance-def}
\end{equation}
Substituting the shot noise from Eq.~\eqref{eq:shot-noise-vector} into this definition yields
\begin{align}
    \hat{\Sigma}_U &= \mathbb{E}_{x, \operatorname{shot}} \left[ (\phi_U(x) + \epsilon(x))(\phi_U(x) + \epsilon(x))^\top \right] =
     \mathbb{E}_{x, \operatorname{shot}} \left[ \phi_U(x)\phi_U(x)^\top + \phi_U(x)\epsilon(x)^\top + \epsilon(x)\phi_U(x)^\top + \epsilon(x)\epsilon(x)^\top \right] \nonumber
    \\ &= \mathbb{E}_{x, \operatorname{shot}}\left[\phi_U(x)\phi_U(x)^\top\right] + \mathbb{E}_{x, \operatorname{shot}}\left[\phi_U(x)\epsilon(x)^\top\right] + \mathbb{E}_{x, \operatorname{shot}}\left[\epsilon(x)\phi_U(x)^\top\right] + \mathbb{E}_{x, \operatorname{shot}}\left[\epsilon(x)\epsilon(x)^\top\right] \nonumber
    \\ &\stackrel{\eqref{eq:joint-product-measure}}{=} \mathbb{E}_x\left[\mathbb{E}_{\operatorname{shot}}\left[\phi_U(x)\phi_U(x)^\top\mid x\right] + \mathbb{E}_{\operatorname{shot}}\left[\phi_U(x)\epsilon(x)^\top\mid x\right] + \mathbb{E}_{\operatorname{shot}}\left[\epsilon(x)\phi_U(x)^\top\mid x\right] + \mathbb{E}_{\operatorname{shot}}\left[\epsilon(x)\epsilon(x)^\top\mid x\right]\right] \nonumber
    \\ &= \underbrace{\mathbb{E}_x\left[\phi_U(x)\phi_U(x)^\top\right]}_{=\Sigma_U} + \mathbb{E}_x[\phi_U(x)\underbrace{\mathbb{E}_{\operatorname{shot}}\left[\epsilon(x)^\top\mid x\right]}_{\stackrel{\eqref{eq:centered-shot-noise}}{=}0} + \underbrace{\mathbb{E}_{\operatorname{shot}}\left[\epsilon(x)\mid x\right]}_{\stackrel{\eqref{eq:centered-shot-noise}}{=}0}\phi_U(x)^\top  + \mathbb{E}_{\operatorname{shot}}\left[\epsilon(x)\epsilon(x)^\top\mid x\right]] \nonumber
    \\ &= \Sigma_U + \mathbb{E}_x\left[\mathbb{E}_{\operatorname{shot}}\left[\epsilon(x)\epsilon(x)^\top\mid x\right]\right], \label{eq:population-covariance-estimator}
\end{align}
applying Fubini's theorem and the fact that the exact readout $\phi_U(x)$ is independent of the shot measure $\mathbb{P}_{\operatorname{shot}}$. 
Let us further analyse the conditional shot noise term $\mathbb{E}_{\operatorname{shot}}\left[\epsilon(x)\epsilon(x)^\top\right]$. 
For each $k,k'\in[d]$ we explicitly compute
\begin{align}
    \mathbb{E}_{\operatorname{shot}}\left[\epsilon(x)\epsilon(x)^\top\mid x \right]_{k,k'} &= \mathbb{E}_{\operatorname{shot}}[\epsilon(x)_k \epsilon(x)_{k'} \mid x] \stackrel{\eqref{eq:shot-noise-components}}{=} \frac{1}{m_0^2} \sum_{m,m'=1}^{m_0} \mathbb{E}_{\operatorname{shot}}\left[ \left(\mu_k^{(m)}(x) - \langle O_k \rangle_x\right)\left(\mu_{k'}^{(m')}(x) - \langle O_{k'} \rangle_x\right) \;\middle|\; x \right]. \label{eq:shot-noise-covariance}
\end{align}
If $k\neq k'$, then
\begin{align}
    &\mathbb{E}_{\operatorname{shot}}\left[ \left(\mu_k^{(m)}(x) - \langle O_k \rangle_x\right)\left(\mu_{k'}^{(m')}(x) - \langle O_{k'} \rangle_x\right) \;\middle|\; x \right] \nonumber
    \\ &= \underbrace{\mathbb{E}_{\operatorname{shot},k}\left[\left(\mu_{k}^{(m)}(x) - \langle O_{k} \rangle_x\right)\;\middle|\;x\right]}_{\stackrel{\eqref{eq:born-expectation}}{=}0}\,\underbrace{\mathbb{E}_{\operatorname{shot},k'}\left[\left(\mu_{k'}^{(m')}(x) - \langle O_{k'} \rangle_x\right)\;\middle|\; x\right]}_{\stackrel{\eqref{eq:born-expectation}}{=}0} = 0 \label{eq:noise-term-off-diagonal}
\end{align}
for all $m,m'\in[m_0]$ assuming a protocol where each observable is measured using a separate, 
independent batch of quantum circuit executions so that the measurements of $O_k$ and $O_{k'}$ are stochastically independent. Notably, this assumed independence is not enough to ensure that the off-diagonal terms of $\mathbb{E}_{\text{shot}}[\epsilon(x)\epsilon(x)^\top\mid x]$ vanish, which is why we require the previously proven conditional unbiasedness in Eq.~\eqref{eq:born-expectation}.
Consequently, the random variables $\mu_k^{(m)}(x)$ and $\mu_{k'}^{(m')}(x)$ are conditionally independent for all shot pairs $m, m'\in[m_0]$. This shows that $\mathbb{E}_{\operatorname{shot}}\left[\epsilon(x)\epsilon(x)^\top\right]$ is diagonal.

Next, we explicitly compute the diagonal elements $k=k$ and see what happens when $m=m'$ and $m\neq m'$, respectively. When measuring the same observable $O_k$, the measurement shots are independent across different shots $m \neq m'$, causing all cross-shot expectations to vanish and only the matched pairs $m = m'$ to survive:
\begin{align}
    &\mathbb{E}_{\operatorname{shot}}\left[ \left(\mu_k^{(m)}(x) - \langle O_k \rangle_x\right)\left(\mu_{k}^{(m')}(x) - \langle O_{k} \rangle_x\right) \;\middle|\; x \right] = \delta_{m,m'}\,\mathbb{E}_{\operatorname{shot}}\left[\left(\mu_k^{(m)}(x) - \langle O_k \rangle_x\right)^2 \;\middle|\; x \right] \nonumber
    \\ &= \delta_{m,m'}\,\mathbb{E}_{\operatorname{shot},k}\left[\left(\mu_k^{(m)}(x)-\mathbb{E}_{\operatorname{shot},k}\left[\mu_k^{(m)}(x)\;\middle|\; x\right]\right)^2\;\middle|\;x\right] = \delta_{m,m'}\, \operatorname{Var}_{\operatorname{shot},k}\left(\mu_k^{(m)}(x)\;\middle|\;x\right) \nonumber
    \\ &\stackrel{\eqref{eq:born-expectation}}{=} \delta_{m,m'}\left(\langle O_k^2\rangle_x - \langle O_k\rangle_x^2\right). \label{eq:shot-covariance-diagonal-elements}
\end{align}
Consequently, for all $k\in[d]$, we explicitly have
\begin{align}
    \mathbb{E}_{\operatorname{shot}}\left[\epsilon(x)\epsilon(x)^\top\mid x \right]_{k,k} &\stackrel{\eqref{eq:shot-noise-covariance}}{=} \frac{1}{m_0^2} \sum_{m,m'=1}^{m_0} \mathbb{E}_{\operatorname{shot}}\left[ \left(\mu_k^{(m)}(x) - \langle O_k \rangle_x\right)\left(\mu_{k'}^{(m')}(x) - \langle O_{k} \rangle_x\right) \;\middle|\; x \right] \nonumber
    \\ &\stackrel{\eqref{eq:shot-covariance-diagonal-elements}}{=} \frac{1}{m_0^2} \sum_{m,m'=1}^{m_0} \delta_{m,m'}\left(\langle O_k^2\rangle_x - \langle O_k\rangle_x^2\right) = \frac{1}{m_0^2}\sum_{m=1}^{m_0} \left(\langle O_k^2\rangle_x - \langle O_k\rangle_x^2\right) \nonumber
    \\ &= \frac{1}{m_0}\left(\langle O_k^2\rangle_x - \langle O_k\rangle_x^2\right). \label{eq:diagonal-noise-term}
\end{align}
In summary, we have established that the updated population covariance 
\begin{equation}
    \hat{\Sigma}_U \stackrel{\eqref{eq:population-covariance-estimator}}{=} \Sigma_U + \mathbb{E}_x\left[\mathbb{E}_{\operatorname{shot}}\left[\epsilon(x)\epsilon(x)^\top\mid x\right]\right] = \Sigma_U + \frac{1}{m_0}\operatorname{diag}\left(\mathbb{E}_x\left[\langle O_k^2\rangle_x - \langle O_k\rangle_x^2\right]\right)_{k=1}^d  = \Sigma_U + D_{\text{shot}},\label{eq:noise-covariance-decomposition}
\end{equation}
adds a diagonal quantum variance term $D_{\text{shot}}$, for each $k\in[d]$ defined as
\begin{equation}
    D_{\text{shot}} := \frac{1}{m_0}\operatorname{diag}(\sigma_k^2)_{k=1}^d\quad\text{where}\quad \sigma_k^2:=\mathbb{E}_x\left[\langle O_k^2\rangle_x - \langle O_k\rangle_x^2\right], \label{eq:diagonal-regularization}
\end{equation}
to the original population covariance that scales inversely to the number $m_0$ of shots. In other words, finite shot noise acts as an intrinsic, data-averaged diagonal regularizer on the population covariance matrix $\Sigma_U$.

\smallskip
\par
\paragraph{Finite Shot Noise Regularization.} In the previous paragraph, we have proven that $\hat{\Sigma}_U = \Sigma_U + D_{\text{shot}}$, where the diagonal matrix $D_{\text{shot}}$ scales inversely with the number $m_0$ of sampling shots. Thus, if $\Sigma_U$ is ill-conditioned, the additional term $D_{\text{shot}}$ can act as a regularization term provided that $m_0$ is sufficiently bounded. We now prove that this regularization improves the condition number $\kappa(\hat{\Sigma}_U)$, and thus generalization by Theorem~\ref{thm:risk-bound-nonisotropic} in Subsection~\ref{sec:non-isotropic-analysis}. By Weyl's inequality~\cite{vershynin2020high} we have the bound
\begin{align}
    \lambda_{\max}(\hat{\Sigma}_U) &\stackrel{\eqref{eq:noise-covariance-decomposition}}{=} \lambda_{\max}(\Sigma_U + D_{\text{shot}}) \leq \lambda_{\max}(\Sigma_U) + \lambda_{\max}(D_{\text{shot}}) \stackrel{\eqref{eq:diagonal-regularization}}{=} \|\Sigma_U\| + \frac{\sigma_{\max}^2}{m_0}
\end{align}
where $\sigma_{\max}^2 := \max_{k\in[d]}\sigma_k^2$. Similarly,
\begin{align}
    \lambda_{\min}(\hat{\Sigma}_U) &\stackrel{\eqref{eq:noise-covariance-decomposition}}{=} \lambda_{\min}(\Sigma_U + D_{\text{shot}}) \geq \lambda_{\min}(\Sigma_U) + \lambda_{\min}(D_{\text{shot}}) \stackrel{\eqref{eq:diagonal-regularization}}{=} \lambda_{\min}(\Sigma_U) + \frac{\sigma_{\min}^2}{m_0} \label{eq:noisy-covariance-spectral-bound}
\end{align}
where $\sigma_{\min}^2:=\min_{k\in[d]}\sigma_k^2$. As a result, we may use the previous bounds to estimate the condition number via
\begin{align}
    \kappa(\hat{\Sigma}_U) &= \frac{\lambda_{\max}(\hat{\Sigma}_U)}{\lambda_{\min}(\hat{\Sigma}_U)} \leq \frac{\|\Sigma_U\| + \frac{\sigma_{\max}^2}{m_0}}{\lambda_{\min}(\Sigma_U)+\frac{\sigma_{\min}^2}{m_0}}, \label{eq:condition-number-improves-with-shots}
\end{align}
which shows that the condition number $\kappa(\hat{\Sigma}_U)$ is indeed regularized with decreasing shot counts $m_0$, which is equivalent to increasing shot noise as per Eqs.~\eqref{eq:shot-noise-vector} and~\eqref{eq:finite-shot-estimator-def}. Importantly, the effect of regularization is only meaningful if $\Sigma_U$ is ill-conditioned, i.e., if $\lambda_{\min}(\Sigma_U) \approx 0$ causing $\kappa(\Sigma_U)$ to explode. More precisely, the bound in Eq.~\eqref{eq:condition-number-improves-with-shots} explicitly tells us that the condition number improves with finite shot noise, i.e. $\kappa(\Sigma_U) > \kappa(\hat{\Sigma}_U)$, if and only if 
\begin{equation}
    \kappa(\Sigma_U) >  \frac{\sigma_{\max}^2}{\sigma_{\min}^2}.
\end{equation}
In particular, if the quantum variance is isotropic among the distinct observables, that is, $\sigma_{\max}^2=\sigma_{\min}^2$, then we always have an improvement of the condition number as long as $\kappa(\Sigma_U)>1$. We have proven the following theorem:

\begin{theorem} \label{thm:condition-number-improves-with-shots} Let $m_0\in\mathbb{N}$ and consider the noisy population covariance
\begin{equation*}
    \hat{\Sigma}_U := \mathbb{E}_{x, \operatorname{shot}} \left[ \hat{\phi}_U(x)\hat{\phi}_U(x)^\top \right] \quad\text{where}\quad \hat{\phi}_U(x) = \phi_U(x)+\epsilon(x)
\end{equation*}
due to finite sampling shot noise $\epsilon(x)$ (cf. Eqs.~\eqref{eq:finite-shot-estimator-def} and~\eqref{eq:joint-product-measure}). Then, we generally have the bound
\begin{equation*}
    \kappa(\hat{\Sigma}_U) \leq \frac{\lambda_{\max}(\Sigma_U) + \frac{\sigma_{\max}^2}{m_0}}{\lambda_{\min}(\Sigma_U)+\frac{\sigma_{\min}^2}{m_0}},
\end{equation*}
where the extreme quantum variances $\sigma_{\min}^2$ and $\sigma_{\max}^2$ are defined by taking the minimum and maximum of $\sigma_k^2:=\mathbb{E}_x\left[\langle O_k^2\rangle_x - \langle O_k\rangle_x^2\right]$ across $k\in[d]$, respectively. In particular, finite shot noise $\epsilon(x)$ improves the noiseless condition number $\kappa(\Sigma_U)$ if and only if $\kappa(\Sigma_U) >  \frac{\sigma_{\max}^2}{\sigma_{\min}^2}$.
\end{theorem}

\smallskip
\par
\paragraph{Shot Noise Limits Correlations and Re-establishes Double Descent.} This is the final section of our contributing theory. It is worth taking a step back and recall what we have shown so far in order to motivate our last main result. 
In Theorem~\ref{thm:risk-bound-nonisotropic} of  Subsection~\ref{sec:non-isotropic-analysis} we have shown that the condition number of the population covariance determines the quality of generalization in the over-parametrized regime. 
Leveraging this result in the context of QELMs, we have then employed the Fourier interpretation of quantum models in Section~\ref{appendix:fourier-interpretation} to explicitly prove that increasing input strength (cf. Section~\ref{subsubsec:input-strength}) and reuploads (cf. Section~\ref{subsubsec:input-reuploads}) both improve the condition number of the population covariance, and thus generalization in the over-parametrized regime. 
Additionally, if the population covariance is ill-conditioned, we have shown in Theorem~\ref{thm:condition-number-improves-with-shots} above that finite sampling shot noise further improves the condition number under certain assumptions of the quantum measurement process. 
Note that all these results only informs us about the shape of the double descent curve in the over-parametrized regime. 
Thus far, we have not established any result in the QELM context describing the risk curve in the under-parametrized curve.

We shall address this problem by revisiting Proposition~\ref{prop:over-fitting} in Subsection~\ref{sec:non-isotropic-analysis}, telling us that the expected risk exhibits an ideal double descent curve if the correlations (cf. Lemma~\eqref{eq:second-order-correlation-measure} in Subsection~\ref{subsubsection:high-dim-probability}) are bounded via
\begin{equation}
    \operatorname{Var}_x(\phi_U(x)A^{(\Sigma)}\phi_U(x))=o(d^2) \label{eq:correlations-bound1i}
\end{equation}
as $d\to\infty$ for all $A\in\mathbb{R}^{d\times d}$ with $\|A\|=\mathcal{O}_d(1)$. Thus, if we are able to prove that the condition in Eq.~\eqref{eq:correlations-bound1i} holds true with the noisy population covariance $\Sigma\equiv \hat{\Sigma}_U$, while it fails to hold true for the noiseless $\Sigma_U$, then, this would confirm that finite sampling noise re-established the double descent curve with an interpolation threshold at $\gamma=1$. 
However, a technical yet important subtlety is that one requires to compute the variance in Eq.~\eqref{eq:correlations-bound1i} with respect to the probability measure of the joint space as per Eq.~\eqref{eq:joint-product-measure}. This results in excessive computations that are not essential for understanding the main idea. For that reason, we shall separately compile these computations in Section~\ref{appendix:others}. Our last main contribution is the following theorem:

\begin{theorem}\label{thm:finite-shot-noise-correlation}
    Assume that $\{O_k\}_{k\in[d]}$ is a rescaled system of observables satisfying $\max_{k\in[d]}\|O_k\|^2=\mathcal{O}(d^{-\frac{1}{2}})$ as $d\to\infty$, and suppose that
    \begin{equation*}
        \mathcal{C}:=\max_{\substack{k,k'\in[d]\\ k\neq k'}}\left|\operatorname{Cov}_x\left(\sigma_k^2(x),\sigma_{k'}^2(x)\right)\right| = o_d(1)\quad\text{where}\quad \sigma_k(x)^2:=\langle O_k^2\rangle_x - \langle O_k\rangle_x^2.
    \end{equation*}
    Then, if we additionally assume that $m_0=o(\sqrt{d})$, we have
    \begin{equation*}
        \sup_{\substack{A\in\mathbb{R}^{d\times d}\\ \|A\|=\mathcal{O}_d(1)}}\operatorname{Var}_{x,\operatorname{shot}}\left(\hat\phi_U(x)^\top A^{(\hat\Sigma_U)}\hat\phi_U(x)\right) = o(d^2)
    \end{equation*}
    as $d\to\infty$, where $A^{(\hat{\Sigma}_U)}:=\hat{\Sigma}_U^{-\frac{1}{2}}A\,\hat{\Sigma}_U^{-\frac{1}{2}}$ with $\hat{\Sigma}_U$ denoting the noisy population covariance defined in Eq.~\eqref{eq:noisy-population-covariance-def}.
\end{theorem}

\noindent
\emph{Proof.} Let $A\in\mathbb{R}^{d\times d}$ be arbitrary with $\|A\|=\mathcal{O}_d(1)$ and define $a := \operatorname{vec}(A^{(\hat{\Sigma}_U)}) \in \mathbb{R}^{d^2 \times 1}$ to be the vectorization of the modified matrix $A^{(\hat{\Sigma}_U)}:=\hat{\Sigma}_U^{-\frac{1}{2}} A\,{\hat{\Sigma}_U}^{-\frac{1}{2}}$ in order to contract tensor representations $\mathcal{T}\in\mathbb{R}^{d^2\times d^2}$ to scalars $a^\top \mathcal{T} a$ in our subsequent analysis. By Lemma~\eqref{eq:second-order-correlation-measure} in Subsection~\ref{subsubsection:high-dim-probability} it suffices to identify the condition on $m_0$ such that the bound
\begin{equation}
    \operatorname{Var}_{x,\text{shot}}\left({\hat{\phi}_U(x)}^\top A^{(\hat{\Sigma}_U)}\hat{\phi}_U(x)\right) = \sum_{k,k',\eta,\eta'=1}^d A^{(\hat{\Sigma}_U)}_{k,k'}A^{(\hat{\Sigma}_U)}_{\eta,\eta'}\,\underbrace{\operatorname{Cov}_{x,\text{shot}}(\hat\phi_U(x)_k\hat\phi_U(x)_{k'},\hat\phi_U(x)_\eta\hat\phi_U(x)_{\eta'})}_{=\mathcal{\hat{V}}_{(k,k'),(\eta,\eta')}} = a^\top \mathcal{\hat{V}} a = o(d^2)
\end{equation}
is satisfied as $d\to\infty$. 
Here we defined $\mathcal{\hat{V}} \in \mathbb{R}^{d^2 \times d^2}$ as the matrix representation of the fourth-order covariance tensor of the full, noisy feature products.
In order to find such a condition on $m_0$, we split
\begin{equation}
    a^\top \mathcal{\hat{V}}a=T_1+T_2+T_3 + T_{\text{special}}
\end{equation}
into four individual terms that we will bound separately in Subsection~\ref{subsubsection:high-dim-probability}. Indeed, following the proof of Lemma~\ref{lemma:correlation-tensor-contraction-decomposition} we have the exact representations
\begin{align}
    T_1 &= \operatorname{Var}_x\left(\phi_U(x)^\top A^{(\hat{\Sigma}_U)}\phi_U(x) + \operatorname{Tr}\left(A^{(\hat{\Sigma}_U)}D(x)\right)\right), \nonumber
    \\ T_2 &= 4\mathbb{E}_x\left[\phi_U(x)^\top A^{(\hat{\Sigma}_U)}D(x)A^{(\hat{\Sigma}_U)}\phi_U(x)\right], \nonumber
    \\ T_3 &= \mathbb{E}_x\left[\operatorname{Var}_{\operatorname{shot}}\left(\epsilon(x)^\top A^{(\hat{\Sigma}_U)}\epsilon(x)\;\middle|\;x\right)\right],
\end{align}
and furthermore $T_\text{special}=\mathcal{O}(d)$ as $d\to\infty$.
Because the special term $T_{\text{special}}$ trivially satisfies $o(d^2)$ as $d\to\infty$, it thus remains to show that the remaining terms $T_i$ for each $i\in\{1,2,3\}$ are also bounded by $o(d^2)$ using by using the stated assumptions of the theorem.

Let us begin with the second term $T_2$ next: We have
\begin{align}
    \frac{1}{4}T_2 &= \mathbb{E}_x\left[\phi_U(x)^\top A^{(\hat{\Sigma}_U)}D(x)A^{(\hat{\Sigma}_U)}\phi_U(x)\right] \leq \mathbb{E}_x\left[\underbrace{\|\phi_U(x)\|^2}_{\leq d M^2}\underbrace{\left\|A^{(\hat\Sigma_U)}\right\|^2}_{\stackrel{\eqref{eq:Frobenius-bound1}}{=} \mathcal{O}_d(m_0^2)}\underbrace{\|D(x)\|}_{\stackrel{\eqref{eq:diagonal-shot-noise-matrix}}{=}\mathcal{O}_d(m_0^{-1})}\right] = \mathcal{O}_d\left(dm_0\right).
\end{align}
This shows that $T_2=\mathcal{O}_d(dm_0)$ as $d\to\infty$ which immediately implies that $m_0 = o(d)$ is a necessary condition for $a^\top\mathcal{\hat{V}}a=o(d^2)$ to hold true.

We continue by bounding the pure shot noise term $T_3$. We first note that
\begin{align}
    T_3 &= \mathbb{E}_x\left[\operatorname{Var}_{\operatorname{shot}}\left(\epsilon(x)^\top A^{(\hat{\Sigma}_U)}\epsilon(x)\;\middle|\;x\right)\right] = \mathbb{E}_x\left[\operatorname{Var}_{\operatorname{shot}}\left( \sum_{k,k'=1}^d A^{(\hat{\Sigma}_U)}_{k,k'} \epsilon(x)_k \epsilon(x)_{k'} \;\middle|\; x \right)\right] \nonumber
    \\ &= \mathbb{E}_x\left[\operatorname{Var}_{\text{shot}}\left(\sum_{k=1}^d A_{k,k}^{(\hat\Sigma_U)}\epsilon(x)_k^2 + \sum_{\substack{k,k'=1 \\ k\neq k'}}^dA_{k,k'}^{(\hat\Sigma_U)}\epsilon(x)_{k}\epsilon(x)_{k'}\,\middle|\;x\right)\right] \nonumber
    \\ &= \mathbb{E}_x\left[\operatorname{Var}_{\text{shot}}\left(\sum_{k=1}^d A_{k,k}^{(\hat\Sigma_U)}\epsilon(x)_k^2\,\middle|\;x\right) + \operatorname{Var}_{\text{shot}}\left(\sum_{\substack{k,k'=1 \\ k\neq k'}}^dA_{k,k'}^{(\hat\Sigma_U)}\epsilon(x)_{k}\epsilon(x)_{k'}\,\middle|\;x\right)\right] \label{eq:T3-term1}
\end{align}
because
\begin{align}
    \operatorname{Cov}_{\text{shot}}\left(\sum_{k=1}^d A_{k,k}^{(\hat\Sigma_U)}\epsilon(x)_k^2, \sum_{\substack{k,k'=1 \\ k\neq k'}}^dA_{k,k'}^{(\hat\Sigma_U)}\epsilon(x)_{k}\epsilon(x)_{k'}\,\middle|\;x\right) = \sum_{\eta=1}^d\sum_{\substack{k,k'=1 \\ k\neq k'}}^dA_{\eta,\eta}^{(\hat\Sigma_U)}A_{k,k'}^{(\hat\Sigma_U)}\underbrace{\operatorname{Cov}_{\text{shot}}\left(\epsilon(x)_\eta^2,\epsilon(x)_k\epsilon(x)_{k'}\mid x\right)}_{=0} = 0.
\end{align}
Indeed, for each $\eta,k,k'\in[d]$ with $k\neq k'$ we have
\begin{align}
    \operatorname{Cov}_{\text{shot}}\left(\epsilon(x)_\eta^2,\epsilon(x)_k\epsilon(x)_{k'}\mid x\right) &= \underbrace{\mathbb{E}_{\text{shot}}\left[\epsilon(x)_\eta^2\epsilon(x)_k\epsilon(x)_{k'}\mid x\right]}_{=0} - \mathbb{E}_{\text{shot}}\left[\epsilon(x)_\eta^2\mid x\right]\underbrace{\mathbb{E}_{\text{shot}}\left[\epsilon(x)_k\epsilon(x)_{k'}\mid x\right]}_{=0}=0
\end{align}
since $\epsilon(x)_k$ and $\epsilon(x)_{k'}$ are independent due to $k\neq k'$ and conditionally centered as per Eq.~\eqref{eq:centered-shot-noise}. Therefore, again by pairwise independence of the shot noise components $\epsilon(x)_k$ and Eq.~\eqref{eq:centered-shot-noise},
\begin{align}
    T_3 &\stackrel{\eqref{eq:T3-term1}}{=} \mathbb{E}_x\left[\operatorname{Var}_{\text{shot}}\left(\sum_{k=1}^d A_{k,k}^{(\hat\Sigma_U)}\epsilon(x)_k^2\,\middle|\;x\right) + \operatorname{Var}_{\text{shot}}\left(\sum_{\substack{k,k'=1 \\ k\neq k'}}^dA_{k,k'}^{(\hat\Sigma_U)}\epsilon(x)_{k}\epsilon(x)_{k'}\,\middle|\;x\right)\right] \nonumber
    \\ &= \mathbb{E}_x\left[\sum_{k=1}^d\left(A_{k,k}^{(\hat\Sigma_U)}\right)^2\underbrace{\operatorname{Var}_{\text{shot}}\left(\epsilon(x)_k^2\mid x\right)}_{=\mathcal{O}_d(m_0^{-3})} + \,2\sum_{\substack{k,k'=1 \\ k\neq k'}}^d\left(A_{k,k'}^{(\hat\Sigma_U)}\right)^2\underbrace{\operatorname{Var}_{\text{shot}}\left(\epsilon(x)_k\mid x\right)\operatorname{Var}_{\text{shot}}\left(\epsilon(x)_{k'}\mid x\right)}_{=\mathcal{O}_d\left(m_0^{-2}\right)}\right] \nonumber
    \\ &= \mathcal{O}_d\left(\frac{1}{m_0^2}\right)\,\left(\sum_{k=1}^d\left(A_{k,k}^{(\hat\Sigma_U)}\right)^2 + \sum_{\substack{k,k'=1 \\ k\neq k'}}^d\left(A_{k,k'}^{(\hat\Sigma_U)}\right)^2\right) = \mathcal{O}_d\left(\frac{1}{m_0^2}\sum_{k,k'=1}^d\left(A_{k,k'}^{(\hat\Sigma_U)}\right)^2\right) \nonumber
    \\ &= \mathcal{O}_d\left(\frac{1}{m_0^2}\operatorname{Tr}\left(\left(A^{(\hat\Sigma_U)}\right)^\top A^{(\hat\Sigma_U)}\right)\right)  \stackrel{\eqref{eq:contraction-cauchy-schwarz-term1}}{=} \mathcal{O}_d\left(\frac{d\|A^{(\hat\Sigma_U)}\|^2}{m_0^2}\right) \stackrel{\eqref{eq:Frobenius-bound1}}{=} \mathcal{O}(d),
\end{align}
where the second equation used both $\operatorname{Var}(XY) = \operatorname{Var}(X)\operatorname{Var}(Y)$ for independent centered random variables $X,Y$, and the fact that
\begin{align}
    \operatorname{Var}_{\text{shot}}\left(\epsilon(x)_k\mid x\right) &\stackrel{\eqref{eq:shot-noise-components}}{=} \operatorname{Var}_{\text{shot}}\left(\frac{1}{m_0}\sum_{m=1}^{m_0}\left\{\mu_k^{(m)}(x) - \langle O_k \rangle_x\right\}\,\middle|\;x\right) \nonumber
    \\ &= \frac{1}{m_0^2}\sum_{m=1}^{m_0}\underbrace{\operatorname{Var}_{\text{shot}}\left(\mu_k^{(m)}(x) - \langle O_k \rangle_x\mid x\right)}_{=\mathcal{O}_d(1)} = \mathcal{O}_d\left(\frac{1}{m_0}\right)
\end{align}
by pairwise independence of $\mu_k^{(m)}(x)$, and similarly, using $\operatorname{Var}(X) + \mathbb{E}[X]^2 =\mathbb{E}[X^2]$,
\begin{align}
    \operatorname{Var}_{\operatorname{shot}}\left(\epsilon(x)_k^2\mid x\right) + \underbrace{\left(\mathbb{E}_{\text{shot}}\left[\epsilon(x)_k^2\mid x\right]\right)^2}_{\stackrel{\eqref{eq:diagonal-noise-term}}{=} \mathcal{O}_d(m_0^{-2})} &= \mathbb{E}_{\text{shot}}\left[\epsilon(x)_k^4\mid x\right] \nonumber
    \\ &\stackrel{\eqref{eq:shot-noise-components}}{=} \frac{1}{m_0^4}\sum_{m=1}^{m_0}\underbrace{\mathbb{E}_{\text{shot}}\left[\left(\mu_k^{(m)}(x) - \langle O_k \rangle_x\right)^4\,\middle|\; x\right]}_{=\mathcal{O}_d(1)} = \mathcal{O}_d\left(\frac{1}{m_0^3}\right).
\end{align}
Therefore, we have in particular shown that $T_3=o(d^2)$ holds true for all $m_0\in\mathbb{N}$.

Finally, we bound the remaining term $T_1$: With the help of Lemma~\ref{lemma:bound-variance-sum-of-RV} in Supplementary Section~\ref{appendix:notations} we decouple
\begin{align}
    \frac{1}{2}T_1 &= \frac{1}{2}\operatorname{Var}_x\left(\phi_U(x)^\top A^{(\hat{\Sigma}_U)}\phi_U(x) + \operatorname{Tr}\left(A^{(\hat{\Sigma}_U)}D(x)\right)\right) \leq \underbrace{\operatorname{Var}_x\left(\phi_U(x)^\top A^{(\hat\Sigma_U)}\phi_U(x)\right)}_{=: T_1^{(\text{I})}} + \underbrace{\operatorname{Var}_x\left(\operatorname{Tr}\left(A^{(\hat\Sigma_U)}D(x)\right)\right)}_{=: T_1^{(\text{II})}}.
\end{align}
We introduce the following quantity that measures correlations of quantum measurement variances,
\begin{equation}
    \mathcal{C}:=\max_{\substack{k,k'\in[d]\\ k\neq k'}}\left|\operatorname{Cov}_x\left(\sigma_k^2(x),\sigma_{k'}^2(x)\right)\right| ,\label{eq:quantum-correlation-constant}
\end{equation}
recalling $\sigma_k^2(x):=\langle O_k^2\rangle_x-\langle O_k\rangle_x^2$, and use it to directly bound the second decoupled term with the help of Cauchy-Schwarz applied with the all-ones vector:
\begin{align}
    T_1^{(\text{II})} &= \operatorname{Var}_x\left(\operatorname{Tr}\left(A^{(\hat{\Sigma}_U)}D(x)\right)\right) \stackrel{\eqref{eq:diagonal-shot-noise-matrix}}{=} \operatorname{Var}_x\left(\frac{1}{m_0}\sum_{k=1}^d \left(A^{(\hat{\Sigma}_U)}\right)_{k,k}\sigma_k(x)^2\right) \nonumber
    \\&= \frac{1}{m_0^2}\left(\sum_{k=1}^d \left(A^{(\hat{\Sigma}_U)}\right)_{k,k}^2 \underbrace{\operatorname{Var}_x\left(\sigma_k(x)^2\right)}_{=\mathcal{O}_d(1)} + \sum_{\substack{k,k'=1 \\ k\neq k'}}^d \left(A^{(\hat{\Sigma}_U)}\right)_{k,k}\left(A^{(\hat{\Sigma}_U)}\right)_{k',k'} \underbrace{\operatorname{Cov}_x\left(\sigma_k(x)^2,\sigma_{k'}(x)^2\right)}_{\stackrel{\eqref{eq:quantum-correlation-constant}}{\leq}\mathcal{C}}\right) \nonumber
    \\ &\leq \frac{\mathcal{O}_d(1)}{m_0^2}\underbrace{\sum_{k=1}^d \left(A^{(\hat{\Sigma}_U)}\right)_{k,k}^2}_{\leq \left\|A^{(\hat{\Sigma}_U)}\right\|_{\text{HS}}^2} + \frac{\mathcal{C}}{m_0^2}\sum_{\substack{k,k'=1 \\ k\neq k'}}^d \left| \left(A^{(\hat{\Sigma}_U)}\right)_{k,k} \right|\,\left| \left(A^{(\hat{\Sigma}_U)}\right)_{k',k'} \right| \nonumber
    \\ &\leq \frac{\mathcal{O}_d(1)}{m_0^2}\left\|A^{(\hat{\Sigma}_U)}\right\|_{\text{HS}}^2 + \frac{\mathcal{C}}{m_0^2}\left(\sum_{k=1}^d \left| \left(A^{(\hat{\Sigma}_U)}\right)_{k,k} \right|\right)^2 \leq \frac{\mathcal{O}_d(1)}{m_0^2}\left\|A^{(\hat{\Sigma}_U)}\right\|_{\text{HS}}^2 + \frac{d\mathcal{C}}{m_0^2}\underbrace{\sum_{k=1}^d \left(A^{(\hat{\Sigma}_U)}\right)_{k,k}^2}_{\leq \left\|A^{(\hat{\Sigma}_U)}\right\|_{\text{HS}}^2} \nonumber
    \\ &= \frac{\mathcal{O}_d(1)}{m_0^2}\left\|A^{(\hat{\Sigma}_U)}\right\|_{\text{HS}}^2 + \frac{\mathcal{O}_d(d\mathcal{C})}{m_0^2}\left\|A^{(\hat{\Sigma}_U)}\right\|_{\text{HS}}^2 = \mathcal{O}_d(d)+\mathcal{O}_d(d^2\mathcal{C}),
\end{align}
where we also bounded the Frobenius norm,
\begin{equation}
    \left\|A^{(\hat{\Sigma}_U)}\right\|_{\text{HS}}^2 = \operatorname{Tr}\left(\left(A^{(\hat{\Sigma}_U)}\right)^\top A^{(\hat{\Sigma}_U)}\right) \leq \|A^{(\hat{\Sigma}_U)}\|^2\operatorname{Tr}(I_d) = d\|A^{(\hat{\Sigma}_U)}\|^2 \stackrel{\eqref{eq:Frobenius-bound1}}{=} \mathcal{O}_d(dm_0^2).
\end{equation}
This shows that $T_1^{(\text{II})} = o(d^2)$ as long as the quantum measurement variance correlations in Eq.~\eqref{eq:quantum-correlation-constant} satisfy $\mathcal{C}=o_d(1)$ as $d\to\infty$.

The last step is to bound the first decoupled term $T_1^{(\text{I})}$. Here, we shall assume that $\{O_k\}_{k\in[d]}\subset\operatorname{Herm}(2^N)$ is a rescaled system of observables such that $\|O_k\|^2 = \mathcal{O}(d^{-\frac{1}{2}})$ as $d\to\infty$ for all $k\in[d]$. Under this assumption we find
\begin{align}
    T_1^{(\text{I})} &= \operatorname{Var}_x\left(\phi_U(x)^\top A^{(\hat\Sigma_U)}\phi_U(x)\right) \leq \mathbb{E}_x\left[\left(\phi_U(x)^\top A^{(\hat{\Sigma}_U)}\phi_U(x)\right)^2\right] \leq \underbrace{\|A^{(\hat\Sigma_U)}\|^2}_{\stackrel{\eqref{eq:Frobenius-bound1}}{=}\mathcal{O}_d(m_0^2)} \,\underbrace{\mathbb{E}_x\left[\|\phi_U(x)\|^4\right]}_{=\mathcal{O}(d)} = \mathcal{O}_d\left(dm_0^2\right), \label{eq:regularized-quadratic-form-variance}
\end{align}
where fourth moment bound is obtained from
\begin{align}
    \|\phi_U(x)\|^2 = \sum_{k=1}^d\langle O_k\rangle_x^2 = \sum_{k=1}^d\operatorname{Tr}(O_k\rho_U(x))^2 \leq \underbrace{\max_{k\in[d]}\|O_k\|^2}_{=\mathcal{O}(d^{-\frac{1}{2}})}\,\sum_{k=1}^d\underbrace{\operatorname{Tr}(\rho_U(x))^2}_{=1} = \mathcal{O}(d^{-\frac{1}{2}})\,d = \mathcal{O}(\sqrt{d})
\end{align}
as $\rho_U(x)\in\mathcal{D}(2^N)$ defines a density matrix. Since $x\in\mathcal{X}$ was arbitrary, this particularly shows that $\mathbb{E}_x[\|\phi_U(x)\|^4]=\mathcal{O}(d)$ as $d\to\infty$. Hence, we have shown that $T_1^{(\text{I})} = o(d^2)$ if $m_0=o(\sqrt{d})$ as $d\to\infty$. The theorem now follows.
\qed

In order to demonstrate that $\mathcal{C}=o_d(1)$ is a non-restrictive, physical assumption, assume that $\{O_k\}_{k\in[d]}$ is a system of non-identity Pauli strings, that $d=\Theta(\operatorname{Poly}(N))$, and further suppose first that $U(x)$ is sampled from a 2-design. In that case, we particularly have $\langle O_k^2\rangle_x=1$ so that $\sigma_k(x)^2=1-\langle O_k\rangle_x^2$. By using the properties $\operatorname{Cov}(aX+b,cY+d)=ac\operatorname{Cov}(X,Y)$ for constants $a,b,c,d\in\mathbb{R}$ and $\operatorname{Cov}(X,Y)^2\leq\operatorname{Var}(X)\operatorname{Var}(Y)$ with $X,Y$ denoting arbitrary random variables, we find
\begin{align}
    \operatorname{Cov}_x\left(\sigma_k^2(x),\sigma_{k'}^2(x)\right) &= \operatorname{Cov}_x\left(1-\langle O_k\rangle_x^2,1-\langle O_{k'}\rangle_x^2\right) = \operatorname{Cov}_x\left(\langle O_k\rangle_x^2,\langle O_{k'}\rangle_x^2\right)  \leq \sqrt{\operatorname{Var}_x\left(\langle O_k\rangle_x^2\right)\operatorname{Var}_x\left(\langle O_{k'}\rangle_x^2\right)}. \label{eq:quantum-variance-covariance-bound}
\end{align}
Next, since $O_k$ is a Pauli string we have $\operatorname{Spec}(O_k)=\{-1,+1\}$ so that $-1\leq\langle \psi |O_k|\psi\rangle\leq 1$ for all quantum states $|\psi\rangle\in\mathcal{H}(2^N)$. In particular, $|\langle O_k\rangle_x|\leq 1$ which implies $\langle O_k\rangle_x^2\leq 1$ and thus $\langle O_k\rangle_x^2\leq \langle O_k\rangle_x^2$. As a result, we may bound the variance via
\begin{equation}
    \operatorname{Var}_x\left(\langle O_k\rangle_x^2\right) \leq \mathbb{E}_x\left[\langle O_k\rangle_x^4\right] \leq \mathbb{E}_x\left[\langle O_k\rangle_x^2\right] = \mathbb{E}_x\left[\langle O_k\rangle_x^2\right] - \underbrace{\mathbb{E}_x\left[\langle O_k\rangle_x\right]^2}_{=0} = \operatorname{Var}_x\left(\langle O_k\rangle_x\right) = \mathcal{O}(2^{-N}), \label{eq:concentrating-quantum-variance}
\end{equation}
where the last two identities follow from a direct application of Lemma~\ref{lemma:quantum-covariance} using that $U(x)$ is sampled from a 2-design. Therefore,
\begin{align}
    \operatorname{Cov}_x\left(\sigma_k^2(x),\sigma_{k'}^2(x)\right) &\stackrel{\eqref{eq:quantum-variance-covariance-bound}}{\leq} \sqrt{\operatorname{Var}_x\left(\langle O_k\rangle_x^2\right)\operatorname{Var}_x\left(\langle O_{k'}\rangle_x^2\right)} \stackrel{\eqref{eq:concentrating-quantum-variance}}{=} \mathcal{O}(2^{-N}) = o(\operatorname{Poly}(N)^{-1}) = o_d(d^{-1})=o_d(1),
\end{align}
where we also used $d=\Theta(\operatorname{Poly}(N))$.

If $U(x)$ is not sampled from a 2-design, we can relax the condition by assuming that the circuit architecture exhibits a moderate concentration of measure, such that $\operatorname{Var}_x\left(\langle O_k\rangle_x^2\right)=o_N(1)$. Under this mild condition, we may similarly derive
\begin{align}
    \operatorname{Cov}_x\left(\sigma_k^2(x),\sigma_{k'}^2(x)\right) &\stackrel{\eqref{eq:quantum-variance-covariance-bound}}{\leq} \sqrt{\operatorname{Var}_x\left(\langle O_k\rangle_x^2\right)\operatorname{Var}_x\left(\langle O_{k'}\rangle_x^2\right)} = o_N(1) = o_d(1),
\end{align}
where the final implication $o_N(1) = o_d(1)$ follows directly from the fact that $d = \Theta(\operatorname{Poly}(N))$, meaning that the limit $N \to \infty$ couples bijectively to the asymptotic limit $d \to \infty$.

As a second remark, note that rescaling the observable such that $\max_{k\in[d]}\|O_k\|=\mathcal{O}(d^{-\frac{1}{2}})$ has no effect in the purely noiseless case since the rescaling $\phi_U(x)\mapsto \mathcal{O}(d^{-\frac{1}{2}})\phi_U(x)$ is perfectly canceled by the matrix $A^{(\Sigma_U)}=\Sigma_U^{-\frac{1}{2}}A\Sigma_U^{-\frac{1}{2}}$ in the quadratic form $\phi_U(x)^\top A^{(\Sigma_U)}\phi_U(x)$ because the previous normalization has the simultaneous effect that $\Sigma_U^{-\frac{1}{2}}\mapsto \mathcal{O}(d^{\frac{1}{2}})\Sigma_U^{-\frac{1}{2}}$ by definition of the noiseless population covariance $\Sigma_U=\mathbb{E}_x[\phi_U(x)\phi_U(x)^\top]$.
Thus, this trick cannot be used to force the desired bound $\phi_U(x)^\top A^{(\Sigma_U)}\phi_U(x)=o(d^2)$ as $d\to\infty$. In contrast to the purely noiseless case, in the noisy setting $\hat{\Sigma}_U=\Sigma_U + D_{\text{shot}}$, the diagonal regularization term $D_{\text{shot}}$ prevents the aforementioned, perfect cancellation due to Eq.~\eqref{eq:regularized-quadratic-form-variance}, thus allowing us to re-establish the bound $\phi_U(x)^\top A^{(\hat\Sigma_U)}\phi_U(x)=o(d^2)$ with the help of the noisy population covariance $\hat\Sigma_U=\mathbb{E}_{x,\text{shot}}[\hat{\phi}_U(x)\hat\phi_U(x)^\top]$. 
Furthermore, it is worth noting that, in practice, it is very well possible that double descent is re-established with shot noise even for non-rescaled observable systems with $\|O_k\|=\Theta_d(1)$ (such as Pauli strings) or larger $m_0=\Omega(\sqrt{d})$ than our theory predicts as we do not claim that the bounds used in our analysis above are tight.

\section{Additional Numerical Diagnostics}
\label{appendix:numerical-experiments}

This section provides additional numerical evidence supporting the main results. We first characterize feature correlations for the synthetic QELM architectures and study how they are modified by finite-shot noise. We then extend the analysis to three real regression datasets, examining both the correlation structure of the resulting feature maps and the associated double-descent behavior across different circuit depths.

\begin{figure}
    \captionsetup{justification=raggedright, singlelinecheck=false}
    \centering
    \includegraphics[width=0.5\linewidth]{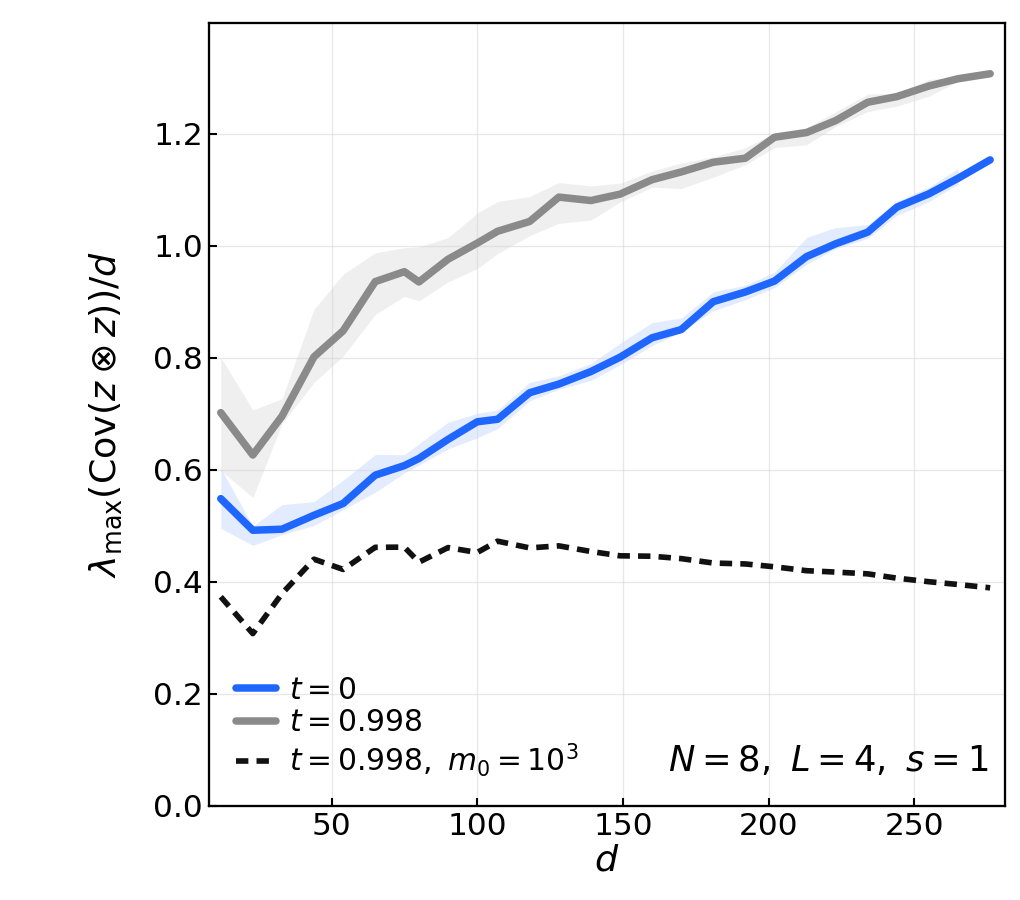}
    \caption{Normalized quadratic-feature covariance statistic $\lambda_{\max}(\operatorname{Cov}(z\otimes z))/d$ as a function of feature dimension $d$ for the $t=0$ reference architecture, the exact $t=0.998$ architecture, and the finite-shot $t=0.998$ architecture with $m_0=10^3$ shots per observable and input. All curves use $N=8$, four input components, $L=4$, $s=1$, and the same pool of 276 weight-one and weight-two Pauli observables. For $d<276$, each point aggregates 20 fixed observable subsets; the full pool is evaluated once. Faint bands show the subset interquartile range.}
    \label{fig:supp-quadratic-variance-synthetic}
\end{figure}

\paragraph{Quadratic correlations in the two synthetic regimes.}
To relate the correlation structure of the $t=0$ and $t=0.998$ QELM regimes to the double-descent behaviour observed in Figure~\ref{fig:experiments}\subref{fig:qelm-g}, we evaluate the normalized quadratic feature covariance measure
$\lambda_{\max}(\operatorname{Cov}(z\otimes z))/d$.
For each exact regime, we generate 32,768 inputs $x\in[0,1)^4$ using an independently scrambled Sobol sequence. Exact-statevector simulation is then used to evaluate all 276 weight-one and weight-two Pauli observables for every input. For each feature dimension $d$ and observable subset, the corresponding $d$ feature coordinates from this common sample are used to estimate the feature second moment $\Sigma$ and construct the whitening map $z=\Sigma^{-1/2}\phi_U(x)$. To estimate $\operatorname{Cov}(z\otimes z)$, we use two independent evaluation sets containing 24,576 scrambled-Sobol inputs each, both disjoint from the inputs used to estimate $\Sigma$, and pool their 49,152 whitened feature vectors. The same input sets are reused across all feature dimensions and observable subsets within each regime. The finite-shot $t=0.998$ curve uses the same underlying inputs and exact feature values as the exact $t=0.998$ curve. Finite-shot features are generated independently for each observable and input according to the estimator in Equation~\eqref{eq:finite-shot-estimator-def}, with $m_0=10^3$ and a fixed random seed; whitening is performed using the corresponding analytic finite-shot second moment.

The resulting behaviour is shown in Figure~\ref{fig:supp-quadratic-variance-synthetic}. Proposition~\ref{prop:over-fitting} applies whenever Equation~\eqref{eq:main-correlation-bound} holds. As shown in Supplementary Section~\ref{subsubsection:high-dim-probability}, the stronger condition
$\lambda_{\max}(\operatorname{Cov}(z\otimes z))=o(d)$
is sufficient to guarantee Equation~\eqref{eq:main-correlation-bound}. Under this sufficient condition, the normalized quantity plotted in Figure~\ref{fig:supp-quadratic-variance-synthetic} must vanish as $d$ increases. The noiseless feature maps do not show this behaviour over the range considered. In particular, the near-dual-unitary $t=0.998$ configuration develops substantially stronger correlations as $d$ grows. This is nevertheless compatible with the interpolation-induced risk divergence observed in Figure~\ref{fig:experiments}\subref{fig:qelm-g}, because the $o(d)$ scaling is sufficient, but not necessary, for Equation~\eqref{eq:main-correlation-bound}. An interpolation peak at $\gamma=1$ can therefore persist even when this particular normalized correlation measure increases with $d$. Finite-shot readout, in contrast, strongly suppresses the correlations of the $t=0.998$ feature map and produces a decreasing trend at larger $d$. This behavior is consistent with the sufficient condition (cf.  Proposition~\ref{prop:spectral-variance-bound}) ensuring the peak at $\gamma=1$ as per Proposition~\ref{prop:over-fitting} in Supplementary Section~\ref{sec:non-isotropic-analysis}, and with the corresponding, double-descent curve with the interpolation peak at $\gamma=1$ in Figure~\ref{fig:experiments}\subref{fig:qelm-g}.

\paragraph{Real datasets.} We consider three regression datasets spanning distinct application domains. 
\begin{itemize}
    \item The \emph{California Housing} dataset contains eight numerical attributes describing 20,640 census block groups, including median income, housing age, occupancy, population, and geographical coordinates, with median house value as the regression target.
    \item The \emph{Diabetes} dataset contains measurements from 442 patients, described by ten baseline variables including age, sex, body mass index, blood pressure, and six blood-serum measurements, with the target given by a quantitative measure of disease progression one year after baseline.
    \item The \emph{AAPL} dataset is constructed from Apple limit-order-book data provided by LOBSTER, which reconstructs timestamped NASDAQ order-book states from Historical TotalView-ITCH messages. The underlying data contain bid and ask prices and volumes together with the sequence of order-book events; in our regression task, these market variables are used to predict the next-minute AAPL log return.
\end{itemize}
For each dataset, we use a permutation generated by NumPy's \texttt{default\_rng(7)} to select $90$ training samples and $1{,}000$, $352$, and $299$ test samples for California Housing, Diabetes, and AAPL, respectively.
For California Housing and Diabetes, input preprocessing is fitted on the training split: each feature is standardized and then mapped to $[0,1]$, with held-out values clipped to this range. For AAPL, we use the most recent order-book state at each one-minute mark on June 21, 2012 and retain the eight level-1 and level-2 ask/bid price and size fields. Prices are transformed as $10^4\log(p/m)$ relative to the contemporaneous level-1 midpoint $m$, sizes as $\log(1+v)$, and the resulting inputs are standardized using the training split. California Housing and AAPL therefore have eight input features and Diabetes has ten, motivating the corresponding choices $N=16$, $N=16$, and $N=20$.

California Housing and Diabetes are obtained through \texttt{sklearn.datasets.fetch\_california\_housing} and \texttt{sklearn.datasets.load\_diabetes}, respectively, while the AAPL order-book files are taken from a \href{https://huggingface.co/datasets/totalorganfailure/lobster-data/tree/1b6ee86b9ec0b462e2a5785bda837294cbfc26be/LOBSTER_SampleFile_AAPL_2012-06-21_10}{public mirror of LOBSTER's level-10 sample for June 21, 2012}.

\begin{figure}
    \captionsetup{justification=raggedright, singlelinecheck=false}
    \centering
    \includegraphics[width=0.99\linewidth]{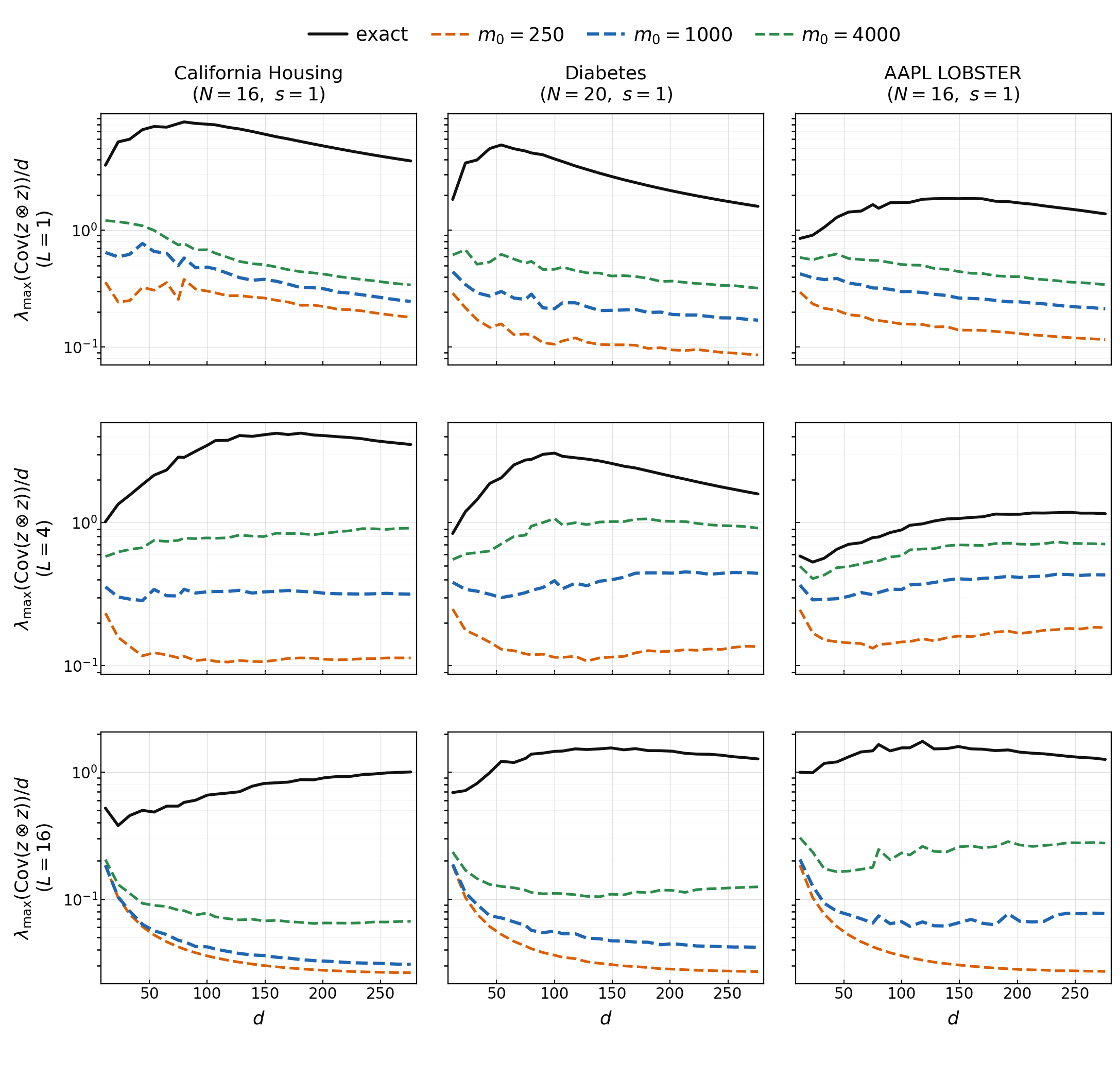}
    \caption{Exact and finite-shot values of $\lambda_{\max}(\operatorname{Cov}(z\otimes z))/d$ for California Housing, Diabetes, and AAPL LOBSTER. Columns identify the datasets and rows identify the frozen reservoirs at $L\in\{1,4,16\}$. Solid curves use exact QELM features; dashed curves use $m_0\in\{250,1000,4000\}$ independent shots per observable and input. Each condition is evaluated on a 45-point grid of feature dimensions $d$; for $d<276$, results are aggregated over 20 fixed observable subsets, while $d=276$ uses the complete observable pool once. The empirical input distributions contain 1{'}090, 442, and 389 preprocessed rows, respectively; labels are not used.}
    \label{fig:supp-quadratic-variance-real-data}
\end{figure}

\paragraph{Quadratic correlations on real-data inputs.}
To assess the correlation structure of the QELM features on the real datasets considered above, we compute the normalized quadratic feature covariance measure
$\lambda_{\max}(\operatorname{Cov}(z\otimes z))/d$
for different circuit depths and readout shot numbers. For each dataset, depth $L$, feature dimension $d$, and observable subset, the whitening map is constructed from the exact QELM features over the complete preprocessed empirical input distribution. The exact statistic is then evaluated after splitting these rows once into two folds. For the finite-shot curves, we instead use two fixed balanced folds of 24,576 rows resampled from the empirical input distribution. Whitening is performed using the corresponding analytic finite-shot second moment, and every observable and input is sampled independently with the specified value of $m_0$. The $m_0=1000$ curve is additionally obtained as the median over five shot-seed-specific subset medians, whereas $m_0=250$ and $m_0=4000$ use a single fixed shot seed.

Figure~\ref{fig:supp-quadratic-variance-real-data} shows that the correlation structure depends strongly on both circuit depth and readout noise. In contrast to the synthetic setting of Figure~\ref{fig:supp-quadratic-variance-synthetic}, where the normalized correlation measure increases with $d$ for the exact feature maps, several of the real-data configurations exhibit a decreasing trend already in the exact simulation, particularly at smaller circuit depths. This behavior is compatible with the sufficient condition
$\lambda_{\max}(\operatorname{Cov}(z\otimes z))=o(d)$, although the finite range of $d$ considered here does not allow us to establish the corresponding asymptotic scaling. For deeper circuits, the normalized correlations tend instead to remain approximately constant or decrease more weakly. Finite-shot readout systematically suppresses the correlation measure across all datasets and depths, with stronger suppression for smaller $m_0$, and in several cases produces a substantially clearer decrease with $d$. Thus, compared with the synthetic experiments, the real-data feature maps can display correlation scaling closer to the sufficient high-dimensional regime even in the absence of shot noise, while finite sampling further enhances this tendency.

\begin{figure}
    \captionsetup{justification=raggedright, singlelinecheck=false}
    \centering
    \includegraphics[width=0.99\linewidth]{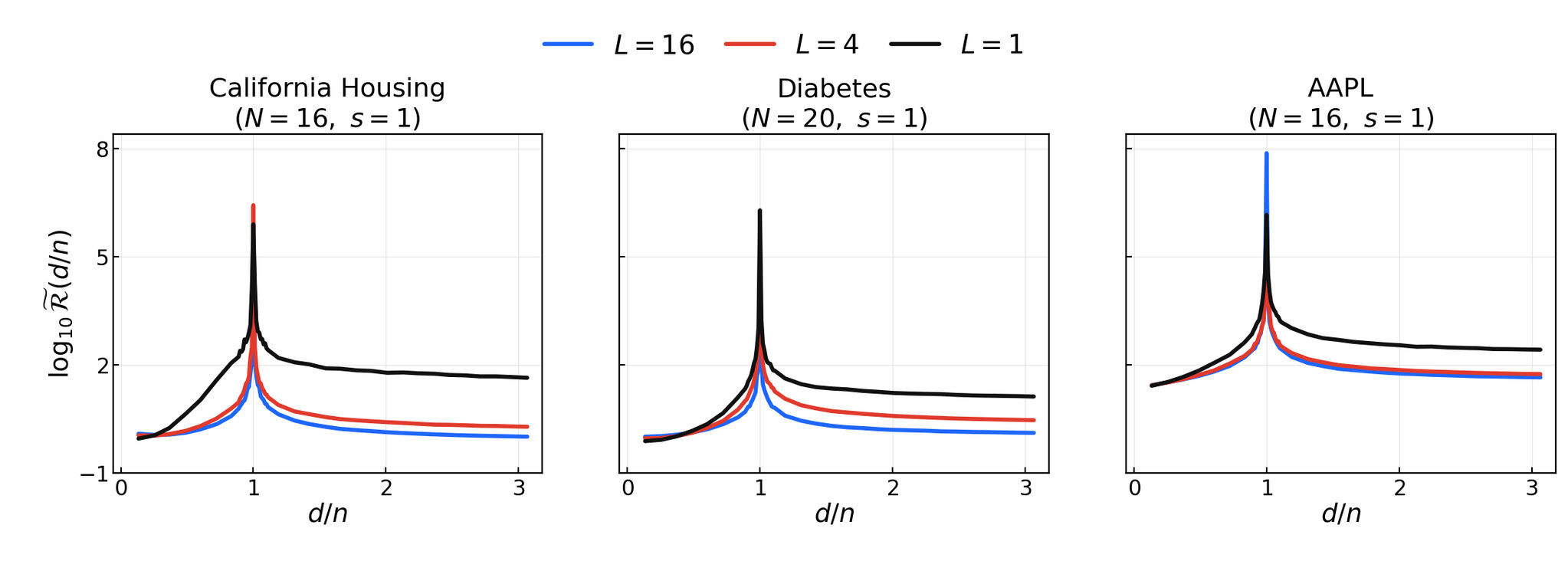}
    \caption{Held-out test mean-squared error as a function of $d/n$ for California Housing, Diabetes, and one-minute AAPL returns. Each panel uses $n=90$, $s=1$, exact-statevector QELM features, $L\in\{1,4,16\}$, a 45-point grid of feature dimensions $d$, and 200 predetermined observable subsets for each $d<276$, with the complete observable pool used at $d=276$. California Housing and Diabetes use $N=16$ and $N=20$, respectively, with targets standardized from the training split. AAPL uses $N=16$ and the raw next-minute return in basis points. Within each panel, the three depth curves are directly comparable, whereas absolute risk magnitudes are not comparable due to different target scalings.}
    \label{fig:supp-real-data-depth}
\end{figure}

\paragraph{Double descent on real-data.}
We next examine whether the correlation behavior discussed above is reflected in the generalization curves of the corresponding QELMs. For each dataset and circuit depth $L\in\{1,4,16\}$, we use a fixed QELM reservoir with input strength $s=1$ and exact-statevector features. We use the dataset splits and qubit numbers described above. For each feature dimension, a minimum-norm linear readout is fitted for each predetermined observable subset using the same 90 training samples, and the curves report the mean held-out test error across subsets.

Figure~\ref{fig:supp-real-data-depth} shows a pronounced interpolation peak around $d/n=1$ for all three datasets and circuit depths. Combined with the correlation analysis in Figure~\ref{fig:supp-quadratic-variance-real-data}, this shows that the peak persists across qualitatively different correlation regimes. In some cases, $\lambda_{\max}(\operatorname{Cov}(z\otimes z))/d$ already decreases with $d$ in the exact simulation, while in others it remains approximately constant or increases over part of the investigated range. This is consistent with Proposition~\ref{prop:over-fitting} of Supplementary Section~\ref{sec:non-isotropic-analysis} and with the fact that $\lambda_{\max}(\operatorname{Cov}(z\otimes z))=o(d)$ is only a sufficient condition for Equation~\eqref{eq:main-correlation-bound}. The circuit depth also changes the height of the interpolation peak and the post-interpolation risk, indicating that feature correlations affect the detailed shape of the double-descent curve.

\section{Technical Details} \label{appendix:techniques}

This section is dedicated to all the technicalities used to prove our main results above which are not essential to understand the key argumentation. We compile these details in order to maintain mathematically rigorous integrity. We divide this part into three groups: First, in Section~\ref{appendix:RMT} we cover the tools used from random matrix theory. 
In Subsection~\ref{appendix:RMT-assumptions} we ensure that the underlying assumptions made in the under- and over-parametrized regime do not contradict each other in order to maintain a valid theory of double descent.
Subsection~\ref{appendix:Stieltjes} then introduces the reader to the Stieltjes formalism used to study the spectral distribution of the random feature matrix in the high-dimensional limit. 
This is essential to apply the Marchenko-Pastur law~\cite{bai2010spectral, adamczak_marchenko-pastur_2011correlation, bryson2021marchenko} in the case of Gaussian features, and to derive the condition number of the population covariance when dealing with more generic, random features allowing limited correlations. Next, we elaborate additional details regarding the Fourier interpretation of quantum models~\cite{schuld2021effect, xiong2025fundamental, mhiri2025constrained} in Section~\ref{appendix:Fourier-analysis}. Section~\ref{appendix:Sobolev-spaces} then briefly introduces Lebesgue spaces~\cite{wheeden1977measure, billingsley2017probability}, including various properties and results used throughout the supplementary material. Not only are these concepts essential to understand the fully rigorous argumentation of the proofs, but they are also the foundation of the Sobolev spaces and weak derivatives~\cite{adams2003sobolev}, the key objects behind our analysis with respect to input strength. Finally, Section~\ref{appendix:others} deals with additional technical details. This involves concentration, sub-Gaussian vectors, correlations and other useful properties used in our computations.

\subsection{Random Matrix Theory} \label{appendix:RMT}

We begin with the additional arguments needed to prove the analytic description of the expected risk in the case of Gaussian random features (Theorem~\ref{thm:double-descent-gaussian-features} in Section~\ref{appendix:gaussian-features}), which made use of a few technical lemmata:

\begin{lemma} [Expected Gaussian Quadratic Form] \label{lemma:expected-gaussian-quadratic-form}
    If $\mathcal{E}\sim\mathcal{N}(0,\sigma^2I_n)$ is a Gaussian random vector and $M\in\mathbb{R}^{n\times n}$ independent of $z$, then
    \begin{equation*}
        \mathbb{E}_\mathcal{E}[\mathcal{E}^\top M \mathcal{E}] = \sigma^2\operatorname{Tr}(M).
    \end{equation*}
\end{lemma}

\noindent
\emph{Proof.} Let $M_{ij}$ be the matrix elements of $M$, and $\mathcal{E}_i$ the components of $\mathcal{E}$, respectively. We directly compute
\begin{align}
    \mathbb{E}_\mathcal{E}[\mathcal{E}^\top M \mathcal{E}] &= \mathbb{E}_\mathcal{E}\left[\sum_{i,j=1}^n M_{ij}\,\mathcal{E}_i  \mathcal{E}_j\right] = \sum_{i=1}^n M_{ii}\,\underbrace{\mathbb{E}_\mathcal{E}[\mathcal{E}_i^2]}_{=\sigma^2} + \sum_{\substack{i,j=1 \\ i\neq j}}^n M_{ij}\,\underbrace{\mathbb{E}_\mathcal{E}[\mathcal{E}_i\mathcal{E}_j]}_{=0} = \sigma^2\sum_{i=1}^n M_{ii} = \sigma^2\,\operatorname{Tr}(M),
\end{align}
where $\mathbb{E}_\mathcal{E}[\mathcal{E}_i\,\mathcal{E}_j]=\mathbb{E}_\mathcal{E}[\mathcal{E}_i]\,\mathbb{E}_\mathcal{E}[\mathcal{E}_j]=0$ for all $i,j\in[n]$ due to $\mathcal{E}\sim\mathcal{N}(0,\sigma^2 I_n)$ and $\mathbb{E}_\mathcal{E}[\mathcal{E}_i^2]=\sigma^2$ being Gaussian.
\qed

\begin{lemma}[Least Square Solution for $\gamma<1$] \label{lemma:least-square-solution}
    Suppose $\gamma<1$. Let $\Phi\in\mathbb{R}^{d\times n}$ denote the random features in Eq.~\eqref{eq:random-features} and assume that $\operatorname{rank}(\Phi\Phi^\top) = d$. Let $Y\in\mathbb{R}^{n\times 1}$ be the training labels satisfying Eq.~\eqref{eq:training-labels}. Then,
    \begin{equation*}
        \hat{w}:=(\Phi\Phi^\top)^{-1}\Phi Y=\arg\min_{w\in\mathbb{R}^d}\|Y-\Phi^\top w\|^2
    \end{equation*}
    defines the unique minimizer.
\end{lemma}

\noindent
\emph{Proof.} Consider the smooth function $J:\mathbb{R}^d\to\mathbb{R}_{\geq0}$ defined as $J(w) := \|Y-\Phi^\top w\|^2$. Rewriting $J(w)$ in terms of an inner product yields
\begin{align}
    J(w) = (Y-\Phi^\top w)^\top(Y-\Phi^\top w) = Y^\top Y - Y^\top\Phi^\top w - w^\top\Phi\,Y + w^\top \Phi\Phi^\top w = \|Y\|^2 +\|\Phi^\top w\|^2 -2Y^\top\Phi^\top w.
\end{align}
Here, we used the fact that the scalar $Y^\top\Phi^\top w$ is equal to its transpose $w^\top\Phi Y$. We proceed by computing the gradient with respect to $w$ (using the standard column-vector convention),
\begin{align}
    \nabla_w J(w) = \nabla_w \{w^\top \Phi\Phi^\top w\} - 2\nabla_w \{Y^\top\Phi^\top w\} = 2\Phi\Phi^\top w - 2\Phi Y.
\end{align}
Hence, setting $\nabla_wJ(w)=0$ results in the equation $\Phi\Phi^\top w = \Phi Y$. Since $\operatorname{rank}(\Phi\Phi^\top)=d$ by assumption, the inverse $(\Phi\Phi^\top)^{-1}$ exists and so does the extremum,
\begin{equation}
    \hat{w} = (\Phi\Phi^\top)^{-1}\Phi Y.
\end{equation}
Since the Hessian,
\begin{equation}
    \nabla_w^2J(w) = 2\nabla_w\{w^\top\Phi\Phi^\top - Y^\top\Phi^\top\} = 2\Phi\Phi^\top
\end{equation}
is strictly positive definite due to $\Phi\Phi^\top$ being invertible, it follows that $\hat{w}$ defines the unique global minimizer.
\qed

\begin{lemma} [Existence of min-Norm Interpolator for $\gamma>1$] \label{lemma:existence-interpolator} Suppose $\gamma>1$. Let $\Phi\in\mathbb{R}^{d\times n}$ denote the random features in Eq.~\eqref{eq:random-features} and assume that $\operatorname{rank}(\Phi^\top\Phi) = n$. Let $Y\in\mathbb{R}^{n\times 1}$ be the training labels satisfying Eq.~\eqref{eq:training-labels}. Then,
\begin{equation*}
    \hat{w}:=\Phi(\Phi^\top\Phi)^{-1}Y
\end{equation*}
defines a min-norm interpolator in the sense of Eq.~\eqref{eq:min-norm-interpolator}.
\end{lemma}

\noindent
\emph{Proof.} Due to the assumption that $\operatorname{rank}(\Phi^\top\Phi)=n$, the matrix $\Phi^\top\Phi$ is invertible and thus
\begin{equation}
    \Phi^\top\hat{w} = \Phi^\top\Phi(\Phi^\top\Phi)^{-1}Y =Y
\end{equation}
immediately yields an exact solution, i.e., $\mathcal{L}_n(\hat{w})=0$. It remains to show that the norm $\|\hat{w}\|$ is minimal. For that purpose, let $w\in\mathbb{R}^d$ be another solution to $\mathcal{L}_n(w)=0$ so that $y=\Phi^\top w$. Then,
\begin{align}
    \langle w-\hat{w},\hat{w}\rangle &= \langle w-\hat{w}, \Phi(\Phi^\top\Phi)^{-1} Y\rangle = \langle \Phi^\top(w-\hat{w}),(\Phi^\top\Phi)^{-1}Y\rangle =0
\end{align}
because of
\begin{equation}
    \Phi^\top(w-\hat{w}) = \Phi^\top w - \Phi^\top\hat{w} = Y - \Phi^\top \Phi(\Phi^\top\Phi)^{-1}Y = Y-Y=0.
\end{equation}
This shows that $w-\hat{w}$ and $\hat{w}$ are orthogonal. As a result,
\begin{align}
    \|w\|^2 = \|w-\hat{w}+\hat{w}\|^2 = \underbrace{\|w-\hat{w}\|^2}_{\geq0} + \|\hat{w}\|^2 + 2\underbrace{\langle w-\hat{w},\hat{w}\rangle}_{=0} \geq \|\hat{w}\|^2.
\end{align}
The lemma follows since $w$ was an arbitrary solution.
\qed

\begin{proposition}[Bias-Variance Decomposition for $\gamma>1$] \label{prop:bias-variance-decomposition}
    Let $\Phi\in\mathbb{R}^{d\times n}$ denote the random features in Eq.~\eqref{eq:random-features} and assume that $\operatorname{rank}(\Phi^\top\Phi) = n$. Let $Y\in\mathbb{R}^{n\times 1}$ be the training labels. Suppose that there exists $w_*\in\mathbb{R}^d$ satisfying $Y=\Phi^\top w_*+\mathcal{E}$ for some Gaussian noise $\mathcal{E}\sim\mathcal{N}(0,\sigma^2I_n)$, and let $\hat{w}\in\mathbb{R}^d$ be the min-norm interpolator from Lemma~\ref{lemma:existence-interpolator}. Then,
    \begin{equation*}
        \|\hat{w}-w_*\|^2 = \underbrace{\|(I-\Phi(\Phi^\top\Phi)^{-1}\Phi^\top)w_*\|^2}_{=:\operatorname{Bias}_\Phi(w_*)^2} + \underbrace{\|\Phi(\Phi^\top\Phi)^{-1}\mathcal{E}\|^2}_{=:\operatorname{Var}_\Phi(\mathcal{E})^2}.
    \end{equation*}
\end{proposition}

\noindent
\emph{Proof.} First, we establish the orthogonality property,
\begin{align}
    \langle (I-\Phi(\Phi^\top\Phi)^{-1}\Phi^\top)w_*,\Phi(\Phi^\top\Phi)^{-1}\mathcal{E} \rangle &= w_*^\top(I-\Phi(\Phi^\top\Phi)^{-1}\Phi^\top)\Phi(\Phi^\top\Phi)^{-1}\mathcal{E} \nonumber
    \\ &= w_*^\top (\Phi-\Phi\underbrace{(\Phi^\top\Phi)^{-1}\Phi^\top\Phi}_{=I_n})(\Phi^\top\Phi)^{-1}\mathcal{E}=0.
\end{align}
By Lemma~\ref{lemma:existence-interpolator}, we know that $\hat{w}=\Phi(\Phi^\top\Phi)^{-1}Y$. Consequently, since $y=\Phi^\top w_*+\mathcal{E}$, the orthogonality property implies
\begin{align}
    \|\hat{w}-w_*\|^2 &= \|\Phi(\Phi^\top\Phi)^{-1}Y-w_*\|^2 = \|\Phi(\Phi^\top\Phi)^{-1}(\Phi^\top w_*+\mathcal{E})-w_*\|^2 = \|(\Phi(\Phi^\top\Phi)^{-1}\Phi^\top-I)w_* + \Phi(\Phi^\top\Phi)^{-1}\mathcal{E}\|^2 \nonumber
    \\ &= \|w_*(I-\Phi(\Phi^\top\Phi)^{-1})\Phi^\top)\|^2 + \|\Phi(\Phi^\top\Phi)^{-1}\mathcal{E}\|^2 +2\langle (I-\Phi(\Phi^\top\Phi)^{-1}\Phi^\top)w_*,\Phi(\Phi^\top\Phi)^{-1}\mathcal{E} \rangle \nonumber
    \\ &= \operatorname{Bias}_\Phi(w_*)^2 + \operatorname{Var}_\Phi(\mathcal{E})^2.
\end{align}
\qed

We also need the following lemma from linear algebra:

\begin{lemma}[Orthogonal Group Commutant] \label{lemma:orthogonal-group-commutant}
    Let $M\in\mathbb{R}^{d\times d}$ be a square matrix. If $[M,Q]=0$ for all $Q\in\operatorname{O}(d)$, then $M=\lambda I_d$ is a multiple of the identity for some $\lambda\in\mathbb{R}$.
\end{lemma}

\noindent
\emph{Proof.} It is sufficient to prove that every unit vector $v\in\mathbb{S}^{d-1}$ defines an eigenvector of $M$ with the same eigenvalue $\lambda\in\mathbb{R}$, i.e., $Mv = \lambda v$ for all $v\in\mathbb{S}^{d-1}$. Indeed, this immediately implies that $M=\lambda I_d$.

Let $v\in\mathbb{S}^{d-1}$ be arbitrary and consider the matrix
\begin{equation}
    Q_v := I-2vv^\top.
\end{equation}
Then, since $v^\top v=\|v\|^2=1$,
\begin{equation}
    Q_v v = (I-2vv^\top)v = v- 2v\underbrace{v^\top v}_{=1} = v - 2v = -v. \label{eq:reflection1}
\end{equation}
On the other hand, for all unit vectors $u\in\mathbb{S}^{d-1}$ orthogonal to $v$,
\begin{equation}
    Q_v u = (I-2vv^\top)u = u-2v\underbrace{v^\top u}_{=0} = u.
\end{equation}
This shows that
\begin{equation}
    \operatorname{span}_\mathbb{R}(v) = \{u\in\mathbb{R}^d:Q_v u=-u\}. \label{eq:span1}
\end{equation}
Moreover, $Q_v^\top = I - 2vv^\top = Q_v$ is symmetric and thus
\begin{equation}
    Q_v^\top Q_v = Q_vQ_v^\top = Q_v^2 = I - 4vv^\top + 4v\underbrace{v^\top v}_{=1} v^\top = I,
\end{equation}
since $v^\top v=\|v\|^2=1$. This shows that $Q_v\in\operatorname{O}(d)$ is orthogonal. Therefore, $[M,Q_v]=0$ by assumption. That is, $Q_v M=MQ_v$, which implies
\begin{equation}
    Q_v(Mv) = M(Q_v v) \stackrel{\eqref{eq:reflection1}}{=} -Mv.
\end{equation}
In particular, $Mv\in\operatorname{span}_{\mathbb{R}}(v)$ by Eq.~\eqref{eq:span1}. In other words, $Mv = \lambda_v v$ for some $\lambda_v\in\mathbb{R}$. Since $v\in\mathbb{S}^{d-1}$ was arbitrary, it remains to prove that $\lambda_v$ is independent of $v$. For that purpose, let $u,u'\in\mathbb{S}^{d-1}$ be two linearly independent unit vectors. By the previous observation, there exist $\lambda_u,\lambda_{u'}\in\mathbb{R}$ such that $Mu=\lambda_u u$ and $Mu' = \lambda_{u'}u'$, respectively. The same observation yields $M(u+u')=\lambda_{u+u'}(u+u')$ for some $\lambda_{u+u'}\in\mathbb{R}$. Therefore,
\begin{equation}
    \lambda_u u+\lambda_{u'}u' = M(u+u') = \lambda_{u+u'}(u+u')
\end{equation}
implying
\begin{equation}
    (\lambda_u - \lambda_{u+u'})u + (\lambda_{u'}-\lambda_{u+u'})u' = 0.
\end{equation}
Since $u$ and $u'$ are linearly independent, we must have that $\lambda_u-\lambda_{u+u'}=0$ and $\lambda_{u'}-\lambda_{u+u'}=0$. Equivalently, $\lambda_u = \lambda_{u+u'} = \lambda_{u'}$. The lemma now follows since $u,u'\in\mathbb{S}^{d-1}$ were arbitrary linearly independent vectors, and due to the fact the linearly dependent eigenvectors share the same eigenvalue.
\qed

Regarding the risk analysis of the anisotropic case in the over-parametrized regime (Theorem~\ref{thm:risk-bound-nonisotropic} in Section~\ref{sec:non-isotropic-analysis}), we also have to establish the corresponding bias-variance decomposition:

\begin{proposition}[General Bias-Variance Decomposition for $\gamma>1$] \label{prop:general-bias-variance-decomposition}
    Let $\Phi\in\mathbb{R}^{d\times n}$ denote the random features in Eq.~\eqref{eq:random-features} and assume that $\operatorname{rank}(\Phi^\top\Phi) = n$. Let $Y\in\mathbb{R}^{n\times 1}$ be the training labels. Suppose that there exists $w_*\in\mathbb{R}^d$ satisfying $Y=\Phi^\top w_*+\mathcal{E}$ for some Gaussian noise $\mathcal{E}\sim\mathcal{N}(0,\sigma^2I_n)$, and let $\hat{w}\in\mathbb{R}^d$ be the min-norm interpolator from Lemma~\ref{lemma:existence-interpolator}. Then,
    \begin{equation*}
        \mathbb{E}_\mathcal{E}\left[\|\hat{w}-w_*\|^2_{ \Sigma}\right] = \underbrace{\|(I-\Phi(\Phi^\top\Phi)^{-1}\Phi^\top)w_*\|^2_{ \Sigma}}_{=:\operatorname{Bias}_\Phi(w_*, \Sigma)^2} + \, \underbrace{\sigma^2\operatorname{Tr}\left( \Sigma\,\Phi(\Phi^\top\Phi)^{-2}\Phi^\top\right)}_{=:\operatorname{Var}_\Phi( \Sigma)},
    \end{equation*}
    where $ \Sigma=\mathbb{E}_{ x}\left[\phi(x)\phi(x)^\top\right]\in\mathbb{R}^{d\times d}$ denotes the true feature covariance in Eq.~\eqref{eq:feature-covariance}.
\end{proposition}

\noindent
\emph{Proof.} First, we note that the cross term 
\begin{align}
    \mathbb{E}_\mathcal{E}\left[\langle (I-\Phi(\Phi^\top\Phi)^{-1}\Phi^\top)w_*,\Phi(\Phi^\top\Phi)^{-1}\mathcal{E} \rangle_{ \Sigma}\right] &= w_*^\top(I-\Phi(\Phi^\top\Phi)^{-1}\Phi^\top) \Sigma\Phi(\Phi^\top\Phi)^{-1}\underbrace{\mathbb{E}_\mathcal{E}[\mathcal{E}]}_{=0} =0
\end{align}
vanishes in expectation over the Gaussian noise $\mathcal{E}$. Furthermore, by Lemma~\ref{lemma:existence-interpolator}, we know that $\hat{w}=\Phi(\Phi^\top\Phi)^{-1}Y$. Consequently, since $Y=\Phi^\top w_*+\mathcal{E}$, the previous observation yields
\begin{align}
    \mathbb{E}_\mathcal{E}\left[\|\hat{w}-w_*\|^2_{ \Sigma}\right] &= \mathbb{E}_\mathcal{E}\left[\|\Phi(\Phi^\top\Phi)^{-1}y-w_*\|^2_{ \Sigma}\right] = \mathbb{E}_\mathcal{E}\left[\|\Phi(\Phi^\top\Phi)^{-1}(\Phi^\top w_*+\mathcal{E})-w_*\|^2_{ \Sigma}\right] \nonumber
    \\ &= \mathbb{E}_\mathcal{E}\left[\|(\Phi(\Phi^\top\Phi)^{-1}\Phi^\top-I)w_* + \Phi(\Phi^\top\Phi)^{-1}\mathcal{E}\|^2_{ \Sigma}\right] \nonumber
    \\ &= \|(I-\Phi(\Phi^\top\Phi)^{-1})\Phi^\top w_*\|^2_{ \Sigma} + \mathbb{E}_\mathcal{E}\left[\|\Phi(\Phi^\top\Phi)^{-1}\mathcal{E}\|^2_{ \Sigma}\right] +2\underbrace{\mathbb{E}_\mathcal{E}\left[\langle (I-\Phi(\Phi^\top\Phi)^{-1}\Phi^\top)w_*,\Phi(\Phi^\top\Phi)^{-1}\mathcal{E} \rangle_{ \Sigma}\right]}_{=0} \nonumber
    \\ &= \operatorname{Bias}_\Phi(w_*)^2 + \operatorname{Var}_\Phi( \Sigma),
\end{align}
where we also applied Lemma~\ref{lemma:expected-gaussian-quadratic-form} to compute
\begin{align}
    \mathbb{E}_\mathcal{E}\left[\|\Phi(\Phi^\top\Phi)^{-1}\mathcal{E}\|^2_{ \Sigma}\right] &= \mathbb{E}_z\left[\mathcal{E}^\top (\Phi^\top\Phi)^{-1}\Phi^\top \Sigma\,\Phi(\Phi^\top\Phi)^{-1}\mathcal{E}\right] = \sigma^2\operatorname{Tr}\left((\Phi^\top\Phi)^{-1}\Phi^\top \Sigma\Phi(\Phi^\top\Phi)^{-1}\right) \nonumber
    \\ &= \sigma^2\operatorname{Tr}\left( \Sigma\,\Phi(\Phi^\top\Phi)^{-2}\Phi^\top\right) = \operatorname{Var}_\Phi( \Sigma),
\end{align}
employing the cyclic property of the trace.
\qed

\subsubsection{Standard Random Matrix Theory Assumptions } \label{appendix:RMT-assumptions}

Let us now take a step back and summarize our standard random matrix theory technique used in the non-isotropic analysis in Section~\ref{sec:non-isotropic-analysis}. Using Eq.~\eqref{eq:isotropic-columns} we established the isotropic random matrix
\begin{equation}
    Z=[z_1\mid \cdots\mid z_n] := \Sigma^{-\frac{1}{2}}\Phi = \left[\Sigma^{-\frac{1}{2}}\phi(x_1)\mid\cdots\mid\Sigma^{-\frac{1}{2}}\phi(x_n)\right] \in\mathbb{R}^{d\times n}
\end{equation}
with independent random column vectors $z_1,\dots,z_n\in\mathbb{R}^d$, where $\Sigma:=\mathbb{E}_x\left[\phi(x)\phi(x)^\top\right]\in\mathbb{R}^{d\times d}$ denotes the positive-definite population covariance. We have employed the random matrix $Z$ to bounds for the expected risk $\mathcal{R}(\gamma)$ in Eq.~\eqref{eq:expected-risk}, namely lower bound it in the under-parametrized regime ($\gamma<1$) and upper bound it in the over-parametrized regime ($\gamma>1$), respectively. In the process of proving those bounds we made use of two assumptions on the random matrix $Z$. For the lower bound in the under-parametrized regime we employed a generalization of the Marchenko-Pastur law that allows correlation among the entries of the columns of $Z$, i.e., Lemma~\ref{lemma:generalized-marchenko-pastur}, and assumed that $\max_{j\in[n]}\operatorname{Var}(z_j^\top A z_j)=o(d^2)$ for all deterministic matrices $A\in\mathbb{R}^{d\times d}$ with $\|A\|=\mathcal{O}(1)$ as $d\to\infty$. On the other hand, in the over-parametrized regime, we additionally assumed that $\|z\|=\sqrt{d}$ almost surely in order to apply a lower bound for the smallest singular value of $Z$ with the help of Lemma~\ref{lemma:singular-value-lower-bound} (observe that $\|z\|=\sqrt{d}$ already implies sub-Gaussianity by Lemma~\ref{lemma:bounded-RV-subGaussian}, shown independently in Supplementary Section~\ref{appendix:others}). The purpose of this section is to show that $\|z\|=\sqrt{d}$ does not necessarily contradict the variance bound $\max_{j\in[n]}\operatorname{Var}(z_j^\top A z_j)=o(d^2)$ by constructing an explicit example.

Consider the scaled sphere $\sqrt{d}\,\mathbb{S}^{d-1}:=\{\sqrt{d}v:v\in\mathbb{S}^{d-1}\}$. We construct a feature map $\Phi$ inducing the uniform distribution $z\sim\operatorname{Unif}(\sqrt{d}\,\mathbb{S}^{d-1})$ on the scaled sphere $\sqrt{d}\,\mathbb{S}^{d-1}$. n particular, $\|z\|=\sqrt{d}$ almost surely by construction. A scalar function $f:\sqrt{d}\,\mathbb{S}^{d-1}\to\mathbb{R}$ on the sphere is said to be \emph{$L$-Lipschitz} for some constant $L>0$ if
\begin{equation}
    |f(z)-f(z')| \leq L\,\|z-z'\|
\end{equation}
for all $z,z'\in\sqrt{d}\,\mathbb{S}^{d-1}$. By Theorem~5.1.3 in Ref.~\cite{vershynin2020high} we have the following concentration of Lipschitz functions on the scaled sphere:
\begin{equation}
    \|f(z)-\mathbb{E}[f(z)]\|_{\psi_2} = \mathcal{O}(L) \label{eq:lipschitz-concentration1}
\end{equation}
for all $L$-Lipschitz functions $f:\sqrt{d}\,\mathbb{S}^{d-1}\to\mathbb{R}$, where $\|x\|_{\psi_2}$ denotes the sub-Gaussian norm (cf. Definition~\eqref{eq:orlicz-space}) of a random vector $x\in\mathbb{R}^d$. 

In order to guarantee the variance bound $\max_{j\in[n]}\operatorname{Var}(z_j^\top A z_j)=o(d^2)$, by Proposition~\ref{prop:spectral-variance-bound} in Supplementary Section~\ref{appendix:others} it suffices to restrict the spectral growth $\lambda_{\max}(\operatorname{Cov}(z\otimes z))=o(d)$ as $d\to\infty$, which may be considered as a measure of the random features correlation strength. Alternatively, by Proposition~\ref{prop:refined-variance-bound} in Supplementary Section~\ref{appendix:others}, it is enough to limit the growth of the sub-Gaussian norm $K:=\max_{j\in[n]}\|z_j\|_{\psi_2}=o(d^{\frac{1}{4}})$ as $d\to\infty$ in order for the variance bound $\max_{j\in[n]}\operatorname{Var}(z_j^\top A z_j)=o(d^2)$ to hold true. Thus, we reduced the problem to upper bounding the sub-Gaussian norm of $z_j$. By definition, the sub-Gaussian norm $\|z\|_{\psi_2}:=\sup_{v\in\mathbb{S}^{d-1}}\|\langle v,z\rangle\|_{\psi_2}$ is the largest sub-Gaussian norm of all possible projections of $z$ on the unit sphere. Thus, it is convenient to consider the projection $f_v:\sqrt{d}\,\mathbb{S}^{d-1}\to\mathbb{R}$ mapping $z\mapsto \langle v,z\rangle$ defined for every unit vector $v\in\mathbb{S}^{d-1}$. In that case, we find with the help of the Cauchy-Schwarz inequality,
\begin{equation}
    |f_v(z)-f_v(z')| =|\langle v,z\rangle - \langle v,z'\rangle| = |\langle v,z-z'\rangle| \leq \|v\|\,\|z-z'\|
\end{equation}
for arbitrary $z,z'\in\sqrt{d}\,\mathbb{S}^{d-1}$. This shows that the projection $f_v:\sqrt{d}\,\mathbb{S}^{d-1}$ is $L$-Lipschitz with $L=\|v\|=1$ and that Eq.~\eqref{eq:lipschitz-concentration1} applies for all $v\in\mathbb{S}^{d-1}$. Furthermore, since $z\sim\operatorname{Unif}(\sqrt{d}\,\mathbb{S}^{d-1})$ is sampled uniformly from a rotationally symmetric domain, we in particular have that $z\stackrel{d}{=}-z$ are equal in distribution. Hence,
\begin{equation}
    \mathbb{E}[f_v(z)]=\mathbb{E}[f_v(-z)] = \mathbb{E}[\langle v,-z\rangle] = -\mathbb{E}[\langle v,z\rangle] = -\mathbb{E}[f_v(z)] \implies \mathbb{E}[f_v(z)] = 0 \label{eq:centered-projection1}
\end{equation}
for all $z\in\mathbb{S}^{d-1}$. As a result,
\begin{equation}
    \|z\|_{\psi_2} = \sup_{v\in\mathbb{S}^{d-1}}\|\langle v,z\rangle\|_{\psi_2} = \sup_{v\in\mathbb{S}^{d-1}}\|f_v(z)\|_{\psi_2} \stackrel{\eqref{eq:centered-projection1}}{=} \sup_{v\in\mathbb{S}^{d-1}}\|f_v(z)-\mathbb{E}[f_v(z)]\|_{\psi_2} \stackrel{\eqref{eq:lipschitz-concentration1}}{=} \sup_{v\in\mathbb{S}^{d-1}}\mathcal{O}(\|v\|) = \mathcal{O}(1).
\end{equation}
In particular, $\|z\|_{\psi_2}=o(d^{\frac{1}{4}})$ as $d\to\infty$ which completes our argumentation.

As mentioned above, in Supplementary Section~\ref{appendix:others} we explicitly demonstrate that the sub-Gaussian norm
\begin{equation}
    K:=\max_{j\in[n]}\left\|\Sigma^{-\frac{1}{2}}\phi(x_j)\right\|_{\psi_2}
\end{equation}
may be understood as a measure of high-dimensional correlation of the random feature matrix $\Phi$. By the previous discussion, we have learned that $K=o(d^{\frac{1}{4}})$ is necessary in order to ensure double descent with an interpolation threshold at $\gamma=1$. This means that the high-dimensional correlations of the features $\Phi$ need to be limited in order to maintain the double descent phenomenon. The coordinates of $z\sim\operatorname{Unif}(\sqrt{d}\mathbb{S}^{d-1})$ must be correlated due to the geometric restriction on the sphere. However, the uniformicity and the rotational symmetry of the distribution minimizes the high-dimensional growth of the sub-Gaussian norm $\|z\|_{\psi_2}=\mathcal{O}(1)$ thus maintaining a dimension-independent scaling of the correlations. Most notably, since a larger growth rate with $\|z\|_{\psi_2}=o(d^{\frac{1}{4}})$ is already sufficient to guarantee double descent with an interpolation threshold at $\gamma=1$, this shows that a small asymptotic gap of order $o(d^{\frac{1}{4}})$ allows slightly larger correlations thus allowing asymmetric distributions with a limited inductive bias.

\subsubsection{Stieltjes Transforms, Deterministic Equivalents and the Marchenko-Pastur Law} \label{appendix:Stieltjes}

In a nutshell, the Marchenko-Pastur law~\cite{bai2010spectral} establishes a closed-form algebraic equation whose solution fully describes the eigenspectrum of certain random matrices in the infinite limit. 
This is achieved with the help of Stieltjes transforms, which bridge the mathematical fields of measure theory and complex analysis. 
Intuitively, the problem of finding the eigendistribution of a random matrix is isomorphically mapped to the problem of finding the solution of an algebraic equation in the complex numbers.
In the special case of the Marchenko-Pastur law, the equation reduces to a simple quadratic equation. We refer to Ref.~\cite{bai2010spectral} for further reading regarding this technique.
The distribution of the eigenspectrum of random matrices arising from machine learning models plays a central role in the analytical description of generalization such as double descent phenomena as elaborated in this work.

\smallskip
\par
\paragraph{The Stieltjes Transform.} For the sake of completeness and the convenience of the reader, we shall briefly introduce all necessary mathematical definitions to understand the Marchenko-Pastur law and additional concepts for its subsequent generalizations. In what follows, we shall denote the upper complex half-plane by
\begin{equation}
    \mathbb{C}_+ := \{z\in\mathbb{C}:\operatorname{Im}(z)>0\}.
\end{equation}
We assume basic knowledge in probability theory and the definition of Lebesgue integral with respect to general measures. The central object in the analysis of random matrix eigenspectra is the so-called Stieltjes transform. Mathematically, it bridges tools from the field of complex analysis and measure theory. We recommend Refs.~\cite{bai2010spectral, couillet2022random} for more details on the Stieltjes transform including applications to machine learning.

\begin{definition}[Stieltjes Transform] \label{def:stieltjes-transform}
    Let $\mu$ be a probability measure on $\mathbb{R}$. The Stieltjes transform of $\mu$ is defined as the complex map $m:\mathbb{C}_+$, mapping to the Lebesgue integral
    \begin{equation*}
        m_\mu(z) := \int_{\mathbb{R}}\frac{1}{\lambda - z}\,\mu(d\lambda).
    \end{equation*}
\end{definition}

The Stieltjes transform defines an analytic function on its domain. The probability measure $\mu$ may always be reconstructed from the Stieltjes transform by the following inversion formula:

\begin{theorem}[Stieltjes Inversion]
Let $a<b$ be continuity points of the cummulative distribution function of $\mu$, and let $m$ be the Stieltjes transform of $\mu$. Then
$$\mu([a,b])=\frac{1}{\pi}\lim_{\varepsilon\to0^+}\int_a^b\operatorname{Im}(m(\lambda+i\varepsilon))\,d\lambda.$$
\end{theorem}

The relationship between sequences of Stieltjes transforms and their corresponding probability measures is characterized by convergence in distribution.

\begin{theorem}[Limiting Measure]
Let $(\mu_n)_{n=1}^\infty$ be a sequence of probability measures with corresponding Stietljes transforms $(m_n)_{n=1}^\infty$. If there exists $m:\mathbb{C}_+\to\mathbb{C}_+$ such that $m$ is the Stieltjes transform of some probability measure $\mu$, and if for all $z\in\mathbb{C}_+$, almost surely as $n\to\infty$, 
$$m_n(z)\to m(z),$$
then $\mu_n\to\mu$ in distribution as $n\to\infty$. The converse statement also holds true.
\end{theorem}

In this context, $\mu$ is defined as the \emph{limiting measure} of the sequence $(\mu_n)_{n=1}^\infty$, and $m$ is the \emph{limiting Stieltjes transform} derived from the sequence $(m_n)_{n=1}^\infty$. The convergence $m_n(z) \to m(z)$ is a fundamental component of random matrix analysis. Consequently, it is useful to examine the structure of the Stieltjes transform $m_n(z)$ associated with the empirical spectral measure $\mu_n$ of a given random matrix.

\begin{definition}[Empirical Spectral Measure] \label{def:empirical-spectral-measure}
Let $X_n\in\mathbb{R}^{n\times n}$ be a random matrix with (random) eigenvalues $\lambda_1,\dots,\lambda_n$. The empirical spectral measure of $X_n$ is defined as
$$\mu_{X_n}(A):=\frac{1}{n}\sum_{j=1}^n\delta_{\lambda_j}(A),\quad A\in\mathcal{B}(\mathbb{R}),$$
where $\delta_x(A):=\mathbf{1}_A(x)$ denotes the Dirac measure.
\end{definition}

This measure can be equivalently represented as the normalized count of eigenvalues within a specific element $A$ of the real Borel $\sigma$-algebra:
\begin{equation}
\mu_{X_n}(A)=\frac{1}{n}\#\{\lambda_j\in A\},\quad A\in\mathcal{B}(\mathbb{R}).
\end{equation}

\smallskip
\par
\paragraph{The Resolvent and Deterministic Equivalents.} The Stieltjes transform of the empirical spectral measure can be expressed through the trace of the \emph{resolvent} matrix, which is a central object in this study.

\begin{definition}[Resolvent] \label{def:resolvent}
Let $X_n\in\mathbb{R}^{n\times n}$ be a random matrix. The (random) resolvent $Q_{X_n}$ of $X_n$ is defined as
$$Q_{X_n}(z):=(X_n-zI_n)^{-1},\quad z\in D(Q_{X_n}),$$
where
$$D(Q_{X_n}):=\{z\in\mathbb{C}_+:(X_n-zI_n)^{-1}\,\text{exists almost surely}\}.$$
\end{definition}

The term $zI_n$ shifts the spectrum of $X_n$. If $(v, \lambda)$ is an eigenpair of $X_n$, then:
\begin{equation}
(X_n-zI_n)v=X_nv-zv=(\lambda-z)v.
\end{equation}
By choosing $z\in\mathbb{C}\setminus\mathbb{R}$ such that $z\in D(Q_{X_n})$, we ensure $X_n-zI_n$ has no zero eigenvalues and is therefore invertible.

\begin{lemma} \label{lemma:Stieltjes-transform}
Let $X_n\in\mathbb{R}^{n\times n}$ be a random matrix, $Q_{X_n}$ its resolvent, and
$$m_{X_n}(z)=\int_{\mathbb{R}}\frac{1}{\lambda-z}\,\mu_{x_n}(d\lambda),\quad z\in\mathbb{C}_+$$
the Stieltjes transform of the empirical spectral measure $\mu_{X_n}$.
Then,
$$m_{X_n}(z)=\frac{1}{n}\operatorname{Tr}Q_{X_n}(z)=\frac{1}{n}\sum_{j=1}^n\frac{1}{\lambda_j-z},$$
where $\lambda_1,\dots,\lambda_n$ are the eigenvalues of $X_n$.
\end{lemma}

\noindent
\emph{Proof.} Let $\lambda_1,\dots,\lambda_n$ be the eigenvalues of $X_n$ so that:
\begin{equation}
\mu_{X_n}(\lambda)=\frac{1}{n}\sum_{j=1}^n\delta(\lambda-\lambda_j).
\end{equation}
The eigenvalues of the resolvent $Q_{X_n}(z)$ are given by $(\lambda_j-z)^{-1}$. For every eigenpair $(\lambda_j,v_j)$, we have:
\begin{align}
v_j=Q_{X_n}(z)Q_{X_n}(z)^{-1}v_j = Q_{X_n}(z)(X_n-zI_n)v_j = Q_{X_n}(z)(\lambda_j-z)v_j,
\end{align}
which yields:
\begin{equation}
Q_{X_n}(z)v_j=\frac{1}{\lambda_j-z}v_j.
\end{equation}
Finally, we conclude:
\begin{align}
m_{X_n}(z) &= \int_{\mathbb{R}}\frac{\mu_{X_n}(d\lambda)}{\lambda-z} = \frac{1}{n}\sum_{j=1}^n\int_{\mathbb{R}}\frac{\delta(\lambda-\lambda_j)}{\lambda-z}\,d\lambda = \frac{1}{n}\sum_{j=1}^n\frac{1}{\lambda_j-z} = \frac{1}{n}\operatorname{Tr}Q_{X_n}(z),
\end{align}
where we utilize the fact that the trace is the sum of the eigenvalues:
\begin{equation}
\operatorname{Tr}Q_{X_n}(z)=\sum_{j=1}^n\frac{1}{\lambda_j-z}.
\end{equation}
\qed

Notably, the random eigenvalues $\lambda_1,\dots,\lambda_n$ of $X_n$ correspond directly to the poles of the complex analytic function $m_{X_n}(z)$. Thus, one can investigate the empirical spectral density in the non-asymptotic regime by analyzing the poles of the empirical Stieltjes transform. This is what esablishes the bridge between complex analysis and measure theory.

The concept of the \emph{deterministic equivalent} $\overline{Q_{X_n}}(z)$ of the resolvent $Q_{X_n}(z)$ is one of various fundamental tools for deriving a self-consistent equation for the limiting Stieltjes transform $m_X(z)$. The core objective is to identify, for a fixed $z\in\mathbb{C}_+$, a deterministic matrix $\overline{Q_{X_n}}(z)\in\mathbb{R}^{n\times n}$ such that, almost surely as $n\to\infty$,
\begin{equation}
\frac{1}{n}\operatorname{Tr}\left(Q_{X_n}(z)- \overline{Q_{X_n}}(z)\right)\to0\quad\text{and}\quad\|\overline{Q_{X_n}}(z)\|\,\leq\mathcal{O}(1).
\end{equation}

The second condition regarding the norm is a necessary technical requirement. The first condition allows us to utilize the convergence of the empirical Stieltjes transform. Specifically, for an arbitrary $z\in\mathbb{C}_+$, we observe that almost surely:
\begin{align}
m_X(z)-\lim_{n\to\infty}\frac{1}{n}\operatorname{Tr}\overline{Q_{X_n}}(z) &= \lim_{n\to\infty}m_{X_n}(z)-\frac{1}{n}\operatorname{Tr}\overline{Q_{X_n}}(z) \notag
\\ &= \lim_{n\to\infty}\frac{1}{n}\operatorname{Tr}\left(Q_{X_n}(z)-\overline{Q_{X_n}}(z)\right) = 0.
\end{align}
This result implies that the limiting Stieltjes transform can be expressed directly in terms of the deterministic equivalent:
\begin{equation}
m_X(z)=\lim_{n\to\infty}\frac{1}{n}\operatorname{Tr}\overline{Q_{X_n}}(z).
\end{equation}

By leveraging this relationship, we could in principle establish a self-consistent equation required for $m_X(z)$. In practice, techniques from random matrix theory and concentration inequalities are employed to derive the explicit form of $\overline{Q_{X_n}}(z)$ that satisfies these asymptotic properties.

\smallskip
\par
\paragraph{The Marchenko-Pastur Law.} The Marchenko-Pastur law establishes a quadratic equation for the limiting Stieltjes transform of a centered sample covariance matrix with independent entries~\cite{bai2010spectral}. We use it to prove the analytic expression of the expected risk in the isotropic case (cf. Theorem~\ref{thm:double-descent-gaussian-features} in Section~\ref{appendix:gaussian-features}) and the risk in the under-parametrized regime under modified assumptions (cf. Proposition~\ref{prop:over-fitting} in Section~\ref{sec:non-isotropic-analysis}), respectively:

\begin{theorem}[Marchenko-Pastur] \label{thm:marchenko-pastur}
    Consider a random matrix $X\in\mathbb{R}^{d\times n}$ with i.i.d. elements $X_{i,j}$ with zero mean $\mathbb{E}[X_{i,j}]=0$ and finite variance $\mathbb{E}[X_{i,j}^2]=\sigma_X^2<\infty$ for all $(i,j)\in[d]\times[n]$. Suppose that $\frac{d}{n}\to\gamma\in(0,\infty)$ as $d,n\to\infty$. Then, the Stieltjes transform $m_{\frac{1}{n}XX^\top}(z)$ of the empirical spectral measure of $\frac{1}{n}XX^\top$ converges almost surely to the Stieltjes transform $S$ satisfying the quadratic equation
    \begin{equation*}
        \gamma z\sigma_X^2 S(z)^2 + (z+\sigma_X^2(\gamma-1))S(z) + 1 = 0.
    \end{equation*}
\end{theorem}

\subsection{Fourier Analysis} \label{appendix:Fourier-analysis}

Recall from Eq.~\eqref{eq:separable-encoding} in Section~\ref{appendix:fourier-interpretation} that we generally assume
\begin{equation}
    \mathcal{G}_\ell(x) = e^{-ix_1H_1^{(\ell)}}\otimes\cdots\otimes e^{-ix_pH_p^{(\ell)}}
\end{equation}
for some generating Hamiltonians $H_1^{(\ell)},\dots,H_p^{(\ell)}$ each with $H_\eta^{(\ell)}\in\operatorname{Herm}(2^{n_\eta})$ with block sizes $n_\eta\in\mathbb{N}$ acting on the accessible qubits such that $n_a=\sum_{\eta=1}^pn_\eta$. Let
\begin{equation}
    \operatorname{Spec}(H_\eta^{(\ell)}):=\{\lambda_{j,\eta}^{(\ell)}\}_{j\in[2^{n_\eta}]}
\end{equation}
denote the eigenspectrum of the generator $H_\eta^{(\ell)}$ acting on the $\eta$-th subsystem at layer $\ell\in[L]$. In Theorem~\ref{thm:general-Fourier-decomposition} we have shown that the readout may be represented as a multi-frequency Fourier decomposition, i.e.,
\begin{equation}
    \phi_U(x) = \sum_{\omega=(\omega_1,\dots,\omega_p)\in\Omega}c_\omega\,e^{i\omega\cdot x}, \label{eq:Fourier-decomposition-2}
\end{equation}
where the total Fourier frequency space is given by
\begin{equation}
    \Omega=\{\Lambda_\gamma-\Lambda_{\gamma'}:\gamma,\gamma'\in\Gamma\}\quad\text{where}\quad \Lambda_\gamma:=\sum_{\ell=1}^L \lambda_{\gamma_\ell}^{(\ell)}
\end{equation}
with $\lambda_{\gamma_\ell}^{(\ell)} := (\lambda_{\gamma_{\ell,1}}^{(\ell)},\dots,\lambda_{\gamma_{\ell,p}}^{(\ell)})\in\mathbb{R}^p$ for each reuploading layer $\ell\in[L]$ and full path space
\begin{equation}
    \Gamma := \left(\prod_{\eta=1}^p [2^{n_\eta}]\right)^L = ([2^{n_1}]\times\cdots\times[2^{n_p}])^L.
\end{equation}
In summary, the total Fourier frequency space is computed by accumulating all possible spectral differences (separately for each input component $\eta\in[p]$ which yields a vector representation) of the generators $H_\eta^{(1)},\dots,H_\eta^{(L)}$ in the circuit. In order to compute this accumulation, we use two ``paths'' $\gamma=(\gamma_1,\dots,\gamma_L),\gamma'=(\gamma'_1,\dots,\gamma'_L)\in\Gamma$ to index all such accumulated spectral differences which allows us to conveniently compute any Fourier frequency
\begin{equation}
    \omega = \Lambda_\gamma-\Lambda_{\gamma'} = \sum_{\ell=1}^L\lambda_{\gamma_\ell}^{(\ell)}-\sum_{\ell=1}^L{\lambda}_{\gamma'_\ell}^{(\ell)} = \sum_{\ell=1}^L\left\{\lambda_{\gamma_\ell}^{(\ell)}-\lambda_{\gamma'_\ell}^{(\ell)}\right\} = \sum_{\ell=1}^L\left(\begin{matrix} \lambda_{\gamma_\ell,1}^{(\ell)}-\lambda_{\gamma'_\ell,1}^{(\ell)} \\ \vdots \\ \lambda_{\gamma_\ell,p}^{(\ell)}-\lambda_{\gamma'_\ell,p}^{(\ell)}\end{matrix}\right) = \left(\begin{matrix} \omega_1 \\ \vdots \\ \omega_p\end{matrix}\right)\in\Omega \label{eq:Fourier-spectrum-original}
\end{equation}
in the total Fourier space, where each individual, multi-index path component $\gamma_\ell\in[2^{n_1}]\times\cdots\times[2^{n_p}]$ is used to index the eigenspectrum $\operatorname{Spec}(H_\eta^{(\ell)})=\{\lambda_{j,\eta}^{(\ell)}\}_{j\in[2^{n_\eta}]}$ of all generators in $\{H_\eta^{(\ell)}\}_{\eta\in[p]}$ simultaneously. In other words, each path $\gamma\in\Gamma$ computes the total phase accumulation $\Lambda_\gamma$ along the path (separately for each generator $\eta\in[p]$ by storing the total phase accumulation of the all the $\eta$-th generators in $\{H_\eta^{(\ell)}:\ell\in[L]\}$ in the $\eta$-th component $\Lambda_{\gamma,\eta}$ of the vector $\Lambda_{\gamma}=(\Lambda_{\gamma,1},\dots,\Lambda_{\gamma,p})\in\mathbb{R}^p$). The accumulation is computed by adding up all spectral differences $\lambda_{\gamma_\ell}^{(\ell)}-\lambda_{\gamma'_{\ell}}^{(\ell)}$ as the two paths $\gamma,\gamma'$ traverse the layers of the circuit. Therefore, by using two paths $\gamma,\gamma'\in\Gamma$ we can compute the total accumulation of spectral differences $\Lambda_\gamma-\Lambda_{\gamma'}=\omega$ along the two paths $\gamma,\gamma'\in\Gamma$ which is precisely the definition of each Fourier frequency $\omega\in\Omega$. 

\smallskip
\par
\paragraph{Bijective Reinterpretation of the Fourier Frequency Space.} Having refreshed the rigorous definition of the Fourier frequency space $\Omega$ above, we now shift our attention to an alternative representation of $\Omega$ which, as demonstrated in Section~\ref{subsubsec:input-reuploads}, is the key in our analysis of the QELM model in terms of the number $L$ of reuploads.

\begin{theorem} \label{thm:equivalent-fourier-space}
    The Fourier frequency space is determined by the input coordinate-wise Cartesian product
    \begin{equation}
        \Omega = \prod_{\eta=1}^p\sum_{\ell=1}^L \{\lambda-\lambda':\lambda,\lambda'\in\operatorname{Spec}(H_\eta^{(\ell)})\}
    \end{equation}
    of the layer-wise Minkowski sum of the set of spectral differences of the generator $H_\eta^{(\ell)}$.
\end{theorem}

\noindent
\emph{Proof.} Consider the mapping
\begin{equation}
    \Psi:\Gamma=\left(\prod_{\eta=1}^p[2^{n_\eta}]\right)^L \to \prod_{\eta=1}^p[2^{n_\eta}]^L=:\Gamma_{\operatorname{decoupled}}
\end{equation}
given by
\begin{equation}
    \Psi(\gamma)=\Psi(\gamma_1,\dots,\gamma_L) := \left(\left(\begin{matrix} \gamma_{1,1} \\ \vdots \\ \gamma_{L,1} \end{matrix}\right),\left(\begin{matrix} \gamma_{1,2} \\ \vdots \\ \gamma_{L,2} \end{matrix}\right),\dots,\left(\begin{matrix} \gamma_{1,p} \\ \vdots \\ \gamma_{L,p} \end{matrix}\right)\right)\quad\text{where}\quad \gamma_\ell=(\gamma_{\ell,1},\dots,\gamma_{\ell,p}) \in \prod_{\eta=1}^p[2^{n_\eta}]. \label{eq:bijection1}
\end{equation}
We will now show that $\Psi^{-1}:\Gamma_{\operatorname{decoupled}}\to\Gamma$, defined as
\begin{equation}
    \Psi^{-1}(\tau)=\Psi^{-1}(\tau_1,\dots,\tau_p) := \left(\left(\begin{matrix} \tau_{1,1} \\ \vdots \\ \tau_{1,p} \end{matrix}\right),\left(\begin{matrix} \tau_{2,1} \\ \vdots \\ \tau_{2,p} \end{matrix}\right),\dots,\left(\begin{matrix} \tau_{L,1} \\ \vdots \\ \tau_{L,p} \end{matrix}\right)\right)\quad\text{where}\quad \tau_\eta =(\tau_{1,\eta},\dots,\tau_{L,\eta}) \in [2^{n_\eta}]^L, \label{eq:bijection2}
\end{equation}
is the inverse of $\Psi$. Indeed, if $\gamma=(\gamma_1,\dots,\gamma_L)\in\Gamma$ with $\gamma_\ell=(\gamma_{\ell,1},\dots,\gamma_{\ell,p})\in\prod_{\eta=1}^p[2^{n_\eta}]$, we have
\begin{align}
    (\Psi^{-1}\circ\Psi)(\gamma) &= (\Psi^{-1}\circ\Psi)(\gamma_1,\dots,\gamma_L) \stackrel{\eqref{eq:bijection1}}{=} \Psi^{-1}\left(\left(\begin{matrix} \gamma_{1,1} \\ \vdots \\ \gamma_{L,1} \end{matrix}\right),\left(\begin{matrix} \gamma_{1,2} \\ \vdots \\ \gamma_{L,2} \end{matrix}\right),\dots,\left(\begin{matrix} \gamma_{1,p} \\ \vdots \\ \gamma_{L,p} \end{matrix}\right)\right) \nonumber
    \\ &\stackrel{\eqref{eq:bijection2}}{=} \left(\left(\begin{matrix} \gamma_{1,1} \\ \vdots \\ \gamma_{1,p} \end{matrix}\right),\left(\begin{matrix} \gamma_{2,1} \\ \vdots \\ \gamma_{2,p} \end{matrix}\right),\dots,\left(\begin{matrix} \gamma_{L,1} \\ \vdots \\ \gamma_{L,p} \end{matrix}\right)\right) = (\gamma_1,\dots,\gamma_L) = \gamma
\end{align}
showing $\Psi^{-1}\circ\Psi=\operatorname{id}_\Gamma$. Conversely, if $\tau=(\tau_1,\dots,\tau_p)\in\Gamma_{\operatorname{decoupled}}$ with $\tau_\eta=(\tau_{1,\eta},\dots,\tau_{L,\eta})\in[2^{n_\eta}]^L$, then
\begin{align}
    (\Psi\circ\Psi^{-1})(\tau) &= (\Psi\circ\Psi^{-1})(\tau_1,\dots,\tau_p) \stackrel{\eqref{eq:bijection2}}{=} \Psi\left(\left(\begin{matrix} \tau_{1,1} \\ \vdots \\ \tau_{1,p} \end{matrix}\right),\left(\begin{matrix} \tau_{2,1} \\ \vdots \\ \tau_{2,p} \end{matrix}\right),\dots,\left(\begin{matrix} \tau_{L,1} \\ \vdots \\ \tau_{L,p} \end{matrix}\right)\right) \nonumber
    \\ &\stackrel{\eqref{eq:bijection1}}{=} \left(\left(\begin{matrix} \tau_{1,1} \\ \vdots \\ \tau_{L,1} \end{matrix}\right),\left(\begin{matrix} \tau_{1,2} \\ \vdots \\ \tau_{L,2} \end{matrix}\right),\dots,\left(\begin{matrix} \tau_{1,p} \\ \vdots \\ \tau_{L,p} \end{matrix}\right)\right) = (\tau_1,\dots,\tau_p) = \tau
\end{align}
showing $\Psi\circ\Psi^{-1}=\operatorname{id}_{\Gamma_{\operatorname{decoupled}}}$.

Having established that $\Psi: \Gamma \to \Gamma_{\operatorname{decoupled}}$ is a bijection, we can reformulate the condition for a frequency vector to belong to $\Omega$. Let $\omega\in\Omega$ be arbitrary. By definition, $\omega = (\omega_1, \dots, \omega_p) \in \Omega$ if and only if there exist paths $\gamma, \gamma' \in \Gamma$ such that $\omega = \Lambda_\gamma - \Lambda_{\gamma'}$ and by Eq.~\eqref{eq:Fourier-spectrum-original}, we can express each coordinate $\omega_\eta$ for $\eta \in [p]$ as
\begin{equation}
    \omega_\eta = \sum_{\ell=1}^L \left( \lambda_{\gamma_{\ell,\eta}}^{(\ell)} - \lambda_{\gamma'_{\ell,\eta}}^{(\ell)} \right). \label{eq:frequency-component1}
\end{equation}
Let $\tau = \Psi(\gamma) \in \Gamma_{\operatorname{decoupled}}$ and $\tau' = \Psi(\gamma') \in \Gamma_{\operatorname{decoupled}}$ be the decoupled image paths, where we write them component-wise as
\begin{equation}
    \tau = (\tau_1, \dots, \tau_p) \quad \text{and} \quad \tau' = (\tau'_1, \dots, \tau'_p)
\end{equation}
with $\tau_\eta = (\tau_{1,\eta}, \dots, \tau_{L,\eta}) \in [2^{n_\eta}]^L$ and $\tau'_\eta = (\tau'_{1,\eta}, \dots, \tau'_{L,\eta}) \in [2^{n_\eta}]^L$. Since $\Psi$ is a bijection, searching over all pairs of paths $\gamma, \gamma' \in \Gamma$ is exactly equivalent to searching over all pairs of decoupled coordinate paths $\tau, \tau' \in \Gamma_{\operatorname{decoupled}}$. From the explicit definition of $\Psi$, the components relate directly via $\gamma_{\ell,\eta} = \tau_{\ell,\eta}$ and $\gamma'_{\ell,\eta} = \tau'_{\ell,\eta}$. Substituting these into our coordinate expression for $\omega_\eta$ in Eq.~\eqref{eq:frequency-component1} yields
\begin{equation}
    \omega_\eta = \sum_{\ell=1}^L \left( \lambda_{\tau_{\ell,\eta}}^{(\ell)} - \lambda_{\tau'_{\ell,\eta}}^{(\ell)} \right). \label{eq:decoupled-fourier-frequency-component}
\end{equation}
Notably, because the target space is the Cartesian product $\Gamma_{\operatorname{decoupled}} = \prod_{\eta=1}^p [2^{n_\eta}]^L$, the choice of the decoupled path component $\tau_\eta$ (and respectively $\tau'_\eta$) for a given coordinate $\eta$ is entirely independent of the choices made for any other coordinate $\eta' \neq \eta$. Consequently, the global set $\Omega$ decomposes cleanly into a Cartesian product of independent coordinate-wise sets, i.e.,
\begin{equation}
    \Omega \stackrel{\eqref{eq:decoupled-fourier-frequency-component}}{=} \left\{(\omega_1,\dots,\omega_p):\omega_\eta = \sum_{\ell=1}^L \left( \lambda_{\tau_{\ell,\eta}}^{(\ell)} - \lambda_{\tau'_{\ell,\eta}}^{(\ell)} \right)\right\} =  \prod_{\eta=1}^p \left\{ \sum_{\ell=1}^L \left( \lambda_{\tau_{\ell,\eta}}^{(\ell)} - \lambda_{\tau'_{\ell,\eta}}^{(\ell)} \right) : \tau_\eta, \tau'_\eta \in [2^{n_\eta}]^L \right\} = \prod_{\eta=1}^p \Omega_{\operatorname{layer}}^{(\eta)}
\end{equation}
recalling the definition $\Omega_{\operatorname{layer}}^{(\eta)}:=\sum_{\ell=1}^L\Omega_{\eta}^{(\ell)}$ where $\Omega_\eta^{(\ell)}:=\{\lambda-\lambda':\lambda,\lambda'\in\operatorname{Spec}(H_\eta^{(\ell)})\}$ with $\operatorname{Spec}(H_\eta^{(\ell)}):=\{\lambda_{j,\eta}^{(\ell)}\}_{j\in[2^{n_\eta}]}$ for each $\eta\in[p]$ and $\ell\in[L]$. Since $\omega\in\Omega$ was arbitrary, this completes the proof.
\qed

\smallskip
\par
\paragraph{Affine Transformation of the Fourier Coefficients.} This section is dedicated to rigorously answering the question of what happens to the Fourier coefficients in Eq.~\eqref{eq:Fourier-coefficients-vector-frequencies} in Section~\ref{appendix:fourier-interpretation} when increasing the number of reuploads $L\mapsto L+\Delta L$ for some $\Delta L\in\mathbb{N}$. Specifically, if $c_\omega^{(L)}$ and $c_\omega^{(L+\Delta L)}$ denote the Fourier coefficients of the $L$- and $(L+\Delta L)$-layered models, respectively, we will explicitly prove that the updated Fourier coefficient
\begin{equation}
    c_\omega^{(L+\Delta L)}= T_0 c_\omega^{(L)}+\tilde{\epsilon}_\omega,
\end{equation}
is an affine transformation of the old, original Fourier coefficient for some certain, explicitly determined matrix $T_0\in\mathbb{C}^{d\times d}$ and vector $\tilde{\epsilon}_\omega\in\mathbb{C}^d$, respectively. Remarkably, this holds true for arbitrary architecture, even if the additional layers are completely different from the original ones.

In order to solve this problem, we need to revisit the double path integral formulation of the Fourier coefficients in Eq.~\eqref{eq:Fourier-coefficients-vector-frequencies} of Section~\ref{appendix:fourier-interpretation} and rigorously study what happens to the structure when we increase $\Delta L$. This requires the introduction of additional notations. Given the increase $L\mapsto L+\Delta L$, we generally have a larger path space
\begin{equation}
    \Gamma_{L+\Delta L}:=\left(\prod_{\eta=1}^p[2^{n_\eta}]\right)^{L+\Delta L} = \quad \underbrace{\left(\prod_{\eta=1}^p[2^{n_\eta}]\right)^{L}}_{=:\Gamma} \times \underbrace{\left(\prod_{\eta=1}^p[2^{n_\eta}]\right)^{\Delta L}}_{=:\Gamma_{\Delta L}} =  \quad \Gamma\times \Gamma_{\Delta L}
\end{equation}
with paths going through additional unitaries $U_\ell$ and $W_\ell$ for $\ell\in\{L+2,\dots,L+\Delta L\}$, respectively. Let $\gamma_{\text{full}}=(\gamma,\mu)\in\Gamma_{L+\Delta L}$ be a full path so that $\gamma\in\Gamma$ is the smaller path existing in the original architecture (i.e. the base reference) and $\mu\in\Gamma_{\Delta}$ the additional contribution to the smaller path due to the deeper circuit. Writing $\mu=(\mu_1,\dots,\mu_{\Delta L})$ to represent the sequence of choices in the extra $\Delta L$ layers, we decompose
\begin{align}
    W_{\gamma_{\text{full}}} &= W_{(\gamma,\mu)} \equiv W_{\gamma_1,\dots,\gamma_L,\mu_1,\dots,\mu_{\Delta L}} = \underbrace{P_{\mu_{\Delta L}}^{(a)}U_{L+\Delta L}\cdots P_{\mu_1}^{(a)}}_{=:V_\mu}U_{L+1}\underbrace{P_{\gamma_L}^{(a)}U_L\cdots P_{\gamma_1}^{(a)} U_1}_{=W_\gamma} \label{eq:fullpath-unitary}
\end{align}
according to Theorem~\ref{thm:general-Fourier-decomposition} in Section~\ref{appendix:fourier-interpretation}. Furthermore, we observe that the spectral accumulation is additive in the sense that
\begin{equation}
    \Lambda_{\gamma_{\text{full}}} = \sum_{\ell=1}^{L+\Delta L}\lambda_{(\gamma_{\text{full}})_\ell}^{(\ell)} = \underbrace{\sum_{\ell=1}^L\lambda_{\gamma_{\ell}}^{(\ell)}}_{=\Lambda_\gamma} + \underbrace{\sum_{j=1}^{\Delta L} \lambda_{\mu_j}^{(j)}}_{=:\Lambda_{\mu}^{(\Delta L)}} = \Lambda_{\gamma} + \Lambda_{\mu}^{(\Delta L)} \label{eq:fullpath-accumulation}
\end{equation}
for each $\gamma_{\text{full}}=(\gamma,\mu)\in\Gamma_{L+\Delta L}$, where $\Lambda_\mu^{(\Delta L)}$ defines the additional spectral accumulation due to the deeper architecture. Thus, together with Eqs.~\eqref{eq:fullpath-accumulation} and~\eqref{eq:fullpath-unitary}, Corollary~\ref{thm:general-Fourier-decomposition} tells us that we explicitly have
\begin{align}
    \left(c_\omega^{(L+\Delta L)}\right)_k &= \sum_{\substack{\gamma_{\text{full}},\gamma'_{\text{full}}\in\Gamma_{L+\Delta L} \\ \Lambda_{\gamma_{\text{full}}}-\Lambda_{\gamma'_{\text{full}}}=\,\omega}}\operatorname{Tr}\left(O_kU_{L+\Delta L+1}W_{\gamma_{\text{full}}}\rho_0W_{\gamma'_{\text{full}}}^\dagger U_{L+\Delta L+1}^\dagger\right) \nonumber
    \\ &\stackrel{\eqref{eq:fullpath-unitary}}{=} \sum_{\substack{(\gamma,\mu),(\gamma',\mu')\in\Gamma\times\Gamma_{\Delta L} \\ (\Lambda_\gamma-\Lambda_{\gamma'})+(\Lambda_\mu^{(\Delta L)}-\Lambda_{\mu'}^{(\Delta L)})=\omega}} \underbrace{\operatorname{Tr}\left(U_{L+\Delta L+1}^\dagger O_k U_{L+\Delta L+1} (V_\mu U_{L+1}W_\gamma)\rho_0(V_{\mu'} U_{L+1}W_{\gamma'})^\dagger\right)}_{=\operatorname{Tr}\left(\left\{V_{\mu'}^\dagger U_{L+\Delta L+1}^\dagger O_k U_{L+\Delta L+1} V_\mu\right\}U_{L+1}W_\gamma\rho_0 W_{\gamma'}^\dagger U_{L+1}^\dagger\right)} \nonumber
    \\ &= \sum_{\substack{(\gamma,\mu),(\gamma',\mu')\in\Gamma\times\Gamma_{\Delta L} \\ (\Lambda_\gamma-\Lambda_{\gamma'})+(\Lambda_\mu^{(\Delta L)}-\Lambda_{\mu'}^{(\Delta L)})=\omega}} \operatorname{Tr}\left(O_k(\mu,\mu')U_{L+1}W_\gamma\rho_0W_{\gamma'}^\dagger U_{L+1}^\dagger\right), \label{eq:deep-Fourier-coefficients}
\end{align}
where we also applied the cyclic property of the trace operator, and defined the path dependent operator $O_k(\mu,\mu'):=V_{\mu'}^\dagger U_{L+\Delta L+1}^\dagger O_k U_{L+\Delta L+1}V_\mu$ (which is not necessarily Hermitian). On the other hand, Corollary~\ref{thm:general-Fourier-decomposition} also provides us with the explicit expression of the original, base reference Fourier coefficients,
\begin{equation}
    \left(c_\omega^{(L)}\right)_k = \sum_{\substack{\gamma,\gamma'\in\Gamma \\ \Lambda_{\gamma}-\Lambda_{\gamma'}=\omega}}\operatorname{Tr}\left(O_k U_{L+1}W_\gamma\rho_0W_{\gamma'}^\dagger U_{L+1}^\dagger\right). \label{eq:original-Fourier-coefficients}
\end{equation}
Hence, it suffices to obtain Eq.~\eqref{eq:original-Fourier-coefficients} directly from Eq.~\eqref{eq:deep-Fourier-coefficients}. 
We do this with the help of a transition matrix and an error term. The key lies in assuming that $\{O_k\}_{k\in[d]}$ defines an orthogonal system
Indeed, with the help of this extra assumption, we can leverage Lemma~\ref{lemma:orthogonal-decomposition} in Section~\ref{appendix:others} to decompose the linear operator $O_k(\mu,\mu')$ such that 
\begin{equation}
    O_k(\mu,\mu') = O_{k}(\mu,\mu')^{\perp} + \sum_{k'=1}^d\underbrace{\left(\frac{\operatorname{Tr}\left(O_{k'}^\dagger O_k(\mu,\mu')\right)}{\operatorname{Tr}(O_{k'}^2)}\right)}_{=:T(\mu,\mu')_{k,k'}}O_{k'} \label{eq:path-operator-orthogonal-decomposition}
\end{equation}
for all $\mu,\mu'\in\Gamma_{\Delta L}$, where $T(\mu,\mu')\equiv(T(\mu,\mu')_{k,k'})_{k,k'\in[d]}\in\mathbb{C}^{d\times d}$ defines the path dependent transition matrix and $O_{k}(\mu,\mu')^\perp$ is the error term such that $\operatorname{Tr}(O_{k'}^\dagger O_k(\mu,\mu')^\perp)=0$ for all $k'\in[d]$.
In other words, the term $\sum_{k'=1}^d T(\mu,\mu')_{k,k'}O_{k'}$ is the orthogonal projection of $O_k(\mu,\mu')$ onto the subspace spanned by the orthogonal system $\{O_k\}_{k\in[d]}$, and $O_k^\perp\in\operatorname{Herm}(2^N)$ the remaining error term that is orthogonal to the system $\{O_k\}_{k\in[d]}$, respectively. Substituting this into Eq.~\eqref{eq:deep-Fourier-coefficients} yields
\begin{align}
    \left(c_\omega^{(L+\Delta L)}\right)_k &\stackrel{\eqref{eq:deep-Fourier-coefficients}}{=} \sum_{\substack{(\gamma,\mu),(\gamma',\mu')\in\Gamma\times\Gamma_{\Delta L} \\ (\Lambda_\gamma-\Lambda_{\gamma'})+(\Lambda_\mu^{(\Delta L)}-\Lambda_{\mu'}^{(\Delta L)})=\omega}} \operatorname{Tr}\left(O_k(\mu,\mu')U_{L+1}W_\gamma\rho_0W_{\gamma'}^\dagger U_{L+1}^\dagger\right) \nonumber
    \\ &\stackrel{\eqref{eq:path-operator-orthogonal-decomposition}}{=} \sum_{\mu,\mu'\in\Gamma_{\Delta L}}\sum_{\substack{\gamma,\gamma\in\Gamma \\ \Lambda_\gamma-\Lambda_{\gamma'}=\omega-(\Lambda_\mu^{(\Delta L)}-\Lambda_{\mu'}^{(\Delta L)})}} \operatorname{Tr}\left(\left(O_k(\mu,\mu')^\perp+\sum_{k'=1}^d T(\mu,\mu')_{k,k'}O_{k'}\right)U_{L+1}W_\gamma\rho_0W_{\gamma'}^\dagger U_{L+1}^\dagger\right) \nonumber
    \\ &= (\epsilon_\omega)_k +  \sum_{\mu,\mu'\in\Gamma_{\Delta L}}\sum_{k'=1}^d T(\mu,\mu')_{k,k'} \underbrace{\sum_{\substack{\gamma,\gamma'\in\Gamma \\ \Lambda_\gamma-\Lambda_{\gamma'}=\omega-(\Lambda_{\mu}^{(\Delta L)}-\Lambda_{\mu'}^{(\Delta L)})}} \operatorname{Tr}\left(O_{k'}U_{L+1}W_\gamma\rho_0 W_{\gamma'}^\dagger U_{L+1}^\dagger\right)}_{\stackrel{\eqref{eq:Fourier-coefficients-vector-frequencies}}{=}\left(c^{(L)}_{\omega-(\Lambda_{\mu}^{(\Delta L)}-\Lambda_{\mu'}^{(\Delta L)})}\right)_k},
\end{align}
introducing the orthogonal error term
\begin{equation}
    (\epsilon_\omega)_k := \sum_{\mu,\mu'\in\Gamma_{\Delta L}} \sum_{\substack{\gamma,\gamma\in\Gamma \\ \Lambda_\gamma-\Lambda_{\gamma'}=\omega-(\Lambda_\mu^{(\Delta L)}-\Lambda_{\mu'}^{(\Delta L)})}} \operatorname{Tr}\left(O_k(\mu,\mu')^\perp U_{L+1} W_\gamma \rho W_{\gamma'}^\dagger U_{L+1}^\dagger\right). \label{eq:orthogonal-leakage-term}
\end{equation}

This establishes the following technical proposition, which tells us that the deep Fourier coefficients are obtained from a discrete convolution between the transition matrix elements and the original Fourier coefficients modulo an additive error term:

\begin{proposition} \label{prop:Fourier-convolution}
    Assume that $\{O_k\}_{k\in[d]}$ defines an orthogonal system of observables and let $\Delta L\in\mathbb{N}$. Then,
    \begin{equation*}
        \left(c_\omega^{(L+\Delta L)}\right)_k = (e_\omega)_k + \sum_{\mu,\mu'\in \Gamma_{\Delta L}}\sum_{k'=1}^d T_{k,k'}(\mu,\mu') \left(c^{(L)}_{\omega-(\Lambda_{\mu}^{(\Delta L)}-\Lambda_{\mu'}^{(\Delta L)})}\right)_{k'}
    \end{equation*}
    for all $k\in[d]$. In particular,
    \begin{equation*}
        c_\omega^{(L+\Delta L)}= T_0 c_\omega^{(L)}+\tilde{\epsilon}_\omega, 
    \end{equation*}
    where for each $k,k'\in[d]$,
    \begin{equation*}
        (T_0)_{k,k'}:=\sum_{\substack{\mu,\mu'\in\Gamma_{\Delta L} \\ \Lambda_{\mu'}^{(\Delta L)}-\Lambda_\mu^{(\Delta L)}=0}}T_{k,k'}(\mu,\mu') \quad\text{and}\quad (\tilde{\epsilon}_\omega)_k:=(\epsilon_\omega)_k \, + \sum_{\substack{\mu,\mu'\in\Gamma_{\Delta L} \\ \Lambda_{\mu'}^{(\Delta L)}-\Lambda_\mu^{(\Delta L)}\neq0}}\sum_{k'=1}^d T_{k,k'}(\mu,\mu') \left(c^{(L)}_{\omega-(\Lambda_{\mu}^{(\Delta L)}-\Lambda_{\mu'}^{(\Delta L)})}\right)_{k'}.
    \end{equation*}
\end{proposition}

\subsection{Lebesgue and Sobolev Spaces} \label{appendix:Sobolev-spaces}

\paragraph{Lebesgue Spaces.} As the mathematical foundations, we begin with the definition of Lebesgue spaces~\cite{wheeden1977measure} (also see Chapter~2 in Ref.~\cite{adams2003sobolev}):

\begin{definition}[Lebesgue Space] \label{def:Lebesgue-space}
    Let $\mathcal{X}\subseteq\mathbb{R}^p$ be open and $q\in[1,\infty)$. The $q$-th Lebesgue space on $\mathcal{X}$ is defined as the quotient space
    \begin{equation*}
        L^q(\mathcal{X}) = \{g:\mathcal{X}\to\mathbb{R}\mid \mathbb{E}[|g(x)|^q] < \infty \} \,/ \sim
    \end{equation*}
    of measurable functions $g:\mathcal{X}\to\mathbb{R}$ with finite $q$-th moments with respect to almost sure equivalence ``$\,\sim$''.
\end{definition}

It is well-known that the Lebesgue space $L^q(\mathcal{X})$ together with the norm
\begin{equation}
    \|g\|_{L^q}:=\left(\mathbb{E}[|g(x)|^q]\right)^{\frac{1}{q}}
\end{equation}
defines a Banach space, i.e., an infinite dimensional complete normed vector space. In fact, in the special case $q=2$, the Lebesgue space $L^2(\mathcal{X})$ equipped with the inner product $\langle g,h\rangle_{L^2}:= \mathbb{E}[g(x)h(x)]$ becomes a complete Hilbert space, and we further have that the mean value $\mathbb{E}[g(x)] = \arg\inf_{c\in\mathbb{R}} \,\|g-c\|_{L^2}$ coincides with the orthogonal projection of $g(x)$ onto the closed subspace of constant functions. In particular, $\operatorname{Var}(g(x)) = \langle g,g\rangle_{L^2}$ if $\mathbb{E}[g(x)]=0$ is centered. The $L^q$-norm may be computed as a standard real Lebesgue integral with the help of the following classical formula which is an immediate consequence of Tonelli's theorem (cf. Theorem~6.10 of Ref.~\cite{wheeden1977measure}). This formula is applied to prove certain concentration inequalities used in our technical argumentation (cf. Subsection~\ref{subsubsection:high-dim-probability}):

\begin{lemma}[Layer Cake Representation] \label{lemma:layer-cake}
    Let $q\in[1,\infty)$ and $g\in L^q(\mathcal{X})$. Then,
    \begin{equation*}
        \|g\|_{L^q}^q = \mathbb{E}[|g(x)|^q] = \int_{\mathcal{X}}|g(x)|^q\,d\mathbb{P}(x) = q\int_0^\infty t^{q-1} \mathbb{P}\left(|g(x)|\geq t\right)\,dt.
    \end{equation*}
\end{lemma}

Another very central tool in measure theory, often used to justify changing limits and integration, is the Lebesgue Dominated Convergence (LDC) theorem. Crucially, this tool is used to justify an interchange between taking limits and integration in the proof of Theorem~\ref{thm:condition-number-input-strength} in Section~\ref{subsubsec:input-strength}:

\begin{lemma}[Lebesgue Dominated Convergence, Theorem~5.36 in Ref.~\cite{wheeden1977measure}] \label{lemma:lebesgue-dominated-convergence}
    Let $\mathcal{X}\subseteq\mathbb{R}^p$ be open and $(g_i)_{i\in\mathcal{I}}$ a sequence of $\mathcal{X}$-measurable functions $g_i:\mathcal{X}\to\mathbb{R}$, indexed by an unbounded subset $\mathcal{I}\subseteq(0,\infty)$, such that $g_i\to f$ almost everywhere as $i\to\infty$ for some $\mathcal{X}$-measurable limit function $g:\mathcal{X}\to\mathbb{R}$ satisfying $|g_i|\leq G$ almost everywhere for almost all $i\in\mathcal{I}$ for some integrable dominant $G\in L^1(\mathcal{X})$. Then, we have $g\in L^1(\mathcal{X})$ and
    \begin{equation}
        \|g_i-g\|_{L^1}\to 0
    \end{equation}
    as $i\to\infty$.
\end{lemma}

To see concretely why the LDC theorem allows one to exchange limits and integration, we note that under the assumptions of Lemma~\ref{lemma:lebesgue-dominated-convergence} we have
\begin{equation}
    \left|\int_{\mathcal{X}} g_i( x)\,d\mathbb{P}(x)-\int_{\mathcal{X}}g( x)\,d\mathbb{P}(x)\right| = \left|\int_{\mathcal{X}}(g_i( x)-g( x))\,d \mathbb{P}(x)\right| \leq \int_{\mathcal{X}}|g_i( x)-g( x)|\,d\mathbb{P}(x) = \|g_i-g\|_{L^1}\to 0
\end{equation}
as $i\to\infty$, which indeed proves
\begin{equation}
    \lim_{i\to\infty}\int_{\mathcal{X}}g_i( x)\,d\mathbb{P}(x) = \int_{\mathcal{X}}\underbrace{\lim_{i\to\infty}g_i( x)}_{=g( x)}\,d\mathbb{P}(x). \label{eq:exchange-limit-integration}
\end{equation}

\smallskip
\par
\paragraph{The Characteristic Function.} Next, we introduce the characteristic function of random vectors and refer to Section~26 in Ref.~\cite{billingsley2017probability} for further details. We use the characteristic function in combination with the Fourier interpretation (cf. Theorem~\ref{thm:general-Fourier-decomposition} in Section~\ref{appendix:fourier-interpretation}) to analytically represent the quantum population covariance to study the asymptotics with respect to input strength (cf. Lemma~\ref{lemma:input-strength-covariance} in Section~\ref{subsubsec:input-strength}). The asymptotic analysis is refined futher below when introducing Sobolev spaces.
The \emph{characteristic function} $\varphi_x$ of a $p$-dimensional random vector $x\in\mathcal{X}$ is defined as the function
\begin{equation}
    \varphi_x:\mathbb{R}^p\to\mathbb{C};\quad \varphi_x(v) := \mathbb{E}[e^{i\langle v,x\rangle}]. \label{eq:characteristic-function-def}
\end{equation}
In particular, $e^{i\langle v, x\rangle}\in L^1(\mathcal{X})$ for all $v\in\mathbb{R}^p$. Moreover, it defines a uniformly continuous function on $\mathbb{R}^p$ satisfying $|\varphi_x(v)| \leq 1$ and $\varphi_x(-v) = \varphi_x(v)^*$ for all $v\in\mathbb{R}^p$. We particularly have the special case $\varphi_x(0)=1$ at $v=0$. Many useful properties follow due to the exponential in the definition: for example, we generally have $\varphi_{ax+b}(v) = e^{i\langle v,b\rangle}\varphi_x(av)$ for all $a\in\mathbb{R}$ and $b,v\in\mathbb{R}^p$. Furthermore, $\varphi_{\sum_{i=1}^n x_i} = \prod_{i=1}^n\varphi_{x_i}$ if $x_1,\dots,x_n\in\mathcal{X}$ are mutually independent random vectors. Notably, for more regular random vectors, specifically if $x\in L^q(\mathcal{X})$ exhibits finite $q$-th moment, then, by Lebesgue dominated convergence, the characteristic becomes $q$-fold differentiable in each dimension $k\in[p]$ with
\begin{equation}
    (\partial_{v_k}^\alpha\varphi_x)(v) = \int_{\mathcal{X}} (ix_k)^\alpha e^{i\langle v,x\rangle}\,d\mathbb{P}(x) \label{eq:moments-property}
\end{equation}
for all $\alpha\in[q]$ and $v\in\mathbb{R}^p$. In particular, $\partial_{v_k}^\alpha\mid_{v=0}\varphi_x(v) = i^\alpha\,\mathbb{E}[x_k^\alpha]$ for all $\alpha\in[q]$ and $k\in[p]$, which establishes a direct link between high-order gradients of the characteristic function and moments of the random vector $x$. However, the core property of the characteristic function is the Fourier inversion formula:
\begin{equation}
    f(x) = \frac{1}{(2\pi)^p}\int_{\mathbb{R}^p} e^{-i\langle v,x\rangle} \varphi_x(v)\,dv,
\end{equation}
which holds true if $d\mathbb{P}(x) = f(x)\,dx$ (i.e., if $x$ has a continuous density). As a corollary of this property we namely have that $\varphi_x = \varphi_y$ for two random vectors $x,y\in\mathcal{X}$ if and only if $x\stackrel{d}{=}y$ are equal in distribution. This is precisely the reason why $\varphi_x$ is called the ``characteristic'' function of $x$ as it fully characterizes its distribution. Table~\ref{tab:characteristic-function} compiles a few relevant examples.

\renewcommand{\arraystretch}{1.5} 
\begin{table}[h]
\centering
\caption{\textbf{Multivariate Distributions and their Characteristic Functions}}
\label{tab:characteristic-function}
\begin{tabular}{l l l} 
\toprule
\textbf{Distribution \& Notation} & \textbf{Parameters \& Support} & \textbf{Characteristic Function } $\varphi_{x}(v)$ \\
\midrule

\makecell[l]{\textbf{Dirac Delta} \\ $x \sim \delta_{a}$} & 
\makecell[l]{$a \in \mathbb{R}^p$ \\ Support: $\{a\}$} & 
$\exp\left(i {v}^T {a}\right)$ \\ \addlinespace[1.5ex]

\makecell[l]{\textbf{Normal} \\ ${x} \sim \mathcal{N}({\mu}, {\Sigma})$} & 
\makecell[l]{${\mu} \in \mathbb{R}^p$ \\ ${\Sigma} \in \mathbb{S}_+^p$ (Symmetric pos.-def.) \\ Support: $\mathbb{R}^p$} & 
$\exp\left(i {v}^T {\mu} - \frac{1}{2} {v}^T {\Sigma} {v}\right)$ \\ \addlinespace[1.5ex]

\makecell[l]{\textbf{Uniform} \\ ${x} \sim \mathcal{U}({a}, {b})$} & 
\makecell[l]{${a}, {b} \in \mathbb{R}^p, \ a_k < b_k\,\forall k\in[p]$ \\ Support: $\prod_{k=1}^p [a_k, b_k]$} & 
$\displaystyle \prod_{k=1}^p \frac{e^{i v_k b_k} - e^{i v_k a_k}}{i v_k (b_k - a_k)}$ \\ \addlinespace[1.5ex]

\makecell[l]{\textbf{Multinom.} \\ ${x} \sim \text{Mult}(n, {\rho})$} & 
\makecell[l]{$n \in \mathbb{N}$ \\ ${\rho} \in [0,1]^p, \ \sum_{k=1}^p \rho_k = 1$ \\ Support: $\{{x} \in \mathbb{N}_0^p : \sum_{k=1}^p x_k = n\}$} & 
$\displaystyle \left( \sum_{k=1}^p \rho_k e^{i v_k} \right)^n$ \\ \addlinespace[1.5ex]

\makecell[l]{\textbf{Cauchy} \\ ${x} \sim \mathcal{C}({\mu}, {\Sigma})$} & 
\makecell[l]{${\mu} \in \mathbb{R}^p$ \\ ${\Sigma} \in \mathbb{S}_+^p$ \\ Support: $\mathbb{R}^p$} & 
$\exp\left(i {v}^T {\mu} - \sqrt{{v}^T {\Sigma} {v}}\right)$ \\ \addlinespace[1.5ex]

\makecell[l]{\textbf{Laplace} \\ ${x} \sim \text{Laplace}({\mu}, {\Sigma})$} & 
\makecell[l]{${\mu} \in \mathbb{R}^p$ \\ ${\Sigma} \in \mathbb{S}_+^p$ \\ Support: $\mathbb{R}^p$} & 
$\displaystyle \frac{\exp(i {v}^T {\mu})}{1 + \frac{1}{2}{v}^T {\Sigma} {v}}$ \\
\bottomrule
\end{tabular}
\end{table}

\smallskip
\par
\paragraph{Input Strength Analysis.} Recall from Section~\ref{subsubsec:input-strength} that we are interesting in bounding the condition number of the rescaled population covariance
\begin{equation}
    \Sigma(s) \equiv  \Sigma_U(s) := \mathbb{E}_{ x}\left[\phi_U(s x)\,\phi_U(s x)^\top\right]
\end{equation}
defined in Eq.~\eqref{eq:rescaled-population-covariance} as it quantifies the generalization performance of the QELM in the over-parametrized regime (cf. Theorem~\ref{thm:risk-bound-nonisotropic} in Section~\ref{appendix:double_descent_linear2}). The following proposition establishes a refined analysis of the rescaled population covariance $\Sigma(s)$ for the movement $s\to0$. Concretely, it uses the asymptotic effect of the input strength to precisely describe how integral moments (cf. Eq.~\eqref{eq:moments-property}) of input-frequency projections $(\omega+\omega')\cdot x$ contribute to the scaled sample covariance $ \Sigma(s)$.

\begin{proposition} \label{prop:covariance-moment-series}
    Let $s>0$. Then, we explicitly have
    \begin{equation*}
         \Sigma(s) = \sum_{m\geq0}\frac{(is)^m}{m!}\sum_{\omega,\omega'\in\Omega} M_{ x}^{(m)}(\omega+\omega') \, c_\omega c_{\omega'}^\top,
    \end{equation*}
    where $c_\omega=(c_\omega^1,\dots, c_\omega^d)\in\mathbb{C}^d$ are explicitly determined in Eq.~\eqref{eq:Fourier-coefficients-vector-frequencies} and independent of the input strength $s>0$, calling $M_{ x}^{(m)}(\omega+\omega') := \mathbb{E}_{ x}[((\omega+\omega')\cdot x)^m]$ the $m$-th frequency projection moment at $\omega+\omega'$ for each $\omega,\omega'\in\Omega$ and $m\in\mathbb{N}_0$.
\end{proposition}

\noindent
\emph{Proof.} Let $s>0$, $m\in\mathbb{N}_0$ and $\omega,\omega'\in\Omega$ be arbitrary. First, note that
\begin{align}
    \partial_{s}^m|_{s=0}\,\varphi_{ x}(s(\omega+\omega')) &= \mathbb{E}_{ x}[\partial_s^m|_{s=0}\,e^{is(\omega+\omega')\cdot x}] = \mathbb{E}_{ x}[(i(\omega+\omega')\cdot x)^m [e^{is(\omega+\omega')\cdot x}]_{s=0}] \nonumber
    \\ &= i^m\mathbb{E}_{ x}[((\omega+\omega')\cdot x)^m] = i^m M_{ x}^{(m)}(\omega+\omega'). \label{eq:moment-step1}
\end{align}
Additionally, since the characteristic function $\varphi_{ x}:\mathbb{R}^p\to[-1,1]$ is analytic, we have
\begin{equation}
    \varphi_{ x}(s(\omega+\omega') = \sum_{m\geq0}\underbrace{\partial_s^m|_{s=0}\,\varphi_{ x}(s(\omega+\omega'))}_{\stackrel{\eqref{eq:moment-step1}}{=} i^mM_{ x}^{(m)}(\omega+\omega')}\frac{s^m}{m!} = \sum_{m\geq0}M_{ x}^{(m)}(\omega+\omega')\frac{(is)^m}{m!}. \label{eq:moment-step2}
\end{equation}
Therefore, we obtain with the help of Lemma~\ref{lemma:input-strength-covariance} that
\begin{align}
     \Sigma(s) &= \sum_{\omega,\omega'\in\Omega} \varphi_{ x}\left(s(\omega+\omega')\right)  c_\omega  c_{\omega'}^\top \stackrel{\eqref{eq:moment-step2}}{=} \sum_{\omega,\omega'\in\Omega}\left(\sum_{m\geq0}M_{ x}^{(m)}(\omega+\omega')\frac{(is)^m}{m!}\right) c_\omega  c_{\omega'}^\top \nonumber
    \\ &= \sum_{m\geq0}\frac{(is)^m}{m!}\sum_{\omega,\omega'\in\Omega} M_{ x}^{(m)}(\omega+\omega') \, c_\omega c_{\omega'}^\top
\end{align}
as claimed.
\qed

\smallskip
\par
\paragraph{Sobolev Spaces and Weak Derivatives.} Having introduced Lebesgue space, we shall extend to Sobolev spaces and weak derivatives~\cite{adams2003sobolev} to characterize more nuanced regularity properties. These regularity properties allow us to upper the bound population covariance condition number with respect to the input strength (cf. Theorem~\ref{thm:condition-number-input-strength} in Section~\ref{subsubsec:input-strength}).
First, we begin with the standard classes of continuously differentiable functions. Let $\mathcal{X}\subseteq\mathbb{R}^p$ be an open domain in $p$-dimensional Euclidean space. For each $K\in\mathbb{N}$ let $\mathcal{C}^K(\mathcal{X})$ denote the class of $K$-fold differentiable functions $g:\mathcal{X}\to\mathbb{R}$, that is, $D_{ x}^\alpha g( x) := \partial_{ x_1}^{\alpha_1}\cdots\partial_{ x_p}^{\alpha_p} g( x)$ exists and is continuous for all multi-indices $\alpha=(\alpha_1,\dots,\alpha_p)\in\mathbb{N}_0^p$ with $\|\alpha\|_1:=\sum_{\eta=1}^p\alpha_\eta\leq K$. The class of smooth functions on $\mathcal{X}$ is then defined as the intersection
\begin{equation}
    \mathcal{C}^\infty(\mathcal{X}) := \bigcap_{K\in\mathbb{N}}\mathcal{C}^K(\mathcal{X})
\end{equation}
of all $K$-fold continuously differentiable classes on $\mathcal{X}$. Next, we introduce the class
\begin{equation}
    \mathcal{C}^\infty_c(\mathcal{X}) := \{g\in\mathcal{C}^\infty(\mathcal{X}):\operatorname{supp}(g)\subseteq\mathcal{X}\text{ is compact}\}
\end{equation}
of smooth functions with compact support in $\mathcal{X}$, recalling that $\operatorname{supp}(g):=\overline{\{ x\in\mathcal{X}:g( x)\neq0\}}$ denotes the support of a function $g:\mathcal{X}\to\mathbb{R}$. We also call $\mathcal{C}^\infty_c(\mathcal{X})$ the space of \emph{test functions} on $\mathcal{X}$. All those function spaces are analogously generalized in the case of complex images by treating real- and imaginary parts accordingly. Test functions are a very useful mathematical tool to conduct refined analysis in function spaces.

Before we may proceed with the definition of Sobolev spaces, we have to introduce a weaker notion of derivatives that are only defined within integrands. In order to do so, we employ test functions to define the so-called \emph{distributions} on $\mathcal{X}$ (not to be confused with probability distributions), i.e., linear and bounded functionals $T:\mathcal{C}_c^\infty(\mathcal{X})\to\mathbb{R}$ that map test functions to the real number line.

\begin{definition}[Distribution] \label{def:distribution1}
    Let $\mathcal{X}\subseteq\mathbb{R}^p$ be open. A map $T:\mathcal{C}_c^\infty(\mathcal{X})\to\mathbb{R}$ is called a distribution on $\mathcal{X}$ if it is linear and bounded, i.e., if
    \begin{itemize}
        \item[(i)] $T(\alpha\varphi+\beta\psi)=\alpha T(\varphi)+\beta T(\psi)$ for all $\varphi,\psi\in\mathcal{C}_c^\infty(\mathcal{X})$ and $\alpha,\beta\in\mathbb{R}$
        \item[(ii)] For all compact $K\subset\mathcal{X}$ there exists a constant $C>0$ and $m\in\mathbb{N}_0$ such that
        \begin{equation*}
            |T(\varphi)| \leq C\sup_{\substack{\alpha\in\mathbb{N}_0^p\\ \|\alpha\|_1\leq m}}\sup_{x\in K}|D_x^\alpha\varphi(x)|
        \end{equation*}
        for every test function $\varphi\in\mathcal{C}_c^\infty(\mathcal{X})$ with $\operatorname{supp}(\varphi)\subseteq K$.
    \end{itemize}
    The vector space of distributions on $\mathcal{X}$ shall be denoted $\mathcal{D}'(\mathcal{X})$.
\end{definition}

Because this work continuously studies expectation values over functions of random variables, including the elements of the population covariance (cf. Eq.~\eqref{eq:rescaled-population-covariance}) and the characteristic function, (cf. Lemma.~\ref{lemma:input-strength-covariance}) in Section~\ref{subsubsec:input-strength}, the class of distributions of our main interest are functionals of the form
\begin{equation}
    T_f:\mathcal{C}_c^\infty(\mathcal{X})\to\mathbb{R},\quad \varphi\mapsto\mathbb{E}_x[\varphi(x)] = \int_{\mathcal{X}}\varphi(x)\,d\mathbb{P}(x) = \int_{\mathcal{X}} \varphi(x)f(x)\,dx, \label{eq:functional-distribution1}
\end{equation}
where $f(x)$ is the Radon-Nikodym derivative of the probability measure $\mathbb{P}$ with respect to the Lebesgue measure (i.e., $d\mathbb{P}(x) = f(x)\,dx$). Indeed, linearity is straightforward by definition of the Lebesgue integral, and boundedness follows due to the fact that, for any compact $K\subset\mathcal{X}$, we have
\begin{equation}
    |T_f(\varphi)| \leq \int_K|\varphi(x)||f(x)|\,dx \leq \sup_{x\in K}|\varphi(x)|\int_{\mathcal{X}}f(x)\,dx = \sup_{x\in K}|\varphi(x)|\underbrace{\int_{\mathcal{X}}d\mathbb{P}(x)}_{=1} = \sup_{x\in K}|\varphi(x)| \label{eq:bounded-distribution-1}
\end{equation}
for all $\varphi\in\mathcal{C}_c^\infty(\mathcal{X})$ with $\operatorname{supp}(\varphi)\subseteq K$. Thus, the property (ii) in Definition~\ref{def:distribution1} holds true by choosing $C=1$ and $m=0$.  Let $q\in[1,\infty)$ and consider the $q$-th order local Lebesgue space
\begin{equation*}
   \mathcal{L}^q_{\operatorname{loc}}(\mathcal{X}) := \left\{g:\mathcal{X}\to\mathbb{R} : \int_K |g(x)|^q\,dx<\infty\text{ for all compact $K\subset\mathcal{X}$}\right\} /\sim
\end{equation*}
of locally Lebesgue integrable functions modulo almost everywhere equivalence ``$\sim$''. In particular, we have that $f\in\mathcal{L}_{\operatorname{loc}}^1(\mathcal{X})$ for any Radon-Nikodym derivative $f$ as per $d\mathbb{P}(x) = f(x)\,dx$. Note that we used ``$\mathcal{L}$'' instead of ``$L$'' to specify that the integral is with respect to the Lebesgue measure, whereas the Lebesgue spaces $L^q(\mathcal{X})$ are defined with respect to the probability measure $\mathbb{P}$.

We now proceed by observing that, if $g\in\mathcal{C}^1(\mathcal{X})$ is continuously differentiable, standard integration by parts immediately yields
\begin{equation}
    T_{\partial_{x_k}g}(\varphi) = \int_{\mathcal{X}} \varphi(x)(\partial_{x_k}g)(x)\,dx = - \int_{\mathcal{X}}  (\partial_{x_k}\varphi)(x)g(x)\,dx= - T_{g}(\partial_{x_k}\varphi)
\end{equation}
for all test functions $\varphi\in\mathcal{C}_c^\infty(\mathcal{X})$ and $k\in[p]$. This motivates the definition of \emph{weak derivatives} (or distributional derivatives):

\begin{definition}[Weak Derivative] \label{def:weak-derivative1}
   Let $\alpha\in\mathbb{N}_0^p$ be a multi-index. A locally Lebesgue integrable function $g\in \mathcal{L}_{\operatorname{loc}}^1(\mathcal{X})$ is called the $\alpha$-order weak derivative of $f\in\mathcal{L}_{\operatorname{loc}}^1(\mathcal{X})$ if
    \begin{equation}
        \int_{\mathcal{X}} (D_x^\alpha\varphi)(x)f(x)\,dx = (-1)^{\|\alpha\|_1}\int_{\mathcal{X}}\varphi(x)g(x)\,dx
    \end{equation}
    for all test functions $\varphi\in\mathcal{C}_c^\infty(\mathcal{X})$, where $\|\alpha\|_1 := \sum_{\eta=1}^p\alpha_\eta$, and derivative operator $D_{ x}^{\alpha}:=\partial_{ x_1}^{\alpha_1}\cdots\partial_{ x_p}^{\alpha_p}$ in the function argument variable $ x=( x_1,\dots, x_p)$. In that case, we write $D_x^\alpha f:=g$.
\end{definition}

In particular, if $f$ is the Radon-Nikodym derivative as per $d\mathbb{P}(x)=f(x)\,dx$, then its weak derivative $D_x^\alpha f$ is defined by setting
\begin{equation}
    \mathbb{E}_x[D_x^\alpha\varphi(x)] = (-1)^{\|\alpha\|_1}\int_{\mathcal{X}} \varphi(x) D_x^\alpha f(x)\,dx
\end{equation}
for all $\varphi\in\mathcal{C}_c^\infty(\mathcal{X})$. We are now finally ready to define Sobolev spaces:

\begin{definition}[Sobolev Space] \label{def:Sobolev-space}
    Let $\mathcal{X}\subseteq\mathbb{R}^p$ be open and $K\in\mathbb{N}$. Let further $q\in[1,\infty)$. The $K$-th order Sobolev space on $\mathcal{X}$ with respect to $q$ is defined as
    \begin{equation*}
        W^{K,q}(\mathcal{X}) := \left\{g\in \mathcal{L}^1_{\operatorname{loc}}(\mathcal{X}): \int_{\mathcal{X}}|D_{ x}^\alpha g(x)|^q\,dx<\infty\text{ for all $\alpha\in\mathbb{N}_0^p$ with 
        $\|\alpha\|_1\leq K$}\right\} / \sim
    \end{equation*}
    modulo almost everywhere equivalence ``$\sim$'', where $\|\alpha\|_1 := \sum_{\eta=1}^p\alpha_\eta$, and derivative operator $D_{ x}^{\alpha}:=\partial_{ x_1}^{\alpha_1}\cdots\partial_{ x_p}^{\alpha_p}$ in the function argument variable $ x=( x_1,\dots, x_p)$, defined for each multi-index $\alpha=(\alpha_1,\dots,\alpha_p)\in\mathbb{N}_0^p$, respectively.
\end{definition}

The Sobolev space $W^{K,q}(\mathcal{X})$ equipped with the norm
\begin{equation}
    \|g\|_{W^{K,q}} := \sum_{\substack{\alpha\in\mathbb{N}_0^p \\ \|\alpha\|_1 \leq K}} \left(\int_{\mathcal{X}}|D_{ x}^\alpha g(x)|^q\,dx\right)^{\frac{1}{q}}
\end{equation}
also defines a Banach space. In particular, we have $g\in W^{K,q}(\mathcal{X})$ if and only if $\|g\|_{W^{K,q}}<\infty$. Remember that we previously claimed test functions being a very useful tool for conducting refined analysis. The defining property of Sobolev functions ensures the existence of weak derivatives as per Definition~\ref{def:weak-derivative1}. The statement is also frequently referred to as ``integration by parts'':

\begin{lemma}[``Integration by Parts''] \label{lemma:sobolev-integration-by-parts}
    Let $K\in\mathbb{N}$ and $\mathcal{X} \subseteq \mathbb{R}^p$ be an open domain, and let $f$ be the Radon-Nikodym derivative of $\mathbb{P}$ with respect to the Lebesgue measure (i.e., $d\mathbb{P}(x) = f(x)\,dx$) such that $f\in W^{K,1}(\mathcal{X})$. Then, for any test function $\varphi\in\mathcal{C}^\infty_c(\mathcal{X})$,
    \begin{equation*}
        \mathbb{E}_x[D_{ x}^\alpha \varphi( x)]=\int_{\mathcal{X}}(D_{ x}^\alpha \varphi)(x)f(x)\,dx = (-1)^K\int_{\mathcal{X}}\varphi( x)D_{ x}^\alpha f( x)dx
    \end{equation*}
    for all multi-indices $\alpha\in\mathbb{N}_0^p$ with $\|\alpha\|_1=K$.
\end{lemma}

\smallskip
\par
\paragraph{Input Strength Analysis Part II.} The following could be considered to be the ``climax'' of the input strength analysis, and thus, one of the more technical parts of the supplementary material. Recall from Eq.~\eqref{eq:asymmetric-remainder-term} in Section~\ref{subsubsec:input-strength} that the rescaled population covariance is generally represented by
\begin{equation}
    \Sigma(s) = \sum_{\omega\in\Omega}c_\omega c_\omega^\dagger + \Delta(s) \quad\text{with}\quad \Delta(s) := \sum_{\substack{\omega,\omega'\in\Omega\\ \omega + \omega'\neq0}} \varphi_{ x}\left(s(\omega+\omega')\right)  c_\omega  c_{\omega'}^\top, \label{eq:population-covariance-decomposition1i}
\end{equation}
where $\Delta(s)$ denotes the asymmetric remainder term. With the help of Sobolev techniques (cf. Section~\ref{appendix:Sobolev-spaces}) we can find a very explicit expression for $\Delta(s)$ in terms of the regularity properties of the input density function:

\begin{proposition} \label{prop:regularity-bound}
    Let $s>0$. Assume that the input $ x$ has a density function $f( x)$ belonging to the $K$-th order Sobolev space $W^{K,1}(\mathcal{X})$ for some $K\in\mathbb{N}$. Then, we explicitly have 
    \begin{equation*}
        \Delta(s) = \sum_{\substack{\omega,\omega'\in\Omega \\ \omega+\omega'\neq0}}\left(\frac{i}{s\|\omega+\omega'\|^2}\right)^K \int_{\mathcal{X}}e^{is(\omega+\omega')\cdot x}D_{\omega+\omega'}^K f( x)\,d x\,c_\omega c_{\omega'}^\top,
    \end{equation*}
    where $c_\omega=(c_\omega^1,\dots,c_\omega^d)\in\mathbb{C}^d$ are explicitly determined in Eq.~\eqref{eq:Fourier-coefficients-vector-frequencies} and independent of the input strength $s>0$, defining directional derivatives (to be understood in the weak sense as per Definition~\ref{def:weak-derivative1})
    \begin{equation*}
        D_{\omega+\omega'}^K f( x) := \sum_{\eta_1,\dots,\eta_K=1}^p \prod_{\kappa =1}^K (\omega_{\eta_\kappa}+\omega'_{\eta_\kappa})\frac{\partial^K f( x)}{\partial_{ x_{\eta_1}}\cdots\,\,\partial_{ x_{\eta_K}}}
    \end{equation*}
    for each $\omega,\omega'\in\Omega$ with $\omega+\omega'\neq0$, respectively.
\end{proposition}

\noindent
\emph{Proof.} Let $s>0$ and $\omega,\omega'\in\Omega$ be arbitrary with $\omega+\omega'\neq0$. We know that the input $ x$ has a density function $f:\mathcal{X}\to[0,1]$ so that $\mathbb{E}_{ x}[g( x)] = \int_{\mathcal{X}}g( x)f( x)\,d x$ for any integrable function $g\in L^1(\mathcal{X})$. In particular,
\begin{equation}
    \varphi_{ x}(s(\omega+\omega')) = \mathbb{E}_{ x}[e^{is(\omega+\omega')\cdot x}] = \int_{\mathcal{X}} e^{is(\omega+\omega')\cdot x} f( x)\,d x. \label{eq:characteristic-function1}
\end{equation}
Next, we evaluate the directional derivative
\begin{align}
    D_{\omega+\omega'}^K e^{is(\omega+\omega')\cdot x} &= \sum_{\eta_1,\dots,\eta_K=1}^p\prod_{\kappa =1}^K (\omega_{\eta_\kappa}+\omega'_{\eta_\kappa})\frac{\partial^K e^{is(\omega+\omega')\cdot x}}{\partial_{ x_{\eta_1}}\cdots\,\,\partial_{ x_{\eta_K}}} = (is)^Ke^{is(\omega+\omega')\cdot x}\sum_{\eta_1,\dots,\eta_K=1}^p \prod_{\kappa=1}^K(\omega_{\eta_{\kappa}}+\omega'_{\eta_\kappa})^2 \nonumber
    \\ &= (is)^K e^{is(\omega+\omega')\cdot x} \left( \sum_{\eta=1}^p (\omega_{\eta}+\omega_{\eta}')^2 \right)^K = (is)^K \|\omega+\omega'\|^{2K} e^{is(\omega+\omega')\cdot x}, \label{eq:directional-derivative1}
\end{align}
where we separately computed the partial derivatives
\begin{equation}
    \frac{\partial^K e^{is(\omega+\omega')\cdot x}}{\partial  x_{\eta_1} \cdots \partial  x_{\eta_K}} = (is)^K \left( \prod_{\kappa=1}^K (\omega_{\eta_\kappa}+\omega_{\eta_\kappa}') \right) e^{is(\omega+\omega')\cdot x}
\end{equation}
for each $(\eta_1,\dots,\eta_K)\in[p]^K$. As a result, rearranging Eq.~\eqref{eq:directional-derivative1} yields
\begin{equation}
    e^{is(\omega+\omega')\cdot x} = \frac{1}{(is)^K \|\omega+\omega'\|^{2K}} D_{\omega+\omega'}^K e^{is(\omega+\omega')\cdot x} = \left( \frac{-i}{s\|\omega+\omega'\|^2} \right)^K D_{\omega+\omega'}^K e^{is(\omega+\omega')\cdot x} \label{eq:exponential-factor1}
\end{equation}
and thus
\begin{align}
     \varphi_{ x}(s(\omega+\omega')) &\stackrel{\eqref{eq:characteristic-function1}}{=} \int_{\mathcal{X}} e^{is(\omega+\omega')\cdot x} f( x)\,d x \stackrel{\eqref{eq:exponential-factor1}}{=} \left( \frac{-i}{s\|\omega+\omega'\|^2} \right)^K\int_{\mathcal{X}} \left( D_{\omega+\omega'}^K e^{is(\omega+\omega')\cdot x} \right) f( x)\,d x
\end{align}
Thus, it suffices to prove
\begin{equation}
    \int_{\mathcal{X}} \left( D_{\omega+\omega'}^K e^{is(\omega+\omega')\cdot x} \right) f( x)\,d x = (-1)^K\int_{\mathcal{X}}e^{is(\omega+\omega')\cdot x}D_{\omega+\omega'}^K f( x)\,d x \label{eq:decomposition-formula-milestone1}
\end{equation}
for the first statement of the proposition. We approach this step with the help of the Sobolev space formalism (cf. Definition~\ref{def:Sobolev-space}) and the integration by parts formula (cf. Lemma~\ref{lemma:sobolev-integration-by-parts}). In the case where $\mathcal{X}\subseteq\mathbb{R}^p$ is unbounded we have to approach this with the help of a cutoff argument to construct valid test functions $g\in\mathcal{C}_c^\infty(\mathcal{X})$. Specifically, for each radius $R>0$, we choose a smooth bump function $b_R\in\mathcal{C}_c^\infty(\mathcal{X})$ satisfying
\begin{equation}
    b_R( x) = \begin{cases}
        1,& \| x\| \leq R \\
        0,& \| x\| \geq 2R \label{eq:bump-function1}
    \end{cases}
\end{equation}
so that $ x\mapsto g_{R,\omega,\omega'}( x):=e^{is(\omega+\omega')\cdot x}b_R( x)$ defines a valid test function $g_{R,\omega,\omega'}\in\mathcal{C}_c^\infty(\mathcal{X})$ for all $R>0$ and $\omega,\omega'\in\Omega$. It shall also be constructed such that it naturally satisfies the scaling property $b_R( x)=b_1\left(\frac{ x}{R}\right)$ for all $ x\in\mathcal{X}$. Explicitly, one may verify that
\begin{equation}
    b_R( x) := \begin{cases} 1, & \| x\| \le R \\ \left(1 + \exp\left(\frac{1}{2 - \frac{\| x\|}{R}} - \frac{1}{\frac{\| x\|}{R} - 1}\right)\right)^{-1}, & R < \| x\| < 2R \\ 0, & \| x\| \ge 2R \end{cases}
\end{equation}
is one possible example of a bump function of the form in Eq.~\eqref{eq:bump-function1}. Therefore, since $f\in W^{K,1}(\mathcal{X})$ defines a $K$-th order Sobolev function, we may apply Lemma~\ref{lemma:sobolev-integration-by-parts} so that
\begin{align}
    \int_{\mathcal{X}} &\left( D_{\omega+\omega'}^K \left\{e^{is(\omega+\omega')\cdot x} b_R( x)\right\}\right) f( x)\,d x = \int_{\mathcal{X}}\left(D_{\omega+\omega'}^Kg_{R,\omega,\omega'}( x)\right)f( x)\,d x \nonumber
    \\ &= \int_{\mathcal{X}}\left(\sum_{\substack{\alpha\in\mathbb{N}_0^p \\ \|\alpha\|_1=K}}\binom{K}{\alpha}(\omega+\omega')^\alpha D_{ x}^\alpha g_{R,\omega,\omega'}( x)\right)f( x)\,d x = \sum_{\substack{\alpha\in\mathbb{N}_0^p \\ \|\alpha\|_1=K}} \binom{K}{\alpha}(\omega+\omega')^\alpha\int_{\mathcal{X}}(D_{ x}^\alpha g_{R,\omega,\omega'}( x)) f( x)\,d x \nonumber
    \\ &\stackrel{\text{Lemma~\ref{lemma:sobolev-integration-by-parts}}}{=} (-1)^K\sum_{\substack{\alpha\in\mathbb{N}_0^p \\ \|\alpha\|_1=K}} \binom{K}{\alpha}(\omega+\omega')^\alpha\int_{\mathcal{X}}g_{R,\omega,\omega'}( x) D_{ x}^\alpha f( x)\,d x \nonumber
    \\ &= (-1)^K\int_{\mathcal{X}}g_{R,\omega,\omega'}( x)\sum_{\substack{\alpha\in\mathbb{N}_0^p \\ \|\alpha\|_1=K}} \binom{K}{\alpha}(\omega+\omega')^\alpha D_{ x}^\alpha f( x)\,d x  =  (-1)^K \int_{\mathcal{X}} g_{R,\omega,\omega'}( x) (D_{\omega+\omega'}^K f( x))\,d x \nonumber
    \\ &= (-1)^K\int_{\mathcal{X}}e^{is(\omega+\omega')\cdot x}b_R( x)(D_{\omega+\omega'}^K f( x))\,d x, \label{eq:decomposition-formula-step1}
\end{align}
for all $R>0$, where we collected the indices via $(\eta_1,\dots,\eta_K)\mapsto \alpha=(\alpha_1,\dots,\alpha_p)\in\mathbb{N}^p$ under the restriction $\|\alpha\|_1=\sum_{\eta=1}^p\alpha_\eta = K$ to obtain the linear operator identity
\begin{align}
    D_{\omega+\omega'}^K = \sum_{\eta_1,\dots,\eta_K=1}^p \prod_{\kappa =1}^K (\omega_{\eta_\kappa}+\omega'_{\eta_\kappa})\frac{\partial^K}{\partial_{ x_{\eta_1}}\cdots\,\,\partial_{ x_{\eta_K}}} = \sum_{\substack{\alpha\in\mathbb{N}_0^p \\ \|\alpha\|_1 = K}}\binom{K}{\alpha} (\omega+\omega')^\alpha D_{ x}^\alpha, \label{eq:derivative-transformation}
\end{align}
where $D_{ x}^\alpha := \partial_{ x_1}^{\alpha_1}\cdots\partial_{ x_p}^{\alpha_p}$, $(\omega+\omega')^\alpha := \prod_{\eta=1}^p(\omega_\eta+\omega'_\eta)^{\alpha_\eta}$ and $\binom{K}{\alpha}:=\frac{K!}{\alpha_1!\cdots\alpha_p!}$. Next, we want to take the limit $R\to\infty$ in order to conclude with Eq.~\eqref{eq:decomposition-formula-milestone1}.

We begin with the right hand side of Eq.~\eqref{eq:decomposition-formula-step1} by employing the Lebesgue dominated convergence theorem (cf. Lemma~\ref{lemma:lebesgue-dominated-convergence}). Applying it with the indexing set $\mathcal{I}=(0,\infty)$ and setting $f_R( x):=e^{is(\omega+\omega')\cdot x}b_R( x)D_{\omega+\omega'}^K f( x)$ for all $R\in\mathcal{I}$, we note that for all $R\in\mathcal{I}$,
\begin{equation}
    |f_R( x)| \leq |D_{\omega+\omega'}^K f( x)| \in L^1(\mathcal{X})
\end{equation}
using $|b_R( x)|\leq 1$ by definition and that $f\in W^{K,1}(\mathcal{X})$ defines a $K$-th order Sobolev function so that $D_{\omega+\omega'}^K f( x)\in L^1(\mathcal{X})$. Therefore, Lemma~\ref{lemma:lebesgue-dominated-convergence} implies $\lim_{R\to\infty}f_R \in L^1(\mathcal{X})$ and allows us to exchange limits and integration as per Eq.~\eqref{eq:exchange-limit-integration}. That is,
\begin{align}
    \lim_{R\to\infty}\int_{\mathcal{X}}e^{is(\omega+\omega')\cdot x}b_R( x)(D_{\omega+\omega'}^K f( x))\,d x &= \lim_{R\to\infty}\int_{\mathcal{X}}f_R( x)\,d x \stackrel{\eqref{eq:exchange-limit-integration}}{=} \int_{\mathcal{X}}\lim_{R\to\infty}f_R( x)\,d x \nonumber
    \\ &= \int_{\mathcal{X}}e^{is(\omega+\omega')\cdot x}\underbrace{\lim_{R\to\infty}b_R( x)}_{=1}D_{\omega+\omega'}^K f( x)\,d x = \int_{\mathcal{X}}e^{is(\omega+\omega')\cdot x}D_{\omega+\omega'}^K f( x)\,d x
\end{align}
by definition of the bump functions $b_R( x)$, since for each fixed $ x\in\mathcal{X}$ we have $b_R( x)=1$ for sufficiently large $R>0$.

Now, we continue with the left hand side of Eq.~\eqref{eq:decomposition-formula-step1} which is slightly more sophisticated to deal with. First, we note that for all $R>0$ and $\omega,\omega'\in\Omega$ with $\omega+\omega'\neq0$, the directional derivative factor in the integrand equals
\begin{align}
    D_{\omega+\omega'}^K \left\{e^{is(\omega+\omega')\cdot x} b_R( x) \right\} &\stackrel{\eqref{eq:derivative-transformation}}{=} \sum_{\substack{\alpha\in\mathbb{N}_0^p \\ \|\alpha\|_1=K}} \binom{K}{\alpha} (\omega+\omega')^\alpha D_{ x}^\alpha\{e^{is(\omega+\omega')\cdot x} b_R( x)\} \nonumber
    \\ &= \sum_{\substack{\alpha\in\mathbb{N}_0^p \\ \|\alpha\|_1=K}}\sum_{\substack{\beta\in\mathbb{N}_0^p \\ \beta_\eta\leq\alpha_\eta\,\forall\eta\in[p]}} \binom{K}{\alpha}\binom{\alpha}{\beta}(\omega+\omega')^\alpha(D_{ x}^\beta e^{is(\omega+\omega')\cdot x}) (D_{ x}^{\alpha-\beta} b_R( x)) \nonumber
    \\ &= \sum_{\substack{\alpha\in\mathbb{N}_0^p \\ \|\alpha\|_1=K}}\sum_{\substack{\beta\in\mathbb{N}_0^p \\ \beta_\eta\leq\alpha_\eta\,\forall\eta\in[p]}} (is)^{\|\beta\|_1}\binom{K}{\alpha}\binom{\alpha}{\beta}(\omega+\omega')^{\alpha+\beta}  e^{is(\omega+\omega')\cdot x} (D_{ x}^{\alpha-\beta} b_R( x))
    \label{eq:decomposition-formula-step2}
\end{align}
where we utilized the general Leibniz formula in the third equality to compute the high-order derivatives with respect to $ x$ of the product $e^{is(\omega+\omega')\cdot x} b_R( x)$, i.e.,
\begin{align}
    D_{ x}^\alpha\{e^{is(\omega+\omega')\cdot x} b_R( x)\} &=  \sum_{\substack{\beta\in\mathbb{N}_0^p \\ \beta_\eta\leq\alpha_\eta\,\forall\eta\in[p]}} \binom{\alpha}{\beta} (D_{ x}^\beta e^{is(\omega+\omega')\cdot x}) (D_{ x}^{\alpha-\beta} b_R( x)),
\end{align}
with $\binom{\alpha}{\beta}:=\prod_{\eta=1}^p\binom{\alpha_\eta}{\beta_\eta}$. In the third equality of Eq.~\eqref{eq:decomposition-formula-step2} we evaluated the derivative
\begin{equation}
    D_{ x}^\beta e^{is(\omega+\omega')\cdot x} = (is)^{\|\beta\|_1} (\omega+\omega')^\beta e^{is(\omega+\omega')\cdot x}. \label{eq:exponential-derivative1}
\end{equation}
At this point, we recall that, by construction, $b_R( x)=1$ is constant on the open ball $B_R(0)=\{ x\in\mathcal{X}:\| x\|<R\}$ of radius $R>0$ centered at the origin. In particular, $D_{ x}^{\alpha-\beta}b_R\equiv 0$ on $B_R(0)$ for all $\alpha,\beta\in\mathbb{N}_0^p$ with $\alpha-\beta\neq0$. Hence, if $ x\in B_R(0)$, the only term in the inner sum over $\beta$ of Eq.~\eqref{eq:decomposition-formula-step2} that survives is when $\beta = \alpha$ in which case we have $\|\beta\|_1 = \|\alpha\|_1 = K$, $\binom{\alpha}{\alpha} = \prod_{\eta=1}^p \binom{\alpha_\eta}{\alpha_\eta} = 1$, $(\omega+\omega')^{\alpha+\beta} = (\omega+\omega')^{2\alpha}$ and $D_{ x}^{\alpha-\beta}b_R( x) = b_R( x) = 1$ on $B_R(0)$. It thus follows from Eq.~\eqref{eq:decomposition-formula-step2} that for any $ x\in B_R(0)$,
\begin{align}
    D_{\omega+\omega'}^K \left\{e^{is(\omega+\omega')\cdot x} b_R( x) \right\} &= \sum_{\substack{\alpha\in\mathbb{N}_0^p \\ \|\alpha\|_1=K}}\sum_{\substack{\beta\in\mathbb{N}_0^p \\ \beta_\eta\leq\alpha_\eta\,\forall\eta\in[p]}} (is)^{\|\beta\|_1}\binom{K}{\alpha}\binom{\alpha}{\beta}(\omega+\omega')^{\alpha+\beta}  e^{is(\omega+\omega')\cdot x} (D_{ x}^{\alpha-\beta} b_R( x)) \nonumber
    \\ &\stackrel{ x \in B_R(0)}{=} \sum_{\substack{\alpha\in\mathbb{N}_0^p \\ \|\alpha\|_1=K}} (is)^K\binom{K}{\alpha} (\omega+\omega')^{2\alpha} e^{is(\omega+\omega')\cdot x} \nonumber
    \\ &= \sum_{\substack{\alpha\in\mathbb{N}_0^p \\ \|\alpha\|_1=K}}  \binom{K}{\alpha}(\omega+\omega')^\alpha\underbrace{(is)^{\|\alpha\|_1}(\omega+\omega')^\alpha e^{is(\omega+\omega')\cdot x}}_{\stackrel{\eqref{eq:exponential-derivative1}}{=} D_{ x}^\alpha e^{is(\omega+\omega')\cdot x}} \stackrel{\eqref{eq:derivative-transformation}}{=} D_{\omega+\omega'}^K e^{is(\omega+\omega')\cdot x}. \label{eq:ball-reduction1}
\end{align}
The next step is to use this for point-wise convergence of the integrand of the left hand side of Eq.~\eqref{eq:decomposition-formula-step1} as $R\to\infty$. Indeed, fix any $ x\in\mathcal{X}$. Then, for all $R>R_0:=\| x\|$ we have $ x\in B_R(0)$ and thus Eq.~\eqref{eq:ball-reduction1} holds true. This establishes
\begin{equation}
    \lim_{R\to\infty} D_{\omega+\omega'}^K \left\{e^{is(\omega+\omega')\cdot x} b_R( x)\right\} = D_{\omega+\omega'}^K e^{is(\omega+\omega')\cdot x} \label{eq:derivative-limit1}
\end{equation}
for all $ x\in\mathcal{X}$. Now, we would like to use the Lebesgue dominated convergence theorem (cf. Lemma~\ref{lemma:lebesgue-dominated-convergence}) again to justify exchanging the limit $R\to\infty$ and the integration over $\mathcal{X}$. To do so, we make use of the scaling property $b_R( x)=b_1\left(\frac{ x}{R}\right)$ of the bump function in Eq.~\eqref{eq:bump-function1} so that, if $R\geq 1$, then for each $\alpha,\beta\in\mathbb{N}_0^p$ with $\beta_\eta\leq\alpha_\eta$ for all $\eta\in[p]$ there exists a constant $C_{\alpha,\beta}>0$ independent of $R>0$ such that
\begin{equation}
    |D_{ x}^{\alpha-\beta} b_R( x)| = \underbrace{R^{-\|\alpha-\beta\|_1}}_{\leq 1} \underbrace{\left|(D_{\frac{ x}{R}}^{\alpha-\beta} b_1)\left(\frac{ x}{R}\right)\right|}_{\leq C_{\alpha,\beta}} \leq C_{\alpha,\beta}
\end{equation}
 relying on the fact that any smooth, compactly supported function $b_1\in\mathcal{C}_c^\infty(\mathcal{X})$ has uniformly bounded derivatives. Indeed, by definition of $b_1\in\mathcal{C}_c^\infty(\mathcal{X})$, the derivative $D^{\alpha-\beta}b_1$ is continuous and vanishes outside the compact ball $\overline{B_2(0)}$, the extreme value theorem guarantees it achieves a finite global maximum $C_{\alpha,\beta}:=\max_{ x\in\overline{B_2(0)}}|D^{\alpha-\beta}_{ x}b_1( x)|$. In particular, we find that the integrand of the left hand side of Eq.~\eqref{eq:decomposition-formula-step1} is upper bounded by
 \begin{align}
     &\left|\left( D_{\omega+\omega'}^K \left\{e^{is(\omega+\omega')\cdot x} b_R( x)\right\}\right) f( x)\right|  \nonumber
     \\ &\stackrel{\eqref{eq:decomposition-formula-step2}}{\leq} \sum_{\substack{\alpha\in\mathbb{N}_0^p \\ \|\alpha\|_1=K}}\sum_{\substack{\beta\in\mathbb{N}_0^p \\ \beta_\eta\leq\alpha_\eta\,\forall\eta\in[p]}} s^{\|\beta\|_1}\binom{K}{\alpha}\binom{\alpha}{\beta}\left|(\omega+\omega')^{\alpha+\beta}\right|\underbrace{|e^{is(\omega+\omega')\cdot x}|}_{=1} \underbrace{\left|D_{ x}^{\alpha-\beta} b_R( x)\right|}_{\leq C_{\alpha,\beta}} |f( x)| \leq C_{K,s,\omega,\omega'}|f( x)|
\end{align}
for some constant $C_{K,s,\omega,\omega'}>0$ independent of $R>0$. But we know that $f\in L^1(\mathcal{X})$ since it defines the probability density of the input $ x$, i.e., $\int_{\mathcal{X}}f( x)\,d x=1$. Therefore, we may apply Lemma~\ref{lemma:lebesgue-dominated-convergence} to justify exchanging the limit and integration:
\begin{align}
    \lim_{R\to\infty}\int_{\mathcal{X}} \left( D_{\omega+\omega'}^K \left\{e^{is(\omega+\omega')\cdot x} b_R( x)\right\}\right) f( x)\,d x &\stackrel{\eqref{eq:exchange-limit-integration}}{=} \int_{\mathcal{X}}\lim_{R\to\infty}\left( D_{\omega+\omega'}^K \left\{e^{is(\omega+\omega')\cdot x} b_R( x)\right\}\right) f( x)\,d x \nonumber
    \\ &\stackrel{\eqref{eq:derivative-limit1}}{=} \int_{\mathcal{X}}D_{\omega+\omega'}^K e^{is(\omega+\omega')\cdot x} f( x)\,d x.
\end{align}
In summary, we have justified that we can simply take the limit $R\to\infty$ in both sides of Eq.~\eqref{eq:decomposition-formula-step1} to conclude with the desired identity in Eq.~\eqref{eq:decomposition-formula-milestone1}, thus completing the proof of the first statement of the proposition. This completes the proof of the proposition.
\qed

\subsection{Other Details} \label{appendix:others}

Finally, we provide the remaining computations and proofs of technical properties used in this supplementary material. 
In summary, this mainly involves concentration inequalities, sub-Gaussian vectors, correlation, and additional quantum and linear algebra results.

\subsubsection{Concentration, sub-Gaussian Vectors and Correlation} \label{subsubsection:high-dim-probability}

In Subsection~\ref{appendix:RMT-assumptions} we ensured that the standard random matrix theory assumptions allow the formulation of a consistent theory of double descent. 
However, it is a priori not clear whether these assumptions are satisfied in the context of QELM feature maps specifically. To justify these assumptions in Section~\ref{appendix:quantum-RMT-assumptions}, we rely on the following concentration inequality involving Lipschitz continuity:

\begin{lemma} \label{lemma:lipschitz-concentration-quantum}
Let $\varepsilon>0$ be arbitrary. Let $z:=\Sigma^{-\frac{1}{2}}\phi_U(x)$ with readout $\phi_U:\mathcal{X}\to\mathbb{R}^d$ as per Eq.~\eqref{eq:quantum-readout}, and population covariance $\Sigma:=\mathbb{E}_x[\phi_U(x)\phi_U(x)^\top]$. Suppose that the function $f(x):=\|\Sigma^{-\frac{1}{2}}\phi_U(x)\|$ is Lipschitz with Lipschitz constant $L_f=\mathcal{O}_d\left(\frac{1}{\sqrt{\lambda_{\min}(\Sigma)}}\right)$, and assume further that for all $t\geq0$,
\begin{equation}
    \mathbb{P}\left(|f(x)-\mathbb{E}_x[f(x)]|\geq t\right) \leq 2\exp\left(-\frac{t^2}{C_1L_f^2}\right) \label{eq:Gaussian-concentration2}
\end{equation}
for some other absolute constant $C_1>0$. Then, if $\lambda_{\min}(\Sigma)=\omega\left(\frac{\log d}{d}\right)$, we have exponential concentration of $\|z\|$ around $\sqrt{d}$ in high dimensions, i.e.,
\begin{align*}
    \mathbb{P}\left(\left|\frac{\|z\|}{\sqrt{d}}-1\right|\geq\varepsilon\right) = \mathcal{O}_d\left(e^{-{d\lambda_{\min}(\Sigma)}}\right)
\end{align*}
as $d\to\infty$.
\end{lemma}

\noindent
\emph{Proof.} Let $\varepsilon>0$ be arbitrary. Taking $t=\varepsilon\sqrt{d}$ yields 
\begin{equation}
    \mathbb{P}\left(|\|z\|-\mathbb{E}[\|z\|]|\geq \varepsilon\sqrt{d}\right) \stackrel{\eqref{eq:Gaussian-concentration2}}{\leq} 2\exp\left(-\frac{\varepsilon^2d}{L_f^2}\right), \label{eq:concentration-bound1}
\end{equation}
where we absorbed the constant $C_1>0$ into $\varepsilon>0$ and recall that $z=\|\Sigma^{-\frac{1}{2}}\phi_U(x)\|$. The layer cake representation (cf. Lemma~\ref{lemma:layer-cake} in Section~\ref{appendix:Sobolev-spaces} applied with $q=2$) allows us to apply the previous sub-Gaussian tails in order to estimate
\begin{align}
    \operatorname{Var}(\|z\|) &= \mathbb{E}\left[(\|z\|-\mathbb{E}[\|z\|])^2\right] = 2\int_0^\infty t\mathbb{P}(|\|z\|-\mathbb{E}[\|z\|]|\geq t)\,dt \stackrel{\eqref{eq:Gaussian-concentration2}}{\leq} 4\int_0^\infty t\exp\left(-\frac{t^2}{C_1L_f^2}\right)\,dt = 2C_1L_f^2, \label{eq:variance-bound-dimensionless}
\end{align}
where we evaluated the Gaussian integral $\int xe^{-ax^2}\,dx = -\frac{1}{2a}e^{-ax^2}+C$ for $a>0$. Furthermore, since $z$ is isotropic by Eq.~\eqref{eq:isotropic-columns} we further find
\begin{equation}
    \mathbb{E}[\|z\|^2] = \mathbb{E}[z^\top z] = \mathbb{E}[\operatorname{Tr}(z^\top z)] = \mathbb{E}[\operatorname{Tr}(zz^\top)] = \operatorname{Tr}(\mathbb{E}[zz^\top]) = \operatorname{Tr}(I_d)=d \label{eq:linear-variance-dimension}
\end{equation}
by the cyclic property of the trace operator. As a result,
\begin{equation}
    0\leq \operatorname{Var}(\|z\|)=\mathbb{E}[\|z\|^2]-\mathbb{E}[\|z\|]^2 \stackrel{\eqref{eq:linear-variance-dimension}}{=} d - \mathbb{E}[\|z\|]^2 = (\sqrt{d}+\mathbb{E}[\|z\|])(\sqrt{d}-\mathbb{E}[\|z\|])
\end{equation}
which implies both $|\mathbb{E}[\|z\|]|\leq \sqrt{d}$ and
\begin{equation}
    |\mathbb{E}[\|z\|]-\sqrt{d}| = \frac{\operatorname{Var}(\|z\|)}{|\sqrt{d}+\mathbb{E}[\|z\|]|} \stackrel{\eqref{eq:variance-bound-dimensionless}}{\leq} \frac{2C_1L_f^2}{|\sqrt{d}+\mathbb{E}[\|z\|]|} = \Theta_d\left(\frac{L_f^2}{\sqrt{d}}\right) = \mathcal{O}_d\left(\frac{1}{\lambda_{\min}(\Sigma)\sqrt{d}}\right) = o\left(\frac{\sqrt{d}}{\log d}\right) = o(\sqrt{d}) \label{eq:expected-difference1}
\end{equation}
where we employed the assumptions that $L_f=\mathcal{O}_d\left(\frac{1}{\sqrt{\lambda_{\min}(\Sigma)}}\right)$ and that $\lambda_{\min}(\Sigma)=\omega\left(\frac{\log d}{d}\right)$ does not decay too fast as $d\to\infty$. We continue with the triangle inequality by noting 
\begin{align}
    |\|z\|-\sqrt{d}| \leq \left|\|z\|-\mathbb{E}[\|z\|]\right|+|\mathbb{E}[\|z\|]-\sqrt{d}| \stackrel{\eqref{eq:expected-difference1}}{=} |\|z\|-\mathbb{E}[\|z\|] | + o(\sqrt{d}) \label{eq:deterministic-difference1}
\end{align}
as $d\to\infty$. Returning to the random event that $|\|z\|-\mathbb{E}[\|z\|]\geq \varepsilon\sqrt{d}$, we thus find
\begin{align}
    \mathbb{P}\left(\left|\frac{\|z\|}{\sqrt{d}}-1\right|\geq 2\varepsilon\right) \leq \mathbb{P}\left(\left|\frac{\|z\|}{\sqrt{d}}-1\right| \geq \varepsilon+o_d(1)\right) = \mathbb{P}(|\|z\|-\sqrt{d}|-o(\sqrt{d})\geq\varepsilon\sqrt{d})  \stackrel{\eqref{eq:deterministic-difference1}}{\leq} \mathbb{P}(|\|z\|-\mathbb{E}[\|z\|]\geq \varepsilon\sqrt{d})
    \label{eq:probability-bound1}
\end{align}
with the first bound holding true for sufficiently large $d$. Therefore, we have shown that
\begin{align}
    \mathbb{P}\left(\left|\frac{\|z\|}{\sqrt{d}}-1\right|\geq\varepsilon\right) \stackrel{\eqref{eq:probability-bound1}}{\leq} \mathbb{P}\left(|\|z\|-\mathbb{E}[\|z\|]|\geq\frac{\varepsilon}{2}\sqrt{d}\right) \stackrel{\eqref{eq:concentration-bound1}}{\leq} 2\exp\left(-\frac{\varepsilon^2d}{4L_f^2}\right) = \mathcal{O}_d\left(e^{-{d\lambda_{\min}(\Sigma)}}\right)
\end{align}
for arbitrarily small but fixed $\varepsilon>0$ as $d\to\infty$, where the last bound again makes use of $L_f=\mathcal{O}_d\left(\frac{1}{\lambda_{\min}(\Sigma)}\right)$.
\qed

Next, we discuss sub-Gaussian vectors, the key concept used to formulate our standard random matrix theory assumptions in Section~\eqref{eq:standard-RMT-model}. To make a direct, conceptual bridge with the previously discussed Lebesgue and Sobolev spaces in Section~\ref{appendix:Sobolev-spaces}, it is worth highlighting that Sub-Gaussian vectors are certain instances of so-called Orlicz spaces~\cite{RaoRen1991} (also see Chapter~8 in Ref.~\cite{adams2003sobolev}), These functional spaces may be understood as a generalization of Lebesgue spaces (cf. Definition~\ref{def:Lebesgue-space}).
However, instead of introducing the most general version, we are instead focusing on $q$-sub-exponential Orlicz spaces, particularly with $q=2$ which is used to characterize sub-Gaussian random vectors~\cite{vershynin2012introduction, vershynin2020high}. Let us begin with the formal definition of Orlicz spaces, which automatically induces the notion of sub-Gaussian vectors:

\begin{definition}[Orlicz Space] \label{eq:orlicz-space}
    Let $\mathcal{X}\subseteq\mathbb{R}^p$ be open and $q\in[1,\infty)$. The $q$-th Orlicz space on $\mathcal{X}$ is defined as the quotient space
    \begin{equation*}
        L_{\psi_q}(\mathcal{X}) := \left\{x\in\mathcal{X} : \|x\|_{\psi_q}<\infty\right\} \, / \sim
    \end{equation*}
    of random vectors $x\in\mathcal{X}$ with finite $q$-th Orlicz norm,
    \begin{equation*}
        \|x\|_{\psi_q}:=\inf\left\{K>0:\sup_{v\in\mathbb{S}^{p-1}}\mathbb{E}\left[\exp\left(\frac{|\langle x,v\rangle|^q }{K^q}\right)\right]\leq 2\right\},
    \end{equation*}
    with respect to almost sure equivalence ``$\,\sim$''.
\end{definition}

It can be shown that the Orlicz space $L_{\psi_q}(\mathcal{X})$ equipped with the norm $\|x\|_{\psi_q}$ defines a Banach space. A random vector $x\in\mathcal{X}$ is said to be \emph{$q$-sub-exponential} if $x\in L_{\psi_q}(\mathcal{X})$ belongs to the $q$-th order Orlicz space. Specifically if $x\in L_{\psi_1}(\mathcal{X})$ (i.e., $q=1$), then we simply say that $x$ is \emph{sub-exponential}. Throughout this work, we are particularly interested in second order Orlicz spaces (i.e., $q=2$). In that case, a random vector $x\in L_{\psi_2}(\mathcal{X})$ is said to be \emph{sub-Gaussian}. It is worthwhile to highlight a few elementary properties of sub-Gaussian random vectors. There is an equivalent characterization of sub-Gaussian random vectors based on the norm appearing in the tail probabilities~\cite{vershynin2020high}:

\begin{lemma} \label{lemma:subgaussian-characterization}
    Let $\mathcal{X}\subseteq\mathbb{R}^p$ be open. Then, a random variable $x\in L_{\psi_2}(\mathcal{X})$ is sub-Gaussian if and only if there exists a $K>0$ such that
    \begin{equation}
        \mathbb{P}\left(|\langle x, v\rangle| \geq t \right) \leq 2\exp\left(-\frac{t^2}{K^2}\right) \label{eq:sub-gaussian-def}
    \end{equation}
    for all $v\in\mathbb{S}^{p-1}$ and $t\geq0$.
\end{lemma}

One may verify that the smallest constant $K>0$ satisfying Eq.~\eqref{eq:sub-gaussian-def} satisfies the lower bound $K\geq \frac{\|x\|_{\psi_2}}{\sqrt{3}}$. Furthermore, from Lemma~2.7.8 in Ref.~\cite{vershynin2020high} we also know that any sub-Gaussian random variable $X\in L_{\psi_2}(\mathbb{R})$ satisfies the centering property,
\begin{equation}
    \|X-\mathbb{E}[X]\|_{\psi_2} \leq C\|X\|_{\psi_2} \label{eq:subgaussian-centering}
\end{equation}
for some absolute constant $C>0$. This property is used to prove Proposition~\ref{prop:refined-variance-bound} further below, which characterizes the feature correlations (the assumption used to ensure the double descent curve as per Proposition~\ref{prop:over-fitting} in Section~\ref{sec:non-isotropic-analysis}) in terms of the sub-Gaussian norm.

In the context of QELM feature maps, we know that the quantum readout elements in Eq.~\eqref{eq:quantum-readout} is uniformly bounded by $\max_{k\in[d]}\|O_k\|$. Asa a result, the following lemma automatically implies that the quantum readout is sub-Gaussian:

\begin{lemma} \label{lemma:bounded-RV-subGaussian}
    Let $x\in\mathbb{R}^p$ be a random vector with $x\in(-M(p),M(p))^p$ almost surely for some $M(p)>0$ that may depend on the dimension $p\in\mathbb{N}$. Then, $x\in L_{\psi_2}(\mathbb{R}^p)$ is sub-Gaussian with
    \begin{equation*}
        \|x\|_{\psi_2} \leq \frac{M(p)\sqrt{p}}{\sqrt{\log 2}} = \mathcal{O}(M(p)\sqrt{p}) 
    \end{equation*}
    as $p\to\infty$. In particular, $\|x\|_{\psi_2}=\mathcal{O}_p(\sqrt{p})$ if $M=\Theta_p(1)$ is dimension independent.
\end{lemma}

\noindent
\emph{Proof.} Let $v=(v_1,\dots,v_p)\in\mathbb{S}^{p-1}$ be arbitrary and consider the random variable $X:=\langle x,v\rangle$ where $x=(x_1,\dots,x_p)\in(-M(p),M(p))^p$ is our bounded random vector (i.e., $|x_k|< M(p)$ for all $k\in[p]$). By the Cauchy-Schwarz inequality,
\begin{equation}
    |X|=|\langle x,v\rangle| \leq \|x\|\|v\| =  \|x\| = \sqrt{\sum_{\eta=1}^px_\eta^2} \leq M(p)\sqrt{p} \label{eq:bounded-RV1}
\end{equation}
since $x_\eta^2\leq p$ almost surely for all $\eta\in[p]$ by assumption and $\|v\|=1$ due to $v\in\mathbb{S}^{p-1}$. Eq.~\eqref{eq:bounded-RV1} immediately implies $X^2\leq M(p)^2p^2$ and thus
\begin{equation}
    \mathbb{E}\left[\exp\left(\frac{|\langle x,v\rangle|^2 }{K^2}\right)\right] = \mathbb{E}\left[\exp\left(\frac{X^2 }{K^2}\right)\right] \leq \exp\left(\frac{M(p)^2p^2 }{K^2}\right) \leq 2 \iff K\geq \frac{M(p)\sqrt{p}}{\sqrt{\log 2}} \label{eq:subgaussian-constant1}
\end{equation}
uniformly in $v\in\mathbb{S}^{p-1}$. Taking the infimum over all valid $K>0$ thus yields, by definition,
\begin{equation}
    \|x\|_{\psi_2} = \inf\left\{K>0 : \sup_{v\in\mathbb{S}^{p-1}}\mathbb{E}\left[\exp\left(\frac{|\langle x,v\rangle|^2 }{K^2}\right)\right]\leq 2\right\} \stackrel{\eqref{eq:subgaussian-constant1}}{\leq} \frac{M(p)\sqrt{p}}{\sqrt{\log 2}} < \infty.
\end{equation}
\qed

Now that we have discussed basic properties of sub-Gaussian random vectors, we return to our previous random matrix theory techniques. Specifically, in the analysis of the non-isotropic case in Section~\ref{sec:non-isotropic-analysis} of Supplementary Section~\ref{appendix:double_descent_linear2} we established that, according to Eq.~\eqref{eq:unit-interpolation-threshold}, the expected risk $\mathcal{R}(\gamma)$ defined in Eq.~\eqref{eq:expected-risk} analytically coincides with the canonical double descent curve
\begin{equation}
    \mathcal{R}(\gamma) = \frac{\sigma^2}{1-\gamma}
\end{equation}
for $\gamma<1$ as long as the variance bound
\begin{equation}
    \sup_{\substack{A\in\mathbb{R}^{d\times d}\\\|A\|=\mathcal{O}(1)}}\operatorname{Var}(z^\top A z) = o(d^2) \label{eq:highdimensional-variance-bound}
\end{equation}
is satisfied as $d\to\infty$, where
\begin{equation}
    \Sigma := \mathbb{E}_{x}\left[\phi(x)\phi(x)^\top\right]\in\mathbb{R}^{d\times d}\quad \text{and}\quad z:= \Sigma^{-\frac{1}{2}}\phi(x)\in\mathbb{R}^d,
\end{equation}
assuming that the population covariance $\Sigma$ is non-singular. Under these assumptions, the violation of the high-dimensional variance bound in Eq.~\eqref{eq:highdimensional-variance-bound} is the only possible cause for a shifted, non-unit interpolation threshold appearing at $\gamma\neq 1$. The following lemma shows that the variance of the quadratic form $z^\top A z$ may be interpreted as a measure of second moment correlations of the features $\phi(x)$. More formally, it is a certain deterministic quadratic form that contracts the $d^2\times d^2$ feature covariance tensor $\operatorname{Cov}(z\otimes z)$ determined by matrix elements $\operatorname{Cov}(\phi_k\phi_{k'},\phi_\eta\phi_{\eta'})$.

\begin{lemma}[Second Moment Covariance Tensor] \label{eq:second-order-correlation-measure}
    Assume that the population covariance $\Sigma$ is invertible and let $z=\Sigma^{-\frac{1}{2}}\phi$ with $\phi\equiv \phi(x)$ where $x\sim\mathbb{P}$ sampled from the input distribution. Then,
    \begin{equation*}
        \operatorname{Var}(z^\top A z) = \operatorname{Var}\left(\phi^\top A^{(\Sigma)}\phi\right) = \sum_{k,k',\eta,\eta'=1}^d A^{(\Sigma)}_{k,k'}A^{(\Sigma)}_{\eta,\eta'}\,\operatorname{Cov}(\phi_k\phi_{k'},\phi_\eta\phi_{\eta'}) = \operatorname{vec}(A)^\top \operatorname{Cov}(z \otimes z) \operatorname{vec}(A)
    \end{equation*}
    for all deterministic $A\in\mathbb{R}^{d\times d}$, where $A^{(\Sigma)}:=\Sigma^{-\frac{1}{2}} A\,\Sigma^{-\frac{1}{2}}$. Taking the supremum over all deterministic $A\in\mathbb{R}^{d\times d}$ with $\|A\|=\mathcal{O}(1)$ thus yields the largest possible linear combination of all possible second moment covariances of the feature elements $\{\phi_k:k\in[d]\}$.
\end{lemma}

\noindent
\emph{Proof.} Let $A\in\mathbb{R}^{d\times d}$ be arbitrary. First, we expand the quadratic form and find
\begin{align}
    z^\top A\, z = \phi^\top\underbrace{\Sigma^{-\frac{1}{2}} A\, \Sigma^{-\frac{1}{2}}}_{=A^{(\Sigma)}}\phi = \phi^\top A^{(\Sigma)}\phi = \sum_{k,k'=1}^d A^{(\Sigma)}_{k,k'}\,\phi_k\phi_{k'}. \label{eq:expanded-quadratic-form1}
\end{align}
Next, we compute the variance,
\begin{align}
    \operatorname{Var}(z^\top A z) &= \mathbb{E}\left[(z^\top A z)^2\right] -\mathbb{E}[z^\top A z]^2 \stackrel{\eqref{eq:expanded-quadratic-form1}}{=} \mathbb{E}\left[\left(\sum_{k,k'=1}^dA^{(\Sigma)}_{k,k'}\,\phi_k\phi_{k'}\right)^2\right] - \mathbb{E}\left[\sum_{k,k'=1}^dA^{(\Sigma)}_{k,k'}\,\phi_k\phi_{k'}\right]^2 \nonumber
    \\ &= \mathbb{E}\left[\sum_{k,k',\eta,\eta'=1}^d A_{k,k'}^{(\Sigma)}A_{\eta,\eta'}^{(\Sigma)}\, \phi_k\phi_{k'}\phi_\eta\phi_{\eta'}\right] - \left(\sum_{k,k'=1}^d A_{k,k'}^{(\Sigma)}\,\mathbb{E}[\phi_k\phi_{k'}]\right)^2 \nonumber
    \\ &= \mathbb{E}\left[\sum_{k,k',\eta,\eta'=1}^d A_{k,k'}^{(\Sigma)}A_{\eta,\eta'}^{(\Sigma)}\, \phi_k\phi_{k'}\phi_\eta\phi_{\eta'}\right] - \sum_{k,k',\eta,\eta'=1}^d A_{k,k'}^{(\Sigma)}A_{\eta,\eta'}^{(\Sigma)}\,\mathbb{E}[\phi_k\phi_{k'}]\mathbb{E}[\phi_\eta\phi_{\eta'}] \nonumber
    \\ &= \sum_{k,k',\eta,\eta'=1}^d A_{k,k'}^{(\Sigma)}A_{\eta,\eta'}^{(\Sigma)}\,\underbrace{\left(\mathbb{E}[\phi_k\phi_{k'}\phi_\eta\phi_{\eta'}]-\mathbb{E}[\phi_k\phi_{k'}]\mathbb{E}[\phi_\eta\phi_{\eta'}]\right)}_{=\operatorname{Cov}(\phi_k\phi_{k'},\phi_\eta\phi_{\eta'})} = \sum_{k,k',\eta,\eta'=1}^dA_{k,k'}^{(\Sigma)}A_{\eta,\eta'}^{(\Sigma)}\,\operatorname{Cov}(\phi_k\phi_{k'},\phi_\eta\phi_{\eta'}).
\end{align}
To show the last equality, we recall the standard definition of the vectorization operator $\operatorname{vec}(A) \in \mathbb{R}^{d^2}$ and the outer product vectorization $z \otimes z \in \mathbb{R}^{d^2}$ by indexing their components using the coordinate pairs $(k, k')$ and $(\eta, \eta')$ for $k, k', \eta, \eta' \in [d]$. Specifically, we establish the entrywise convention,
\begin{equation}
    \operatorname{vec}(A)_{(k,k')} = A_{k,k'} \quad \text{and} \quad (z \otimes z)_{(k,k')} = z_k z_{k'}.
\end{equation}
First, we express the scalar quadratic form as a standard vector inner product in $\mathbb{R}^{d^2}$,
\begin{align}
    z^\top A z = \sum_{k,k'=1}^d A_{k,k'} z_k z_{k'} = \sum_{k,k'=1}^d (\operatorname{vec}(A))_{(k,k')} (z \otimes z)_{(k,k')} = \operatorname{vec}(A)^\top (z \otimes z). \label{eq:expanded-quadratic-form-vec}
\end{align}
We proceed, we compute the variance,
\begin{align}
    \operatorname{Var}(z^\top A z) &\stackrel{\eqref{eq:expanded-quadratic-form-vec}}{=} \operatorname{Var}\left(\operatorname{vec}(A)^\top (z \otimes z)\right) = \mathbb{E}\left[\left(\operatorname{vec}(A)^\top (z \otimes z)\right)^2\right] - \mathbb{E}\left[\operatorname{vec}(A)^\top (z \otimes z)\right]^2 \nonumber
    \\ &= \mathbb{E}\left[\left(\sum_{k,k'=1}^d \operatorname{vec}(A)_{(k,k')} (z \otimes z)_{(k,k')}\right)^2 \right] - \left(\sum_{k,k'=1}^d \operatorname{vec}(A)_{(k,k')}\,\mathbb{E}\left[(z \otimes z)_{(k,k')}\right]\right)^2 \nonumber 
    \\&= \mathbb{E}\left[\sum_{k,k',\eta,\eta'=1}^d \operatorname{vec}(A)_{(k,k')} \operatorname{vec}(A)_{(\eta,\eta')} (z \otimes z)_{(k,k')} (z \otimes z)_{(\eta,\eta')}\right]  \nonumber
    \\ &\qquad- \sum_{k,k',\eta,\eta'=1}^d \operatorname{vec}(A)_{(k,k')} \operatorname{vec}(A)_{(\eta,\eta')}\,\mathbb{E}\left[(z \otimes z)_{(k,k')}\right]\mathbb{E}\left[(z \otimes z)_{(\eta,\eta')}\right] \nonumber
    \\&= \sum_{k,k',\eta,\eta'=1}^d \operatorname{vec}(A)_{(k,k')} \operatorname{vec}(A)_{(\eta,\eta')}\underbrace{\left(\mathbb{E}\left[(z \otimes z)_{(k,k')} (z \otimes z)_{(\eta,\eta')}\right]-\mathbb{E}\left[(z \otimes z)_{(k,k')}\right]\mathbb{E}\left[(z \otimes z)_{(\eta,\eta')}\right]\right)}_{=\operatorname{Cov}\left((z \otimes z)_{(k,k')}, (z \otimes z)_{(\eta,\eta')}\right)} \nonumber
    \\ &= \sum_{k,k',\eta,\eta'=1}^d \operatorname{vec}(A)_{(k,k')} \operatorname{vec}(A)_{(\eta,\eta')}\,\operatorname{Cov}\left((z \otimes z)_{(k,k')}, (z \otimes z)_{(\eta,\eta')}\right). \label{eq:cov-sum-vec}
\end{align}
By definition, the pairwise covariances between the elements of the random vector $z \otimes z$ constitute the entries of its $d^2 \times d^2$ covariance matrix $\operatorname{Cov}(z \otimes z)$. The entry located at row $(k,k')$ and column $(\eta,\eta')$ is precisely given by
\begin{align}
   \operatorname{Cov}(z \otimes z)_{(k,k'), (\eta,\eta')} = \operatorname{Cov}\left((z \otimes z)_{(k,k')}, (z \otimes z)_{(\eta,\eta')}\right) = \operatorname{Cov}(z_k z_{k'}, z_\eta z_{\eta'}).
\end{align}
Substituting this matrix representation back into the index summation in Eq.~\eqref{eq:cov-sum-vec} allows us to collect the terms into a clean quadratic form in $\mathbb{R}^{d^2}$, i.e.,
\begin{align}
    \operatorname{Var}(z^\top A z) &= \sum_{k,k',\eta,\eta'=1}^d \operatorname{vec}(A)_{(k,k')} \operatorname{vec}(A)_{(\eta,\eta')}\,\operatorname{Cov}(z \otimes z)_{(k,k'), (\eta,\eta')} = \operatorname{vec}(A)^\top \operatorname{Cov}(z \otimes z) \operatorname{vec}(A).
\end{align}

\qed

In our theoretical framework, we always assume that $z=\Sigma^{-\frac{1}{2}}\phi(x)\in[-M,M]^d$ is bounded. Hence, $\|z\|^2 = \sum_{k=1}^d z_k^2 \leq dM^2=\mathcal{O}(d^2)$. By the Cauchy-Schwarz inequality, we generally have
\begin{equation}
    \operatorname{Var}(z^\top A z) \leq \mathbb{E}\left[(z^\top A z)^2\right] \leq \mathbb{E}\left[|\langle z,Az\rangle|^2\right] \leq \mathbb{E}\left[||z||^2||Az\|^2\right] \leq \underbrace{\|A\|^2}_{=\mathcal{O}(1)} \underbrace{\mathbb{E}[\,||z||^4}_{\leq d^2}] = \mathcal{O}(d^2)
\end{equation}
as $d\to\infty$. In particular, it is sufficient to find one constant-scaling matrix $A_0\in\mathbb{R}^{d\times d}$ with $\|A_0\|=\mathcal{O}(1)$ such that $\operatorname{Var}(z^\top A_0 z) = \Theta(d^2)$ as $d\to\infty$ in order to know that double descent with an interpolation threshold at $\gamma=1$ could potentially fail due to Eq.~\eqref{eq:unit-interpolation-threshold}. On the other hand, to guarantee double descent with an interpolation threshold at $\gamma=1$, one has to verify the bound $\operatorname{Var}(z^\top A z)=o(d^2)$ for all matrices $A\in\mathbb{R}^{d\times d}$ with $\|A\|=\mathcal{O}(1)$ as $d\to\infty$, which is of course totally infeasible to verify in practice. By leveraging loose bounds, we may derive a practically implementable condition in terms of the largest eigenvalue of the covariance tensor $\operatorname{Cov}(z\otimes z)$.

\begin{proposition}[Spectral Measure of Feature Correlations] \label{prop:spectral-variance-bound}
Let $z = \Sigma^{-\frac{1}{2}}\phi(x) \in \mathbb{R}^d$ be the whitened feature vector, and let $\operatorname{Cov}(z \otimes z) \in \mathbb{R}^{d^2 \times d^2}$ denote the covariance matrix of its vectorized outer product. Then, for all deterministic matrices $A \in \mathbb{R}^{d \times d}$ satisfying $\|A\| = \mathcal{O}_d(1)$, the variance of the quadratic form satisfies
\begin{equation*}
    \operatorname{Var}(z^\top A z) = \mathcal{O}_d\left( d \,\lambda_{\max}(\operatorname{Cov}(z \otimes z)) \right)
\end{equation*}
as $d\to\infty$. Consequently, to establish that $\operatorname{Var}(z^\top A z) = o(d^2)$ holds uniformly for all valid $A$, it suffices to check the dimension-independent spectral growth condition, as $d\to\infty$,
\begin{equation*}
\lambda_{\max}(\operatorname{Cov}(z \otimes z)) = o(d).
\end{equation*}
\end{proposition}

\noindent
\emph{Proof.} Let $A \in \mathbb{R}^{d \times d}$ be an arbitrary deterministic matrix such that $\|A\| = \mathcal{O}_d(1)$. By Lemma~\eqref{eq:second-order-correlation-measure} we know that
\begin{align}
    \operatorname{Var}(z^\top A z) = \operatorname{vec}(A)^\top \operatorname{Cov}(z \otimes z) \operatorname{vec}(A). \label{eq:prop-quad-form}
    \end{align}
Since $\operatorname{Cov}(z \otimes z)$ is a real symmetric positive semi-definite matrix, we may upper Eq.~\eqref{eq:prop-quad-form} using its maximum eigenvalue,
\begin{align}
    \operatorname{vec}(A)^\top \operatorname{Cov}(z \otimes z) \operatorname{vec}(A) \leq \|\operatorname{vec}(A)\|^2 \, \lambda_{\max}(\operatorname{Cov}(z \otimes z)), \label{eq:rayleigh-bound}
\end{align}
By the definition of the vectorization, the squared Euclidean norm of $\operatorname{vec}(A)$ is identical to the squared Frobenius norm of the original matrix $A$. It follows that
\begin{align}
    \|\operatorname{vec}(A)\|^2 = \sum_{k,k'=1}^d A_{k,k'}^2 = \|A\|_{\operatorname{HS}}^2 \leq d\|A\|^2 = \mathcal{O}(d) \label{eq:frobenius-equivalence}
\end{align}
using the assumption that $\|A\|=\mathcal{O}_d(1)$. Substituting Eq.~\eqref{eq:frobenius-equivalence} into the bound established in Eq.~\eqref{eq:rayleigh-bound} yields
\begin{align}
    \operatorname{Var}(z^\top A z) = \mathcal{O}_d\left(d\,\lambda_{\max}(\operatorname{Cov}(z \otimes z)\right). 
\end{align}
This completes the proof.
\qed

On the other hand, it is worth remarking that the measure, $\operatorname{Var}(z^\top A z)$, of second moment feature correlations is also characterized by the sub-Gaussian norm $\|z\|_{\psi_2}$ of the whitened features $z=\Sigma^{-\frac{1}{2}}\phi(x)$. To prove this insight, we employ a concentration inequality: We consider the following generalization of the Hanson-Wright inequality~\cite{rudelson2013hanson, vershynin2020high} that allows dependent coordinates in the underlying random vector:

\begin{lemma}[Zajkowski~\cite{zajkowski2020bounds}, Corollary~2.8] \label{lemma:general-hanson-wright}
    Let $\xi\in L_{\psi_2}(\mathbb{R}^d)$ be a sub-Gaussian random vector with $\mathbb{E}[\xi]=0$. Then, for all deterministic $A\in\mathbb{R}^{d\times d}$ and $t\geq0$, we have
    \begin{equation*}
        \mathbb{P}\left(|\xi^\top A \xi - \mathbb{E}[\xi^\top A \xi]|\geq t\right) \leq 2\exp\left(-\min\left\{\frac{t^2}{c^2\|A\|_{\operatorname{HS}}^2K^4},\frac{t}{c\|A\|_{\operatorname{HS}}K^2}\right\}\right)
    \end{equation*}
    for some absolute constant $c>0$ and $K:=\|\xi\|_{\psi_2}$.
\end{lemma}

If $x_1,\dots,x_n\sim\mathbb{P}$ are i.i.d. samples drawn from the input distribution $\mathbb{P}$, then $\phi(x)\stackrel{d}{=}\phi(x_j)=\Phi_j$ are equal in distribution for all samples $j\in[n]$. In particular, random vectors of the form $z_j:=\Sigma^{-\frac{1}{2}}\Phi_j = \Sigma^{-\frac{1}{2}}\phi(x_j)$ are mutually independent and isotropic by Eq.~\eqref{eq:isotropic-columns}. Therefore, the following result only requires sub-Gaussian tails as an additional condition:

\begin{proposition} \label{prop:refined-variance-bound}
    Let $Z=[z_1|\cdots|z_n]\in\mathbb{R}^{d\times n}$ be a random matrix with independent, isotropic, sub-Gaussian columns $z_j\in L_{\psi_2}(\mathbb{R}^d)$. Then,
    \begin{equation*}
        \max_{j\in[n]}\operatorname{Var}(z_j^\top A z_j) = \mathcal{O}(K^4d)
    \end{equation*}
    as $d\to\infty$ for all deterministic $A\in\mathbb{R}^{d\times d}$ with $\|A\| = \mathcal{O}(1)$, where $K:=\max_{j\in[n]}\|z_j\|_{\psi_2}$ denotes the maximal sub-Gaussian norm of the columns of $Z$.
\end{proposition}

\noindent
\emph{Proof.} Let $A\in\mathbb{R}^{d\times d}$ with $\|A\|=\mathcal{O}(1)$ and $j\in[n]$ be arbitrary. Since $z_j^\top A z_j = z_j^\top\left(\frac{A+A^\top}{2}\right)z_j$ we may assume without loss of generality that $A\in\mathbb{R}^{d\times d}$ is symmetric. Let $\mu = \mathbb{E}[z_j] \in \mathbb{R}^d$ be the deterministic mean vector of the $j$-th column $z_j$. Consider the random vector $\xi_j := z_j - \mu$. Then,
$\mathbb{E}[\xi_j]=\mathbb{E}[z_j]-\mu=0$ by definition and, for each fixed $u\in\mathbb{S}^{d-1}$, we have
\begin{equation}
    \|\langle \xi_j, u \rangle\|_{\psi_2} = \|\langle z_j-\mu,u\rangle\|_{\psi_2} = \|\langle z_j, u \rangle - \mathbb{E}[\langle z_j, u \rangle]\|_{\psi_2} \stackrel{\eqref{eq:subgaussian-centering}}{\leq} C_0 \|\langle z_j, u \rangle\|_{\psi_2} \leq C_0 K \label{eq:sub-gaussianity1}
\end{equation}
for some absolute constant $C_0>0$, using that $z_j\in L_{\psi_2}(\mathbb{R}^d)$ is sub-Gaussian by assumption (i.e., the random variable $X\equiv \langle z_j,u\rangle\in L_{\psi_2}(\mathbb{R})$ is sub-Gaussian). In particular, $\max_{j\in[n]}\|\xi_j\|_{\psi_2}\leq C_0K$. Now, we substitute $z_j = \mu + \xi_j$ into our quadratic form,
\begin{equation}
    z_j^\top A z_j = (\mu + \xi_j)^\top A (\mu + \xi_j) = \mu^\top A \mu + \xi_j^\top A \xi_j + 2\mu^\top A \xi_j
\end{equation}
using that $A$ is symmetric. As a result, since $\mu^\top A\mu$ is a deterministic constant, it does not contribute to the variance so that
\begin{align}
    \operatorname{Var}(z_j^\top A z_j) = \operatorname{Var}(\mu^\top A \mu + \xi_j^\top A \xi_j + 2\mu^\top A \xi_j) = \operatorname{Var}(\xi_j^\top A \xi_j + 2\mu^\top A \xi_j) \label{eq:variance-quadratic-form-bound1}
\end{align}
Utilizing Lemma~\ref{lemma:bound-variance-sum-of-RV} yields the bound
\begin{equation}
    \operatorname{Var}(\xi_j^\top A \xi_j + 2\mu^\top A \xi_j) \leq 2\operatorname{Var}(\xi_j^\top A \xi_j) + \operatorname{Var}(2\mu^\top A \xi_j) = 2\operatorname{Var}(\xi_j^\top A \xi_j) + 8\operatorname{Var}(\mu^\top A \xi_j). \label{eq:variance-quadratic-form-bound2}
\end{equation}

Let us first analyse the linear term $\operatorname{Var}(\mu^\top A \xi_j)$. Since $\xi_j$ is centered, we find
\begin{align}
    \operatorname{Var}(\mu^\top A \xi_j) &= \mathbb{E}[(\mu^\top A \xi_j)^2] - \mathbb{E}[\mu^\top A \xi_j]^2 = \mathbb{E}[(\mu^\top A \xi_j)^2] = \mathbb{E}[(\mu^\top A \xi_j)(\mu^\top A \xi_j)^\top] \nonumber
    \\ &= \mathbb{E}[\mu^\top A\xi_j\xi_j^\top A^\top \mu] = \mu^\top A \,\mathbb{E}[\xi_j\xi_j^\top] A^\top \mu \label{eq:linear-term-bound1}
\end{align}
because $\mathbb{E}[\mu^\top A \xi_j] = \mu^\top A\mathbb{E}[\xi_j]=0$ due to $\mathbb{E}[\xi_j]=0$ by construction. Note that we also used $\mu^\top A\xi_j = (\mu^\top A\xi_j)^\top$ which trivially holds true since it is simply a scalar. Since $z_j$ is isotropic by assumption, we further have
\begin{align}
    I_d &= \mathbb{E}[z_j z_j^\top] = \mathbb{E}[(\mu + \xi_j)(\mu + \xi_j)^\top] = \mathbb{E}\left[\mu\mu^\top + \xi_j\xi_j^\top + \mu\xi_j^\top + \xi_j\mu^\top\right] \nonumber
    \\ &= \mu\mu^\top + \mathbb{E}[\xi_j\xi^\top] + \mu\underbrace{\mathbb{E}[\xi_j]^\top}_{=0} + \underbrace{\mathbb{E}[\xi_j]}_{=0}\mu^\top = \mu\mu^\top + \mathbb{E}[\xi_j \xi_j^\top] \label{eq:isotropic-columns-1}
\end{align}
which implies $\mathbb{E}[\xi_j \xi_j^\top] = I_d - \mu\mu^\top$. Since $\mathbb{E}[\xi_j\xi_j^\top] \succeq 0$ must be positive semi-definite by definition, we have $I_d\succeq \mu\mu^\top$. In particular, $\|\mu\|^2\leq 1$. In other words, the isotropy of $z_j$ ensures that  that the Euclidean norm of the deterministic mean vector $\mu=\mathbb{E}[z_j]$ is bounded by 1. Consequently,
\begin{align}
    \operatorname{Var}(\mu^\top A \xi_j) &\stackrel{\eqref{eq:linear-term-bound1}}{=} \mu^\top A \,\underbrace{\mathbb{E}[\xi_j\xi_j^\top]}_{\stackrel{\eqref{eq:isotropic-columns-1}}{=}I_d-\mu\mu^\top} A^\top \mu = \mu^\top A A^\top \mu - \mu^\top A \,\mu\mu^\top A^\top \mu \nonumber
    \\ &= \|A^\top\mu\|^2 - \underbrace{(\mu^\top A \mu)^2}_{\geq 0} \leq \|A^\top \mu\|^2 \leq \|A\|^2\underbrace{\|\mu\|^2}_{\leq 1} = \mathcal{O}(1) \label{eq:linear-variance-constant-bonud1}
\end{align}
since $\|A\|=\mathcal{O}(1)$ by assumption. This shows that the linear term $\operatorname{Var}(\mu^\top A \xi_j)$ scales with $\mathcal{O}(1)$.

Next, we deal with the quadratic term $\operatorname{Var}(\xi_j^\top A \xi_j)$. This part is slightly more technical as we shall employ tail integration using a generalization of the Hanson-Wright inequality (cf. Lemma~\ref{lemma:general-hanson-wright}). For that purpure, we first introduce the centered random variable $W:=\xi_j^\top A\xi_j - \mathbb{E}[\xi_j^\top A\xi_j]$ so that $\operatorname{Var}(\xi_j^\top A \xi_j) = \mathbb{E}[W^2]$. By the layer cake representation (cf. Lemma~\ref{lemma:layer-cake}) applied with $q=2$ and $f(x) \equiv W$ we thus obtain
\begin{equation}
    \operatorname{Var}(\xi_j^\top A \xi_j) = \mathbb{E}[W^2] = 2\int_0^\infty t\, \mathbb{P}\left(|W|>t\right)\,dt. \label{eq:variance-integral1}
\end{equation}
In order to bound the probability in the integrand, we may directly apply the aforementioned generalized Hanson-Wright inequality (cf. Lemma~\ref{lemma:general-hanson-wright}) to the quadratic form $\xi_j^\top A\xi_j$:
\begin{equation}
    \mathbb{P}\left(|W|>t\right) = \mathbb{P}\left(|\xi_j^\top A\xi_j - \mathbb{E}[\xi_j^\top A\xi_j]|\geq t\right) \leq 2\exp\left(-\min\left\{\frac{t^2}{c^2\|A\|_{\operatorname{HS}}^2K^4},\frac{t}{c\|A\|_{\operatorname{HS}}K^2}\right\}\right) \label{eq:hanson-wright1}
\end{equation}
for some absolute constant $c>0$. Note that we absorbed the absolute constant $C_0>0$ from Eq.~\eqref{eq:sub-gaussianity1} (which states $\max_{j\in[n]}\|\xi_j\|_{\psi_2}\leq C_0K$) into $c>0$. Furthermore, observe that we are indeed in a position to apply Lemma~\ref{lemma:general-hanson-wright} because $z_j\in L_{\psi_2}(\mathbb{R}^d)$ is sub-Gaussian by assumption and we have already seen that $\mathbb{E}[\xi_j]=0$. Hence,
\begin{equation}
    \int_0^\infty t\, \mathbb{P}\left(|W|>t\right)\,dt \leq 2 \int_0^\infty t\,\exp\left(-\min\left\{\frac{t^2}{c^2\|A\|_{\operatorname{HS}}^2K^4},\frac{t}{c\|A\|_{\operatorname{HS}}K^2}\right\}\right) \, dt. \label{eq:HW-integral1}
\end{equation}
To evaluate this integral cleanly, we find where the two terms in the minimum cross over. We have
\begin{equation}
    \frac{t^2}{c^2K^4 \|A\|_{\operatorname{HS}}^2} \leq \frac{t}{cK^2 \|A\|_{\operatorname{HS}}} \iff t\leq t_0 := cK^2 \|A\|_{\operatorname{HS}} \label{eq:integral-split-threshold}
\end{equation}
for all $t\in\mathbb{R}$. In other words, the cross over happens exactly at the threshold $t_0=cK^2\|A\|_{\operatorname{HS}}$. We use this threshold  to split the integral in Eq.~\eqref{eq:HW-integral1} into two terms,
\begin{equation}
    \int_0^\infty t\,\exp\left(-\min\left\{\frac{t^2}{c^2\|A\|_{\operatorname{HS}}^2K^4},\frac{t}{c\|A\|_{\operatorname{HS}}K^2}\right\}\right) \, dt \stackrel{\eqref{eq:integral-split-threshold}}{=} \underbrace{\int_0^{t_0} t\,\exp\left(-\frac{t^2}{t_0^2}\right)\,dt}_{=:I_1} + \underbrace{\int_{t_0}^\infty t\,\exp\left(-\frac{t}{t_0}\right)\,dt}_{=:I_2}.
\end{equation}
We evaluate the first integral $I_1$ by performing the change of variables $y=t^2$ (i.e., $dy=2t\,dt$) so that
\begin{align}
    I_1 &= \frac{1}{2}\int_0^{t_0} \exp\left(-\frac{t^2}{t_0^2}\right) 2t\,dt = \frac{1}{2}\int_0^{t_0^2}\exp\left(-\frac{y}{t_0^2}\right)\,dy = -\frac{t_0^2}{2}\left[\exp\left(-\frac{y}{t_0^2}\right)\right]_0^{t_0^2} = -\frac{t_0^2}{2}\left(\exp\left(-\frac{t_0^2}{t_0^2}\right)-1\right) \nonumber
    \\ &= \frac{t_0^2}{2}\left(1-\frac{1}{e}\right) =\frac{c^2K^4\|A\|^2_{\operatorname{HS}}}{2}\left(1-\frac{1}{e}\right) = \mathcal{O}_d(K^4\|A\|_{\operatorname{HS}}^2)
\end{align}
as $d\to\infty$ since the constant $c>0$ is absolute. We have employed the elementary integration formula $\int e^{-at}\,dt=-\frac{1}{a}e^{-at}+C$ for $a>0$. Next, we evaluate the second integral $I_2$,
\begin{align}
    I_2 &= \int_{t_0}^\infty t\,\exp\left(-\frac{t}{t_0}\right)\,dt = -\left[\left(t_0t+t_0^2\right)\exp\left(-\frac{t}{t_0}\right)\right]_{t_0}^\infty = \left(t_0^2+t_0^2\right)\exp\left(-\frac{t_0}{t_0}\right)-\underbrace{\lim_{t\to\infty}\left(t_0t+t_0^2\right)\exp\left(-\frac{t}{t_0}\right)}_{=0} \nonumber
    \\ &= \frac{2t_0^2}{e} = \frac{2c^2K^4\|A\|_{\operatorname{HS}}^2}{e} = \mathcal{O}_d(K^4\|A\|_{\operatorname{HS}}^2),
\end{align}
as $d\to\infty$, by utilizing integration by parts
\begin{equation}
    \int t e^{-at}\,dt = -\frac{t}{a}e^{-at}-\underbrace{\int e^{-at}\,dt}_{=\frac{1}{a}e^{-at}+\,C_2}  +\, C_1 = -\left(\frac{t}{a}+\frac{1}{a^2}\right)e^{-at} + C
\end{equation}
for $a>0$, again using the fact that $c>0$ is absolute.

Therefore, the quadratic term is bounded by
\begin{align}
    \operatorname{Var}(\xi_j^\top A \xi_j) &\stackrel{\eqref{eq:variance-integral1}}{=} 2\int_0^\infty t\, \mathbb{P}\left(|W|>t\right)\,dt \stackrel{\eqref{eq:HW-integral1}}{\leq} 4 \underbrace{\int_0^\infty t\,\exp\left(-\min\left\{\frac{t^2}{c^2\|A\|_{\operatorname{HS}}^2K^4},\frac{t}{c\|A\|_{\operatorname{HS}}K^2}\right\}\right) \, dt}_{= I_1 + I_2 = \mathcal{O}_d(K^4\|A\|_{\operatorname{HS}}^2)} = \mathcal{O}_d(K^4\|A\|_{\operatorname{HS}}^2) \label{eq:quadratic-variance-bound1}
\end{align}
as $d\to\infty$. By definition of the Hilbert-Schmidt norm, we have
\begin{equation}
    \|A\|_{\operatorname{HS}}^2 = \operatorname{Tr}(AA^\top) = \sum_{k=1}^d \lambda_k(AA^\top) \leq d\|A\|^2 = \mathcal{O}(d) \label{eq:HS-bound1}
\end{equation}
by using the fact that the trace of an operator equals the sum of its eigenvalues, and that $\|A\|=\mathcal{O}(1)$ as $d\to\infty$ by assumption. In particular,
\begin{equation}
    \operatorname{Var}(z_j^\top A z_j) \stackrel{\eqref{eq:variance-quadratic-form-bound1}}{=} \operatorname{Var}(\xi_j^\top A \xi_j + 2\mu^\top A \xi_j) \stackrel{\eqref{eq:variance-quadratic-form-bound2}}{\leq} 2\underbrace{\operatorname{Var}(\xi_j^\top A \xi_j)}_{\stackrel{\eqref{eq:quadratic-variance-bound1}}{\leq} \mathcal{O}(K^4\|A\|_{\operatorname{HS}}^2)} + 8\underbrace{\operatorname{Var}(\mu^\top A \xi_j)}_{\stackrel{\eqref{eq:linear-variance-constant-bonud1}}{=}\mathcal{O}(1)} = \mathcal{O}(K^4\|A\|_{\operatorname{HS}}^2) \stackrel{\eqref{eq:HS-bound1}}{=} \mathcal{O}(K^4 d)
\end{equation}
as $d\to\infty$. Since both  $j\in[n]$ and $A\in\mathbb{R}^{d\times d}$ with $\|A\|=\mathcal{O}(1)$ were arbitrary, this completes the proof.
\qed

Here, we shift our focus to the proof of Theorem~\ref{thm:finite-shot-noise-correlation} of Section~\ref{subsec:finite-shot-noise}, The key idea of the proof was to decompose the involved high-dimensional tensor contractions into four separate terms. The following lemma elaborates these details:

\begin{lemma} \label{lemma:correlation-tensor-contraction-decomposition}
Let $A\in\mathbb{R}^{d\times d}$ be arbitrary with $\|A\|=\mathcal{O}_d(1)$ as $d\to\infty$. Define the vectorized contraction $a^\top \hat{\mathcal{V}} a$ of the matrix $\hat{\mathcal{V}}\in\mathbb{R}^{d^2\times d^2}$ via
    \begin{equation*}
         a^\top \mathcal{\hat{V}} a := \sum_{k,k',\eta,\eta'=1}^d A^{(\hat{\Sigma}_U)}_{k,k'}A^{(\hat{\Sigma}_U)}_{\eta,\eta'}\,\underbrace{\operatorname{Cov}_{x,\text{shot}}(\hat\phi_U(x)_k\hat\phi_U(x)_{k'},\hat\phi_U(x)_\eta\hat\phi_U(x)_{\eta'})}_{=\mathcal{\hat{V}}_{(k,k'),(\eta,\eta')}},
    \end{equation*}
    where $\hat{\Sigma}_U:=\mathbb{E}_{x,\text{shot}}[\hat{\phi}(x)\hat{\phi}(x)^\top]$ defines the noisy population covariance in Eq.~\eqref{eq:noisy-population-covariance-def}, and $a:=\operatorname{vec}(A^{(\hat{\Sigma}_U)}) \in \mathbb{R}^{d^2 \times 1}$ where $A^{(\hat{\Sigma}_U)}:=\hat{\Sigma}_U^{-\frac{1}{2}}A\,\hat{\Sigma}_U^{-\frac{1}{2}}$. Then, we have $a^\top\hat{\mathcal{V}}a=T_1 + T_2 + T_3 + T_{\text{special}}$, where
    \begin{align*}
        T_1 &= \operatorname{Var}_x\left(\phi_U(x)^\top A^{(\hat{\Sigma}_U)}\phi_U(x) + \operatorname{Tr}\left(A^{(\hat{\Sigma}_U)}D(x)\right)\right), \nonumber
        \\ T_2 &= 4\mathbb{E}_x\left[\phi_U(x)^\top A^{(\hat{\Sigma}_U)}D(x)A^{(\hat{\Sigma}_U)}\phi_U(x)\right], \nonumber
        \\ T_3 &= \mathbb{E}_x\left[\operatorname{Var}_{\operatorname{shot}}\left(\epsilon(x)^\top A^{(\hat{\Sigma}_U)}\epsilon(x)\;\middle|\;x\right)\right],
    \end{align*}
    and $T_{\text{special}}=\mathcal{O}(d)$ as $d\to\infty$.
\end{lemma}

\noindent
\emph{Proof.} Let $A\in\mathbb{R}^{d\times d}$ be arbitrary with $\|A\|=\mathcal{O}_d(1)$.
We begin by fixing a tensor index $(k,k'),(\eta,\eta')\in[d]\times[d]$. By the law of total covariance we find
\begin{align}
    \mathcal{\hat{V}}_{(k,k'),(\eta,\eta')} &= \operatorname{Cov}_x\left(\mathbb{E}_{\operatorname{shot}}[\hat{\phi}_U(x)_k\hat{\phi}_U(x)_{k'}\mid x],\,\mathbb{E}_{\operatorname{shot}}[\hat{\phi}_U(x)_\eta\hat{\phi}_U(x)_{\eta'}\mid x]\right) \nonumber
    \\ &\qquad+ \mathbb{E}_x\left[\operatorname{Cov}_{\operatorname{shot}}\left(\hat{\phi}_U(x)_k\hat{\phi}_U(x)_{k'},\,\hat{\phi}_U(x)_\eta\hat{\phi}_U(x)_{\eta'}\mid x\right)\right] =: \alpha + \beta. \label{eq:tensor-decomposition}
\end{align}
We begin by evaluating the conditional expectation of the noisy feature product within the first covariance term $\alpha$,
\begin{align}
    \mathbb{E}_{\operatorname{shot}}[\hat{\phi}_U(x)_k\hat{\phi}_U(x)_{k'}\mid x] &\stackrel{\eqref{eq:shot-noise-vector}}{=} \mathbb{E}_{\operatorname{shot}}[(\phi_U(x)_k + \epsilon(x)_k)(\phi_U(x)_{k'} + \epsilon(x)_{k'})\mid x] \nonumber 
    \\&= \phi_U(x)_k \phi_U(x)_{k'} + \phi_U(x)_k \underbrace{\mathbb{E}_{\operatorname{shot}}[\epsilon(x)_{k'}\mid x]}_{\stackrel{\eqref{eq:centered-shot-noise}}{=}0} + \phi_U(x)_{k'} \underbrace{\mathbb{E}_{\operatorname{shot}}[\epsilon(x)_k\mid x]}_{\stackrel{\eqref{eq:centered-shot-noise}}{=}0} + \mathbb{E}_{\operatorname{shot}}[\epsilon(x)_k \epsilon(x)_{k'}\mid x] \nonumber 
    \\&= \phi_U(x)_k \phi_U(x)_{k'} + D(x)_{k,k'}
\end{align}
using that the exact readouts $\phi_U(x)$ are independent of the shot noise. The last equality is obtained by via Eq.~\eqref{eq:noise-term-off-diagonal} from
\begin{align}
    \mathbb{E}_{\operatorname{shot}}[\epsilon(x)_k \epsilon(x)_{k'}\mid x] &\stackrel{\eqref{eq:diagonal-noise-term}}{=} \delta_{k,k'}\,\frac{1}{m_0} \underbrace{\left(\langle O_k^2\rangle_x - \langle O_k\rangle_x^2\right)}_{=:\sigma_k^2(x)=\mathcal{O}_d(1)} =: D(x)_{k,k'} \label{eq:diagonal-shot-noise-matrix}
\end{align}
which defines a (random) diagonal matrix $D(x)\in\mathbb{R}^{d\times d}$ with elements $D(x)_{k,k'}=\mathcal{O}_d(m_0^{-1})$ as $d\to\infty$. This establishes
\begin{equation}
    \alpha = \operatorname{Cov}_x\left(\phi_U(x)_k \phi_U(x)_{k'} + D(x)_{k,k'}, \phi_U(x)_{\eta} \phi_U(x)_{\eta'} + D(x)_{\eta,\eta'}\right).
\end{equation}
Contracting the terms of the form of $\alpha$ with the vectorized matrix coefficients $A^{(\hat{\Sigma}_U)}_{k,k'}A^{(\hat{\Sigma}_U)}_{\eta,\eta'}$ and utilizing the bilinearity of covariance, we synthesize our first scalar term $T_1$:
\begin{align}
    T_1 &:= \sum_{k,k',\eta,\eta'=1}^d A^{(\hat{\Sigma}_U)}_{k,k'}A^{(\hat{\Sigma}_U)}_{\eta,\eta'}\underbrace{\operatorname{Cov}_x\left(\phi_U(x)_k \phi_U(x)_{k'} + D(x)_{k,k'}, \phi_U(x)_{\eta} \phi_U(x)_{\eta'} + D(x)_{\eta,\eta'}\right)}_{=\alpha\equiv \alpha_{k,k',\eta,\eta'}} \nonumber
    \\ &= \operatorname{Var}_x\left( \sum_{k,k'=1}^d A^{(\hat{\Sigma}_U)}_{k,k'} \left(\phi_U(x)_k \phi_U(x)_{k'} + D(x)_{k,k'}\right) \right) = \operatorname{Var}_x\left(\phi_U(x)^\top A^{(\hat{\Sigma}_U)}\phi_U(x) + \operatorname{Tr}\left(A^{(\hat{\Sigma}_U)}D(x)\right)\right)
\end{align}
using that $D(x)$ is diagonal. This establishes the first term $T_1$.

Next, we are going to further analyse the second term $\beta$ in our first decomposition of Eq.~\eqref{eq:tensor-decomposition}. For that purpose,  we introduce the conditionally centered random variable $\Delta_{k,k'}(x)$ representing the shot-to-shot fluctuations of the feature products:
\begin{equation}
    \Delta_{k,k'}(x) := \hat{\phi}_U(x)_k\hat{\phi}_U(x)_{k'} - \mathbb{E}_{\operatorname{shot}}[\hat{\phi}_U(x)_k\hat{\phi}_U(x)_{k'}\mid x] = L_{k,k'}(x) + Q_{k,k'}(x),
\end{equation}
where, using $\hat{\phi}_U(x) = \phi_U(x) + \epsilon(x)$ in combination with Eq.~\eqref{eq:diagonal-shot-noise-matrix}, we partition $\Delta_{k,k'}(x)$ into a linear noise component $L_{k,k'}(x)$ and a quadratic noise component $Q_{k,k'}(x)$ defined as
\begin{equation}
    L_{k,k'}(x) := \phi_U(x)_k \epsilon(x)_{k'} + \phi_U(x)_{k'} \epsilon(x)_k \quad\text{and}\quad Q_{k,k'}(x) := \epsilon(x)_k \epsilon(x)_{k'} - D(x)_{k,k'}, \label{eq:L-Q-random-variables}
\end{equation}
respectively. The conditional covariance tensor can now be fully written as the expectation of the product of these shot-to-shot fluctuations:
\begin{align}
    &\operatorname{Cov}_{\operatorname{shot}}\left(\hat{\phi}_U(x)_k\hat{\phi}_U(x)_{k'},\hat{\phi}_U(x)_{\eta}\hat{\phi}_U(x)_{\eta'}\mid x\right) =  \mathbb{E}_{\operatorname{shot}}[\Delta_{k,k'}(x)\Delta_{\eta,\eta'}(x) \mid x] \nonumber \\
     &= \mathbb{E}_{\operatorname{shot}}[L_{k,k'}(x)L_{\eta,\eta'}(x) \mid x] + \mathbb{E}_{\operatorname{shot}}[L_{k,k'}(x)Q_{\eta,\eta'}(x) \mid x] +\mathbb{E}_{\operatorname{shot}}[Q_{k,k'}(x)L_{\eta,\eta'}(x) \mid x] + \mathbb{E}_{\operatorname{shot}}[Q_{k,k'}(x)Q_{\eta,\eta'}(x) \mid x] \label{eq:covariance-decomposition2}
\end{align}
That is, we have linear-linear terms, quadratic-quadratic terms, and mixed cross-terms.
We first examine the these cross-terms containing one linear and one quadratic component, such as $\mathbb{E}_{\operatorname{shot}}[L_{k,k'}(x)Q_{\eta,\eta'}(x) \mid x]$. Expanding this product yields terms of the form
\begin{equation}
    \mathbb{E}_{\operatorname{shot}}[L_{k,k'}(x)Q_{\eta,\eta'}(x) \mid x] = \phi_U(x)_k \mathbb{E}_{\operatorname{shot}}[\epsilon(x)_{k'} \epsilon(x)_{\eta} \epsilon(x)_{\eta'} \mid x] + \phi_U(x)_{k'} \mathbb{E}_{\operatorname{shot}}[\epsilon(x)_k \epsilon(x)_{\eta} \epsilon(x)_{\eta'} \mid x] \label{eq:mixed-covariance-tensor-constributions}
\end{equation}
Following an analogous argumentation as in Eq.~\eqref{eq:centered-shot-noise}, the joint expectation of any odd-powered combination of $\epsilon(x)$ components vanishes identically because the quantum measurements are conditionally independent: If all three indices are distinct, independence simplifies the expression to a product of individual zero-mean expectations. Even if two indices match (e.g., $k'=\eta$), the third distinct index ($\eta'$) remains independent and conditionally centered, forcing the expectation to zero so that
\begin{equation}
    \mathbb{E}_{\operatorname{shot}}[\epsilon(x)_{k'} \epsilon(x)_{\eta} \epsilon(x)_{\eta'} \mid x] = 0 \quad \implies \quad \mathbb{E}_{\operatorname{shot}}[L_{k,k'}(x)Q_{\eta,\eta'}(x) \mid x] = 0.
\end{equation}
because mixed terms of the form $\mathbb{E}_{\text{shot}}[\phi_U(x)_k \epsilon(x)_{k'} D(x)_{\eta,\eta'}\mid x]$ disappear since $\varepsilon(x)$ is conditionally centered by Eq.~\eqref{eq:centered-shot-noise}. By symmetry, $\mathbb{E}_{\operatorname{shot}}[Q_{k,k'}(x)L_{\eta,\eta'}(x) \mid x] = 0$.

However, the main technical caveat is the special case where all three indices match $k'=\eta=\eta'$ since the third conditional moments of $\epsilon(x)_k'$ do not zero in general without additional symmetry assumptions on the observable eigenspectrum. We are going to isolate this special case into the separate term $T_{\text{special}}$, which contracts all terms of the form $\mathbb{E}_{\operatorname{shot}}[L_{k,k'}(x)Q_{k',k'}(x) \mid x]$, and show that these terms scale as $\mathcal{O}(d)$ regardless of $m_0$. In that case, by Eq.~\eqref{eq:mixed-covariance-tensor-constributions} it is enough to directly bound contractions of the form 
\begin{align}
    T_{\text{special}} := \sum_{k,k'=1}^d A_{k,k'}^{(\hat\Sigma_U)}A_{k',k'}^{(\hat\Sigma_U)} \phi_U(x)_k\,\mathbb{E}_{\text{shot}}\left[\epsilon(x)_{k'}^3\mid x\right] = \sum_{k'=1}^d A_{k',k'}^{(\hat\Sigma_U)}\,\mathbb{E}_{\text{shot}}\left[\epsilon(x)_{k'}^3\mid x\right]\underbrace{\left(\sum_{k=1}^d A_{k,k'}^{(\hat\Sigma_U)} \phi_U(x)_k \right)}_{=\left(A^{(\hat\Sigma_U)}\phi_U(x)\right)_{k'}} \label{eq:T-special}
\end{align}
to handle this special case of conditional third noise moments. We may assume without loss of generality that $A^{(\hat\Sigma_U)}\in\mathbb{R}^{d\times d}$ is symmetric because of $z^\top A^{(\hat\Sigma_U)} z = z^\top \left(\frac{A^{(\hat\Sigma_U)}+{A^{(\hat\Sigma_U)}}^\top}{2}\right) z$ for all $z\in\mathbb{R}^d$. First, note that
\begin{align}
    \mathbb{E}_{\text{shot}}\left[\epsilon(x)_{k'}^3\mid x\right] &\stackrel{\eqref{eq:shot-noise-components}}{=} \mathbb{E}_{\text{shot}}\left[\left(\frac{1}{m_0^3}\sum_{m=1}^{m_0}\left\{\mu_{k'}^{(m)}(x)-\langle O_{k'}\rangle_x\right\}\right)^3\;\middle|\;x\right] \nonumber
    \\ &= \frac{1}{m_0^3} \sum_{m=1}^{m_0}\underbrace{\mathbb{E}_{\text{shot}}\left[\left(\mu_{k'}^{(m)}(x)-\langle O_{k'}\rangle_x\right)^3\;\middle|\; x\right]}_{=\mathcal{O}_d(1)} = \mathcal{O}_d\left(\frac{1}{m_0^2}\right), \label{eq:third-moment-bound}
\end{align}
where used the fact that the measurements are independent and that each $\mu_{k'}^{(m_i)}(x)$ is an (conditionally) unbiased estimator of $\langle O_{k'}\rangle_x$ as per Eq.~\eqref{eq:born-expectation}. We also used the uniform bound $|\mu_{k'}^{(m)}(x)-\langle O_{k'}\rangle_x| \leq 2\max_{k\in[d]}\|O_k\|=\mathcal{O}_d(1)$ due to Eq.~\eqref{eq:bounded-estimator}.
Next, we apply Cauchy-Schwarz to bound the expression
\begin{align}
    \left|\sum_{k'=1}^d A_{k',k'}^{(\hat\Sigma_U)} \left(A^{(\hat\Sigma_U)}\phi_U(x)\right)_{k'}\right| &\leq \sqrt{\sum_{k'=1}^d \left(A_{k',k'}^{(\hat\Sigma_U)}\right)^2}\sqrt{\sum_{k'=1}^d \left(A^{(\hat\Sigma_U)}\phi_U(x)\right)_{k'}^2} \label{eq:contraction-cauchy-schwarz}
\end{align}
and separately establish
\begin{align}
    \sqrt{\sum_{k'=1}^d \left(A_{k',k'}^{(\hat\Sigma_U)}\right)^2} &\leq \sqrt{\operatorname{Tr}\left(\left(A^{(\hat\Sigma_U)}\right)^\top A^{(\hat\Sigma_U)}\right)} = \sqrt{\sum_{k=1}^d\lambda_{k}\left(A^{(\hat\Sigma_U)}\right)^2} \leq \sqrt{d}\lambda_{\max}\left(A^{(\hat\Sigma_U)}\right) = \sqrt{d}\|A^{(\hat\Sigma_U)}\|, \label{eq:contraction-cauchy-schwarz-term1}
\end{align}
since $A^{(\hat\Sigma_U)}$ is symmetric, and
\begin{align}
    \sqrt{\sum_{k'=1}^d \left(A^{(\hat\Sigma_U)}\phi_U(x)\right)_{k'}^2} &= \|A^{(\hat\Sigma_U)}\phi_U(x)\| \leq \|A^{(\hat\Sigma_U)}\|\,\underbrace{\|\phi_U(x)\|}_{\leq\sqrt{d}M} \leq \sqrt{d} M \|A^{(\hat\Sigma_U)}\|, \label{eq:contraction-cauchy-schwarz-term2}
\end{align}
since $|\phi_U(x)|_k\leq M$ for all $k\in[d]$. Hence, since $M=\mathcal{O}_d(1)$, we may derive the desired bound
\begin{align}
    T_{\text{special}} &\stackrel{\eqref{eq:T-special}}{=} \sum_{k'=1}^d A_{k',k'}^{(\hat\Sigma_U)}\left(A^{(\hat\Sigma_U)}\phi_U(x)\right)_{k'}\, \underbrace{\mathbb{E}_{\text{shot}}\left[\epsilon(x)_{k'}^3\mid x\right]}_{\stackrel{\eqref{eq:third-moment-bound}}{=} \mathcal{O}_d\left(m_0^{-2}\right)} \stackrel{\eqref{eq:contraction-cauchy-schwarz}}{=} \mathcal{O}_d\left(\frac{1}{m_0^2}\underbrace{\sqrt{\sum_{k'=1}^d \left(A_{k',k'}^{(\hat\Sigma_U)}\right)^2}}_{\stackrel{\eqref{eq:contraction-cauchy-schwarz-term1}}{\leq}\sqrt{d}\|A^{(\hat\Sigma_U)}\|}\underbrace{\sqrt{\sum_{k'=1}^d \left(A^{(\hat\Sigma_U)}\phi_U(x)\right)_{k'}^2}}_{\stackrel{\eqref{eq:contraction-cauchy-schwarz-term2}}{\leq \sqrt{d} M \|A^{(\hat\Sigma_U)}\|}}\right)
    \nonumber
    \\ &= \mathcal{O}_d\left(\frac{d\|A^{(\hat\Sigma_U)}\|^2}{m_{\max^2}}\right) = \mathcal{O}(d),
\end{align}
where the last identity follows from the observation that
\begin{align}
    \|A^{(\hat\Sigma_U)}\| =\left\|\left(\hat\Sigma_U^{-\frac{1}{2}}\right)^\top A \,\hat\Sigma_U^{-\frac{1}{2}}\right\|\leq \underbrace{\|A\|}_{=\mathcal{O}_d(1)}\|\hat\Sigma_U^{-1}\| \stackrel{\eqref{eq:noisy-covariance-spectral-bound}}{=} \mathcal{O}_d\left(\frac{1}{\lambda_{\min}(\hat\Sigma_U)}\right) \stackrel{\eqref{eq:noise-covariance-decomposition}}{=} \mathcal{O}_d\left(\frac{m_0}{\sigma_{\min}^2}\right) =   \mathcal{O}_d\left(m_0\right) \label{eq:Frobenius-bound1}
\end{align}
since we may assume that $\sigma_{\min}^2=\min_{k\in[d]}\mathbb{E}_x\left[\langle O_k^2\rangle_x - \langle O_k\rangle_x^2\right] = \Theta_d(1)$ is independent of the dimension $d$ under standard quantum measurement protocol assumptions.

This proves the claimed scaling behaviour of the term $T_{\text{special}}$. Therefore, returning to Eq.~\eqref{eq:mixed-covariance-tensor-constributions}, we may assume of the form $\mathbb{E}_{\operatorname{shot}}[L_{k,k'}(x)Q_{\eta,\eta'}(x) \mid x] = \mathbb{E}_{\operatorname{shot}}[Q_{k,k'}(x)L_{\eta,\eta'}(x) \mid x] = 0$ vanish by absorbing the special case where three indices match into the term $T_{\text{special}}$ as elaborated above. This means that only the linear-linear and quadratic-quadratic terms remain so that
\begin{align}
    \beta &= \mathbb{E}_x\left[\operatorname{Cov}_{\operatorname{shot}}\left(\hat{\phi}_U(x)_k\hat{\phi}_U(x)_{k'},\,\hat{\phi}_U(x)_\eta\hat{\phi}_U(x)_{\eta'}\mid x\right)\right] \nonumber
    \\ &\stackrel{\eqref{eq:covariance-decomposition2}}{=} \mathbb{E}_x\left[\mathbb{E}_{\operatorname{shot}}[L_{k,k'}(x)L_{\eta,\eta'}(x) \mid x]\right] +  \mathbb{E}_x\left[\mathbb{E}_{\operatorname{shot}}[Q_{k,k'}(x)Q_{\eta,\eta'}(x) \mid x]\right]. \label{eq:second-term-covariance-decomposition}
\end{align}
By contracting these two terms of $\beta$ individually we aim to identify $T_2$ and $T_3$.

To find $T_2$, we focus on contracting the linear-linear terms $\mathbb{E}_{\operatorname{shot}}[L_{k,k'}(x)L_{\eta,\eta'}(x) \mid x]$ in Eq.~\eqref{eq:covariance-decomposition2}. Employing the diagonal structure $D(x)_{k,k'} = \delta_{k,k'}D(x)_{k,k}$ of $D(x)$, we find
\begin{align}
    \mathbb{E}_{\operatorname{shot}}[L_{k,k'}(x)L_{\eta,\eta'}(x) \mid x] &= \mathbb{E}_{\operatorname{shot}}\left[ (\phi_U(x)_k \epsilon(x)_{k'} + \phi_U(x)_{k'} \epsilon(x)_k)(\phi_U(x)_{\eta} \epsilon(x)_{\eta'} + \phi_U(x)_{\eta'} \epsilon(x)_{\eta}) \;\middle|\; x \right] \nonumber \\
    &\stackrel{\eqref{eq:diagonal-shot-noise-matrix}}{=} \phi_U(x)_k \phi_U(x)_{\eta} \delta_{k',\eta'} D(x)_{k',k'} + \phi_U(x)_k \phi_U(x)_{\eta'} \delta_{k',\eta} D(x)_{k',k'} \nonumber \\
    &\quad + \phi_U(x)_{k'} \phi_U(x)_{\eta} \delta_{k,\eta'} D(x)_{k,k} + \phi_U(x)_{k'} \phi_U(x)_{\eta'} \delta_{k,\eta} D(x)_{k,k}
\end{align}
Now, contracting this tensor entry-by-entry with $A^{(\hat{\Sigma}_U)}_{k,k'}A^{(\hat{\Sigma}_U)}_{\eta,\eta'}$, we evaluate each of the four latter addends using the symmetry property $A^{(\hat{\Sigma}_U)}_{k,k'} = A^{(\hat{\Sigma}_U)}_{k',k}$, i.e., contracting the first addend where only ($k' = \eta'$)-terms contribute, yields
\begin{equation}
   \sum_{k,k',\eta=1}^d A^{(\hat{\Sigma}_U)}_{k,k'}A^{(\hat{\Sigma}_U)}_{\eta,k'} \phi_U(x)_k \phi_U(x)_{\eta} D(x)_{k',k'} = \sum_{k'=1}^d D(x)_{k',k'} \left( \sum_{k=1}^d A^{(\hat{\Sigma}_U)}_{k,k'}\phi_U(x)_k \right)^2 = \phi_U(x)^\top A^{(\hat{\Sigma}_U)} D(x) A^{(\hat{\Sigma}_U)} \phi_U(x).
\end{equation}
Analogously, the remaining addends contribute exactly the same way as $\phi_U(x)^\top A^{(\hat{\Sigma}_U)} D(x) A^{(\hat{\Sigma}_U)} \phi_U(x)$.
Since all four addends contract to the identical quadratic form, we may thus identify
\begin{equation}
    T_2 :=  \sum_{k,k',\eta,\eta'=1}^d A^{(\hat{\Sigma}_U)}_{k,k'}A^{(\hat{\Sigma}_U)}_{\eta,\eta'} \mathbb{E}_x\left[\mathbb{E}_{\operatorname{shot}}[L_{k,k'}(x)L_{\eta,\eta'}(x) \mid x]\right] = 4\mathbb{E}_x\left[\phi_U(x)^\top A^{(\hat{\Sigma}_U)}D(x)A^{(\hat{\Sigma}_U)}\phi_U(x)\right]
\end{equation}
by the previous computation. This identifies the term $T_2$.

Finally, it remains to derive the last term $T_3$ which is obtained by contracting the quadratic-quadratic terms $\mathbb{E}_{\operatorname{shot}}[Q_{k,k'}(x)Q_{\eta,\eta'}(x) \mid x]$ in Eq.~\eqref{eq:covariance-decomposition2}.  By definition, the conditional expectation of the product of these quadratic centered variables represents the conditional covariance of the noise self-products, i.e.,
\begin{equation}
    \mathbb{E}_{\operatorname{shot}}[Q_{k,k'}(x)Q_{\eta,\eta'}(x) \mid x] \stackrel{\eqref{eq:L-Q-random-variables}}{=} \operatorname{Cov}_{\operatorname{shot}}\left(\epsilon(x)_k \epsilon(x)_{k'}, \epsilon(x)_{\eta} \epsilon(x)_{\eta'} \;\middle|\; x\right).
\end{equation}
Contracting this block with $A^{(\hat{\Sigma}_U)}_{k,k'}A^{(\hat{\Sigma}_U)}_{\eta,\eta'}$ and taking the outer data expectation yields the final term
\begin{align}
    T_3 &:= \mathbb{E}_x\left[\sum_{k,k',\eta,\eta'=1}^d A^{(\hat{\Sigma}_U)}_{k,k'}A^{(\hat{\Sigma}_U)}_{\eta,\eta'} \operatorname{Cov}_{\operatorname{shot}}\left(\epsilon(x)_k \epsilon(x)_{k'}, \epsilon(x)_{\eta} \epsilon(x)_{\eta'} \;\middle|\; x\right) \right]
    = \mathbb{E}_x\left[\operatorname{Var}_{\operatorname{shot}}\left( \sum_{k,k'=1}^d A^{(\hat{\Sigma}_U)}_{k,k'} \epsilon(x)_k \epsilon(x)_{k'} \;\middle|\; x \right)\right] \nonumber
    \\ &= \mathbb{E}_x\left[\operatorname{Var}_{\operatorname{shot}}\left(\epsilon(x)^\top A^{(\hat{\Sigma}_U)}\epsilon(x)\;\middle|\;x\right)\right].
\end{align}

In summary, we have thus established that
\begin{equation}
     a^\top \hat{\mathcal{V}} a = T_1 + T_2 + T_3 + \underbrace{T_{\text{special}}}_{=\mathcal{O}(d)},
\end{equation}
which completes the proof of the lemma.
\qed

\subsubsection{Useful Quantum and Linear Algebra Properties}

Here, we prove useful mathematical properties arising due to quantum mechanics used throughout the supplementary material. The following lemma was used to prove Lipschitz regularity of the quantum readout in Section~\ref{appendix:quantum-RMT-assumptions}:

\begin{lemma}[Lipschitz Regularity of the Readout] \label{lemma:lipschitz-readout}
    Assume that the system $\{O_k:k\in[d]\}$ of observables defines 1-tight frame, i.e.,
    \begin{equation}
        \sum_{k=1}^d\left|\operatorname{Tr}\left(O_kA\right)\right|^2 \leq \|A\|_{\operatorname{HS}}^2 \label{eq:1-tight-frame}
    \end{equation}
    for all Hermitian operators $A\in\operatorname{Herm}(2^N)$, where $\|A\|_{\operatorname{HS}}:=\sqrt{\langle A,A\rangle_{\operatorname{HS}}}$. Then, the readout $\phi_U:\mathcal{X}\to\mathbb{R}^d$ of the QELM model as per Eq.~\eqref{eq:quantum-readout} is Lipschitz continuous with Lipschitz constant $L_{\phi_U}=\Theta_d(1)$ as $d\to\infty$.
\end{lemma}

\noindent
\emph{Proof.} First, we recall from Eq.~\eqref{eq:quantum-readout} that $\langle O_k\rangle = \operatorname{Tr}(O_k\rho_U(x))$ for all $k\in[d]$, where $\rho_U(x)=U(x)\rho_0 U(x)^\dagger$ and $U(x)=U_{L+1}\prod_{\ell=1}^L\left\{(\mathcal{G}_\ell(x)\otimes I_{2^{n_h}})U_\ell\right\}$ for some input-independent unitaries $U_{L+1},\dots,U_1\in\operatorname{U}(2^N)$ as per Eq.~\eqref{eq:data-uploading}. 
For each layer $\ell\in[L]$, the data uploading PQC $\mathcal{G}_\ell(x)$ is generated by Hermitian generators $\{H_{\eta}^{(\ell)}:\eta\in[p]\}\subset\operatorname{Herm}(2^N)$ (that is, one generator for each input coordinate $x_\eta$ of $x=(x_1,\dots,x_p)$) such that
\begin{equation}
    \partial_{x_\eta}\mathcal{G}_\ell(x)=-iH_\eta^{(\ell)}\mathcal{G}_\ell(x) \label{eq:generator-derivative}
\end{equation}
for all $\eta\in[p]$, where the spectrum of each generator is bounded by the dimension-independent constant
\begin{equation}
    \lambda_0 := \max_{(\ell,\eta)\in[L]\times[p]}\|H_\eta^{(\ell)}\| = \Theta_d(1). \label{eq:dimension-independent-generator-bound}
\end{equation}
We are interested in quantifying the sensitivity of the readout $\phi_U(x)$ with respect to the input $x\in\mathcal{X}$. Let $S_\ell(x):=\mathcal{G}_\ell(x)\otimes I_{2^{n_h}}$. Then, Eq.~\eqref{eq:generator-derivative} implies
\begin{equation}
    \partial_{x_\eta} S_\ell(x) = -i(H_\eta^{(\ell)}\otimes I_{2^{n_h}})S_\ell(x) \label{eq:generator-derivative2}
\end{equation}
for all layers $\ell\in[L]$ and input components $\eta\in[p]$. In particular, by the product rule,
\begin{align}
    \partial_{x_\eta}U(x) &= U_{L+1}\partial_{x_\eta}\prod_{\ell=1}^L\left\{S_\ell(x)U_\ell\right\} = \sum_{\ell=1}^LU_{L+1}\left(\prod_{\ell'=\ell+1}^L\{S_{\ell'}(x)U_{\ell'}\}\right)\partial_{x_\eta}\left\{S_\ell(x)U_\ell\right\}\left(\prod_{\ell'=1}^{\ell-1}\{S_{\ell'}(x)U_{\ell'}\}\right) \nonumber
    \\ &\stackrel{\eqref{eq:generator-derivative2}}{=} -i\sum_{\ell=1}^L\underbrace{U_{L+1}\left(\prod_{\ell'=\ell+1}^L \{S_{\ell'}(x)U_{\ell'}\}\right)}_{=:U_{>\ell}(x)}(H_\eta^{(\ell)}\otimes I_{2^{n_h}}) \underbrace{S_\ell(x)U_\ell\prod_{\ell'=1}^{\ell-1}\{S_{\ell'}(x)U_{\ell'}\}}_{=:U_{\leq \ell}(x)} \nonumber
    \\ &= -i\sum_{\ell=1}^L U_{>\ell}(x) (H_\eta^{(\ell)}\otimes I_{2^{n_h}})U_{\leq\ell}(x)
\end{align}
for all $\eta\in[p]$. In particular, $U(x) = U_{>\ell}(x)U_{\leq\ell}(x)$ and $\partial_{x_\eta}U(x)=-iV_\eta(x)U(x)$, where we introduced the Hermitian operator
\begin{equation}
    V_\eta(x) := \sum_{\ell=1}^L U_{>\ell}(x)(H_\eta^{(\ell)}\otimes I_{2^{n_h}})U_{>\ell}(x)^\dagger\in\operatorname{Herm}(2^N)
\end{equation}
for each $\eta\in[p]$. Let $\eta\in[p]$ be arbitrary. We bound in the operator norm of the latter operator and find
\begin{equation}
    \|V_\eta(x)\| \leq \sum_{\ell=1}^L\|U_{>\ell}(x)(H_\eta^{(\ell)}\otimes I_{2^{n_h}})U_{>\ell}(x)^\dagger\| \leq \sum_{\ell=1}^L\|H_\eta^{(\ell)}\| \stackrel{\eqref{eq:dimension-independent-generator-bound}}{\leq} L\lambda_0 \label{eq:dimensionless-bound1}
\end{equation}
using that $U_{>\ell}(x)\in\operatorname{U}(2^N)$ for all $\ell\in[L]$ and the operator norm remains invariant under unitary transformations. We once again apply the product rule and show that the derivative $\partial_{x_\eta}\rho_U(x)$ of the pre-measurement state $\rho_U(x)$ reduces to the commutator of $V_\eta(x)$ and $\rho_U(x)$ (which is related to the \emph{parameter-shift rule}~\cite{schuld2019evaluating} for PQC gradients):
\begin{align}
    \partial_{x_\eta}\rho_U(x) &= \partial_{x_\eta}\{U(x)\rho_0U(x)^\dagger\} = \underbrace{\{\partial_{x_\eta}U(x)\}}_{=-iV_\eta(x)U(x)}\rho_0U(x)^\dagger + U(x)\rho_0\underbrace{\{\partial_{x_\eta}U(x)\}^\dagger}_{=+iU(x)^\dagger V_\eta(x)} \nonumber
    \\ &= -i V_\eta(x)\underbrace{U(x)\rho_0 U(x)^\dagger}_{=\rho_U(x)} +i\underbrace{U(x)\rho_0U(x)^\dagger}_{=\rho_U(x)} V_\eta(x) = -i\left\{ V_\eta(x)\rho_U(x)- \rho_U(x)V_\eta(x)\right\} = -i[V_\eta(x),\rho_U(x)] \label{eq:parameter-shift}
\end{align}
using $V_\eta(x)^\dagger = V_\eta(x)$ since $V_\eta(x)\in\operatorname{Herm}(2^N)$.

To show that the readout $\phi_U:\mathcal{X}\to\mathbb{R}^d$ is Lipschitz continuous, we bound the operator norm of the Jacobian $J_{\phi_U}(x)\in\mathbb{R}^{d\times p}$ of $\phi_U(x)$. For that purpose, let $v\in\mathbb{S}^{p-1}$ be an arbitrary unit vector. Then, recalling that $\phi_U(x)_k=\operatorname{Tr}(O_k(\rho_U(x))$ for each component $k\in[d]$,
\begin{align}
    (J_{\phi_U}(x)v)_k &= v^\top\nabla_x\phi_U(x)_k = \sum_{\eta=1}^p v_\eta\partial_{x_\eta}\phi(x)_k = \sum_{\eta=1}^pv_\eta\operatorname{Tr}(O_k\{\partial_{x_\eta}\rho_U(x)\}) \stackrel{\eqref{eq:parameter-shift}}{=} -i\sum_{\eta=1}^p v_\eta \operatorname{Tr}(O_k[V_\eta(x),\rho_U(x)]) \nonumber
    \\ &= -i\operatorname{Tr}\left(O_k\left[\sum_{\eta=1}^p v_\eta V_\eta(x),\rho_U(x)\right]\right) = -i\operatorname{Tr}(O_k [V_v(x),\rho_U(x)]), \label{eq:Jacobian-projection1}
\end{align}
where we introduced the directional operator
\begin{equation}
    V_v(x):=\sum_{\eta=1}^p v_\eta V_\eta(x) \in\operatorname{Herm}(2^N)
\end{equation}
defined for each unit vector $v\in\mathbb{S}^{p-1}$. Hence, under the assumption that the system $\{O_k:k\in[d]\}$ defines a 1-tight frame as per Eq.~\eqref{eq:1-tight-frame}, then we automatically have the bound
\begin{align}
    \|J_{\phi_U}(x)v\|^2 &= \sum_{k=1}^d |(J_{\phi_U}(x)v)_k|^2 \stackrel{\eqref{eq:Jacobian-projection1}}{=} \sum_{k=1}^d|\operatorname{Tr}(O_k[V_v(x),\rho_U(x)])|^2 \stackrel{\eqref{eq:1-tight-frame}}{\leq} \|[V_v(x),\rho_U(x)]\|^2_{\operatorname{HS}} \nonumber
    \\ &\leq 4\|V_v(x)\|^2\|\rho_U(x)\|^2_{\operatorname{HS}} \leq 4L^2\lambda_0^2p, \label{eq:Jacobian-bound}
\end{align}
where the last inequality follows from $\|\rho_U(x)\|^2_{\operatorname{HS}} = \operatorname{Tr}(\rho_U(x)^2)\leq 1$ since $\rho_U(x)\in\mathcal{D}(2^N)$ defines a density matrix, and due to the estimation
\begin{equation}
    \|V_v(x)\| \leq \sum_{\eta=1}^p|v_\eta|\underbrace{\|V_\eta(x)\|}_{\stackrel{\eqref{eq:dimensionless-bound1}}{\leq} L\lambda_0} \leq L\lambda_0\sum_{\eta=1}^p|v_\eta|\leq L\lambda_0\sqrt{p},
\end{equation}
where, by Cauchy-Schwarz applied to $\tilde{v}:=(|v_1|,\dots,|v_p|)\in\mathbb{R}^p$ and $\vec{1}:=(1,\dots,1)\in\mathbb{R}^p$,
\begin{equation}
    \sum_{\eta=1}^p|v_\eta| = \langle \tilde{v},\vec{1}\rangle\leq\|\tilde{v}\|\,\|\vec{1}\| = \sqrt{\sum_{\eta=1}^pv_\eta^2}\sqrt{\sum_{\eta=1}^p1} = \sqrt{p}\underbrace{\|v\|}_{=1}=\sqrt{p}
\end{equation}
using $v\in\mathbb{S}^{p-1}$. Since $v\in\mathbb{S}^{p-1}$ was arbitrary, this shows that for all $x\in\mathcal{X}$, the Jacobian obeys the spectral bound
\begin{equation}
    \|J_{\phi_U}(x)\|=\sup_{v\in\mathbb{S}^{p-1}}\|J_{\phi_U}(x)v\| \stackrel{\eqref{eq:Jacobian-bound}}{\leq} 2L\lambda_0\sqrt{p}. \label{eq:readout-Lipschitz-bound}
\end{equation}
This bound holds true uniformly on the input domain $\mathcal{X}$, thus proving that the readout $\phi_U:\mathcal{X}\to\mathbb{R}^d$ is $L_{\phi_U}$-Lipschitz with $L_{\phi_U}\leq 2L\lambda_0\sqrt{p}=\Theta_d(1)$. This completes the proof.
\qed

As a small remark, note that the number of reuploads $L$ dries a linear penalty into the input sensitivity but the bound remains completely independent of the feature dimension $d=\mathcal{O}(\operatorname{Poly}(N))$ (and the total qubit count $N$ assuming that $L$ scales independently of $N$). The following lemma establishes the existence of non-restrictive 1-tight observable systems, used to justify the random matrix theory assumption in the setting of unitary 2-designs in Section~\ref{appendix:quantum-RMT-assumptions}:

\begin{lemma}[Existence of 1-Tight Observable Systems] \label{lemma:1-tight-system}
    Let $\{O_k\}_{k\in[d]}$ be an orthonormal system spanning a subspace of the Hilbert-Schmidt space $\operatorname{Herm}(2^N)$. Then,
    \begin{equation*}
        \sum_{k=1}^d|\operatorname{Tr}(O_kA)|^2\leq\|A\|_{\operatorname{HS}}^2
    \end{equation*}
    for all $A\in\operatorname{Herm}(2^N)$.
\end{lemma}

\noindent
\emph{Proof.} The space $\operatorname{Herm}(2^N)$ of $2^N\times 2^N$ Hermitian operators defines a Hilbert space with inner product $\langle A,B\rangle_{\operatorname{HS}}=\operatorname{Tr}(AB)$ and induces the Schatten 2-norm $\|A\|_{\operatorname{HS}}:=\sqrt{\langle A,A\rangle_{\operatorname{HS}}}$. Let $A\in\operatorname{Herm}(2^N)$ be arbitary and decompose it into
\begin{equation}
    A = A_{\parallel} + A_{\perp}
\end{equation}
where $A_{\parallel}\in\operatorname{Herm}(2^N)$ is the orthogonal projection of $A$ onto the $d$-dimensional linear subspace $\mathcal{H}:=\operatorname{span}_{\mathbb{R}}\{O_k:k\in[d]\}$ of $\operatorname{Herm}(2^N)$, and $A_{\perp}\in\operatorname{Herm}(2^N)$ the left-over component in the orthogonal complement $\mathcal{H}^\perp$ of $\mathcal{H}$ such that $\langle A_{\parallel},A_{\perp}\rangle_{\operatorname{HS}}=0$. In that case, we find for all $k\in[d]$ that
\begin{align}
    \operatorname{Tr}(O_kA)=\langle O_k,A\rangle_{\operatorname{HS}} = \langle O_k,A_\parallel+A_\perp\rangle_{\operatorname{HS}} = \langle O_k,A_\parallel\rangle_{\operatorname{HS}} + \underbrace{\langle O_k,A_\perp\rangle_{\operatorname{HS}}}_{=0} = \langle O_k,A_\parallel\rangle_{\operatorname{HS}} = \operatorname{Tr}(O_kA_\parallel)
\end{align}
using that $A_\perp\in\mathcal{H}^\perp$ is in the orthogonal complement of $\mathcal{H}$. Furthermore, we have
\begin{align}
    \|A_\parallel\|^2_{\operatorname{HS}} = \langle A_\parallel,A_\parallel\rangle_{\operatorname{HS}} = \sum_{k,k'=1}^d\operatorname{Tr}(O_kA_\parallel)\operatorname{Tr}(O_{k'}A_\parallel) \underbrace{\operatorname{Tr}(O_kO_{k'})}_{=\delta_{k,k'}} = \sum_{k=1}^d\operatorname{Tr}(O_kA_\parallel)^2 \label{eq:schatten-2-norm-bound}
\end{align}
using that $\{O_k\}_{k\in[d]}$ defines an orthonormal system and substituting the orthogonal projection
\begin{equation}
    A_{\parallel}=\sum_{k=1}^d\operatorname{Tr}(O_kA_{\parallel})O_k.
\end{equation}
Consequently, we may directly estimate
\begin{align}
    \sum_{k=1}^d|\operatorname{Tr}(O_kA)|^2 \stackrel{\eqref{eq:schatten-2-norm-bound}}{=} \|A_\parallel\|^2_{\operatorname{HS}} \leq \|A_\parallel\|^2_{\operatorname{HS}} + \underbrace{\|A_\perp\|_{\operatorname{HS}}^2}_{\geq0} = \|A\|^2_{\operatorname{HS}},
\end{align}
where the last identity follows from the Pythagorean theorem. Since $A\in\operatorname{Herm}(2^N)$ was arbitrary, this completes the proof.
\qed

Last but not least, we employed the following lemma to show explicitly how the Fourier coefficients transform under additional reuploads (cf. Proposition~\ref{prop:Fourier-convolution} in Section~\ref{appendix:Fourier-analysis}):

\begin{lemma}[Orthogonal Decomposition of General Operators] \label{lemma:orthogonal-decomposition}
    Let $\{O_k\}_{k\in[d]}$ be an orthogonal system of non-zero operators spanning a subspace of the Hilbert-Schmidt space $\operatorname{Herm}(2^N)$. Then, any linear operator $A \in \mathbb{C}^{2^N \times 2^N}$ can be uniquely decomposed as
    \begin{equation*}
        A = A^\perp + \sum_{k'=1}^d \left( \frac{\operatorname{Tr}(O_{k'}^\dagger A)}{\operatorname{Tr}(O_{k'}^2)} \right) O_{k'}
    \end{equation*}
    where $A^\perp \in \mathbb{C}^{2^N \times 2^N}$ is an error operator that is orthogonal to the subspace $\mathcal{H} := \operatorname{span}_{\mathbb{C}}\{O_k : k \in [d]\}$, satisfying $\operatorname{Tr}(O_k^\dagger A^\perp) = 0$ for all $k \in [d]$.
\end{lemma}

\noindent
\emph{Proof.} The space $\mathbb{C}^{2^N \times 2^N}$ of complex $2^N \times 2^N$ matrices forms a complex Hilbert space when equipped with the standard Hilbert-Schmidt inner product $\langle B, C \rangle_{\operatorname{HS}} = \operatorname{Tr}(B^\dagger C)$, which induces the Schatten 2-norm $\Vert{}B\Vert{}_{\operatorname{HS}} := \sqrt{\langle B, B \rangle_{\operatorname{HS}}}$. Let $\mathcal{H} := \operatorname{span}_{\mathbb{C}}\{O_k : k \in [d]\}$ be the $d$-dimensional complex linear subspace of $\mathbb{C}^{2^N \times 2^N}$ spanned by our system. 
Since the constituent operators are elements of $\operatorname{Herm}(2^N)$, they are self-adjoint, meaning $O_k^\dagger = O_k$ for all $k \in [d]$. 
The mutual orthogonality condition implies that $\langle O_k, O_{k'} \rangle_{\operatorname{HS}} = \operatorname{Tr}(O_k^\dagger O_{k'}) = \operatorname{Tr}(O_k O_{k'}) = 0$ for all $k \neq k'$.
Let $A \in \mathbb{C}^{2^N \times 2^N}$ be an arbitrary operator. 
By the projection theorem in Hilbert spaces (cf. Theorem~1 in Section~3.3 and Theorem~1 in Section~3.5 of Ref.~\cite{luenberger1997optimization}), the linear operator $A$ can be uniquely decomposed into
\begin{equation}
    A = A_\parallel + A^\perp
\end{equation}
where $A_\parallel \in \mathcal{H}$ is the unique orthogonal projection of $A$ onto the subspace $\mathcal{H}$, and $A^\perp \in \mathcal{H}^\perp$ is the leftover component belonging to the orthogonal complement, satisfying $\langle O_k, A^\perp \rangle_{\operatorname{HS}} = \operatorname{Tr}(O_k^\dagger A^\perp) = 0$ for all $k \in [d]$.
Because $A_\parallel \in \mathcal{H}$, it can be written as a linear combination of the underlying basis elements:
\begin{equation}
    A_\parallel = \sum_{k'=1}^d \alpha_{k'} O_{k'}
\end{equation}
for some set of expansion coefficients $\alpha_{k'} \in \mathbb{C}$. 
To identify these coefficients, we compute the Hilbert-Schmidt inner product of $A$ with an arbitrary basis element $O_k$,
\begin{equation}
    \operatorname{Tr}(O_k^\dagger A) = \langle O_k, A \rangle_{\operatorname{HS}} = \langle O_k, A_\parallel + A^\perp \rangle_{\operatorname{HS}}
\end{equation}
Using the linearity of the inner product and the fact that $\langle O_k, A^\perp \rangle_{\operatorname{HS}} = 0$, this simplifies directly to
\begin{equation}
    \operatorname{Tr}(O_k^\dagger A) = \langle O_k, A_\parallel \rangle_{\operatorname{HS}} = \left\langle O_k, \sum_{k'=1}^d \alpha_{k'} O_{k'} \right\rangle_{\operatorname{HS}} = \sum_{k'=1}^d \alpha_{k'} \operatorname{Tr}(O_k^\dagger O_{k'}).
\end{equation}
Applying the orthogonality condition of the system, all terms in the summation where $k' \neq k$ vanish, yielding $\operatorname{Tr}(O_k^\dagger O_{k'}) = \delta_{k,k'} \operatorname{Tr}(O_k^2)$ (recalling that $O_k^\dagger = O_k$). The summation collapses to a single term,
\begin{equation}
    \operatorname{Tr}(O_k^\dagger A) = \alpha_k \operatorname{Tr}(O_k^2).
\end{equation}
Since the operators in the system are non-zero by definition, the normalization factor satisfies $\operatorname{Tr}(O_k^2) = \Vert{}O_k\Vert{}_{\operatorname{HS}}^2 > 0$. Solving for $\alpha_k$ yields
\begin{equation}
    \alpha_k = \frac{\operatorname{Tr}(O_k^\dagger A)}{\operatorname{Tr}(O_k^2)}.
\end{equation}
Since $A$ was chosen arbitrarily, the proof is complete.\qed

\end{document}